\documentclass{article}

\usepackage{epubtk}
\usepackage{booktabs}
\usepackage{graphicx}
\usepackage{rotating}

\begin{document}

\title{Modified Newtonian Dynamics (MOND):\\ Observational Phenomenology and Relativistic Extensions}

\author{%
\epubtkAuthorData{Benoit Famaey}{%
Observatoire Astronomique de Strasbourg \\ CNRS, UMR 7550, France \\
and AIfA, University of Bonn, Germany}{%
benoit.famaey@astro.unistra.fr}{%
http://astro.u-strasbg.fr/~famaey/}%
\and
\epubtkAuthorData{Stacy McGaugh}{%
Department of Astronomy \\ University of Maryland, USA \\
and Case Western Reserve University, USA}{%
ssm@astro.umd.edu}{%
http://www.astro.umd.edu/~ssm/}%
}

\date{}
\maketitle

\begin{abstract}
A wealth of astronomical data indicate the presence of mass
discrepancies in the Universe.  The motions observed in a variety of
classes of extragalactic systems exceed what can be explained by the
mass visible in stars and gas.  Either (i) there is a vast amount of
unseen mass in some novel form -- dark matter -- or (ii) the data
indicate a breakdown of our understanding of dynamics on the relevant
scales, or (iii) both. Here, we first review a few outstanding
challenges for the dark matter interpretation of mass discrepancies in
galaxies, purely based on observations and independently of any
alternative theoretical framework. We then show that many of these
puzzling observations are predicted by one single relation
-- Milgrom's law -- involving an acceleration constant $a_0$ (or a
characteristic surface density $\Sigma_\dagger = a_0/G$) of the order of the
square-root of the cosmological constant in natural units. This
relation can at present most easily be interpreted as the effect of a
single universal force law resulting from a modification of Newtonian
dynamics (MOND) on galactic scales. We exhaustively review the current
observational successes and problems of this alternative paradigm at
all astrophysical scales, and summarize the various theoretical
attempts (TeVeS, GEA, BIMOND, and others) made to effectively embed
this modification of Newtonian dynamics within a relativistic
theory of gravity.
\end{abstract}

\epubtkKeywords{astronomical observations, astrophysics,
extragalactic astronomy, fundamental physics, 
cosmology, equations of motion, Newtonian limit, theories of gravity
}

\newpage

%\begin{quote}
%\raggedleft
%Failure is not falling down, but refusing to get up. \\
%--- Chinese proverb
%\end{quote}
%\newpage

\tableofcontents

\newpage

\section{Introduction}
\label{section:introduction}

Two of the most tantalizing mysteries of modern astrophysics are known as the \textit{dark matter} and \textit{dark energy} problems. These problems come from the discrepancies between, on one side, the observations of galactic and extragalactic systems (as well as the observable Universe itself in the case of dark energy) by astronomical means, and on the other side, the predictions of General Relativity from the observed amount of matter-energy in these systems. In short, what astronomical observations are telling us is that the dynamics of galactic and extragalactic systems as well as the expansion of the Universe itself do not correspond to the observed mass-energy as they should if our understanding of gravity is complete.  This thus indicates \emph{either} (i) the presence of unseen (and yet unknown) mass-energy, \emph{or} (ii) a failure of our theory of gravity, \emph{or} (iii) both.

The third case is \textit{a priori} the most plausible, as there are good reasons for there being more particles than those of the standard model of particle physics \cite{supersym} (actually, even in the case of baryons, we suspect that a lot of them have not been seen yet and thus literally make up \textit{unseen mass}, in the form of ``missing baryons''), and as there is \textit{a priori} no reason that General Relativity should be valid over a wide range of scales where it has never been tested \cite{Bertolami1}, and where the need for a dark sector actually prevents the theory from being tested until this sector has been detected by other means than gravity itself\epubtkFootnote{Up to now, all the dark matter particle candidates still elude both direct and indirect non-gravitational detection.}. However, either of the first two cases could be the \textit{dominant} explanation of the discrepancies in a given class of astronomical systems (or even in all astronomical systems), and this \textit{is} actually testable. 

For instance, as far as (ii) is concerned, if the mass discrepancies in a class of systems are mostly caused by some subtle change in gravitational physics, then there should be a clear signature of a single, universal force law at work in this whole class of systems. If instead there is a distinct dark matter component in these, the kinematics of any given system should then depend on the particular distribution of both dark and luminous mass.  This distribution would vary from system to system, depending on their environment and past history of formation, and should in principle not result in anything like an apparent universal force law\epubtkFootnote{However, a way to effectively reproduce an apparent universal force law from an exotic dark component could be to enforce an intimate connection between the distribution of baryons, the dark component, and the gravitational field through, e.g., a fifth force effect. This possibility will be extensively discussed in Section~\ref{sec:covariant}, notably Section~\ref{sec:dipolar}}.

Over the years, there have been a large variety of such attempts to alter the theory of gravity in order to remove the need for dark matter and/or dark energy. In the case of dark energy, there is some wiggle room, but in the case of dark matter, most of these alternative gravity attempts fail very quickly, and for a simple reason: once a force law is specified, it must fit all relevant kinematic data in a given class of systems, with the mass distribution specified by the visible matter only. This is a tall order with essentially zero wiggle room: at most one particular force law can work. However, among all these attempts, there is one survivor: the \textit{Modified Newtonian Dynamics} (MOND) hypothesized by Milgrom almost 30 years ago~\cite{Milgrom1, 22, original} seems to come close to satisfying the criterion of a universal force law in a whole class of systems, namely \textit{galaxies}. This success implies a unique relationship between the distribution of baryons and the gravitational field in galaxies and is extremely hard to understand within the present dominant paradigm of the concordance cosmological model, hypothesizing that General Relativity is correct on every relevant scale in cosmology including galactic scales, and that the dark sector in galaxies is made of non-baryonic dissipationless and collisionless particles. Even if such particles are detected directly in the near to far future, the success of MOND on galaxy scales as a phenomenological law, as well as the associated appearance of a universal critical acceleration constant $a_0 \simeq 10^{-10}\mathrm{\ m\, s}^{-2}$ in various seemingly unrelated aspects of galaxy dynamics, will anyway have to be explained and understood by \textit{any} successful model of galaxy formation and evolution. Previous reviews of various aspects of MOND, at an observational and theoretical level, can be found in~\cite{Bek06,Bruneton07,Clifton,Ferreira09,MdB98,Blois08,MDorDM,SM02,Scarpa06,Skordis09}. A blog-website dedicated to this topic is also maintained, with all the relevant literature as well as introductory level articles~\cite{MONDpages} (see also~\cite{Scilogs}).

Here, we first review the basics of the dark matter problem (Section~\ref{sec:missing-mass}) as well as the basic ingredients of the present-day concordance model of cosmology (Section~\ref{sec:LCDM}). We then point out a few outstanding challenges for this model (Section~\ref{sec:challenges}), both from the point of view of unobserved predictions of the model, and from the point of view of unpredicted observations (all uncannily involving a common acceleration constant $a_0$). Up to that point, the challenges presented are purely based on observations, and are fully independent of any alternative theoretical framework\epubtkFootnote{The first four sections provide the observational evidence for the MOND phenomenology through the different appearances of $a_0$ in galactic dynamics, but they are actually independent of any specific theory, while the reader more specifically interested into MOND \textit{per se} could go directly to Section~\ref{sec:Kepler}.}. We then show that, surprisingly, many of these puzzling observations can be summarized within one single empirical law, Milgrom's law (Section~\ref{sec:Kepler}), which can be most easily (although not necessarily uniquely) interpreted as the effect of a single universal force law resulting from a modification of Newtonian dynamics (MOND) in the weak-acceleration regime $a < a_0$, for which we present the current observational successes and problems (Section~\ref{sec:MOND}). We then summarize the various attempts currently made to embed this modification in a generally covariant relativistic theory of gravity (Section~\ref{sec:covariant}) and how such theories allow new predictions on gravitational lensing (Section~\ref{sec:lensing}) and cosmology (Section~\ref{sec:cosmology}). We finally draw conclusions in Section~\ref{sec:summary}.

\newpage

\section{The Missing Mass Problem in a Nutshell}
\label{sec:missing-mass}

There exists overwhelming evidence for mass discrepancies in the Universe from multiple independent observations.  
This evidence involves the dynamics of extragalactic systems: the motions of stars and gas in galaxies and clusters
of galaxies.  Further evidence is provided by gravitational lensing, the temperature of hot, X-ray emitting gas in clusters
of galaxies, the large scale structure of the Universe, and the gravitating mass density of the Universe itself
(Figure~\ref{figure:evidencetree}).
For an exhaustive historical review of the problem, we refer the reader to~\cite{Sandersbook}.

\epubtkImage{evidencetree.png}{%
\begin{figure}[htbp]
  \centerline{\includegraphics[angle=90,width=14.5cm]{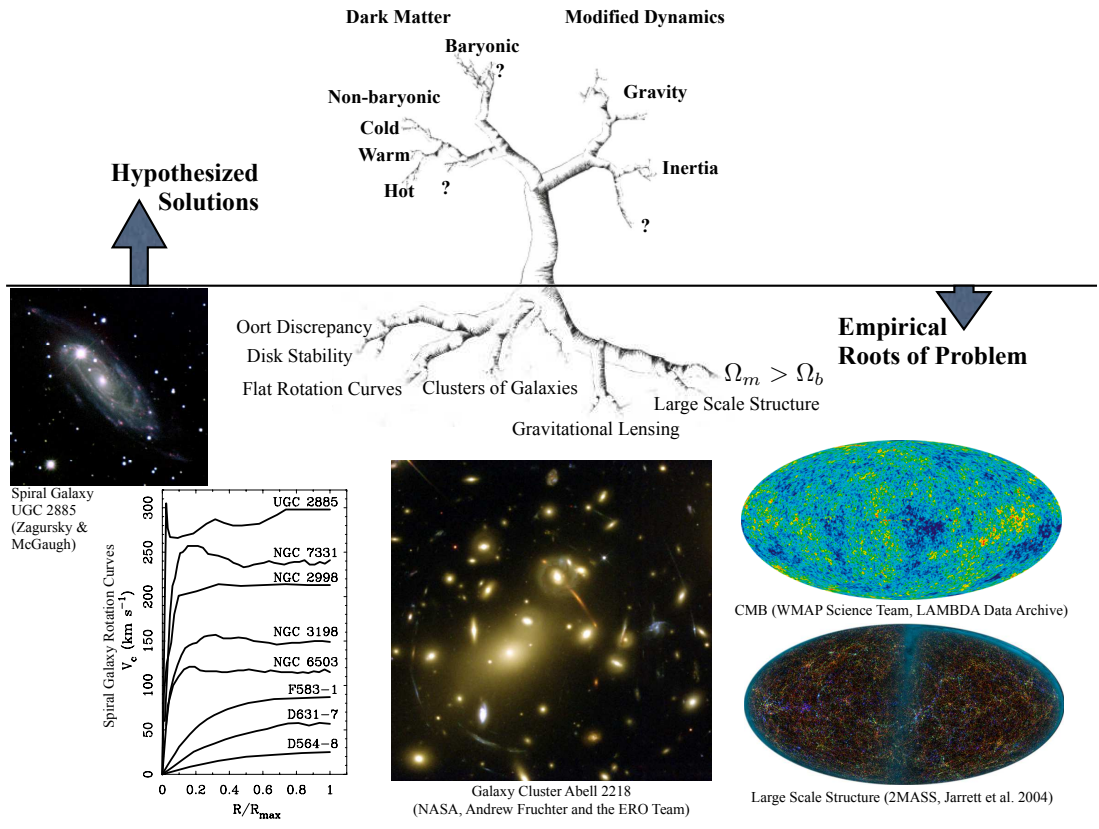}}
  \caption{Summary of the empirical roots of the missing mass problem (below line) and the generic possibilities for
  its solution (above line).  Illustrated lines of evidence include the approximate flatness of the rotation curves of spiral
  galaxies, gravitational lensing in a cluster of galaxies, and the growth of large scale structure from an initially 
  very nearly homogeneous early Universe.  Other historically important lines of evidence include the Oort discrepancy,
  the need to stabilize galactic disks, motions of galaxies within clusters of galaxies and the hydrodynamics of hot,
  X-ray emitting gas therein, and the apparent excess of gravitating mass density over the mass density of baryons
  permitted by big bang nucleosynthesis.  From these many distinct problems grow several possible solutions.
  Generically, the observed discrepancies either imply the existence of dark matter, or the necessity to modify
  dynamical laws.  Dark matter could in principle be any combination of non-luminous baryons and/or some non-baryonic
  form of mass like neutrinos (hot dark matter) or some new particle whose mass makes it dynamically cold or perhaps warm. Alternatively, the observed discrepancies might point to the need to modify the equation of gravity that is employed to  infer the existence of dark matter, or perhaps some other fundamental dynamical assumption like the equivalence of
  inertial mass and gravitational charge.  Many specific ideas of each of these types have been considered over the years. Note that none of these ideas are mutually exclusive, and that some form or the other of dark matter could happily cohabit with a modification of the gravitational law, or could even be itself the cause of an effective modification of the gravitational law. Question marks on some tree branches represent the fruit of ideas yet to be had.  Perhaps these might also address the dark energy problem, with the most satisfactory result being a theory that would simultaneously explain the acceleration scale in the dark matter problem as well as the accelerating expansion of the Universe, and explain the coincidence of scales between these two problems, a coincidence exhibited in Sect.~4.1.}
  \label{figure:evidencetree}
\end{figure}}

The data leave no doubt that when the law of gravity as currently known is applied to extragalactic systems,
it fails if only the observed stars and gas are included as sources in the stress-energy tensor.  This leads to a 
stark choice:  either the Universe is pervaded by some unseen form of mass -- dark matter -- or the dynamical
laws that lead to this inference require revision.  Though the mass discrepancy problem is now well established
\cite{Sandersbook,TrimbleDM}, such a dramatic assertion warrants a brief review of the evidence.

Historically, the first indications of the modern missing mass problem came in the 1930s shortly after galaxies were 
recognized to be extragalactic in nature.  Oort~\cite{oortdiscrepancy} noted that the sum of the observed stars in
the vicinity of the sun fell short of explaining the vertical motions of stars in the disk of the
Milky Way.  The luminous matter did not provide a sufficient restoring force for the observed stellar vertical oscillations.
This became known as the Oort discrepancy.  Around the same time, Zwicky~\cite{zwicky} reported that the velocity
dispersion of galaxies in clusters of galaxies was far too high for these objects to remain bound for a substantial
fraction of cosmic time.  The Oort discrepancy was approximately a factor of two in amplitude, and confined to the
Galactic disk -- it required local dark matter, not necessarily the quasi-spherical halo we now envision.  It was long 
considered a serious problem, but has now largely (though perhaps not fully) gone away~\cite{HF04,KG89}.
The discrepancy Zwicky reported was less subtle, as the required dark mass outweighed the visible stars by
a factor of at least 100.  This result was apparently not taken seriously at the time.

One of the first indications of the need for dark matter in modern times came from the stability of galactic disks.
Stars in spiral galaxies like the Milky Way are predominantly on approximately circular orbits, with relatively few on
highly eccentric orbits~\cite{ELS}.  The small velocity dispersion of stars relative to their circular velocities makes galactic
disks dynamically cold.  Early simulations~\cite{OstPeeb} revealed that cold, self-gravitating disks were subject to
severe instabilities.  In order to prevent the rapid, self-destructive growth of these instabilities, and hence preserve
the existence of spiral galaxies over a sizable fraction of a Hubble time, it was found to be necessary to embed the
disk in a quasi-spherical potential well -- a role that could be played by a halo of dark matter, as first proposed in 1973 by Ostriker \& Peebles~\cite{OstPeeb}.

Perhaps the most persuasive piece of evidence was then provided, notably through the seminal works of Bosma and Rubin, by establishing that the rotation curves of spiral galaxies are approximately flat~\cite{bosma,rubin}. A system obeying Newton's law of gravity should have a rotation curve that, like the Solar system, declines in a Keplerian manner once the bulk of the mass is enclosed: $V_c \propto r^{-1/2}$.  Instead, observations 
indicated that spiral galaxy rotation curves tended to remain approximately flat with increasing radius: $V_c \sim$ constant.
This was shown to happen over and over and over again~\cite{rubin} with the approximate flatness of the rotation curve
persisting to the largest radii observable~\cite{bosma}, well beyond where the details of each galaxy's mass distribution
mattered so that Keplerian behavior \textit{should} have been observed.  Again, a quasi-spherical halo of dark matter as proposed by Ostriker and Peebles was implicated.

Other types of galaxies exhibit mass discrepancies as well.  Perhaps most notable are the dwarf spheroidal galaxies that are
satellites of the Milky Way~\cite{SG,Walker09} and of Andromeda~\cite{jason}.  These satellites are tiny by galaxy standards, possessing only millions, or in the case of the so-called ultrafaint dwarfs, thousands, of individual stars.  They are close enough that the line-of-sight velocities of individual stars can be measured, providing for a precise measurement of the system's velocity dispersion.  The mass inferred from these motions (roughly, $M \sim r \sigma^2/G$) greatly exceeds the mass visible in luminous stars.  Indeed, these dim satellite galaxies exhibit some of the largest mass discrepancies observed.  In contrast, bright
giant elliptical galaxies (often composed of much more than the $\sim 10^{11}$ stars of the Milky Way) exhibit remarkably modest
and hard to detect mass discrepancies~\cite{romanowsky}.  It is thus inferred that fainter galaxies are progressively more
dark matter dominated than bright ones.  However, as we shall expand on in Sect.~4.3, the primary correlation is not with luminosity, but with surface brightness: the lower the surface brightness of a system, the larger its mass discrepancy~\cite{MdB98}.  

On larger scales, groups and clusters of galaxies also show mass discrepancies, just as individual galaxies do.
One of the earliest lines of evidence comes from the so-called ``timing argument'' in the Local Group~\cite{KahnW}.
Presumably the material that was to become the Milky Way and Andromeda (M31) was initially expanding apart with
the general Hubble expansion.  Currently they are approaching one another at $\sim 100\mathrm{\ km\, s}^{-1}$.
In order for the Milky Way and M31 to have overcome the initial expansion and fallen back towards one another,
there must be a greater than average gravitating mass between the two.  To arrive
at their present separation with the observed blueshifted line of sight velocity after a 
Hubble time requires a dynamical mass-to-light ratio $M/L > 80$.  This greatly exceeds the mass-to-light ratio of the 
stars themselves, which is of order unity in Solar units~\cite{Bel03} (the Sun is a fairly average star, so averaged over many
stars each Solar mass produces roughly one Solar luminosity).

Rich clusters of galaxies are rare structures containing dozens or even hundreds of bright galaxies.
These objects exhibit mass discrepancies in several distinct ways. 
Measurements of the redshifts of individual cluster members give velocity dispersions in the vicinity of
$1,000\mathrm{\ km\, s}^{-1}$ typically implying dynamical mass-to-light ratios in excess of 100~\cite{Bahcall}.
The actual mass discrepancy is not this large, as most of the detected baryonic mass in clusters is in
a diffuse intracluster gas rather than in the stars in the galaxies (something Zwicky was not aware of back in 1933).  This gas is heated to the virial temperature and emits X-rays.  Mapping the temperature and emission of this X-ray gas provides another probe of the cluster mass through the equation of hydrostatic equilibrium.  In order to hold the gas in the clusters at the observed temperatures requires there to be dark matter that outweighs the gas by a factor of $\sim$~8~\cite{giodini}.
Furthermore, some clusters are observed to gravitationally lens background galaxies (Figure~\ref{figure:evidencetree}).
Once again, mass above and beyond that observed is required to explain this phenomenon~\cite{abell2218}.
Thus three independent methods all imply the need for about the same amount of dark matter in clusters of galaxies.

In addition to the abundant evidence for mass discrepancies in the dynamics of extragalactic systems,
there are also strong motivations for dark matter in cosmology.  Two observations are particularly important:
(i) the small baryonic mass density $\Omega_b$ inferred from big bang nucleosynthesis (and from the measured Hubble parameter), and (ii) the growth of large scale structure by a factor of $\sim 10^5$ from the surface of last scattering of the cosmic microwave background at redshift $z \sim 1000$ until present-day $z=0$, implying $\Omega_m > \Omega_b$. Together, these observations thus imply not only the need for dark matter, but for some exotic new form of non-baryonic cold dark matter. Indeed, observational estimates of the gravitating mass density of the Universe $\Omega_m$, measured for instance from peculiar galaxy (or large-scale) velocity fields, have for several decades persistently returned values in the range $1/4 < \Omega_m < 1/3$~\cite{DavisOm}.  While shy of the value needed for a flat Universe, this mass density is well in excess of the baryon density inferred from big bang nucleosynthesis (BBN).  The observed abundances of
the light isotopes deuterium, helium, and lithium are consistent with having been produced in the first few minutes after
the big bang if the baryon density is just a few percent of the critical value: $\Omega_b < 0.05$
\cite{BBNcomp1,BBNcomp2}.
Thus $\Omega_m > \Omega_b$.  Consequently, we don't just need dark matter, we need the dark matter to be
non-baryonic.

Another early Universe constraint is provided by the Cosmic Microwave Background (CMB).  The small (microKelvin) amplitude of the temperature fluctuations at the time of baryon-photon decoupling ($z \sim 1000$) indicates that the Universe was initially very homogeneous, roughly to one part in $10^5$. The Universe today ($z=0$) is very inhomogeneous, at least on ``small'' scales of less than $\sim$~100~Mpc ($\sim 3 \times 10^8 \,$light-years), with huge density
contrasts between planets, stars, galaxies, clusters, and empty intergalactic space.  The only attractive long-range force
acting on the entire Universe, that can make such structures, is gravity.  In a rich-get-richer while the poor-get-poorer process, the small initial over-densities attract more mass and grow into structures like galaxies while under-dense regions become
less dense, leading to voids.  The catch is that gravity is rather weak, so this process takes a long time.  If the baryon
density from BBN is all we have to work with, we can only obtain a growth factor of $\sim 10^2$ in a Hubble time~\cite{joesilkcmb}, orders of magnitude short of the observed $10^5$.  The solution is to boost the growth rate with
extra invisible mass displaying larger density fluctuations:  dark matter.  In order not to make the same mark on the CMB that baryons would, this dark matter must not interact with the photons. So, in effect, the density fluctuations in the dark matter can already be very large at the epoch of baryon-photon decoupling, especially if the dark matter is cold (i.e., with effectively zero Jeans length). The baryons fall into the already deep dark matter potential wells only after that, once released from their electromagnetic link to the photon bath.  Before decoupling, the fluctuations in the baryon-photon fluid did not grow but were oscillating in the form of acoustic waves, maintaining the same amplitude as when they entered the horizon; actually they were even slightly diffusion-damped. In principle, at baryon-photon decoupling, CMB fluctuations on smaller angular scales, having entered the horizon earlier, would thus have been damped with respect to those on larger scales (Silk damping). Nevertheless, the presence of decoupled non-baryonic dark matter would provide a net forcing term countering the damping of the oscillations at recombination, meaning that the second and third acoustic peaks of the CMB could then be of equal amplitude rather than exhibiting a damping tail. The actual observation of a high third-peak in the CMB angular power spectrum is thus another compelling evidence for non-baryonic dark matter (see, e.g.,~\cite{Komatsu11}). Both BBN and the CMB thus drive us to consider a form of mass that is non-baryonic and which does not interact electromagnetically.  Moreover, in order to form structure (see Sect.~3.2), the mass must be dynamically cold (i.e., moving much slower than the speed of light when it decouples from the photon bath), and is known as \textit{cold dark matter} (CDM). 

Now, in addition to CDM, modern cosmology also requires something even more mysterious, dubbed \textit{dark energy}.  The fact that the baryon fraction in clusters of galaxies was such that $\Omega_m$ was implied to be much smaller than 1 -- the value needed for a flat Euclidean Universe favored by inflationary models -- , as well as tensions between the measured Hubble parameter and independent estimates of the age of the Universe, led Ostriker \& Steinhardt~\cite{OstStein} to propose in 1995 a so-called ``concordance model of cosmology'' or $\Lambda$CDM model, where a cosmological constant $\Lambda$ -- supposed to represent vacuum energy or \textit{dark energy} -- provided the major contribution to the Universe's energy density. Three years later, the observations of SNIa~\cite{Perlmutter99,Riess98} indicating late-time acceleration of the Universe's expansion, led most people to accept this model. This concordance model has since been refined and calibrated through subsequent large-scale observations of the CMB and of the matter power spectrum, to lead to the favored cosmological model prevailing today (see Sect.~3). However, as we shall see, curious coincidences of scales between the dark matter and dark energy sectors (see Sect.~4.1) have prompted the question of whether these two sectors are really physically independent, and the existence of dark energy itself has led to a renewed interest in modified gravity theories as a possible alternative to this exotic fluid~\cite{Clifton}.

\newpage

\section{A Brief Overview of the $\Lambda$CDM Cosmological Model}
\label{sec:LCDM}

General relativity provides a clear and compelling cosmology, the Friedmann--Lema\^itre--Robertson--Walker (FLRW) model. The expansion of the Universe discovered by Hubble and Slipher found a natural explanation\footnote{Arguably a non-static, expanding or contracting Universe was an \textit{a priori} prediction of General Relativity in its original form lacking the cosmological constant.} in this context. The picture of a hot big bang cosmology that emerged from this model famously predicted the existence of the 3 degree CMB and the abundances of the light isotopes via BBN.  

Within the FLRW framework, we are inexorably driven to infer the existence of both non-baryonic cold dark matter and a non-zero cosmological constant as discussed in Sect.~\ref{sec:missing-mass}.  The resulting concordance $\Lambda$CDM model -- first proposed in 1995 by Ostriker and Steinhardt \cite{OstStein} -- is encouraged by a wealth of observations: the consistency of the Hubble parameter with the ages of the oldest stars \cite{OstStein}, the consistency between the dynamical mass density of the Universe, that of baryons from BBN (see also discussion in Sect.~9.2), and the baryon fraction of clusters \cite{baryoncatastrophe}, as well as the power spectrum of density perturbations~\cite{Cole05,Tegmark04}.  A prediction of the concordance model is that the expansion rate of the Universe should be accelerating; this was confirmed by observations of high redshift Type Ia supernovae \cite{Perlmutter99,Riess98}. Another successful prediction was the scale of the baryonic acoustic oscillation \cite{BAOdetect}. Perhaps the most emphatic support for $\Lambda$CDM comes from fits to the acoustic power spectrum of temperature fluctuations in the CMB \cite{Komatsu11}.

For a brief review of the basics and successes of the concordance cosmological model we refer the reader to, e.g., \cite{Smoot,Peebrecent} and all references therein.  We note that, while most of the cosmological probes in the above list are not uniquely fit by the $\Lambda$CDM model on their own, when tey are taken together they provide a remarkably tight set of constraints. The success of this now favoured cosmological model on large scales is thus remarkable indeed, as there was {\it a priori} no reason that such a parameterized cosmology could explain all these completely independent data sets with such outstanding consistency. 

In this model, the Hubble constant is $H_0 = 70\mathrm{\ km\, s^{-1}\, Mpc}^{-1}$ (i.e., $h=0.7$), the amplitude of density fluctuations within a top-hat sphere of $8h^{-1}$~Mpc is $\sigma_8=0.8$, the optical depth to reionization is $\tau = 0.08$, the spectral index measuring how fluctuations change with scale is $n_s = 0.97$, and the price we pay for the outstanding success of the model is new physics in the form of a dark sector. This dark sector is making up 95\% of the mass-energy content of the Universe in $\Lambda$CDM: it is composed separately of a \textit{dark energy sector} and a \textit{cold dark matter sector}, which we briefly describe below.

\subsection{Dark Energy ($\Lambda$)}

In $\Lambda$CDM, dark energy is a non-vanishing vacuum energy represented by the cosmological constant $\Lambda$ in the field equations of General Relativity.  Einstein's cosmological constant is equivalent to vacuum energy with equation of state $p/\rho=w=-1$.  In principle, the equation of state could be merely close to, but not exactly $w=-1$.  In this case, the dark energy could evolve and clump, depending on the value of $w$ and its evolution $\dot w$. However, to date, there is no compelling observational reason to require any form of dark energy more complex than the simple cosmological constant introduced by Einstein.

The various observational datasets discussed above constrain the ratio of the dark energy density to the critical density to be $\Omega_{\Lambda} = \Lambda/3H_0^2=0.73$, where $H_0$ is Hubble's constant and $\Lambda$ is expressed in $s^{-2}$. This value, together with the matter density $\Omega_m$ (see below), leads to a total $\Omega=\Omega_{\Lambda}+\Omega_m=1$, i.e., a spatially flat Euclidean geometry in the Robertson-Walker sense that is nicely consistent with the expectations of inflation. It is important to stress that this model relies on the cosmological principle, i.e., that our observational location in the Universe is not special, and on the fact that on large scales, the Universe is isotropic and homogeneous. For possible challenges to these assumptions and their consequences, we refer the reader to, e.g., \cite{Buchert08, Wiltshire07, Wiltshire08}.

\subsection{Cold Dark Matter (CDM)}

In $\Lambda$CDM, dark matter is assumed to be made of non-baryonic dissipationless massive particles~\cite{Bertone}, the so-called ``cold dark matter'' (CDM).  This dark matter outweighs the baryons that participate in BBN by about 5:1.  The density of baryons from the CMB is $\Omega_b=0.046$, grossly consistent with BBN \cite{Komatsu11}.  This is a small fraction of the critical density; with the non-baryonic dark matter the total matter density is $\Omega_m = \Omega_{CDM}+\Omega_b=0.27$.

The ``cold'' in cold dark matter means that CDM moves slowly so that it is non-relativistic when it decouples from photons. This allows it to condense and begin to form structure while the baryons are still electromagnetically coupled to the photon fluid.  After recombination, when protons and electrons first combine to form neutral atoms so that the cross-section for interaction with the photon bath suddenly drops, the baryons can fall into the potential wells already established by the dark matter, leading to a hierarchical scenario of structure formation with the repeated merger of smaller CDM clumps to form ever larger clumps. 

Particle candidates for the CDM must be massive, non-baryonic, and immune to electromagnetic interactions. The currently preferred CDM candidates are Weakly Interacting Massive Particles (WIMPs,~\cite{Bertonenature,BertoneLHC,Bertone}) that condensed from the thermal bath of the early Universe. These should have masses on the order of about 100~GeV so that (i) the free-streaming length is small enough to create small-scale structures as observed (e.g., dwarf galaxies), and (ii) that thermal relics with cross-sections typical for weak nuclear reactions account for the right amount of matter density $\Omega_m$ (see, e.g., Eq.~28 of~\cite{Bertone}). This last point is known as the WIMP miracle\footnote{The WIMP miracle seems however to fade away with modern particle physics constraints \cite{Baer}.}. 

For lighter particle candidates (e.g., ordinary neutrinos or light sterile neutrinos), the damping scale becomes too large. For instance, a hot dark matter (HDM) particle candidate with mass of a few to 15~eV would have a free-streaming length of about $\sim100$~Mpc, leading to too little power at the small-scale end of the matter power spectrum. The existence of galaxies at redshift $z \sim 6$ implies that the coherence length should have been smaller than 100 kpc or so, meaning that even warm dark matter (WDM) particles with masses between 1 and 10~keV are close to being ruled out as well (see, e.g.,~\cite{Peacock03}).  $\Lambda$CDM thus presently remains the state-of-the-art in Cosmology, although some of the challenges listed below in Sect.~\ref{sec:challenges} are leading to a slow drift of the standard concordance model from CDM to WDM~\cite{Lovell}, but this drift brings along its own problems, and fails to address most of the current observational challenges summarized in the following section, which might thus perhaps point to a more radical alternative to the model.

\newpage

\section{Some Challenges for the $\Lambda$CDM Model}
\label{sec:challenges}

The great concordance of independent cosmological observables from Gpc to Mpc scales lends a certain air of inevitability to the $\Lambda$CDM model.  If we accept these observables as sufficient to prove the model, then any discrepancy appears as trivia that will inevitably be explained away.  If instead we require a higher standard, such as positive laboratory evidence for the dark sectors, then $\Lambda$CDM appears as a yet unproven hypothesis that relies heavily on two potentially fictitious invisible entities. An important test of $\Lambda$CDM as a scientific hypothesis is thus the existence of dark matter. By this we mean not just unseen mass, but specifically CDM: some novel form of particle with the right microscopic properties and correct cosmic mass density.  Searches for WIMPs are now rather mature and not particularly encouraging.  Direct detection experiments have as yet no positive detections, and have now excluded \cite{XENON100} the bulk of the parameter space (interaction cross-section and particle mass) where WIMPs were expected to reside.  Indirect detection through the observation of $\gamma$-rays produced by the self-annihilation\footnote{The simplest WIMPs are their own antiparticle.} of WIMPs in the Galactic halo and in nearby satellite galaxies have similarly returned null results \cite{Ferminot1,Ferminot2,Ferminot3} at interestingly restrictive levels.  For the most plausible minimally supersymmetric models, particle colliders should already have produced evidence for WIMPs \cite{CDFdm2,CDFdm1,Baer}. The right model need not be minimal.  It is always possible to construct a more complicated model that manages to evade all experimental constraints.  Indeed, it is readily possible to imagine dark matter candidates that do not interact at all with the rest of the Universe except through gravity.  Though logically possible, such dark matter candidates are profoundly unsatisfactory in that they could not be detected in the laboratory:  their hypothesized existence could neither be confirmed nor falsified.

Apart from this current non-detection of CDM candidates, there also exists prominent observational challenges for the $\Lambda$CDM model, which might point towards the necessity of an alternative model (or, at the very least, an improved one). These challenges are that (i) some of the parameters of the model appear fine-tuned (Sect~4.1), and that (ii) as far as galaxy formation and evolution are concerned (mainly processes happening on kpc scales so that the predictions are more difficult to make because the baryon physics should play a more prominent role), many predictions that have been made were not successful (Sect.~4.2); (iii) what is more, a number of observations on these galactic scales do exhibit regularities that are fully unexpected in any CDM context without a substantial amount of fine-tuning in terms of baryon feedback (Sect.~4.3).  

\subsection{Coincidences}

What is generally considered as the biggest problem for the $\Lambda$CDM model is that it requires a large and still unexplained fine-tuning to reduce by 120 orders of magnitude the theoretical expectation of the vacuum energy to yield the observed cosmological constant value, and, even more importantly, that it faces a coincidence problem to explain why the dark energy density $\Omega_\Lambda$ is precisely of the same order of magnitude as the other cosmological components \textit{today}\footnote{In addition, the time-averaged value of the deceleration parameter $q$ over the present age of the Universe is quite consistently $\langle q \rangle =0$ \cite{Lewisq}, another currently unexplained coincidence.}. This uncanny coincidence is generally seen as evidence for some yet-to-discover underlying cosmological mechanism ruling the evolution of dark energy (such as quintessence or generalized additional fluid components, see, e.g.,~\cite{Copeland}). But it could also indicate that the effect attributed to dark energy is rather due to a breakdown of General Relativity (GR) on the largest scales~\cite{Frieman}.

Then, as we shall see in more detail in Sect.~4.3, another coincidence, which is central to this whole review, is the appearance of a characteristic scale -- dubbed $a_0$ -- in the behavior of the dark matter sector, a scale with units of acceleration. This acceleration scale appears in various seemingly unrelated galactic scaling relations, mostly unpredicted by the $\Lambda$CDM model (see Sect.~4.3). The value of this scale is $a_0 \simeq  10^{-10}\mathrm{\ m\, s}^{-2}$, which yields in natural units\epubtkFootnote{$c=G= \hbar =1$.}, $a_0 \sim H_0$ (or, more precisely, $a_0 \approx cH_0/2 \pi$). It is perhaps even more meaningful~\cite{Binney,Milcomments,unruh} to note that, in these same units:
\begin{equation}
a_0^2 \sim \Lambda,
\label{eq:a0}
\end{equation}
where $\Lambda$ is the currently favored value of the cosmological constant\epubtkFootnote{We have that $\Lambda \sim a_0^2/c^2$ if expressed in inverse time-squared or $\Lambda \sim a_0^2/c^4$ if expressed in inverse length-squared (more precisely, the natural scale associated with the cosmological constant is $a_0 \approx (c^2/2 \pi) \sqrt(\Lambda/3)$). Another way of expressing this coincidence is thus to say that predictions of GR from visible matter alone always break down for physics involving a length-scale constant of the order of the Hubble radius $l \sim \Lambda^{-1/2} \sim c^2/a_0$. This scale $l$ could perhaps play a similar role to the Planck scale $l_P$~\cite{smolin,bernal}, at the other end of the ladder (as we have $l \approx 10^{60}l_P)$. This is however {\it not} the length at which the modification would be seen, exactly as quantum mechanics does not depart from classical physics at a given length.}. Whether these numerical coincidences are physically relevant or just true (insignificant) coincidences remains an open question, closely related to the nature of the dark sector, which we are going to elaborate on in the next sections. But, at this stage, it is in any case striking that the dark matter and dark energy sectors do have such a \textit{common scale}. This coincidence of scales, together with the coincidence of energy densities at redshift zero, might perhaps be a strong indication that one should cease to consider dark energy as an additional component physically independent from the dark matter sector~\cite{AWE}, and/or cease to consider that GR correctly describes gravity on the largest scales and in extremely weak gravitational fields, in order to perhaps address the two above coincidence problems at the same time.

Finally, let us note that the existence of the $a_0$-scale is actually not the only dark matter-related coincidence, as there is also in principle absolutely no reason why the mechanism leading to the baryon asymmetry (between baryonic matter and antimatter) would simultaneously leave both the baryon and dark matter densities with a similar order of magnitude ($\Omega_{\rm DM}/\Omega_b=5$). If the effects we attribute to dark matter are actually also due to a breakdown of GR on cosmological scales, then such a coincidence might perhaps appear more natural as the baryons would then be the actual source of the effect attributed to the dark matter sector.

\subsection{Unobserved predictions}

Apart from the above puzzling coincidences, the concordance $\Lambda$CDM model also has a few more concrete empirical challenges to address, in the sense of having made a few predictions in contradiction with observations (with the caveat in mind that the model itself is not always that predictive on small scales). These include the following non-exhaustive list:

\begin{itemize}
\item[{\bf 1.}] \textbf{The bulk flow challenge.} Peculiar velocities of galaxy clusters are predicted to be of the order of 200~km/s in the $\Lambda$CDM model. These can actually be measured by studying the fluctuations in the CMB generated by the scattering of the CMB photons by the hot X-ray-emitting gas inside clusters (the kinematic SZ effect). This yields an observed coherent bulk flow of order 1000~km/s (5 times more than predicted) on scales out to at least 400~Mpc~\cite{Kash}. This bulk flow challenge appears not only in SZ studies but also in galaxy studies \cite{bulkgalaxies}. A related problem is the collision velocity larger than 3100~km/s for the merging bullet cluster 1E0657-56 at $z=0.3$, much too high to be accounted for by $\Lambda$CDM~\cite{Lee,Thompson}. These observations would seem to indicate that the attractive force between DM particles is enhanced compared to what $\Lambda$CDM predicts, and changing CDM into WDM would not solve the problem.

\item[{\bf 2.}] \textbf{The high-z clusters challenge.} Observation of even a single massive cluster at high redshift can falsify $\Lambda$CDM~\cite{Huterer}. In this respect the existence of the galaxy cluster XMMU~J2235.3-2557~\cite{Rosati} with a mass of of $\sim 4 \times 10^{14}\,M_{\odot}$ at $z = 1.4$, even though not sufficient to rule out the model, is very surprising and could indicate that structure formation is actually taking place earlier and faster than in $\Lambda$CDM (see also~\cite{Sheth} on the Shapley supercluster and the Sloan Great Wall).

\item[{\bf 3.}] \textbf{The Local Void challenge.} The Local Volume is composed of 562 known galaxies at distances smaller than 8~Mpc from the center of the Local Group, and the region known as the ``Local Void'' hosts only 3 of them. This is much less than the expected $\sim 20$ for a typical similar void in $\Lambda$CDM~\cite{PeeblesNusser}. What is more, in the Local Volume, large luminous galaxies are over-represented by a factor of 6 in the underdense regions, exactly opposite to what is expected from $\Lambda$CDM. This could mean that the Local Volume is just a statistical anomaly, but it could also point,  in line with the two previous challenges, towards more rapid structure formation, allowing sparse regions to more quickly form large galaxies cleaning their environment, making the galaxies larger and the voids emptier at early times~\cite{PeeblesNusser}.

\item[{\bf 4.}] \textbf{The missing satellites challenge.} It has long
  been known that the model predicts an overabundance of dark subhalos
  orbiting Milky Way-sized galaxies compared to the observed number of
  satellite galaxies around the Milky Way~\cite{Moore}. This is a
  different problem from the above predicted overabundance of small
  galaxies in voids.  It has subsequently been suggested that stellar
  feedback and heating processes limit baryonic growth, that
  re-ionisation prevents low-mass dark halos from forming stars, and
  that tidal forces from the host halo limit growth of the dark matter
  sub-halos and lead to their truncation. This important theoretical
  effort has led recent semi-analytic models to predict a reduced
  number of $\sim$~100 to 600 faint satellites rather than the
  original thousands. Moreover, during the past 15~years 13 ``new''
  and mostly ultra-faint satellite galaxies have been found in
  addition to the 11 previously known classical bright ones. Since
  these new galaxies have been largely discovered with the Sloan
  Digital Sky Survey (SDSS), and since this survey covered only one
  fifth of the sky, it has been argued that the problem was
  solved. However, there are actually still missing satellites on the
  low mass and high mass end of the mass function predicted by
  ``$\Lambda$CDM+re-inoisation'' semi-analytic models. This is best
  illustrated on Figure~2 of~\cite{Kroupa} showing the cumulative
  distribution for the predicted and observationally derived masses
  within the central 300~pc of Milky Way satellites. A lot of low-mass
  satellites are still missing, and the most massive predicted
  subhaloes are also incompatible with hosting any of the known Milky
  Way satellites~\cite{Bovill,Bullock2,Bullock}. This is thus the
  modern version of the missing satellites challenge. An obvious but
  rather discomforting way-out would be to simply state that the Milky
  Way must be a statistical outlier, but this is contradicted by the
  study of~\cite{strigari} on the abundance of bright satellites
  around Milky Way-like galaxies in SDSS. Another solution would be to
  change from CDM to WDM~\cite{Lovell} (it is actually one of the only
  listed challenges that such a change would probably immediately
  solve).

\item[{\bf 5.}] \textbf{The satellites phase-space correlation challenge.} In addition to the above challenge, the distribution of dark subhalos around the Galaxy is also predicted by $\Lambda$CDM to be isotropic, or quasi-isotropic. However, the Milky Way satellites are currently observed to be correlated in phase-space: they lie within a seemingly rotation-supported disk~\cite{Kroupa}. Young halo globular clusters define the same disk, and streams of stars and gas, tracing the orbits of the objects from which they are stripped, preferentially lie in this disk, too~\cite{pawlow}. Since SDSS covered only one fifth of the sky, it will be interesting to see whether future surveys such as Pan-Starrs will confirm this state of affairs. Whether or not this  phase-space correlation would be unique to the Milky Way should also be carefully checked, the evidence in M31 being currently much less convincing, with a richer and more complex satellite population~\cite{metz07}. But in any case, the current distribution of satellites around the Milky Way is statistically incompatible with the predictions of $\Lambda$CDM at a very high level of confidence even when taking into account the observational bias from SDSS~\cite{Kroupa}. While this might perhaps have been explained by the infall of a small group of galaxies that would have retained correlated orbits, this solution is ruled out by the fact that no nearby groups are observed to be anywhere near as spatially small as the disk of satellites~\cite{metz09}. Another solution might be that most Milky Way satellites are actually not primordial galaxies but old tidal dwarf galaxies created in an early major merger event, accounting for their presently correlated phase-space distribution~\cite{pawlowtdg}. Note in passing that if only one or two long-lived tidal dwarfs are created in each gas-dissipational galaxy encounter, they could probably account for most of the dwarf galaxy population in the Universe, leaving no room for small CDM subhalos to create galaxies, which would transform the missing satellites challenge into a missing satellites catastrophe~\cite{Kroupa}.

\item[{\bf 6.}] \textbf{The cusp-core challenge.}  Another long-standing problem of $\Lambda$CDM is the fact that the simulations of the collapse of CDM halos lead to a density distribution as a function of radius, $\rho(r)$, which is well fitted by a smooth function asymptoting to a central \textit{cusp} with slope $d{\rm ln} \rho/d{\rm ln} r = -1$ in the central parts~\cite{Diemand,NFW}, while observations clearly point towards large constant density cores in the central parts~\cite{deblok, Gentilecore, walker}. Even though the latest simulations~\cite{Navarro} rather point towards Einasto~\cite{Einasto} profiles with $d{\rm ln} \rho/d{\rm ln} r \propto -r^{(1/n)}$ (with $n$ slightly varying with halo mass, and $n \sim 6$ for a Milky Way-sized halo, meaning that the slope is zero only {\it very} close to the nucleus \cite{Graham1}, and is still $\sim -1$ at 200~pc from the center), fitting such profiles to observed galactic kinematical data such as rotation curves~\cite{Chemin} leads to values of $n$ that are much smaller than simulated values (meaning that they have much larger cores), which is another way of re-assessing the old cusp problem of $\Lambda$CDM. Note that a change from CDM to WDM could solve the problem in dwarf galaxies, by leading to the formation of small cores, but certainly not in large galaxies where large cores are needed from observations. One thus has to rely on baryon feedback to erase the cusp from all galaxies. But this is not easily done, as the adiabatic cooling of baryons in the centre of dark matter halos should lead to an even more concentrated dark matter distribution. A possibility would be that angular momentum transfer from a rotating stellar bar destroys dark-matter cusps: however, significant cusp destruction requires substantially more angular momentum than is realistically available in stellar bars~\cite{Cheminher, McMillan}. Note also that not all galaxies are barred (e.g., M33 is not). The state-of-the-art solution nowadays is to enforce strong supernovae outflows that move large amounts of low-angular-momentum gas from the central parts and that ``pull'' on the central dark matter concentration to create a core~\cite{governato}, but this is still a highly fine-tuned process which fails to address the baryon fraction problem (see challenge 10 below).

\item[{\bf 7.}] \textbf{The angular momentum challenge.}  As a consequence of the merger history of galaxy disks in a hierarchical formation scenario, as well as of the associated transfer of angular momentum from the baryonic disk to the dark halo, the specific angular momentum of the baryons ends up being much too small in simulated disks, which in turn end up much smaller than the observed ones~\cite{Abadi}. Similarly, elliptical systems end up too concentrated too. Addressing this challenge within the standard paradigm essentially relies on forming disks through late-time quiescent gas accretion from large-scale filaments, with much less late-time mergers than presently predicted in $\Lambda$CDM.

\item[{\bf 8.}] \textbf{The pure disk challenge.} Related to the previous challenge, large \textit{bulgeless} thin disk galaxies are extremely difficult to produce in simulations. This is because major mergers, at any time in the galaxy formation process, typically create bulges, so bulgeless galaxies would represent the quiescent tail of a distribution of merger histories for galaxies of the Local Volume. However, these bulgeless disk galaxies represent more than half of large galaxies (with $V_c > 150\mathrm{\ km/s}$) in the Local Volume~\cite{Graham2,Kormendy}. Solving this problem  would rely, e.g., on suppressing central spheroid formation for mergers with mass ratios lower than 30\% \cite{Koda}.

\item[{\bf 9.}] \textbf{The stability challenge.}  Round CDM halos tend to stabilize very low surface density disks against the formation of bars and spirals, due to a lack of disk self-gravity~\cite{Mihos}. The observation~\cite{McG95} of Low Surface Brightness (LSB) disk galaxies with strong bars and spirals is thus challenging in the absence of a significant disk component of dark matter. What is more, in the absence of such a disk DM component, the lack of disk self-gravity prevents the creation of very large razor thin LSB disks, but these are observed~\cite{superthin1,superthin2}. In the standard context, these observations would tend to point towards an additional disk DM component, either a CDM-one linked to in-plane accretion of satellites or a baryonic one in the form of molecular gas.

\item[{\bf 10.}] \textbf{The missing baryons challenge(s).}  As mentioned above, constraints from the CMB imply $\Omega_m = 0.27$ and $\Omega_b=0.046$. However, our inventory of known baryons in the local Universe, summing over all observed stars, gas, etc., comes up short of the total. For example,~\cite{Bel03} estimate that the sum of stars and cold gas is only $\sim$~5\% of $\Omega_b$. While there now seems to be a good chance that many of the missing baryons are in the form of highly ionized gas in the warm-hot intergalactic medium (WHIM), we are still far from being able to give a confident account of where all the baryons reside. Indeed, there could be multiple distinct reservoirs in addition to the WHIM, each comparable to the mass in stars, within the current uncertainties. But there is another missing baryons challenge, namely the halo-by-halo missing baryons. Indeed, each CDM halo can, to a first approximation, be thought of as a microcosm of the Whole. As such, one would naively expect each halo to have the same baryon fraction as the whole Universe, $f_b=\Omega_b/\Omega_m=0.17$. On the scale of clusters of galaxies, this is approximately true (but still systematically low), but for individual galaxies, observations depart from this in a systematic way which we have yet to understand, and which has nothing to do with the truncation radius. The ratio of the galaxy-detected baryon fraction over the cosmological one, $f_d$, is plotted as a function of the potential well of the systems in Figure~\ref{figure:fdV}~\cite{barcontent}. There is a clear correlation, less massive objects being much more dark matter dominated than massive ones. This correlation is \textit{a priori} not predicted at all by $\Lambda$CDM, at least not with the correct shape~\cite{halobyhalo}. This missing baryons challenge is actually closely related to the baryonic Tully--Fisher relation, which we expand on in the following Sect.~4.3.1.

\end{itemize}

\epubtkImage{fdV_LR.png}{%
\begin{figure}[htbp]
  \centerline{\includegraphics[width=0.9\textwidth]{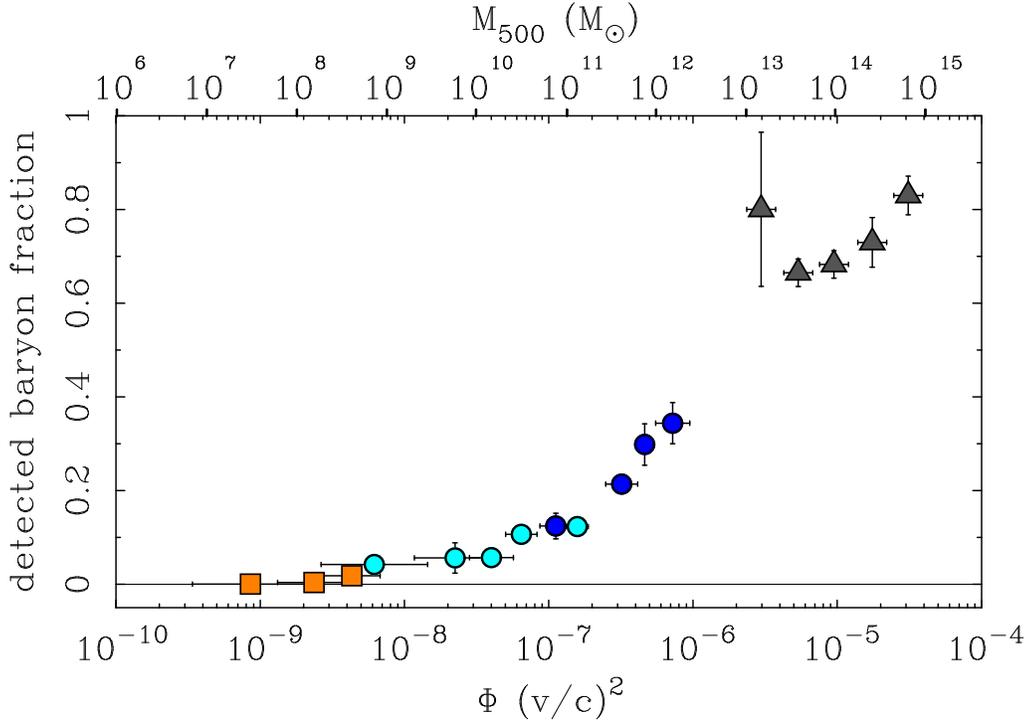}}
  \caption{The fraction of the expected baryons that are detected as a function of potential well 
depth (bottom axis) and mass (top). Measurements are referenced to the radius $R_{500}$
  where the enclosed density is 500 times the cosmic mean \cite{barcontent}.  The 
detected baryon fraction
  $f_d= M_b/(0.17 M_{500})$ where $M_b$ is the detected baryonic mass, 0.17 is the 
universal baryon fraction \cite{Komatsu11},
  and $M_{500}$ is the dynamical mass (baryonic + dark mass) enclosed by $R_{500}$.  Each point is a bin 
representing many objects.
  Gray triangles represent galaxy clusters, which come close to containing the cosmic 
fraction.  The detected baryon
  fraction declines systematically for smaller systems.  Dark blue circles represent star 
dominated spiral galaxies.
  Light blue circles represent gas dominated disk galaxies. Orange squares represent 
Local Group dwarf satellites for which
  the baryon content can be less than 1\% of the cosmic value.  Where these missing 
baryons reside is one of the challenges
  currently faced by $\Lambda$CDM.}
  \label{figure:fdV}
\end{figure}}

Let us however note that, while challenges 1 to 3 are not real smoking guns yet for the $\Lambda$CDM model, challenges 4 to 10 are concerned with processes happening on kpc scales, for which it is fair to consider that the model is not very predictive because the baryon physics should play a more important role, and this is hard to take into account rigorously. However, it is not sufficient to qualitatively invoke handwavy baryon physics to avoid confronting predictions of $\Lambda$CDM with observations. It is also mandatory to show that the feedback from the baryons which is needed to solve the observational problems is what would quantitatively happen in a physical galaxy. This, presently, is not the case yet for the aforementioned challenges. However, these challenges are ``model-dependent problems'', in the sense of being failed predictions of a given model, but would not have appeared \textit{a priori} surprising without the standard concordance model at hand. This means that subtly changing some parameters of the model (like, e.g., swapping CDM for WDM, making DM more self-interacting, etc.) might help solving at least a few of them. But what is even more challenging is a set of observations that appear surprising \textit{independently} of any specific dark matter model, as they involve a fine-tuned relation between the distribution of visible and dark matter. These are what we call hereafter ``unpredicted observations''.

\subsection{Unpredicted observations}

There are several important examples of systematic relations between the dynamics of galaxies (in theory presumed to be dominated
by dark matter) and their baryonic content.  These relations are fully empirical, and as such must be explained by \textit{any} viable theory.
As we shall see, they inevitably involve a critical acceleration scale, or equivalently, a critical surface density of baryonic matter.  

\subsubsection{Baryonic Tully--Fisher relation}

One of the strongest correlations in extragalactic astronomy is the Tully--Fisher relation~\cite{TForig}.
Originally identified as an empirical relation between a galaxy's luminosity and its HI line-width, it has been extensively
employed as a distance indicator.  Though extensively studied for decades, the physical basis of the relation remains unclear.

Luminosity and line-width are readily accessible observational quantities.
The optical luminosity of a galaxy is a proxy for its stellar mass, and the HI line-width is a proxy for its rotation velocity.
The quality of the correlation improves as more accurate indicators of these quantities are employed.  
For example, resolved rotation curves where the flat portion of the rotation curve $V_f$ or the maximum peak velocity $V_p$
can be measured give relations that are tighter than those utilizing only line-width information~\cite{courteauscatter}.  
Similarly, the scatter declines as we shift from optical luminosities to those in the near-infrared~\cite{verheijen}
as the latter are expected to give a more reliable mapping of starlight to stellar mass~\cite{Bel03}.

It was then realized~\cite{braun,freemanconf,btforig} that a more fundamental relation was that between the total observed baryonic mass and the rotation velocity.
In most bright galaxies, the stars harbor the majority of the detected baryonic mass, so luminosity suffices as a proxy for mass.
The next most important known reservoir of baryons is the neutral atomic hydrogen (HI) of the interstellar medium.  As
studies have probed down the mass spectrum to lower mass, more slowly rotating systems, a higher preponderance of
gas rich galaxies is found.  The luminous Tully--Fisher relation breaks down~\cite{btforig,M05}, but a tight relation persists
if instead of luminosity, the detected baryonic mass $M_b = M_*+M_g$ is 
used~\cite{btforig,verheijen,Bel03,M05,pfennBTF,begum,stark,trach,McGgasrich}. 
This is the Baryonic Tully--Fisher Relation (BTFR), plotted on Figure~\ref{figure:btf}.

\epubtkImage{BTF_LR.png}{%
\begin{figure}[htbp]
  \centerline{\includegraphics[width=9cm]{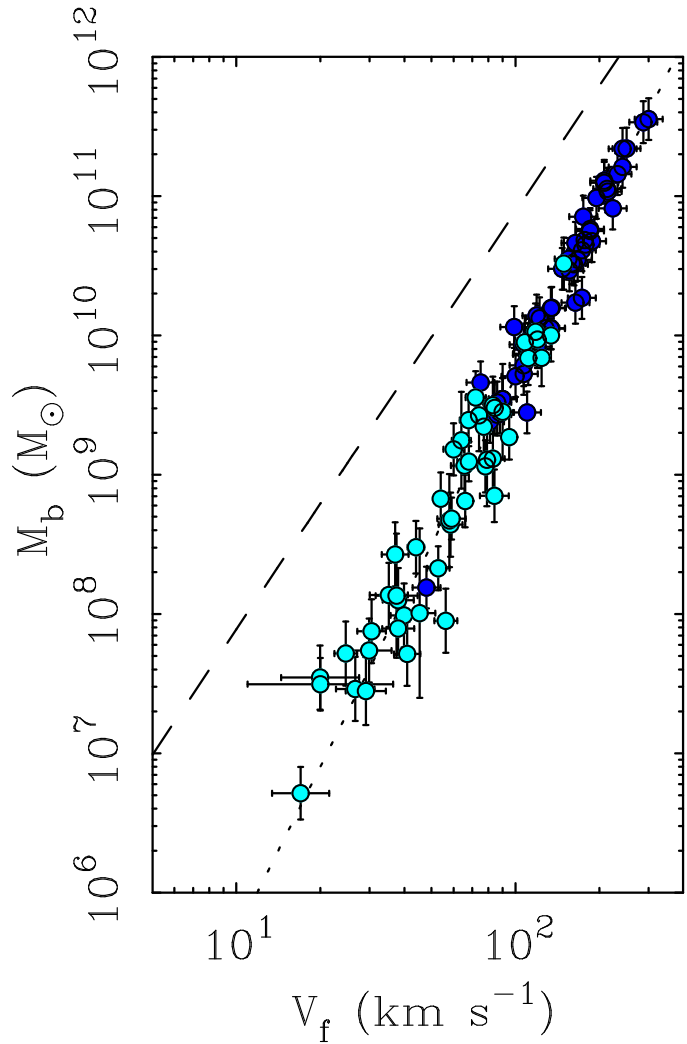}}
  \caption{The Baryonic Tully--Fisher (mass--rotation velocity)
    relation for galaxies with well measured outer velocities $V_f$.
    The baryonic mass is the combination of observed stars and gas:
    $M_b = M_* + M_g$.  Galaxies have been selected that have well observed, extended rotation curves from   
   21~cm interferrometric observations providing a good measure of the outer, flat rotation 
   velocity. The dark blue points are galaxies with $M_* >
    M_g$~\cite{M05}.  The light blue points have $M_* <
    M_g$~\cite{McGgasrich} and are generally less precise in velocity,
    but more accurate in terms of the harmlessness on the result of
    possible systematics on the stellar mass-to-light ratio. For a
    detailed discussion of the stellar mass-to-light ratios used here,
    see~\cite{M05,McGgasrich}. The dotted line has slope 4
    corresponding to a constant acceleration parameter, $1.2 \times
    10^{-10}\mathrm{\ m\, s}^{-2}$. The dashed line has slope 3 as
    expected  in $\Lambda$CDM with the normalization expected if all
    of the baryons associated with dark matter halos are detected.
    The difference between these two lines is the origin of the
    variation in the detected baryon fraction in
    Figure~\ref{figure:fdV}.}
  \label{figure:btf}
\end{figure}}

The luminous Tully--Fisher relation extends over about two decades in luminosity.  Recent work extending the relation
to low mass, typically low surface brightness and gas rich galaxies~\cite{begum,stark,trach}
extends the dynamic range of the BTFR to five decades in baryonic mass.  Over this range, the BTFR has remarkably little intrinsic scatter (consistent with zero given the observational errors) and is well described as a power law, or equivalently, as a straight line in log-log space:
\begin{equation}
\log M_b = \alpha \log V_f - \log \beta
\end{equation}
with slope $\alpha=4$~\cite{M05,stark,McGgasrich}.  
This slope is consistent with a constant acceleration scale $\mathrm{a} = V_f^4/(GM_b)$ such 
that\epubtkFootnote{As we shall see (Sect.~5 and 6), MOND was constructed to predict a relation $a_0 = V_f^4/(GM)$ for a point mass $M$ (note that the slope of 4 is however a pure consequence of the acceleration base, it is not possible to get an arbitrary slope from such an idea).
Since spiral galaxies are not point masses but rather flattened mass distributions 
that rotate faster than the equivalent spherical mass distribution~\cite{BTbook}, the empirical acceleration $\mathrm{a}$ is
close to but not identical to $a_0$ in MOND.  The geometric correction is about 20\% so that $a_0 = 0.8 \mathrm{a}$~\cite{M05}.}
the normalization constant $\beta = G\mathrm{a}$.

The acceleration scale $\mathrm{a} \approx 10^{-10}\;\mathrm{m}\,\mathrm{s}^{-2} \sim \Lambda^{1/2}$ (Eq.~\ref{eq:a0}) is thus present in the data.
Figure~\ref{figure:azero} shows the distribution of this acceleration $V_f^4/M_b$, around the best fit line in Figure~\ref{figure:btf}, strongly peaked around $\sim 2 \times 10^{-62}$ in natural units.  As we shall see, this acceleration scale arises empirically in a variety of distinct situations involving the mass discrepancy problem.

\epubtkImage{histazero_LR.png}{%
\begin{figure}[htb]
  \centerline{\includegraphics[width=9cm]{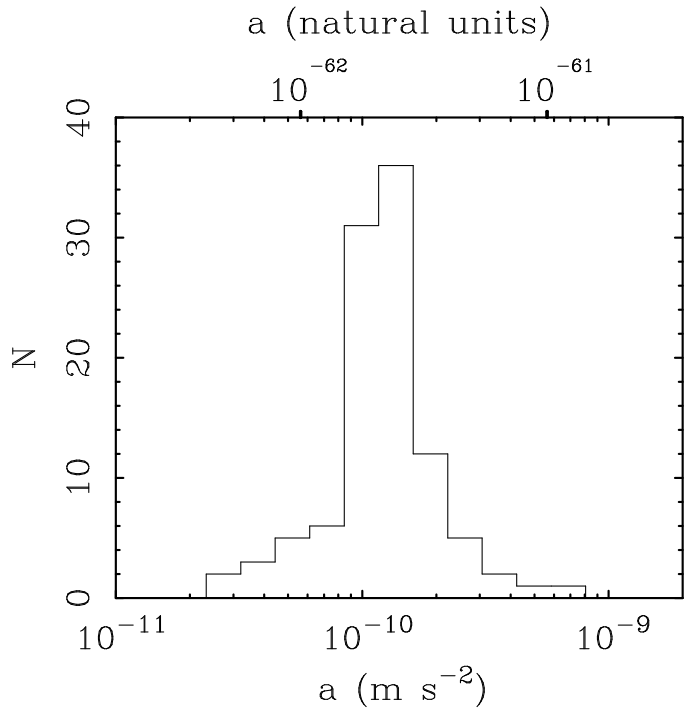}}
  \caption{Histogram of the accelerations $\mathrm{a} = V_f^4/(GM_b)$ in m s$^{-2}$ (bottom axis) and natural units [$c^4/(Gm_P)$ where $m_P$ is the Planck mass]
  for galaxies with well measured $V_f$.  The data are peaked around a characteristic value of $\sim 
10^{-10}\;\mathrm{m}\,\mathrm{s}^{-2}$ ($\sim 2 \times 10^{-62}$ in natural units).}
  \label{figure:azero}
\end{figure}}

A BTFR of the observed form does not arise \textit{naturally} in $\Lambda$CDM.  The naive expectation is
$\alpha = 3$ and $\beta = 10 f_V^3 G H_0$~\cite{SNonTF}\footnote{The factor 10 arises from the commonly adopted definition of the virial radius of the dark matter halo at an overdensity of 200 times the critical density of the Universe \cite{NFW}.} where $H_0$ is the Hubble constant and $f_V$ is a factor
of order unity (currently estimated to be $\approx 1.3$~\cite{reyes}) that relates the observed $V_f$ to the circular velocity
of the potential at the virial radius\epubtkFootnote{Note that~\cite{Gurovich} claimed to measure a slope of 3 for the BTFR, but they relied on unresolved line-widths from single dish 21 cm observations to estimate rotation velocity rather than measuring $V_f$ from resolved rotation curves. Line-widths give a systematically different estimate of the slope of the BTFR than $V_f$, even for the same galaxies~\cite{testbtf,NorVer,verheijen}, and they cannot be related at all to the circular velocity of the potential at the virial radius, nor to the prediction of MOND (Sect. 5 and 6).}.  This modest fudge factor is necessary because $\Lambda$CDM does not explicitly
predict either axis of the observed BTFR.  Rather, there is a relationship between total (baryonic plus dark) mass and
rotation velocity at very large radii.  This simple scaling fails (dashed line in Figure~\ref{figure:btf}), obliging us to
introduce an additional fudge factor $f_d$~\cite{halobyhalo,barcontent} that relates the detected baryonic mass to the total mass of
baryons available in a halo.  This mismatch drives the variation in the detected baryon fraction $f_d$ seen in 
Figure~\ref{figure:fdV}.  A constant $f_d$ is excluded by the difference between the observed and predicted slopes;
$f_d$ must vary with $V_f$, or $M$, or the gravitational potential $\Phi$.

This brings us to the first fine-tuning problem posed by the data.
There is essentially zero intrinsic scatter in the BTFR~\cite{McGgasrich} while the detected baryon fraction $f_d$ could
in principle obtain any value between zero and unity.  Somehow galaxies must ``know'' what the circular velocity of the
halo they reside in is so that they can make observable the correct fraction of baryons.

Quantitatively, in the $\Lambda$CDM picture, the baryonic mass plotted in the BTFR (Figure~\ref{figure:btf}) is $M_b = M_*+M_g$ while the total baryonic mass
available in a halo is $f_b M_{\mathrm{tot}}$.  The difference between these quantities implies a reservoir of dark baryons in some
undetected form, $M_{\mathrm{other}}$.  It is commonly speculated that the undetected baryons could be in a hard-to-detect hot,
diffuse, ionized phase mixed in with the dark matter halo (and extending to comparable radius), or that the missing baryons
have been entirely blown away by winds from supernovae.  For the purposes of this argument, it does not matter which
form the dark baryons take.  All that matters is that a substantial mass of them are required so that~\cite{btforig}
\begin{equation}
f_d = \frac{M_b}{f_b M_{\mathrm{tot}}} = \frac{M_*+M_g}{M_*+M_g+M_{\mathrm{other}}}.
\label{eqn:fdtuning}
\end{equation}
Since there is negligible intrinsic scatter in the observed BTFR, there must be effectively zero scatter in $f_d$.
By inspection of Eq.~\ref{eqn:fdtuning}, it is apparent that small scatter in $f_d$ can only be obtained naturally
in the limits $M_*+M_g \gg M_{\mathrm{other}}$ so that $f_d \rightarrow 1$ or $M_*+M_g \ll M_{\mathrm{other}}$ so that $f_d \rightarrow 0$.
Neither of these limits apply.  We require not only an appreciable mass in dark baryons $M_{\mathrm{other}}$, but we need
the fractional mass of these missing baryons to vary in lockstep with the observed rotation velocity $V_f$.  
Put another way, for any given galaxy, we know not only how many baryons we see, but also how many we do not see ---
a remarkable feat of non-observation.

\epubtkImage{btf_resid_forLR.png}{%
\begin{figure}[htbp]
  \centerline{\includegraphics[width=14.5cm]{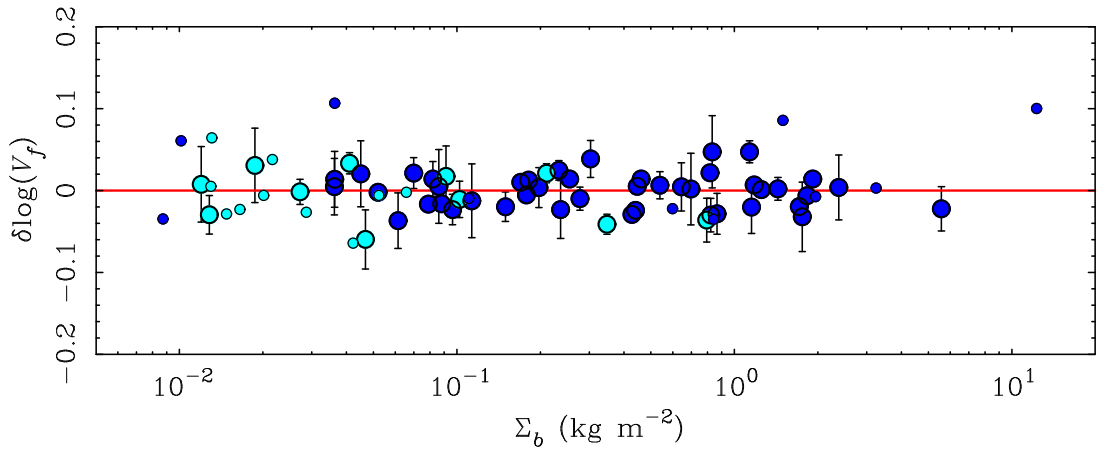}}
  \caption{Residuals ($\delta \log V_f$) from the baryonic Tully--Fisher relation as a function of a galaxy's characteristic 
  baryonic surface density ($\Sigma_b = 0.75 M_b/R_p^2$~\cite{McGbalance}, $R_p$ being the radius at which the contribution of baryons to the rotation curve peaks).  
  Color differentiates between star (dark blue) and gas (light blue) dominated galaxies as in Figure~\ref{figure:btf},
  but not all galaxies there have sufficient data (especially of $R_p$) to plot here.  Stellar masses have been estimated
  with stellar population synthesis models~\cite{Bel03}.  More accurate data, with uncertainty on rotation velocity less than 
  5\%, are shown as larger points; less accurate data are shown as smaller points. The rotation velocity of galaxies shows
  no dependence on the distribution of baryons as measured by $\Sigma_b$ or $R_p$. This is puzzling in the conventional 
  context, where $V^2 = GM/r$ should lead to a strong systematic residual~\cite{CRix}.}
%  This puzzling behavior~\cite{MdB98,McGbalance} is expected from Milgrom's formula (see Sect.~5).   }
  \label{figure:btfresiduals}
\end{figure}}

Another remarkable fact about the BTFR is that it shows no residuals with variations in the distribution of 
baryons~\cite{ZwaanTF,sprayTF,CRix,McGbalance}.  Figure~\ref{figure:btfresiduals} shows deviations from the BTFR
as a function of the characteristic baryonic surface density of the galaxies, as defined in~\cite{McGbalance}, i.e., $\Sigma_b = 0.75 M_b/R_p^2$ where $R_p$ is the radius at which the rotation curve $V_b(r)$ of baryons peaks. Over several decades in surface density, the BTFR is completely insensitive to variations in the mass distribution of the baryons.  
This is odd because, \textit{a priori}, $V^2 \sim M/R$, and thus $V^4 \sim M \Sigma$.  Yet the BTFR is $M_b \sim V_f^4$ with no dependence
on $\Sigma$. This brings us to a second fine-tuning problem.  For some time, it was thought~\cite{freemanlaw} that spiral galaxies all had
very nearly the same surface brightness (a condition formerly known as ``Freeman's Law'').  If this is indeed the case, the observed
BTFR naturally follows from the constancy of $\Sigma$.  However, there do exist many low surface brightness galaxies~\cite{McG96}
that violate the constancy of surface brightness implied in Freeman's Law.  One would thus expect them to deviate systematically
from the Tully--Fisher relation, with lower surface brightness galaxies having lower rotation velocities at a given mass.
Yet they do not.  Thus one must fine-tune the mass surface density of the dark matter to precisely make up for that of
the baryons~\cite{MdB98}.  As the surface density of baryons declines, that of the dark matter must increase
\textit{just so} as to fill in the difference (Figure~\ref{fig:vbvpsd}~\cite{McGbalance}). The relevant quantity is the dynamical surface density enclosed within the radius where the velocity is measured.
The latter matters little along the flat portion of the rotation curve, but the former is the sum of dark and baryonic matter.

One might be able to avoid fine-tuning if all galaxies are dark matter dominated~\cite{CRix}.  In the 
limit $\Sigma_{DM} \gg \Sigma_b$, the dynamics are entirely dark matter dominated and the distribution of the
baryons is irrelevant.  There is some systematic uncertainty in the mass-to-light ratios of stellar populations~\cite{Bel03},
making such an approach \textit{a priori} tenable.  
In effect, we return to the interpretation of $\Sigma \sim$ constant originally made by~\cite{aaronson}
in the context of Freeman's Law, but now we invoke a constant surface density of CDM rather than of baryons.
But as we will see, such an interpretation, i.e., that $\Sigma_b \ll \Sigma_{DM}$ in all disk galaxies, is flatly contradicted by other observations (e.g., Figure~\ref{figure:XiSd} and Figure~\ref{fig:4panel}).

\epubtkImage{VbVpSd_LR.png}{%
\begin{figure}[htbp]
  \centerline{\includegraphics[width=14.5cm]{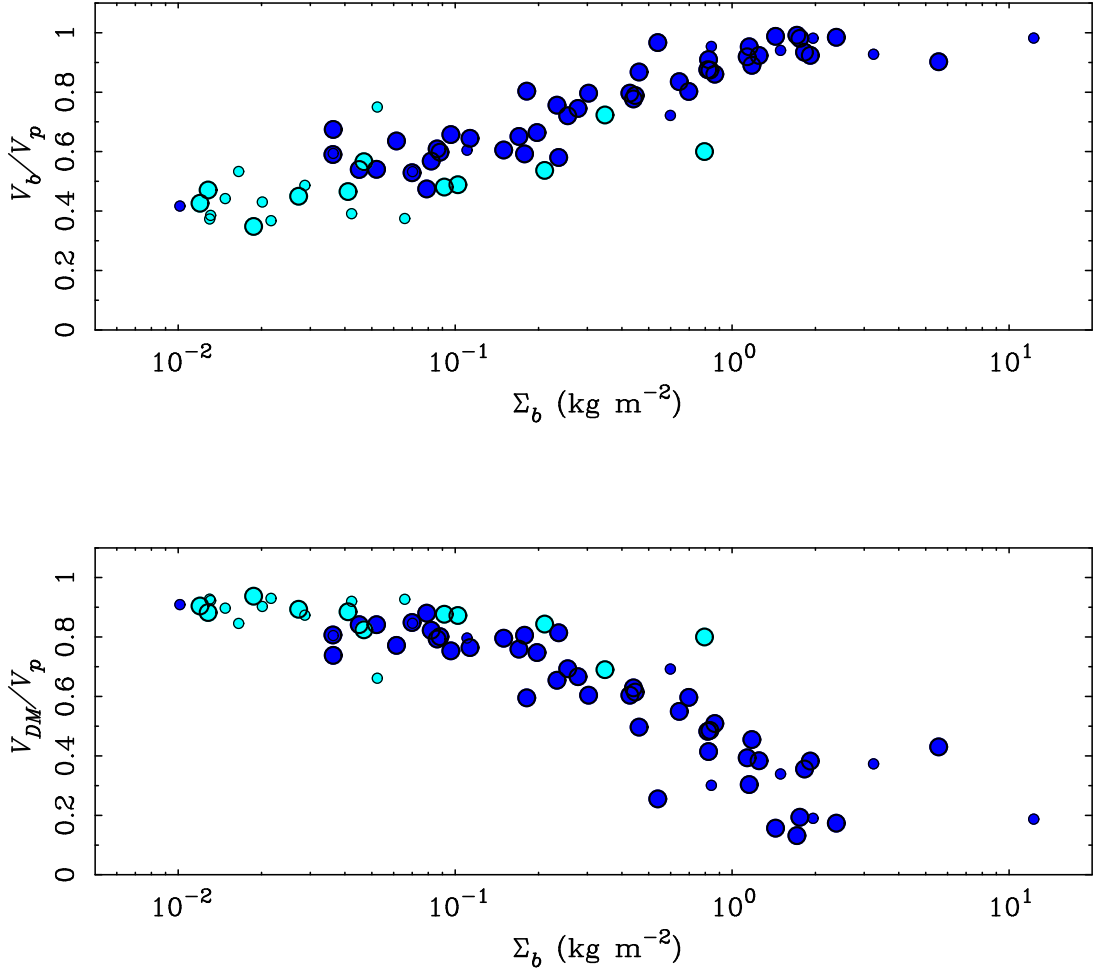}}
  \caption{ The fractional contribution to the total velocity $V_p$ at the
radius $R_P$ where the contribution of the baryons peaks for both
baryons ($V_b/V_p$, top) and dark matter ($V_{DM}/V_p$, bottom).  
Points as per Figure~\ref{figure:btfresiduals}.
As the baryonic surface density increases, the contribution of the baryons to the total gravitating
mass increases.  The dark matter contribution declines in compensation, maintaining a 
see-saw balance that manages to leave no residual in the BTFR (Figure~\ref{figure:btfresiduals}).  
The absolute amplitude of $V_b$ and $V_{DM}$ depends on choice of stellar mass estimator,
but the fine-tuning between them must persist for any choice of $M_*/L$. }
  \label{fig:vbvpsd}
\end{figure}}

The Tully--Fisher relation is remarkably persistent.  Originally posited for bright spirals, it applies to galaxies that one would naively
expected to deviate from it.  This includes low luminosity, gas dominated irregular galaxies~\cite{stark,trach,McGgasrich}, 
low surface brightness galaxies of all luminosities~\cite{ZwaanTF,sprayTF}, and even
tidal dwarfs formed in the collision of larger galaxies~\cite{GentileTDG}.  Such tidal dwarfs may be especially important in this context (see also Sect.~6.5.4).
Galactic collisions should be very effective at segregating dark and baryonic matter.  The rotating gas disks of galaxies that provide
the fodder for tidal tails and the tidal dwarfs that form within them initially have nearly circular, coplanar orbits.  In contrast, the
dark matter particles are on predominantly radial orbits in a quasi-spherical distribution.  This difference in phase space leads
to tidal tails that themselves contain very little dark matter~\cite{Bournaud07}.  When tidal dwarfs form from tidal debris, they should be
largely devoid\epubtkFootnote{The difference in phase space between gas and dark matter also prevents the accretion of tidal 
gas onto any dark matter sub-halos that may be present.  It does not suffice for a tidal tail to intersect the location of a sub-halo
in coordinate space, they must also dock in velocity space.  The gas is moving at the characteristic velocity of the entire system
(typically $\sim 200\mathrm{\ km\, s}^{-1}$) which by definition exceeds the escape speed of typical sub-halos
(usually $< 100\mathrm{\ km\, s}^{-1}$).  The odds of capture are therefore effectively zero unless the tail and sub-halo
happen to be on very nearly the same orbit initially, which is itself very unlikely because of the initial difference in their
phase space distribution.} of dark matter.  Nevertheless, tidal dwarfs do appear to contain dark matter~\cite{Bournaud07}
and obey the BTFR~\cite{GentileTDG}. 

The critical acceleration scale of Eq.~\ref{eq:a0} also appears in non-rotating galaxies.  Elliptical galaxies are three-dimensional stellar systems supported
more by random motions than organized rotation.  First of all, in such systems of measured velocity dispersion $\sigma$, the typical acceleration $\sigma^2/R$ is also of the order of $a_0$ within a factor of a few, where $R$ is the effective radius of the system~\cite{SM02}. Moreover, they obey an analogous relation to the Tully--Fisher one, known as the Faber-Jackson relation
(Figure~\ref{figure:faberjackson}). In bulk, the data for these star-dominated galaxies follow the relation $\sigma^4/(GM_*) \propto a_0$ (dotted line in Figure~\ref{figure:faberjackson}). This is not strictly analogous to the flat part of the rotation curves of spiral galaxies, the dispersion typically being measured at smaller radii where the equivalent circular velocity curve is often falling~\cite{romanowsky,dearth}, or in a temporary plateau before falling again (see also Sect.~6.6.1).  Indeed, unlike the case in spiral galaxies where the
distribution of stars is irrelevant, it clearly does matter in elliptical galaxies (the Faber--Jackson relation is just one projection
of the ``fundamental plane'' of elliptical galaxies~\cite{SAURON}).  This is comforting:  at small radii in dense stellar systems
where the baryonic mass of stars is clearly important, the data behave as Newton predicts.

\epubtkImage{FaberJackson_LR.png}{%
\begin{figure}[htbp]
  \centerline{\includegraphics[width=9cm]{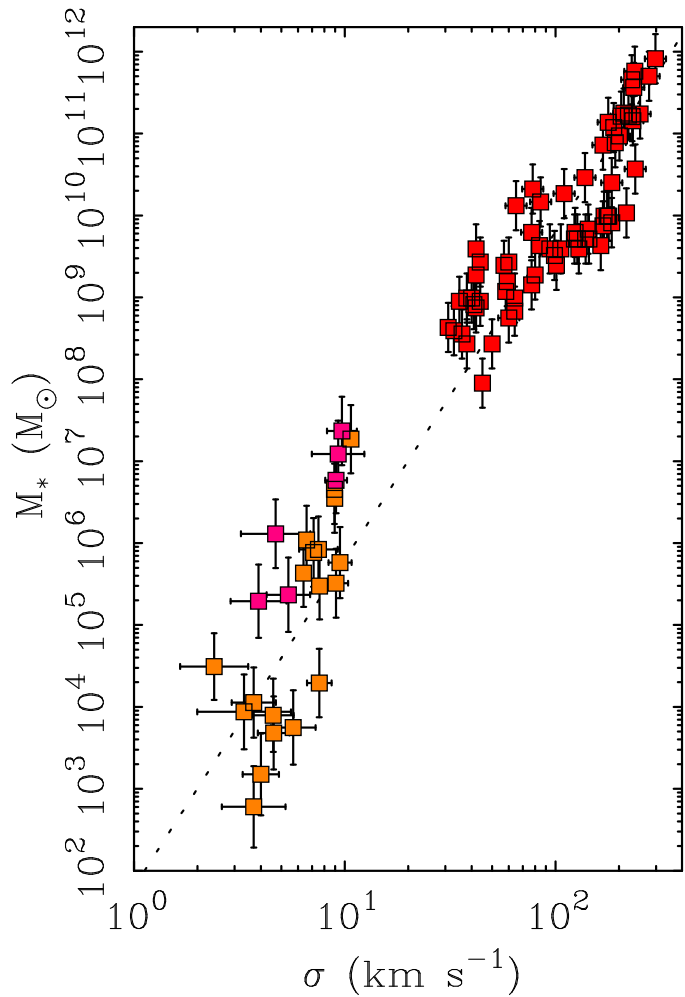}}
  \caption{The Faber--Jackson relation for spheroidal galaxies,
    including both elliptical galaxies (red squares, \cite{SAURON,comadwarf}) and Local Group dwarf satellites~\cite{MWolf}
  (orange squares are satellites of the Milky Way; pink squares are satellites of M31).  In analogy with the
  Tully--Fisher relation for spiral galaxies, spheroidal galaxies follow a relation between stellar mass and line of sight
  velocity dispersion ($\sigma$).  The dotted line represents a constant value of the acceleration parameter $\sigma^4/(GM_*)$. Note however that this relation is different from the BTFR because it applies to the {\it bulk} velocity dispersion while the BTFR applies to the asymptotic circular velocity. In the context of Milgrom's law (Sect.~5 hereafter) the Faber--Jackson relation is predicted only when relying on assumptions such as isothermality, isotropy, and the slope of the baryonic density distribution (see 3rd law of motion in Sect.~5.2). In addition, not all pressure-supported systems are in the weak-acceleration regime. So, in the context of Milgrom's law, deviations from the weak-field regime, from isothermality and from isotropy, as well as variations in the baryonic density distribution slope, would thus explain the scatter in this relation.}
  \label{figure:faberjackson}
\end{figure}}

The acceleration scale $a_0$ is clearly imprinted on the data for local galaxies.  This is an empirical statement that might
not hold at all times, perhaps evolving over cosmic time or evaporating altogether.  Substantial efforts have been made to
investigate the Tully--Fisher relation to high redshift.  To date, there is no persuasive evidence of evolution in the zero point
of the BTFR out to $z = 0.6$~\cite{PuechTF,PuechBTF} and perhaps even to $z = 1$~\cite{Weiner}.  One must exercise caution
in interpreting such results given the difficulty inherent in peering many Gyr back in cosmic time.  Nonetheless, it appears that
the scale $a_0$ remains present in the data and has not obviously changed over the more recent half of the age of the Universe.

\subsubsection{The role of surface density}

The Freeman limit~\cite{freemanlaw} is the maximum central surface brightness in the distribution of galaxy surface brightnesses.
Originally thought to be a universal surface brightness, it has since become clear that instead 
galaxies exist over a wide range in surface brightness~\cite{McG96}.  In the absence of a perverse and fine-tuned anti-correlation between
surface brightness and stellar mass-to-light ratio~\cite{ZwaanTF}, this implies a comparable range in baryonic surface density 
(Figure~\ref{fig:freeman}).  

\epubtkImage{RpSd_LR.png}{%
\begin{figure}[htbp]
  \centerline{\includegraphics[width=14.5cm]{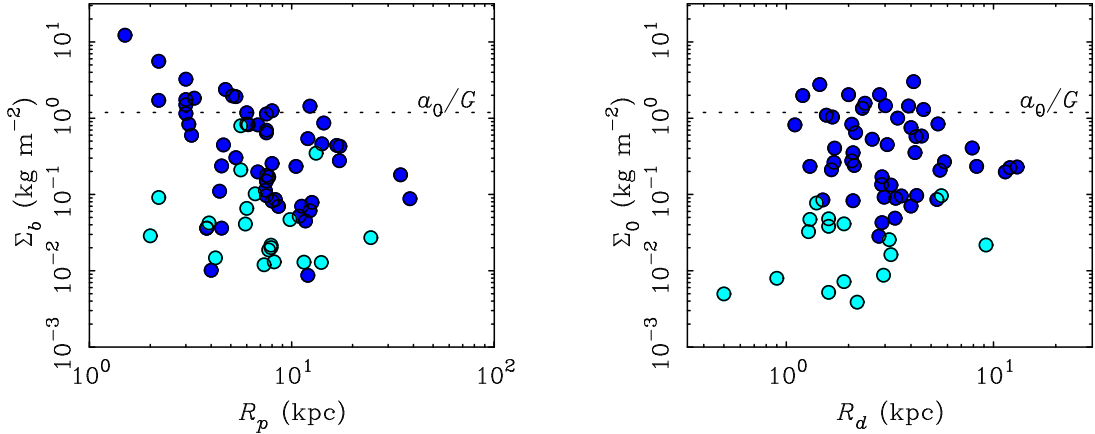}}
  \caption{Size and surface density.
  The characteristic surface density of baryons as defined in Figure~\ref{figure:btfresiduals} 
  is plotted against their dynamical scale length $R_p$ in the left panel. The dark blue points are star-dominated galaxies and the light-blue ones gas-dominated. High characteristic surface densities at low $R_p$ in the left panel are typical of bulge-dominated galaxies. The stellar disk component of most spiral galaxies is well approximated by the exponential disk 
  with $\Sigma(R) = \Sigma_0 e^{-R/R_d}$.  This disk-only 
   central surface density and the exponential scale length of the stellar disk are plotted in the right panel.  
  Galaxies exist over a wide range in both size and surface density.  There is a maximum surface density threshold 
  (sometimes referred to as Freeman's limit) above which disks become very rare~\cite{McG96}.  This is presumably a
  stability effect, as purely Newtonian disks are unstable~\cite{OstPeeb,sellwoodevans}.  Stable disks only appear below a
  critical surface density $\Sigma_{\dagger} \approx a_0/G$~\cite{Milstab,BradaM}.
  }
  \label{fig:freeman}
\end{figure}}

An upper limit to the surface brightness distribution is interesting in the context of disk stability.
Recall that dynamically cold, purely Newtonian disks are subject to potentially self destructive instabilities,
one cure being to embed them in the potential wells of spherical dark matter halos~\cite{OstPeeb}.
While the proper criterion for stability is much debated~\cite{ELN,sellwoodevans}
it is clear that the dark matter halo moderates the growth of instabilities and that the ratio  of halo to disk
self gravity is a relevant quantity.  The more self-gravitating a disk is, the more likely it is to suffer undamped
growth of instabilities. But in principle, galaxies with a baryonic disk and a dark matter halo are totally scalable: if a galaxy model has a certain dynamics, and one multiplies all densities by any (positive) constant (and also scales the velocities appropriately) one gets another galaxy with exactly the same dynamics (with scaled time scales). So if one is stable, so is the other. In turn, the mere fact that there might be an upper limit to $\Sigma_b$ is \textit{a priori} surprising, and even more so that there might be a coincidence of this upper limit with the acceleration scale $a_0$ identified dynamically.

The scale $\Sigma_{\dagger} = a_0/G$ is clearly present in the data (Figure~\ref{fig:freeman}).  Selection effects
make high surface brightness galaxies easy to detect and hence discover, but their intrinsic numbers appear to decline exponentially
when the central surface density of the stellar disk $\Sigma_0 > \Sigma_{\dagger}$~\cite{McG96}.   
It seems natural to associate the dynamical scale $a_0$ with the disk stability scale $\Sigma_{\dagger}$ since they are
numerically indistinguishable and both arise in the context of the mass discrepancy.  However, there is no reason
to expect this in $\Lambda$CDM, which predicts denser dark matter halos than 
observed~\cite{M07,Gentilecore,gentilenature,KdN08,KdN09,Walker10,deblok}.
Such dense dark matter halos could stabilize much higher density disks than are observed to exist.
Lacking a clear mechanism to specify this scale, it is introduced into models by hand~\cite{DSS97}.

\epubtkImage{XiSd_LR.png}{%
\begin{figure}[htb]
  \centerline{\includegraphics[width=0.9\textwidth]{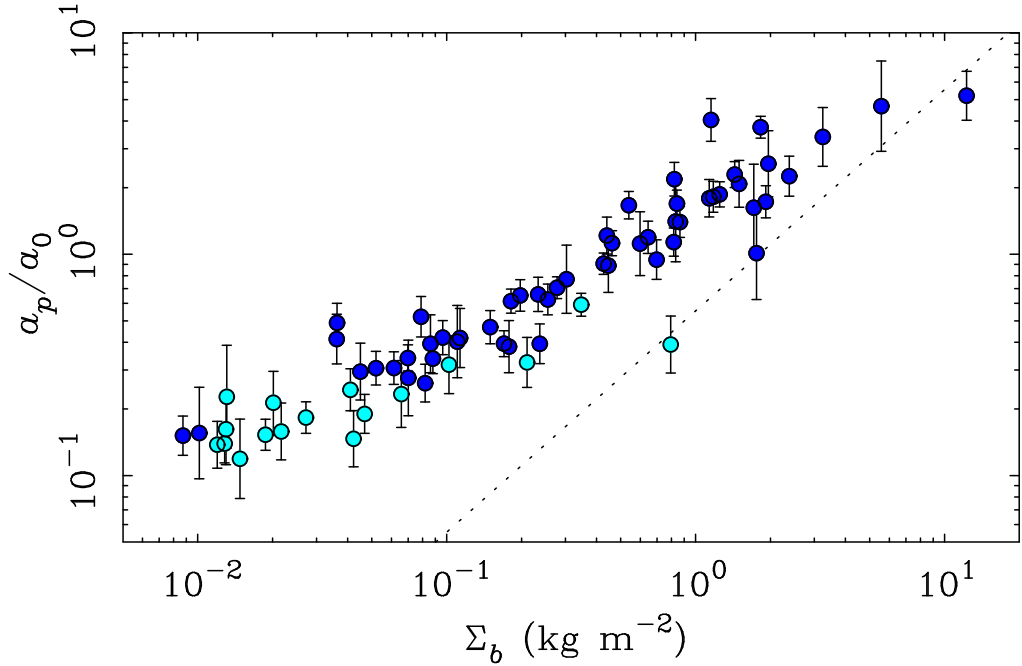}}
  \caption{The dynamical acceleration $a_p = V_p^2/R_p$ in units of $a_0$ 
  plotted against the characteristic baryonic surface density~\cite{McGmalta}.  Points as per Figure~\ref{figure:btfresiduals}.
  The dotted line shows the relation $a_p = G \Sigma_b$ that would be 
%  expected in the absence of either dark matter or modified gravity.
  obtained if the visible baryons sufficed to explain the observed velocities in Newtonian dynamics.
  Though the data do not follow this line, they do show a correlation ($a_p \propto \Sigma_b^{1/2}$).
  This clearly indicates a dynamical role for the baryons, in contradiction to the 
  simplest interpretation~\cite{CRix} of Figure~\ref{figure:btfresiduals} that
  dark matter completely dominates the dynamics.
  }
  \label{figure:XiSd}
\end{figure}}

Poisson's equation provides a direct relation between the force per unit mass (centripetal acceleration in the case of circular
orbits in disk galaxies), the gradient of the potential, and the surface density of gravitating mass.  If there is no dark matter,
the observed surface density of baryons must correlate perfectly with the dynamical acceleration.  If, on the other hand, 
dark matter dominates the dynamics of a system, as we might infer from Figure~\ref{figure:btfresiduals}~\cite{MdB98,CRix}, 
then there is no reason to expect a correlation between acceleration and the dynamically insignificant baryons.  Figure~\ref{figure:XiSd} shows the dynamical acceleration as a function of baryonic surface density in disk galaxies. The acceleration $a_p = V_p^2/R_p$ is measured at the radius $R_p$ where the rotation curve $V_b(r)$ of baryons peaks.
Given the systematic variation of rotation curve shape~\cite{SalucciURC,Ysalucci}, the specific choice of radii is unimportant.
Nevertheless, this radius is advocated to be used by~\cite{CRix} since this maximizes the possibility of perceiving the baryonic 
contribution in the plot of Figure~\ref{figure:btfresiduals}.  That this contribution is not present leads to the inference that $\Sigma_b \ll \Sigma_{DM}$ in all disk galaxies~\cite{CRix}.  This is directly contradicted by Figure~\ref{figure:XiSd}, which shows a clear correlation between $a_p$ and $\Sigma_b$.  

The higher the surface density of baryons is, the higher the observed acceleration.  The slope of the
relation is not unity, $a_p \propto \Sigma_b$, as we would expect in the absence of a mass discrepancy, but rather
$a_p \propto \Sigma_b^{1/2}$.  To simultaneously explain Figure~\ref{figure:btfresiduals} and Figure~\ref{figure:XiSd},
there must be a strong fine-tuning between dark and baryonic surface densities (i.e., Figure~\ref{fig:vbvpsd}), a sort of repulsion between them, a repulsion which is however contradicted by the correlations between baryonic and dark matter bumps and wiggles in rotation curves (see Sect.~4.3.4).

\subsubsection{Mass discrepancy-acceleration relation}

So far we have discussed total quantities.
For the BTFR, we use the total observed mass of a galaxy and its characteristic rotation velocity.
Similarly, the dynamical acceleration--baryonic surface density relation uses a single characteristic value for each galaxy.
These are not the only ways in which the ``magical'' acceleration constant $a_0$ appears in the data.  In general, the mass discrepancy only
appears at very low accelerations $a < a_0$ and not (much) above $a_0$.  Equivalently, the need for dark matter only
becomes clear at very low baryonic surface densities $\Sigma < \Sigma_{\dagger} = a_0/G$.  Indeed, the amplitude of the mass
discrepancy in galaxies anti-correlates with acceleration~\cite{MDA}.

\epubtkImage{MDaccRgn_LR.png}{%
\begin{figure}[htbp]
  \centerline{\includegraphics[width=14.5cm]{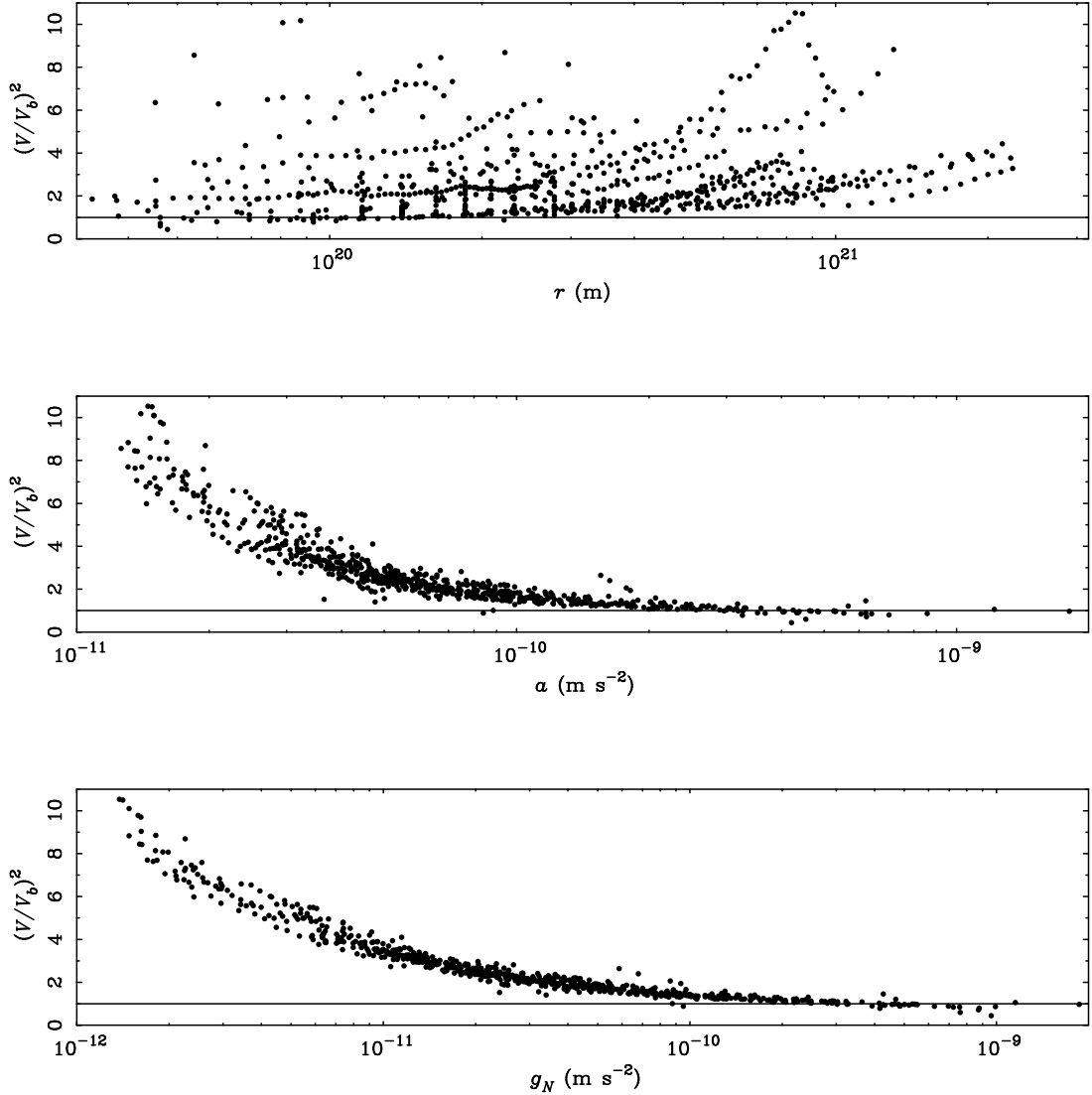}}
  \caption{The mass discrepancy in spiral galaxies.  The mass discrepancy
  is defined~\cite{MDA}
  as the ratio $V^2/V_b^2$ where $V$ is the observed velocity and
  $V_b$ is the velocity attributable to visible baryonic matter.  The ratio of squared
  velocities is equivalent to the ratio of total to baryonic enclosed mass for spherical
  systems.  No dark matter is required when $V = V_b$, only when $V>V_b$.
  Many hundreds of individual resolved measurements along the rotation curves  
  of nearly one hundred spiral galaxies are plotted.
  The top panel plots the mass discrepancy as a function
  of radius.  No particular linear scale is favored.  Some galaxies exhibit mass
  discrepancies at small radii while others do not appear to need dark matter
  until quite large radii.  The middle panel plots the mass discrepancy as a function
  of centripetal acceleration $a = V^2/r$, while the bottom panel plots it against the acceleration 
  $g_N = V_b^2/r$ predicted by Newton from the observed baryonic surface density $\Sigma_b$. Note that the correlation appears a little better with $g_N$ because the data are strecthed out over a wider range in $g_N$ than in $a$. Note also that systematics on the stellar mass-to-light ratios can make this relation slightly more blurred than shown here, but the relation is nevertheless always present irrespective of the assumptions on stellar mass-to-light ratios~\cite{MDA}. There is thus a clear organization: the amplitude of the mass discrepancy increases systematically with decreasing acceleration and baryonic surface density.}
  \label{figure:MDRA}
\end{figure}}

In~\cite{MDA}, one examined the role of various possible scales, as well as the effects of different stellar mass-to-light ratio estimators, on the mass discrepancy problem.
The amplitude of the mass discrepancy, as measured by $(V/V_b)^2$, the ratio of observed velocity to that predicted by the 
observed baryons, depends on the choice of estimator for stellar $M_*/L$.  However, for any plausible (non-zero) $M_*/L$, 
the amplitude of the mass discrepancy correlates with acceleration (Figure~\ref{figure:MDRA}) and baryonic surface density, as originally noted in \cite{Bobmda,mdaconf,mdascarpa}. It does \textit{not} correlate with radius and only weakly with orbital frequency\footnote{Note that this correlation with acceleration was looked at notably because it was pointed to by Milgrom's law (see Sect.~5).}.

There is no reason in the dark matter picture why the mass discrepancy should correlate with any physical scale.  Some systems might happen to contain lots of dark matter; others very little.  In order to make a prediction with a dark matter model, it is necessary to model the formation of the dark matter halo, the condensation of gas within it, the formation of stars therefrom, and any feedback processes whereby the formation of some stars either enables or suppresses the formation of further stars.  This complicated sequence of events is challenging to model.  Baryonic ``gastrophysics'' is particularly difficult, and has thus far precluded the emergence of a clear prediction for galaxy dynamics from $\Lambda$CDM.

$\Lambda$CDM does make a prediction for the distribution of mass in baryonless dark matter halos: the NFW halo~\cite{NFW, Navarro}.  These are remarkable for being scale free.  Small halos have a profile similar to large halos.  No feature stands out that marks a unique physical scale as observed. Galaxies do not resemble pure NFW halos \cite{SM05}, even when dark matter dominates the dynamics as in low surface brightness galaxies \cite{KdN08,KdN09,deblok}.  The inference in $\Lambda$CDM is that gastrophysics, especially the energetic feedback from stellar winds and supernova explosions, plays a critical role in sculpting observed galaxies.  This role is not restricted to the minority baryonic constituents; it must also affect the majority dark matter \cite{governato}.  Simulations incorporating these effects in a quasi-realistic way are extremely expensive computationally, so a comprehensive survey of the plausible parameter space occupied by such models has yet to be made.  We have no reason to expect that a particular physical scale will generically emerge as the result of baryonic gastrophysics.  Indeed,  feedback from star formation is inherently a random process.  While it is certainly possible for simple laws to emerge from complicated physics (e.g., the fact that SNIa are standard candles despite the complicated physics involved), the more common situation is for chaos to beget chaos.  It therefore seems unnatural to imagine feedback processes leading to the orderly behavior that is observed (Figure~\ref{figure:MDRA}), nor is it obvious how they would implicate any particular physical scale.  Indeed, the dark matter halos formed in $\Lambda$CDM simulations~\cite{NFW, Navarro} provide an initial condition with greater scatter than the final observed one~\cite{M07,Walker10}, so we must imagine that the chaotic processes of feedback not only impart order, but do so in a way that cancels out some of the scatter in the initial conditions. 

\epubtkImage{MDacc_solarsystem.png}{%
\begin{figure}[htbp]
  \centerline{\includegraphics[width=14.5cm]{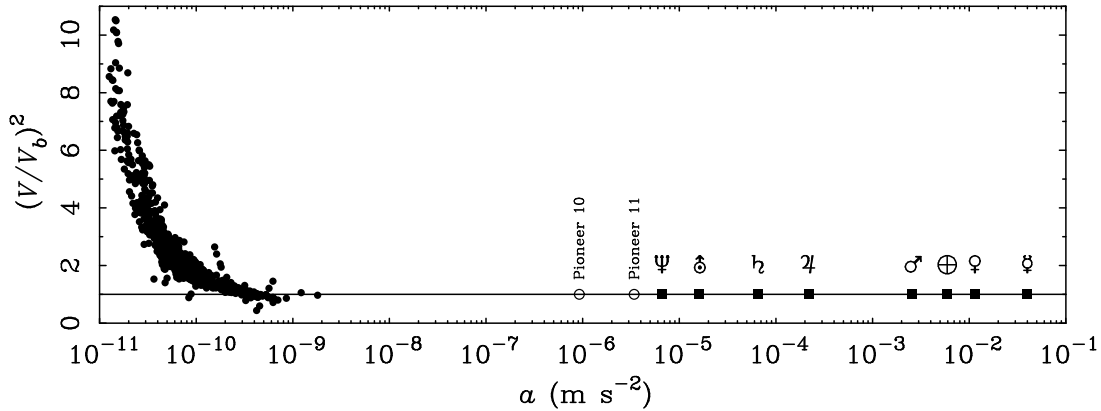}}
  \caption{The mass discrepancy--acceleration relation from Figure~\ref{figure:MDRA} extended to Solar system scales
  (each planet is labelled).
  This illustrates the large gulf in scale between galaxies and the Solar system where high precision tests are possible.
  The need for dark matter only appears at very low accelerations.    }
  \label{fig:mdasolsys}
\end{figure}}

In any case, and whatever the reason for it, a physical scale is clearly observationally present in the data:
$a_0$ (Eq.~\ref{eq:a0}). At high accelerations $a \gg a_0$, there is no indication of the need for dark matter.  Below this acceleration, the mass discrepancy appears. It cannot be emphasized enough that the role played by $a_0$ in the BTFR and this role as a transition acceleration have strictly no intrinsic link with each other, they are fully independent of each other. There is \textit{nothing} in $\Lambda$CDM that stipulates that these two relations (the existence of a transition acceleration and the BTFR) should exist at all, and even less that these should harbour an \textit{identical} acceleration scale.

It is thus important to realize not only that the relevant dynamical scale is one of acceleration, not size, but also that the mass discrepancy appears only at extremely low accelerations.  Just as galaxies are much bigger than the Solar system, so too are the centripetal accelerations experienced by stars orbiting within a galaxy much smaller than those experienced by planets in the Solar system.  Many of the precise tests of gravity that have been made in the Solar system do not explore the relevant regime of physical parameter space.  This is emphasized in Figure~\ref{fig:mdasolsys}, which extends the mass discrepancy--acceleration relation to Solar system scales.  Many decades in acceleration separate the Solar system from galaxies.  Aside from the possible exception of the Pioneer anomaly, there is no hint of a discrepancy in the Solar system:  $V = V_b$.  Even the Pioneer anomaly\footnote{The Pioneer anomaly has an amplitude of the order of $\sim 
10^{-9}\;\mathrm{m}\,\mathrm{s}^{-2}$ but appears at a location in the 
solar system where the total gravitational acceleration is $\sim 
10^{-6}\;\mathrm{m}\,\mathrm{s}^{-2}$.  The discrepancy in 
Figure~\ref{fig:mdasolsys} is thus $(V/V_b)^2 \approx 1.001$.}
is well removed from the regime where the mass discrepancy manifests in 
galaxies, and is itself much too subtle to be perceptible in Figure~\ref{fig:mdasolsys}.  Indeed, to within of a factor of $\sim 2$, no system exhibits a mass discrepancy at accelerations $a \gg a_0$.

The systematic increase in the amplitude of the mass discrepancy with decreasing acceleration and baryonic surface density has
a remarkable implication.  Even though the observed velocity is not correctly predicted by the observed baryons, it is
predictable from them.  Independently of any theory, we can simply fit a function $D(G\Sigma)$ to then describe the variation of the discrepancy $(V/V_b)^2$ with baryonic surface density~\cite{MDA}.  We can then apply it to any new system we encounter to predict $V = D^{1/2} V_b$.  In effect, $D$ boosts the velocity already predicted by the observed baryons.  While this is a purely empirical exercise with no underlying theory, it is quite remarkable that the distribution of dark
matter required in a galaxy is entirely predictable from the distribution of its luminous mass (see also~\cite{gentilenature}).  In the conventional picture, dark matter outweighs baryonic matter by a factor of five, and more in individual galaxies given the halo-by-halo missing baryon problem (Figure~\ref{figure:fdV}), but apparently the baryonic tail wags the dark matter dog. And it does so again through the acceleration scale $a_0$. Indeed, at very low accelerations, the mass discrepancy is precisely defined by the inverse of the square-root of the gravitational acceleration generated by the baryons \textit{in units of} $a_0$. This actually asymptotically leads to the BTFR.

So, up to now, we have seen five roles of $a_0$ in galaxy dynamics. (i) It defines the zero point of the Tully--Fisher relation, (ii) it appears as the characteristic acceleration at the effective radius of spheroidal systems, (iii) it defines the Freeman limit for the maximum surface density of pure disks, (iv) it appears as a transition-acceleration above which no dark matter is needed, and below which it appears, and (v) it defines the amplitude of the mass-discrepancy in the weak-field regime (this last point is not a fully independent role as it leads to the Tully--Fisher relation). Let us eventually note that there is yet a final role played by $a_0$, which is that it defines the central surface density of all dark matter halos as being of the order of $a_0/(2 \pi G)$~\cite{donato,gentilenature,Mildark}.

\subsubsection{Renzo's rule}

The relation between dynamical and baryonic surface densities appears as a global scaling relation in disk galaxies (Figure~\ref{figure:XiSd}) and as a local correspondence within each galaxy (Figure~\ref{figure:MDRA}). When all galaxies are plotted together as in Figure~\ref{figure:MDRA}, this connection appears as a single smooth function $D(a)$.  This does not suffice to illustrate that individual galaxies have features in their baryon distribution that are reflected in their dynamics. While the above correlations could be interpreted as a sort of repulsion between dark and baryonic matter, the following rather indicates closer than natural attraction. 

Figure~\ref{fig:coolpicture} shows the spiral galaxy NGC~6946.  Two multi-color images of the stellar component are given.
The optical bands provide a (nearly) true color picture of the galaxy, which is perceptibly redder near the center and becomes
progressively more blue further out.  This is typical of spiral galaxies and reflects real differences in stellar content:  the
stars towards the center tend to be older an more dominated by the light of red giants, while those further out are younger
on average so the light has a greater fractional contribution from bright but short-lived main sequence stars.  The near-infrared
bands~\cite{2MASS} give a more faithful map of stellar mass, and are less affected by dust obscuration.  Radio synthesis
imaging of the 21 cm emission from the hydrogen spin-flip transition maps the atomic gas in the interstellar medium, which
typically extends to rather larger radii than the stars.  

Surface density profiles of galaxies are constructed by fitting ellipses to images like those illustrated in Figure~\ref{fig:coolpicture}.
The ellipses provide an axisymmetric representation of the variation of surface brightness with radius.  This is shown
in the top panels of Figure~\ref{fig:4panel} for NGC~6946 (Figure~\ref{fig:coolpicture}) and the nearby, gas rich, low surface
brightness galaxy NGC 1560.  The $K$-band light distribution is thought to give the most reliable mapping of observed
light to stellar mass~\cite{Bel03}, and has been used to trace the run of stellar surface density in Figure~\ref{fig:4panel}.
The sharp feature at the center is a small bulge component visible as the red central region in Figure~\ref{fig:coolpicture}.
The bulge contains only 4\% of the $K$-band light.  The remainder is the stellar disk; a straight line fit to the data outside
the central bulge region gives the parameters of the exponential disk approximation, $\Sigma_0$ and $R_d$.
Similarly, the surface density of atomic gas is traced by the 21 cm emission, with a correction for the cosmic abundance of
helium -- the detected hydrogen represents 75\% of the gas mass believed to be present, with most of the rest being
helium, in accordance with big bang nucleosynthesis.

\epubtkImage{ngc6946picture.png}{%
\begin{figure}[htbp]
  \centerline{\includegraphics[width=14.5cm]{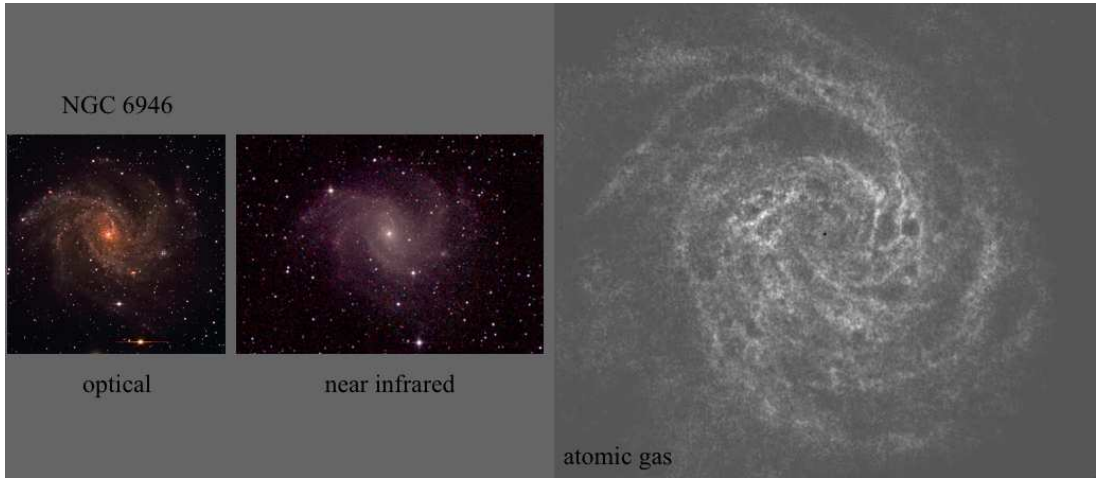}}
  \caption{The spiral galaxy NGC~6946 as it appears in the optical (color composite from the
  $BVR$ bands, left; image obtained
  by SSM with Rachel Kuzio de Naray using the Kitt Peak 2.1 m telescope), near-infrared 
  ($JHK$ bands, middle~\cite{2MASS}), and in atomic gas (21 cm radiaiton, right~\cite{FTHINGS}). 
  The images are shown at the same physical scale, illustrating how the atomic gas typically extends 
  to greater radii than the stars.  Images like these are used to construct mass models representing
  the observed distribution of baryonic mass.
  }
  \label{fig:coolpicture}
\end{figure}}

Mass models (bottom panels of Figure~\ref{fig:4panel}) are constructed from the surface density profiles by numerical 
solution of the Poisson equation~\cite{BTbook,GIPSY}.  No approximations (like sphericity or an exponential disk) are made at this step.
The disks are assumed to be thin, with radial scale length exceeding their vertical scale by 8:1, as is typical of 
edge-on disks~\cite{kregel}.  Consequently, the computed rotation curves (various broken lines in Figure~\ref{fig:4panel})
are not smooth, but reflect the observed variations in the observed surface density profiles of the various components.
The sum (in quadrature) leads to the total baryonic rotation curve $V_b(r)$ (the solid lines in Figure~\ref{fig:4panel}):
this is what would be observed if no dark matter were implicated.  Instead, the observed rotation (data points in
Figure~\ref{fig:4panel}) exceeds that predicted by $V_b(r)$:  this is the mass discrepancy.

It is often merely stated that flat rotation curves require dark matter.
But there is considerably more information in rotation curve data than asymptotic flatness.
For example, it is common that the rotation curve in the inner parts of high surface brightness galaxies like NGC~6946
is well described by the baryons alone.  The data are often consistent with a very low density of dark matter at small
radii with baryons providing the bulk of the gravitating mass.  This condition is referred to as maximum disk~\cite{vAS86},
and also runs contrary to our inferences of dark matter dominance from Figure~\ref{figure:btfresiduals}~\cite{maximumjerry}.
More generally, features in the baryonic rotation curve $V_b(r)$ often correspond to features in the total rotation $V_c(r)$.

\epubtkImage{4panel_SdVr_LR.png}{%
\begin{figure}[htbp]
  \centerline{\includegraphics[width=14.5cm]{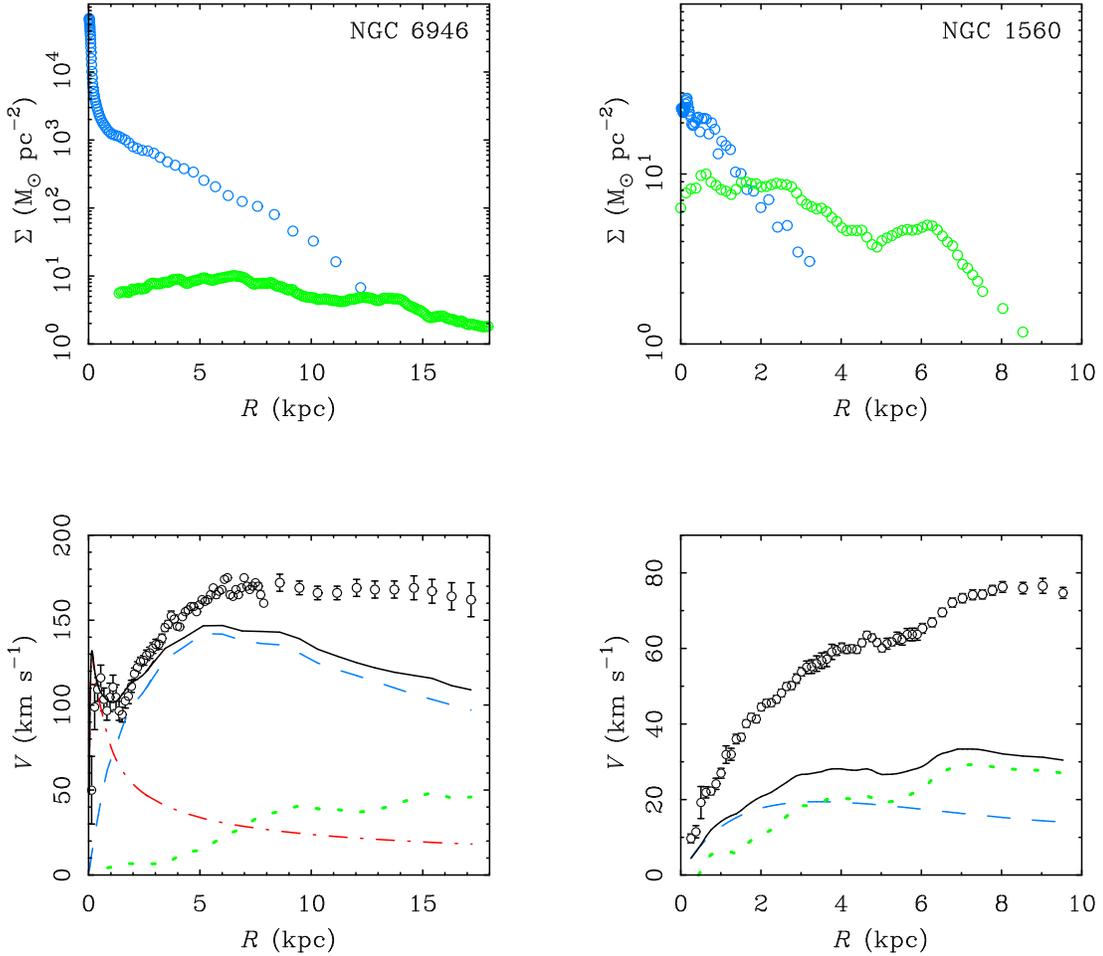}}
  \caption{Surface density profiles (top) and rotation curves (bottom) of two galaxies:  the high surface brightness
  spiral NGC~6946 (Figure~\ref{fig:coolpicture}, left) and the low surface brightness galaxy NGC 1560 (right).
  The surface density of stars (blue circles) is estimated by azimuthal averaging in ellipses fit to the $K$-band ($2.2\mu$m) light
  distribution.  Similarly, the gas surface density (green circles) is estimated by applying the same procedure to the
  21 cm image.  Note the different scale between low and high surface brightness galaxies.  Also note features like
  the central bulge of NGC~6946, which corresponds to a sharp increase in stellar surface density at small radius. In the lower panels, the observed rotation curves (data points)
are shown together with the baryonic mass models (lines) constructed
from the observed distribution of baryons.
Velocity data for NGC 6946 include both HI data that define the outer, flat portion of 
the rotation curve \cite{boomsma} and H$\alpha$ data from two independent observations 
\cite{BOAC,SINGSFP} that define the shape of the inner rotation curve.  Velocity data for 
NGC 1560 come from two independent interferometric HI observations 
\cite{BBS,gentile1560}. Baryonic mass models are constructed from the surface density 
profiles by numerical solution of the Poisson equation using GIPSY 
\cite{GIPSY}. The dashed blue line is the stellar disk, the red dot-dashed line is the central bulge, and the green dotted line is the gas.
  The solid black line is the sum of all baryonic components.  This provides a decent match to the rotation curve at small radii
  in the high surface brightness galaxy, but fails to explain the flat portion of the rotation curve at large radii.  This discrepancy,
  and its systematic ubiquity in spiral galaxies, ranks as one of the primary motivations for dark matter.  
  Note that the mass discrepancy is large at all radii in the low surface brightness galaxy.
  }
  \label{fig:4panel}
\end{figure}}

Perhaps the most succinct empirical statement of the detailed connection between baryons and dynamics has been given
by Renzo Sancisi, and known as Renzo's rule~\cite{sancisi}: 
\textit{``For any feature in the luminosity profile there is a corresponding feature in the rotation curve.''}
Both galaxies illustrated in Figure~\ref{fig:4panel} illustrate this statement.  
In the inner region of NGC~6946, the small but compact bulge component causes a sharp feature in $V_b(r)$ that declines
rapidly before the rotation curve rises again as mass from the disk begins to contribute.  The up-down-up morphology
predicted by the observed distribution of the baryons is observed in high resolution observations~\cite{BOAC,SINGSFP}.
A dark matter halo with a monotonically varying density profile cannot produce such a morphology; the stellar bulge must be
the dominant mass component at small radii in this galaxy.   

A surprising aspect of Renzo's rule is that it applies to low surface brightness galaxies as well as those of high surface brightness.
That the baryons should have some dynamical impact where their surface density is highest is natural, though there is no
reason to demand that they become competitive with dark matter.  What is distinctly unnatural is for the baryons to have a
perceptible impact where dark matter must clearly dominate.  NGC 1560 provides an example where they appear to do just that.
The gas distribution in this galaxy shows a substantial kink in its surface density profile~\cite{BBS} (recently confirmed by 
\cite{gentile1560}) that has a distinct impact on $V_b(r)$.  This occurs at a radius where $V \gg V_b$, so dark matter should
be dominant.  A spherical dark matter halo with particles on randomly oriented, highly radial orbits cannot support the same sort of structure as seen in the gas disk, and the spherical geometry, unlike a disk geometry, would smear the effect on the local acceleration. And yet the wiggle in the baryonic rotation curve is reflected in the total, as per Renzo's rule\epubtkFootnote{Note that such wiggles are often associated with spiral arm features (the existence of which in LSB galaxies being itself challenging in the presence of a massive dark matter halo, see Sect.~4.2), and hence associated with non-circular motions. It is conceivable that such observed wiggles are partly due to these, but the effect of local density contrasts due to spiral arms on the tangential velocity should be damped by the global effect of the spherical dark matter halo, which is apparently not the case.}.

One inference that might be made from these observations is that the dark matter is baryonic.  This is unacceptable from
a cosmological perspective, but it is possible to have a multiplicity of dark matter components.  That is, we could have
baryonic dark matter in the disks of galaxies in addition to a halo of non-baryonic cold dark matter.  It is often possible to
scale up the atomic gas component to fit the total rotation~\cite{hoekstraHI}.  That implies a component of mass that is traced
by the atomic gas -- presumably some other dynamically cold gas component -- that outweighs the observed hydrogen
by a factor of six to ten~\cite{hoekstraHI}.  One hypothesis for such a component is very cold molecular gas~\cite{Pfen}.
It is difficult to exclude such a possibility, though it also appears to be hard to sustain in LSB galaxies\cite{spaans}.  Dynamically, one might expect the extra mass to destabilize the LSB disk.  One also returns to a fine-tuning between baryonic surface density and mass-to-light ratio.  In order to maintain the balance observed in Figure~\ref{figure:btfresiduals}, relatively
more dark molecular gas will be required in lower surface brightness galaxies so as to maintain a constant surface
density of gravitating mass, but given the interactions at hand, this might be at least a bit more promising than explaining it with CDM halos.

As a matter of fact, low surface brightness galaxies play a critical role in testing many of the existing models for dark matter.  This happens in part
because they were appreciated as an important population of galaxies only after many relevant hypotheses were established,
and thus provide good tests of their \textit{a priori} expectations.  Observationally, we infer that low surface brightness disks
exhibit large mass discrepancies down to small radii~\cite{dBM97}.  Conventionally, this means that dark matter completely
dominates their dynamics:  the surface density of baryons in these systems is never high enough to be relevant.
Nevertheless, the observed distribution of baryons suffices to predict the total rotation~\cite{MdB98,dBM98}.
Once again, the baryonic tail wags the dark matter dog, with the observations of the minority baryonic component sufficing to predict the distribution of the dominant dark matter.  Note that, reversely, \textit{nothing} is ``observable'' about the dark matter, in present-day simulations, that predicts the distribution of baryons.

We thus see that there are many observations, mostly on galaxy scales, that are unpredicted, and perhaps unpredictable, in the standard dark matter context. They mostly involve a unique relationship between the distribution of baryons and the gravitational field, as well as an acceleration constant $a_0$ of the order of the square-root of the cosmological constant, and they represent the most significant challenges to the current $\Lambda$CDM model.

\newpage

\section{Milgrom's Empirical Law and ``Kepler Laws'' of Galactic Dynamics}
\label{sec:Kepler}

Up to this point in this review, the challenges that we have presented have been \textit{purely} based on observations, and fully independent of any alternative theoretical framework. However, at this point, it would obviously be a step forward if at least some of these puzzling observations could be summarized and empirically unified in some way, as such a unifying process is largely what physics is concerned with, rather than simply exposing a jigsaw of apparently unrelated empirical observations. And such an empirical unification \textit{is} actually feasible for many of the unpredicted observations presented in the previous section, and goes back to a rather old idea of the Israeli physicist Mordehai Milgrom.

Almost 30 years ago, back in 1983 (and thus before most of the aforementioned observations had been carried out), simply prompted by the question of whether the missing mass problem could perhaps reflect a breakdown of Newtonian dynamics in galaxies, Milgrom~\cite{original} devised a formula linking the Newtonian gravitational acceleration $g_N$ to the true gravitational acceleration $g$ in galaxies. Such attempts to rectify the mass discrepancy by gravitational means often begin by noting that galaxies are much larger than the Solar system. It is easy to imagine that at some suitably large scale, let's say of the order of 1~kpc, there is a transition from
the usual dynamics applicable in the comparatively tiny Solar system
to some more general theory that applies on the scale of galaxies
in order to explain the mass discrepancy problem.  If so, we would expect
the mass discrepancy to manifest itself at a particular length scale
in all systems. However, as already noted hereabove, there is \textit{no} universal length scale apparent in the data (Figure~\ref{figure:MDRA})
\cite{Bobmda, mdaconf, mdascarpa, MdB98, MDA}.  The mass discrepancy appears already at small
radii in some galaxies; in others there is no apparent need for dark matter
until very large radii.  This now observationally excludes all hypotheses that simply alter the force law at a linear length-scale.

\subsection{Milgrom's law and the dielectric analogy}

Before such precise data were available, Milgrom~\cite{original} already noted that other scales were also possible, and that one that is as unique to galaxies as size is acceleration. The typical centripetal acceleration of a star in a galaxy is of order $\sim 10^{-10} \textrm{m}\textrm{s}^{-2}$.  This is eleven orders of magnitude less than the surface gravity of the Earth. 
As we have seen in the previous section, this acceleration constant appears ``miraculously'' in very different scaling relations that should in principle not be related with each other\footnote{Note that many of these relations were scrutinized during the last 30 years because they were pointed to by Milgrom's law. This law thus already achieved an important role of a theoretical idea, i.e. to point an direct observations and their arrangement}. This observational evidence for the universal appearance of $a_0 \simeq 10^{-10} \textrm{m}\textrm{s}^{-2}$ in galactic scaling relations was not at all observationally evident back in 1983. What Milgrom~\cite{original} then hypothesized was a modification of Newtonian dynamics below this acceleration constant $a_0$, appropriate to the tiny accelerations encountered in galaxies\epubtkFootnote{Of course, there is also a natural length scale associated with this acceleration constant, $l = c^2/a_0$, but this length scale will enter the modification nonlinearly, and is thus \textit{not} the length at which the modification would be seen in galaxies, as it is rather of the order of the Hubble radius}. This new constant $a_0$ would then play a similar role as the Planck constant $h$ in quantum physics or the speed of light $c$ in special relativity. For large acceleration (or force per unit mass), $F/m=g \gg a_0$, everything would be normal and Newtonian, i.e., $g=g_N$. Or, put differently, formally taking $a_0 \rightarrow 0$ should make the theory tend to standard physics, just like recovering classical mechanics for $h \rightarrow 0$. On the other hand, formally taking $a_0 \rightarrow \infty$ (and $G \rightarrow 0$), or equivalently, in the limit of small accelerations $g \ll a_0$, the modification would apply in the form:
\begin{equation}
g = \sqrt{g_N a_0},
\label{equation:asymplimit}
\end{equation}
where $g = |\mathbf{g}|$ is the true gravitational acceleration, and $g_N = |\mathbf{g}_N|$ the Newtonian one as calculated from the observed distribution of visible matter. Note that this limit follows naturally from the scale-invariance symmetry of the equations of motion under transformations $(t,r) \rightarrow (\lambda t, \lambda r)$~\cite{Milinvar}. This particular modification was only suggested in 1983 by the asymptotic flatness of rotation curves and the slope of the Tully--Fisher relation. It is indeed trivial to see that the desired behavior follows from equation~(\ref{equation:asymplimit}). For a test particle in circular motion around a point mass $M$, equilibrium between the radial component of the force and the centripetal acceleration yields $V_c^2/r = g_N = GM/r^2$.  In the weak-acceleration limit 
this becomes 
\begin{equation}
\frac{V_c^2}{r} = \sqrt{\frac{G M a_0}{r^2}}.
\label{equation:pointmass}
\end{equation}
The terms involving the radius $r$ cancel, simplifying to
\begin{equation}
V_c^4(r) = V_f^4 = a_0 G M.
\label{equation:TF}
\end{equation}
The circular velocity no longer depends on radius, asymptoting to
a constant $V_f$ that depends only on the mass of the central object and fundamental constants. The equation above is the equivalent of the observed baryonic Tully--Fisher relation. It is often wrongly stated that Milgrom's formula was constructed in an \textit{ad hoc} way in order to reproduce galaxy rotation curves, while this statement is only true of these two observations: (i) the asymptotic flatness of the rotation curves, and (ii) the slope of the baryonic Tully--Fisher relation (but note that, at the time, it was not clear at all that this slope would hold, nor that the Tully--Fisher relation would correlate with baryonic mass rather than luminosity, and even less clear that it would hold over orders of magnitude in mass). All the other successes of Milgrom's formula related to the phenomenology of galaxy rotation curves were pure \textit{predictions} of the formula made \textit{before} the observational evidence. The predictions that are encapsulated in this simple formula can be thought of as sort of ``Kepler-like laws'' of galactic dynamics. These various laws only make sense once they are unified within their parent formula, exactly as Kepler's laws only make sense once they are unified under Newton's law.

In order to ensure a smooth transition between the two regimes $g \gg a_0$ and $g \ll a_0$, Milgrom's law is written in the following way:
\begin{equation}
\mu\left(\frac{g}{a_0}\right){\bf g}={\bf g_N}, 
\label{moti}
\end{equation}
where the interpolating function
\begin{equation}
\mu(x) \rightarrow 1 \; {\rm for} \; x \gg 1 \; {\rm and} \; \mu(x) \rightarrow x \; {\rm for} \; x \ll 1.
\label{muasympt}
\end{equation}
Written like this, the analogy between Milgrom's law and Coulomb's law in a dielectric medium is clear, as noted in \cite{Blanchet2}. Indeed, inside a dielectric medium, the amplitude of the electric field $E$ generated by an external point charge $Q$ located at a distance $r$ obeys the following equation:
\begin{equation}
\mu(E) E = \frac{Q}{4 \pi \epsilon_0 r^2},
\label{coulomb}
\end{equation}
where $\mu$ is the relative permittivity of the medium, and can depend on $E$. In the case of a gravitational field generated by a point mass $M$, it is then clear that Milgrom's interpolating function plays the role of `` gravitational permittivity''. Since it is smaller than 1, it makes the gravitational field stronger than Newtonian (rather than smaller in the case of the electric field in a dielectric medium, where $\mu >1$). In other words, the gravitational susceptibility coefficient $\chi$ (such that $\mu$=1+$\chi$) is negative, which is correct for a force law where like masses attract rather than repel \cite{Blanchet2}. This dielectric analogy has been explicitly used in devising a theory\cite{Blanchet} where Milgrom's law arises from the existence of a ``gravitationally polarizable'' medium (see Sect.~7).

Of course, inverting the above relation, Milgrom's law can also be written as
\begin{equation}
{\bf g} = \nu\left(\frac{g_N}{a_0}\right){\bf g_N}, 
\label{inversemoti}
\end{equation}
where 
\begin{equation}
\nu(y) \rightarrow 1 \; {\rm for} \; y \gg 1 \; {\rm and} \; \nu(y) \rightarrow y^{-1/2} \; {\rm for} \; y \ll 1.
\end{equation}
However, as we shall see in the next section, in order for ${\bf g}$ to remain a conservative force field, these expressions (Eqs. \ref{moti} and \ref{inversemoti}) cannot be rigorous outside of highly symmetrical situations. It nevertheless allows one to make numerous very general predictions for galactic systems, or in other words, to derive ``Kepler-like laws'' of galactic dynamics, unified under the banner of Milgrom's law. As we shall see, many of the observations unpredicted by $\Lambda$CDM on galaxy scales naturally ensue from this very simple law. However, even though Milgrom originally devised this as a modification of dynamics, this law is \textit{a priori} nothing more than an algorithm which allows one to calculate the distribution of force in an astronomical object from the observed distribution of baryonic matter. Its success would simply mean that the observed gravitational field in galaxies is mimicking a universal force law generated by the baryons alone, meaning that (i) either the force law itself is modified, or that (ii) there exists an intimate connection between the distribution of baryons and dark matter in galaxies.

It was for instance suggested~\cite{Kaplinghat} that such a relation might arise naturally in the CDM context, if halos possess a one-parameter density profile that leads to a characteristic acceleration profile that is only weakly dependent upon the mass of the halo. Then with a fixed collapse factor for the baryonic material, the transition from dominance of dark over baryonic occurs at a universal acceleration, which by numerical coincidence, is of the order of $cH_0$ and thus of $a_0$ (see also~\cite{Scott}). While, still today, it remains to be seen whether this scenario would quantitatively hold in numerical simulations, it was noted by Milgrom~\cite{rebut} that this scenario only explained the role of $a_0$ as a transition radius between baryon and dark matter dominance in high-surface brightness (HSB) galaxies, precluding \textit{altogether} the existence of low-surface brightness (LSB) galaxies where dark matter dominates everywhere. The real challenge for $\Lambda$CDM is rather to explain \textit{all} the different roles played by $a_0$ in galaxy dynamics, different roles that can all be summarized within the single law proposed by Milgrom, just like Kepler's laws are unified under Newton's law. We list these Kepler-like laws of galactic dynamics hereafter, and relate each of them with the unpredicted observations of Sect.~4, keeping in mind that these were mostly \textit{a priori} predictions of Milgrom's law, made before the data were as good as today, not ``postdictions'' like we are used to in modern cosmology.

\subsection{Galactic Kepler-like laws of motion}

\begin{itemize}
\item[{\bf 1.}] \textbf{Asymptotic flatness of rotation curves.} The rotation curves of galaxies are \textit{asymptotically} flat, even though this flatness is not always attained at the last observed point (see point hereafter about the \textit{shapes} of rotation curves as a function of baryonic surface density). What is more, Milgrom's law can be thought of as including the \textit{total} acceleration with respect to a preferred frame, which can lead to the prediction of asymptotically falling rotation curves for a galaxy embedded in a large external gravitational field (see Sect.~6.3).

\item[{\bf 2.}] \textbf{\boldmath$Ga_0$ defining the zero-point of the baryonic Tully--Fisher relation.} The plateau of a rotation curve is $V_f = (GMa_0)^{1/4}$. The true Tully--Fisher relation is predicted to be a relation between this asymptotic velocity and \textit{baryonic mass}, not luminosity. Milgrom's law yields immediately the slope (precisely 4) and zero-point of this baryonic Tully--Fisher law.  The observational baryonic Tully--Fisher relation should thus be consistent with zero scatter around this prediction of Milgrom's law (the dotted line of Figure~\ref{figure:btf}). And indeed it is. All rotationally supported systems in the weak acceleration limit should fall on this relation, irrespective of their formation mechanism and history, meaning that completely isolated galaxies or tidal dwarf galaxies formed in interaction events all behave as every other galaxy in this respect. 

\item[{\bf 3.}] \textbf{\boldmath$Ga_0$ defining the zero-point of the Faber--Jackson relation.} For quasi-isothermal systems~\cite{Milisotherm}, such as elliptical galaxies, the bulk velocity dispersion depends only on the total baryonic mass via $\sigma^4 \sim GMa_0$. Indeed, since the equation of hydrostatic equilibrium for an isotropic isothermal system in the weak field regime reads $d(\sigma^2 \rho)/dr = -\rho (G M a_0)^{1/2}/r$, one has $\sigma^4 = \alpha^{-2} \times GMa_0$ where $\alpha = d{\rm ln}\rho/d{\rm ln}r$. This underlies the Faber-Jackson relation for elliptical galaxies (Figure~\ref{figure:faberjackson}), which is however \textit{not} predicted by Milgrom's law to be as tight and precise (because it relies e.g., on isothermality and on the slope of the density distribution) as the BTFR.

\item[{\bf 4.}] \textbf{Mass discrepancy defined by the inverse of the acceleration in units of \boldmath$a_0$.} Or alternatively, defined by the inverse of the square-root of the gravitational acceleration generated by the baryons in units of $a_0$. The mass discrepancy is precisely equal to this in the very low-acceleration regime, and leads to the baryonic Tully--Fisher relation. In the low-acceleration limit, $g_N/g = g/a_0$, so in the CDM language, inside the virial radius of any system whose virial radius is in the weak acceleration regime (well below $a_0$), the baryon fraction is given by the acceleration in units of $a_0$. If we adopt a rough relation $M_{500} \simeq 1.5 \times 10^{5}\ M_{\odot} \times V_c^3 ({\rm km/s})^{-3}$, we get that the acceleration at $R_{500}$, and thus the system baryon fraction predicted by Milgrom's formula, is $M_b/M_{500} = a_{500}/a_0 \simeq 4 \times 10^{-4} \times V_c ({\rm km/s})^{-1}$. Divided by the cosmological baryon fraction, this explains the trend for $f_d= M_b/(0.17 M_{500})$ with potential ($\Phi = V_c^2$) in Figure~\ref{figure:fdV}, thereby naturally explaining the halo-by-halo missing baryon challenge in galaxies.  No baryons are actually missing; rather, we infer their existence because the natural scaling between mass and circular velocity $M_{500} \propto V_{c}^3$ in $\Lambda$CDM differs by a factor of $V_c$ from the observed scaling $M_b \propto V_c^4$.

\item[{\bf 5.}] \textbf{\boldmath$a_0$ as the characteristic acceleration at the effective radius of isothermal spheres.} As a corollary to the Faber-Jackson relation for isothermal spheres, let us note that the baryonic isothermal sphere would not require any dark matter up to the point where the internal gravity falls below $a_0$, and would thus resemble a purely baryonic Newtonian isothermal sphere up to that point. But at larger distances, in the presence of the added force due to Milgrom's law, the baryonic isothermal sphere would rather fall~\cite{Milisotherm} as $r^{-4}$, thereby making the radius at which the gravitational acceleration is $a_0$ the effective baryonic radius of the system, thereby explaining why, at this radius $R$ in quasi-isothermal systems, the typical acceleration $\sigma^2/R$ is almost always observed to be of the order of $a_0$. Of course, this is valid for systems where such a transition radius does exist, but going to very low surface brightness systems, if the internal gravity is everywhere below $a_0$, one can then have typical accelerations as low as one wishes.

\item[{\bf 6.}] \textbf{\boldmath$a_0/G$ as a critical mean surface density for stability.} Disks with mean surface density $\langle \Sigma \rangle \leq \Sigma_\dagger=a_0 /G$ have added stability. Most of the disk is then in the weak-acceleration regime, where accelerations scale as $a \propto \sqrt{M}$, instead of $a \propto M$. Thus $\delta a/a = (1/2)\delta M/M$ instead of $\delta a/a = \delta M/M$, leading to a weaker response to small mass perturbations~\cite{Milstab}. This explains the Freeman limit (Figure~\ref{fig:freeman}).

\item[{\bf 7.}] \textbf{\boldmath$a_0$ as a transition acceleration.} The mass discrepancy in galaxies always appears (transition from baryon dominance to dark matter dominance) when $V_c^2/R \sim a_0$, yielding a clear mass-discrepancy acceleration relation (Figure~\ref{figure:MDRA}). This, again, is the case for every single rotationally supported system irrespective of its formation mechanism and history. For HSB galaxies, where there exists two distinct regions where $V_c^2/R>a_0$ in the inner parts and $V_c^2/R<a_0$ in the outer parts, locally measured mass-to-light ratios should show no indication of hidden mass in the inner parts, but rise beyond the radius where $V_c^2/R \approx a_0$ (Figure~\ref{figure:MLRSd}). Note that this is the only role of $a_0$ that the scenario of~\cite{Kaplinghat} was poorly trying to address (forgetting, e.g., about the existence of LSB galaxies).

\item[{\bf 8.}] \textbf{\boldmath$a_0/G$ as a transition central surface density.} The acceleration $a_0$ defines the transition from HSB galaxies to LSB galaxies: baryons dominate in the inner parts of galaxies whose \textit{central} surface density is higher than some critical value of the order of $\Sigma_{\dagger}=a_0/G$, while in galaxies whose central surface density is much smaller (LSB galaxies), DM dominates everywhere, and the magnitude of the mass discrepancy is given by the inverse of the acceleration in units of $a_0$, see (5) below. The mass discrepancy thus appears at smaller radii and is more severe in galaxies of lower baryonic surface densities (Figure~\ref{figure:MLRSd}). The shapes of rotation curves are predicted to depend on surface density: HSB galaxies are predicted to have rotation curves that rise steeply then become flat, or even fall somewhat to the not-yet-reached asymptotic flat velocity, while LSB galaxies are supposed to have rotation curves that rise slowly to the asymptotic flat velocity. This is precisely what is observed (Figure~\ref{figure:RCshape}), and is in accordance~\cite{GentileURC} with the more complex empirical parametrization of observed rotation curves that has been proposed in~\cite{SalucciURC}. Finally the \textit{total} (baryons+DM) acceleration is predicted to decline with the mean \textit{baryonic} surface density of galaxies, exactly as observed (Figure~\ref{figure:ARSd}), in the form $a \propto \Sigma_b^{1/2}$ (see also Figure~\ref{figure:XiSd}).

\item[{\bf 9.}] \textbf{\boldmath$a_0/2 \pi G$ as the central surface density of dark halos.} Provided they are mostly in the Newtonian regime, galaxies are predicted to be embedded in dark halos (whether real or virtual, i.e., ``phantom'' dark matter) with a central surface density of the order of $a_0/(2 \pi G)$ as observed\footnote{Note that the denominator $2 \pi G$ comes from integrating the phantom dark matter density along a vertical line as per \cite{Mildark}, which leads to a slightly smaller characteristic surface density for phantom dark matter than the $\Sigma_\dagger$ defining Freeman limit in the 6th law hereabove}.  LSBs should have a halo surface density scaling as the square-root of the baryonic surface density, in a much more compressed range than for the HSB ones, explaining the consistency of observed data with a constant central surface density of dark matter~\cite{gentilenature,Mildark}.

\item[{\bf 10.}] \textbf{Features in the baryonic distribution imply features in the rotation curve.} Because a small variation in $g_N$ will be directly translated into a similar one in $g$, Renzo's rule  (Sect.~4.3.4) is explained naturally.

\end{itemize}

As a conclusion, \textit{all} the apparently independent roles that the characteristic acceleration $a_0$ plays in the unpredicted observations of Sect.~4.3 (see end of Sect.~4.3.3 for a summary), as well as Renzo's rule (Sect.~4.3.4), have been elegantly unified by the single law proposed by Milgrom~\cite{original} in 1983 as a unique scaling relation between the gravitational field generated by observed baryons and the total observed gravitational force in galaxies.

\epubtkImage{MLRSdlog_pop.png}{%
\begin{figure}[htbp]
  \centerline{\includegraphics[width=14.5cm]{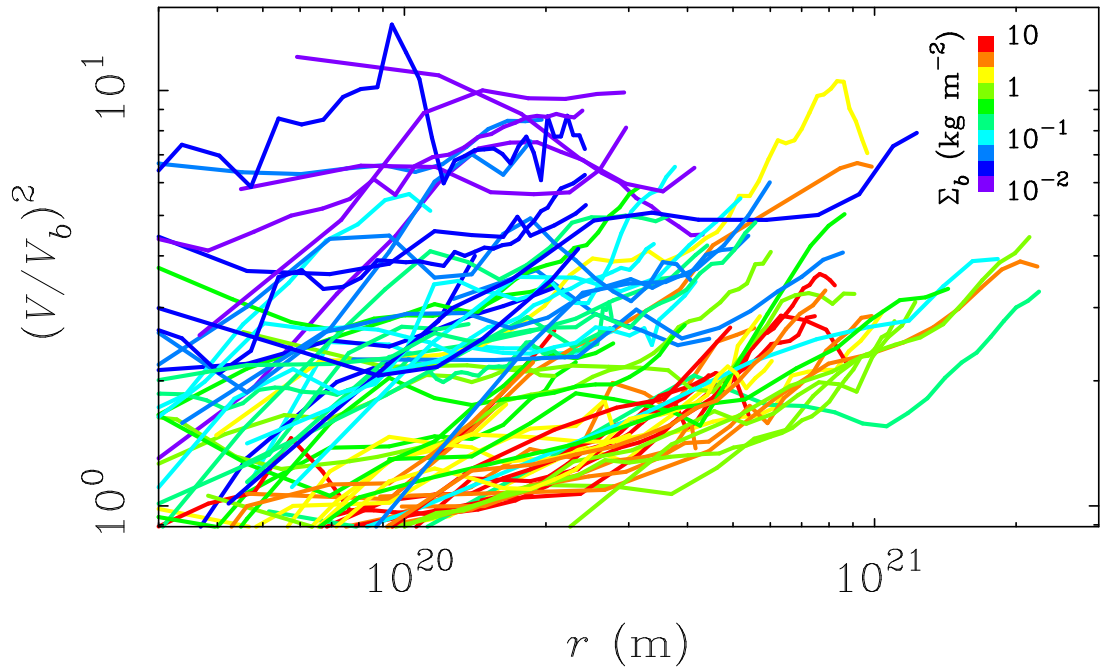}}
  \caption{The mass discrepancy (as in Figure~\ref{figure:MDRA})
  as a function of radius in observed spiral galaxies.  The curves for individual
  galaxies (lines) are color-coded by their characteristic baryonic surface
  density (as in Figure~\ref{figure:btfresiduals}).  In order to be completely empirical and fully
  independent of any assumption such as maximum disk, stellar masses have been estimated with population synthesis models~\cite{Bel03}.
  The amplitude of the mass discrepancy is initially small in high surface
  density galaxies, and grows only slowly at large radii.  As the baryonic
  surface densities of galaxies decline, the mass discrepancy becomes more
  severe and appears at smaller radii.  This trend confirms one of the
  \textit{a priori} predictions of Milgrom's law~\cite{Milgrom1}.
  }
  \label{figure:MLRSd}
\end{figure}}

\epubtkImage{VRSd_pop.png}{%
\begin{figure}[htbp]
  \centerline{\includegraphics[width=14.5cm]{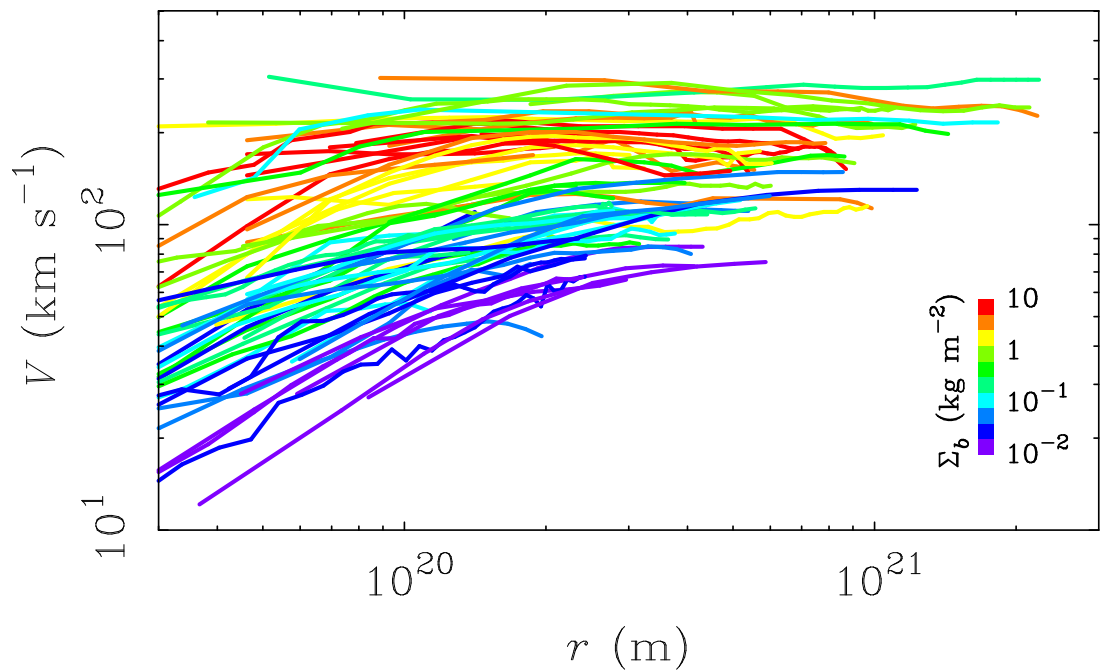}}
  \caption{The shapes of observed rotation curves depend on baryonic
  surface density (color coding as per Figure~\ref{figure:MLRSd}).
  High surface density galaxies have rotation curves that rise steeply
  then become flat, or even fall somewhat to the asymptotic flat velocity.
  Low surface density galaxies have rotation curves
  that rise slowly to the asymptotic flat velocity.
  This trend confirms one of the \textit{a priori} predictions of Milgrom's law~\cite{Milgrom1}.
   }
  \label{figure:RCshape}
\end{figure}}

\epubtkImage{ARSd_pop.png}{%
\begin{figure}[htbp]
  \centerline{\includegraphics[width=14.5cm]{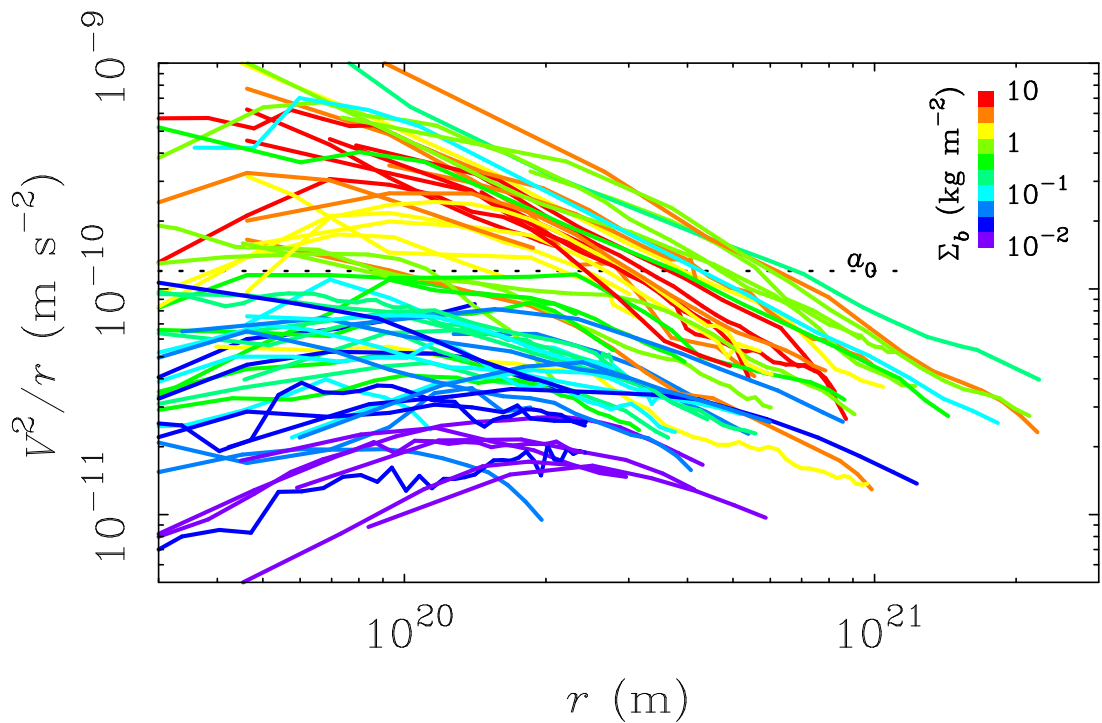}}
  \caption{Centripetal acceleration as a function of radius and surface
  density (color coding as per Figure~\ref{figure:MLRSd}).
  The critical acceleration $a_0$ is denoted by the dotted line.
  Milgrom's formula predicts that acceleration should decline with 
  baryonic surface density, as observed.  Moreover, high surface
  density galaxies transition from the Newtonian regime at small
  radii to the weak-field regime at large radii, whereas low surface
  density galaxies fall entirely in the regime of low acceleration
  $a < a_0$, as anticipated by Milgrom~\cite{Milgrom1}.
  }
  \label{figure:ARSd}
\end{figure}}

\newpage

\section{Milgrom's Law as a Modification of Classical Dynamics: MOND}
\label{sec:MOND}

It thus appears that many puzzling observations, that are difficult to understand in the $\Lambda$CDM context (and/or require an extreme fine-tuning of the DM distribution), are well summarized by a single heuristic law. It would therefore appear natural that this law derives from a universal force law, and would reflect a modification of dynamics rather than the addition of massive particles interacting (almost) only gravitationally with baryonic matter\epubtkFootnote{Note that the main motivation for modifying dynamics is thus \textit{not} to get rid of DM, but to explain why the observed gravitational field in galaxies is apparently mimicking a universal force law generated by the baryons alone. The simplest explanation is of course \textit{a priori} not that DM arranges itself by chance to mimick this force law, but rather that the force law itself is modified. Note that at a fundamental level, relativistic theories of modified gravity often will have to include new fields to reproduce this force law, so that dark matter is effectively replaced by ``dark fields'' in these theories, or even by dark matter exhibiting a new interaction with baryons (one could speak of ``dark matter" if the stress-energy tensor of the new fields is numerically comparable to the density of baryons): this makes the confrontation between modified gravity and dark matter less clear than often believed. The actual confrontation is rather that between all sorts of theories embedding the phenomenology of Milgrom's law vs. theories of DM made of simple self-uninteracting billiard balls assembling themselves in galactic halos under the sole influence of unmodified gravity, theories which currently appear unable to explain the \textit{observed} phenomenology of Milgrom's law.}. However, applying blindly Eq.~\ref{moti} to a set of massive bodies directly leads to serious problems~\cite{Felten, original} such as the non-conservation of momentum. In a two-body configuration, as the implied force is not symmetric in the two masses, Newton's third law (action and reaction principle) does not hold, so the momentum is not conserved. Consider a translationally invariant isolated system of such two masses $m_1$ and $m_2$ small enough to be in the very weak acceleration limit, and placed at rest on the $x$-axis. The amplitude of the Newtonian force is then $F_N=Gm_1m_2/(x_2-x_1)^2$, and applying blindly Eq.~\ref{moti}, would lead to individual accelerations $|{\bf a}_i| = \sqrt{F_N a_0/m_i}$. This then immediately leads to 
\begin{equation}
\dot{p}=\sqrt{a_0 F_N} (\sqrt{m_1}-\sqrt{m_2}) \neq 0 \; {\rm if} \; m_1\neq m_2 ,
\end{equation}
meaning that for different masses, the momentum of this isolated system is not conserved. This thus means that Eq.~\ref{moti} cannot truly represent a universal force law. If Eq.~\ref{moti} is to be more than just a heuristic law summarizing how dark matter is arranged in galaxies with respect to baryonic matter, it must then be an approximation (valid only in highly symmetric configurations) of a more general force law deriving from an action and a variational principle. Such theories at the classical level can be classified under the acronym MOND, for \textit{Modified Newtonian Dynamics}\epubtkFootnote{Generally covariant theories approaching these classical theories in the weak-field limit will then also be classified under this same MOND acronym, even if they really are \textit{Modified Einsteinian Dynamics} (see Sect.~7)}. In this section, we sketch how to devise such theories at the classical level, and list detailed tests of these theories at all astrophysical scales.

\subsection{Modified inertia or modified gravity: Non-relativistic actions}

If one wants to modify dynamics in order to reproduce Milgrom's heuristic law while still benefiting from usual conservation laws such as the conservation of momentum, one can start from the action at the classical level. Clearly such theories are only toy-models until they become the weak-field limit of a relativistic theory (see Sect.~7), but they are useful both as targets for such relativistic theories, and as internally consistent models allowing to make predictions at the classical level (i.e., neither in the relativistic or quantum regime).

A set of particles of mass $m_i$ moving in a gravitational field generated by the matter density distribution $\rho=\sum_i m_i \delta({\bf x} - {\bf x}_i)$ and described by the Newtonian potential $\Phi_N$ has the following action\footnote{The Newtonian mass density also satisfies the continuity equation $\partial \rho /\partial t + \nabla . (\rho {\bf v})=0$}:
\begin{equation}
S_N=S_{\rm kin}+S_{\rm in}+S_{\rm grav}=\int \frac{\rho {\bf v}^2}{2} d^3x \, dt \, - \int \rho \Phi_N d^3x \, dt - \int \frac{|\nabla \Phi_N|^2}{8 \pi G} d^3x \, dt.
\label{Newtonac}
\end{equation}
Varying this action with respect to configuration space coordinates yields the equations of motion ${\rm d}^2\textbf{x}/{\rm d}t^2 = -\nabla \Phi_N$, while varying it with respect to the potential leads to Poisson equation $\nabla^2 \Phi_N = 4 \pi G \rho$. Modifying the first (kinetic) term is generally referred to as ``modified inertia'' and  modifying the last term as ``modified gravity''\epubtkFootnote{In General Relativity, the first two terms $\int \rho ({\bf v}^2/2 -  \Phi_N) d^3x \, dt$ are lumped together into the matter action (also containing the rest mass contribution in GR), and the last term is generalized by the Einstein--Hilbert action}. 

\subsubsection{Modified inertia}

The first possibility, modified inertia, has been investigated by Milgrom~\cite{Mil94,MIrecent}, who constructed modified kinetic actions\epubtkFootnote{Let us note in passing that it would not be the first time that the kinetic action would be modified as special relativity does just this too, changing for a single particle $mv^2/2 \rightarrow -mc^2 \gamma^{-1}$ in $S_{\rm kin}$ (where $\gamma(v)=1/\sqrt{1-(v/c)^2}$), leading for a moving body to a redefinition of the effective mass as $m_{\rm eff} = m \gamma(v)$. With this analogy in mind, a rather simplified view of the Lorentz-breaking modification of inertia needed in order to reproduce MOND would be that $m_{\rm eff} \simeq m \mu(a)$, where $a$ is the amplitude of the acceleration with respect to an absolute preferred inertial frame.} (the first term $S_{\rm kin}$ in Eq.~\ref{Newtonac}) that are functionals depending on the trajectory of the particle as well as on the acceleration constant $a_0$. By construction, the gravitational potential is then still determined from the Newtonian Poisson equation, but the particle equation of motion becomes, instead of Newton's second law:
\begin{equation} 
\textbf{A}[\{{\bf x}(t)\}, a_0]=-\nabla \Phi_N,
\label{modin}
\end{equation}
where $\textbf{A}$ is a functional of the whole trajectory $\{{\bf x}(t)\}$, with the dimensions of acceleration. The Newtonian and MOND limits correspond to $[a_0 \rightarrow 0, \textbf{A} \rightarrow {\rm d}^2\textbf{x}/{\rm d}t^2]$ and $[a_0 \rightarrow \infty, \textbf{A}[\{{\bf x}(t)\}, a_0] \rightarrow a_0^{-1}\textbf{Q}(\{{\bf x}(t)\})]$ where $\textbf{Q}$  has dimensions of acceleration squared. 

Milgrom~\cite{Mil94} investigated theories of this vein and rigorously showed that they always had to be time-nonlocal (see also Sect.~7.10) to be Galilean invariant\epubtkFootnote{Such non-local theories, which also have to be nonlinear (like any MOND theory) are not easy to construct, and there is presently no real fully-fledged theory which has been developed in this vein, although hints in this direction are summarized in Sect.~7.10.}. Interestingly, he also showed that quantities such as energy and momentum had to be redefined but were then enjoying conservation laws: this even leads to a generalized virial relation for bound trajectories, and in turn to an important and robust prediction for circular orbits in an axisymmetric potential, shared by all such theories. Eq.~\ref{modin} becomes for such trajectories:
\begin{equation}
\mu\left(\frac{V_c^2}{R a_0}\right)\frac{V_c^2}{R}=-\frac{\partial\Phi_N}{\partial R}, 
\end{equation}
where, $V_c$ and $R$ are the orbital speed and radius, and $\mu(x)$ is
universal for each theory, and is derived from the expression of the
action specialized to circular trajectories. Thus, for circular trajectories, these theories recover \textit{exactly} the heuristic Milgrom's law. Interestingly, it is this law which is used to fit galaxy rotation curves, while in the modified gravity framework of MOND (see hereafter), one should actually calculate the exact predictions of the modified Poisson formulations which can differ a little bit from Milgrom's law. However, for orbits other than circular, it becomes very difficult to make predictions in modified inertia, as the time non-locality can make the anomalous acceleration at any location depend on properties of the whole orbit. For instance, if the accelerations are small on some segments of a trajectory, MOND effects can be felt also on segments where the accelerations are high, and conversely~\cite{MIrecent}. This can thus give rise to different effects on bound and unbound orbits, as well as on circular and highly elliptic orbits, meaning that ``predictions'' of modified inertia in pressure-supported systems could differ significantly from those derived from Milgrom's law \textit{per se}. Let us finally note that testing modfied inertia on Earth would need to properly define an inertial reference frame, contrary to what has been done in \cite{Abramovici,Gundlach} where the laboratory itself was not an inertial frame. Proper set-ups for testing modified inertia on Earth have been described, e.g. in \cite{Ignatiev1,Ignatiev2}: under the circumstances described in these papers, modified inertia would inevitably predict a departure from Newtonian dynamics, even if the exact departure cannot be predicted at present, except for circular motion.

\subsubsection{Bekenstein--Milgrom MOND}

The idea of modified gravity is to preserve the particle equation of motion by preserving the kinetic action, but to change the gravitational action, and thus modify the Poisson equation. In that case, all the usual conservation laws will be preserved by construction.

A very general way to do so is to write~\cite{BM84}:
\begin{equation}
S_{{\rm grav} \, {\rm BM}} \equiv - \int \frac{a_0^2 F(|\nabla \Phi|^2/a_0^2)}{8 \pi G} d^3x \, dt,
\label{aqualaction}
\end{equation}
where $F$ can be any dimensionless function. The Lagrangian being non-quadratic in $|\nabla \Phi|$, this has been dubbed by Bekenstein \& Milgrom~\cite{BM84} \textit{Aquadratic Lagrangian} theory (AQUAL). Varying the action with respect to $\Phi$ then leads to a non-linear generalization of the Newtonian Poisson equation\epubtkFootnote{Following the dielectric analogy (Sect.~5.1), this is akin to Maxwell's first equation, Gauss' law, in terms of free charge density $\rho_f$, i.e., $\nabla . [\mu \epsilon_0 {\bf E}] = \rho_f$, where ${\bf E}$  is the electric field and $\mu \epsilon_0 {\bf E} = {\bf D}$ is the electric displacement field. See \cite{Blanchet2} for a thorough discussion of the analogy.}:
\begin{equation}
\nabla . \left[\mu \left( \frac{|\nabla \Phi|}{a_0}  \right) \nabla \Phi \right] = 4 \pi G \rho
\label{BM}
\end{equation}
where $\mu(x) = F'(z)$ and $z=x^2$. In order to recover the $\mu$-function behavior of Milgrom's law (Eq.~\ref{moti}), i.e., $\mu(x) \rightarrow 1$ for $x \gg 1$ and  $\mu(x) \rightarrow x$ for $x \ll 1$, one needs to choose:
\begin{equation}
F(z) \rightarrow z \; {\rm for} \; z \gg 1 \; {\rm and} \; F(z) \rightarrow \frac{2}{3}z^{3/2} \; {\rm for} \; z \ll 1.
\end{equation}
The general solution of the boundary value problem for Eq.~\ref{BM} leads to the following relation between the acceleration $\textbf{g}=-\nabla \Phi$ and the Newtonian one, $\textbf{g}_N=- \nabla \Phi_{N}$
\begin{equation}
\mu\left(\frac{g}{a_{0}}\right)\textbf{g} = \textbf{g}_N +\textbf{S},
\label{curl}
\end{equation}
where $g=|\textbf{g}|$, and $\textbf{S}$ is a solenoidal vector field with no net flow across any closed surface (i.e., a curl field $ \textbf{S}=\nabla \times  \textbf{A}$ such that $\nabla . \textbf{S} = 0$). It is thus equivalent to Milgrom's law (Eq.~\ref{moti}) up to a curl field correction, and is precisely equal to Milgrom's law in highly symmetric one-dimensional systems, such as spherically symmetric systems or flattened systems for which the isopotentials are locally spherically symmetric. For instance, the Kuzmin disk \cite{BTbook} is an example of a flattened axisymmetric configuration for which Milgrom's law is precisely valid, as its Newtonian potential $\Phi_N=-GM/\sqrt{R^2+(b+|z|)^2}$ is equivalent on both sides of the disk to that of a point mass above or below the disk respectively.

In vacuum and at very large distances from a body of mass $M$, the isopotentials always tend to become spherical and the curl field tends to zero, while the gravitational acceleration falls well below $a_0$ (a regime known as the ``deep-MOND'' regime), so that:
\begin{equation}
\Phi(r) \sim \sqrt{GMa_0} \, {\rm ln}(r). 
\label{potentielMondien}
\end{equation} 

An important point, demonstrated by Bekenstein \& Milgrom~\cite{BM84}, is that a system with a low center-of-mass acceleration with respect to a larger (more massive) system, sees the motion of its constituents combine to give a MOND motion for the center-of-mass even if it is made of constituents whose internal accelerations are above $a_0$ (for instance a compact globular cluster moving in the outer Galaxy). The center-of-mass acceleration is independent of the internal structure of the system (if the mass of the system is small), namely the Weak Equivalence Principle is satisfied.

In a modified gravity theory, any time-independent system must still satisfy the virial theorem:
\begin{equation}
2 K + W = 0.
\end{equation}
where $K=M\langle v^2 \rangle/2$ is the total kinetic energy of the system, $M=\sum_i m_i$ being the total mass of the system, $\langle v^2 \rangle$ the second moment of the velocity distribution, and $W= - \int \rho {\bf x} \, . \nabla \Phi d^3x$ is the ``virial'', proportional to the total potential energy. Milgrom~\cite{virial1,virial2} showed that, in Bekenstein--Milgrom MOND, the virial is given by:
\begin{equation}
W = -\frac{2}{3}\sqrt{GM^3a_0} - \frac{1}{4 \pi G} \int \left[ \frac{3}{2} a_0^2 F(|\nabla \Phi|^2/a_0^2) - \mu(|\nabla \Phi|/a_0)|\nabla \Phi|^2 \right] d^3x .
\end{equation}
For a system entirely in the extremely weak field limit (the ``deep-MOND'' limit $x=g/a_0 \ll 1$) where  $\mu(x) = x$ and $F(z) = (2/3) z^{3/2}$, the second term vanishes and we thus get $W=(-2/3)\sqrt{GM^3a_0}$ (see~\cite{virial1} for the specific conditions for this to be valid). In this case, we can get an analytic expression for the two-body force under the approximation that the two bodies are very far apart compared to their internal sizes~\cite{virial1,virial3,zhaokepler}. Since the kinetic energy $K=K_{\rm orb}+K_{\rm int}$ can be separated into the orbital energy $K_{\rm orb}=m_1m_2 v^2_{\mathrm{rel}}/(2M)$ 
and the internal energy of the bodies 
$K_{\rm int}=\sum (1/3) \sqrt{Gm_i^3a_0}$ , we get from the scalar virial theorem of a stationary system:
\begin{equation}
\frac{m_1m_2 v^2_{\mathrm{rel}}}{M} = \frac{2}{3} \left[ \sqrt{GM^3a_0} - \sum_i \sqrt{Gm_i^3a_0} \right].
\end{equation}
We can then assume an approximately circular velocity such that the two-body force (satisfying the action and reaction principle) can be written analytically in the deep-MOND limit as :
\begin{equation}
F_{\rm 2body} 
= \frac{m_1m_2}{m_1+m_2} \frac{v^2_{\mathrm{rel}}}{r}
= \frac{2}{3} \left[ (m_1+m_2)^{3/2} - m_1^{3/2} - m_2^{3/2} \right] \frac{\sqrt{Ga_0}}{r}.
\label{2body}
\end{equation}

The latter equation is not valid for N-body configurations, for which the Bekenstein--Milgrom (BM) modified Poisson equation (Eq.~\ref{BM}) must be solved numerically (apart from highly symmetric N-body configurations). This equation is a non-linear elliptic partial differential equation. It can be solved numerically using various methods~\cite{Bienayme,BradaM,Ciotticode,Feix,Llinares,Tiret}. One of them~\cite{BradaM, Tiret} is to use a multigrid algorithm to solve the discrete form of Eq.~\ref{BM} (see also Figure~\ref{fig:discr}):
\begin{eqnarray}
\label{eq:monddiscret}
&\ &4\pi G \rho_{i,j,k}= \\ \nonumber
&\ &[(\Phi_{i+1,j,k}-\Phi_{i,j,k})\mu_{M_1}-(\Phi_{i,j,k}-\Phi_{i-1,j,k})\mu_{L_1} \\ \nonumber
&\ &+(\Phi_{i,j+1,k}-\Phi_{i,j,k})\mu_{M_2}-(\Phi_{i,j,k}-\Phi_{i,j-1,k})\mu_{L_2} \\ \nonumber
&\ &+(\Phi_{i,j,k+1}-\Phi_{i,j,k})\mu_{M_3}-(\Phi_{i,j,k}-\Phi_{i,j,k-1})\mu_{L_3}] /h^2
\end{eqnarray}
where
\begin{itemize}
\item $\rho_{i,j,k}$ is the density discretized on a grid of step $h$,
\item $\Phi_{i,j,k}$ is the MOND potential discretized on the same grid of step $h$, 
\item $\mu_{M_1}$, and $\mu_{L_1}$, are the values of $\mu(x)$ at points $M_1$ and $L_1$ corresponding to $(i+1/2,j,k)$ and  $(i-1/2,j,k)$ respectively (Figure~\ref{fig:discr}).
\end{itemize}
The gradient component $(\partial /\partial x,\partial /\partial y,\partial /\partial z)$, in $\mu(x)$, are approximated in the case of $\mu_{M_l}$ by $([\Phi(B)-\Phi(A)]/ h , [\Phi(I)+\Phi(H)-\Phi(K)-\Phi(J)]/(4h) , [\Phi(C)+\Phi(D)-\Phi(E)-\Phi(F)] / (4h))$ (see Figure~\ref{fig:discr}).

\epubtkImage{olivier.png}{%
\begin{figure}[htbp]
  \centerline{\includegraphics[width=10.5cm]{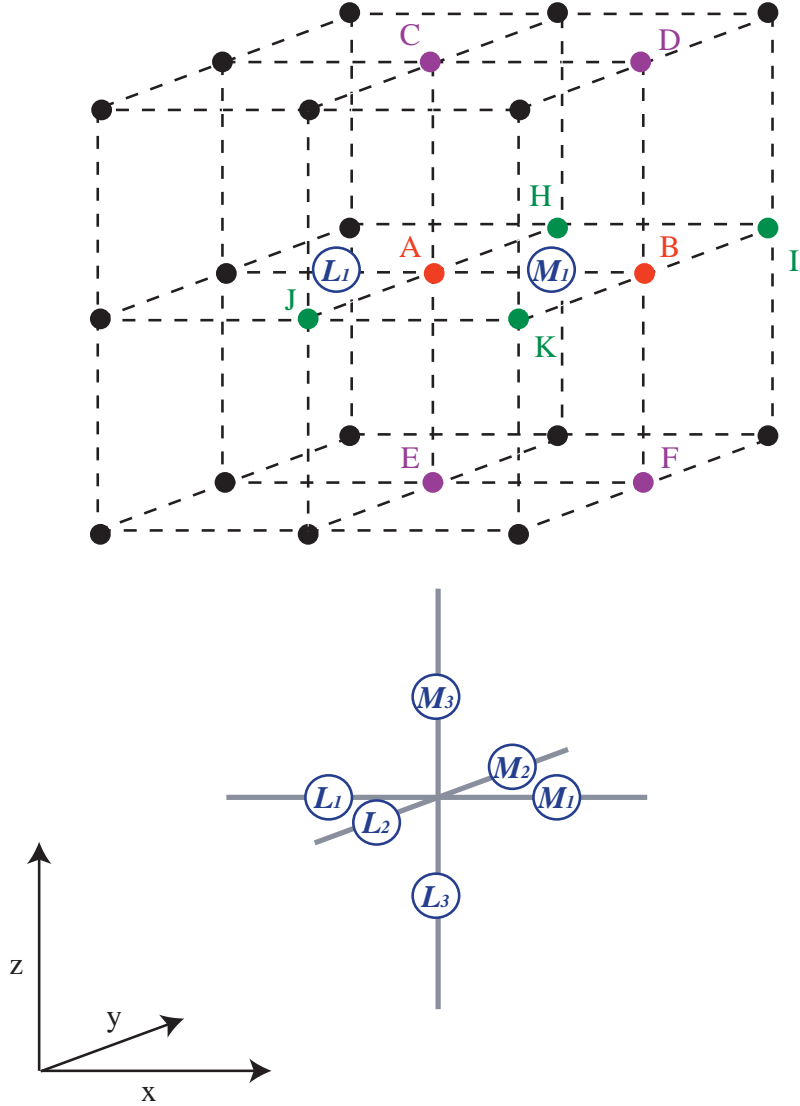}}
  \caption{Discretisation scheme of the BM modified Poisson equation (Eq.~\ref{BM}) and of the phantom dark matter derivation in QUMOND. The node $(i,j,k)$ corresponds to $A$ on the upper panel. The gradient components in $\mu(x)$ (for Eq.~\ref{eq:monddiscret}) and $\nu(y)$ (for Eq.~\ref{numpdm}) are estimated at the $L_i$ and $M_i$ points. (Figure courtesy of O. Tiret)}
  \label{fig:discr}
\end{figure}}

In~\cite{Tiret}, the Gauss--Seidel relaxation with red and black ordering is used to solve this discretized equation, with the boundary condition for the Dirichlet problem given by Eq.~\ref{potentielMondien} at large radii. It is obvious that subsequently devising an evolving N-body code for this theory can only be done using particle-mesh techniques rather than the gridless multipole expansion treecode schemes widely used in standard gravity. 

Finally, let us note that it could be imagined that MOND, given some of its observational problems (developed in Sect.~6.6), is incomplete and needs a new scale in addition to $a_0$. There are several ways to implement such an idea, but for instance, Bekenstein~\cite{bektalk} proposed in this vein a generalization of the AQUAL formalism by adding a velocity scale $s_0$, in order to allow for \textit{effective} variations of the acceleration constant as a function of the deepness of the potential, namely:
\begin{equation}
S_{{\rm grav} \, {\rm Bek}} \equiv - \frac{1}{8 \pi G} \int a_0^2 e^{-2 \Phi/s_0^2} F(|\nabla \Phi|^2 e^{2 \Phi/s_0^2}/a_0^2) d^3x \, dt,
\end{equation}
leading to
\begin{equation}
\nabla . \left[ \mu \left( \frac{|\nabla \Phi|}{a_{0{\rm eff}}} \right) \nabla \Phi \right] - \frac{|\nabla \Phi|^2}{s_0^2} \mu \left( \frac{|\nabla \Phi| }{a_{0{\rm eff}}} \right) +  \frac{a_{0{\rm eff}}^2}{s_0^2} F \left( \frac{|\nabla \Phi|^2}{a_{0{\rm eff}}^2} \right) = 4 \pi G \rho ,
\label{Bekmod}
\end{equation}
where $a_{0{\rm eff}} = a_0 e^{-\Phi/s_0^2}$. Interestingly, with this ``modified MOND'', Gauss' theorem (or Newton's second theorem) would no longer be valid in spherical symmetry. A suitable choice of $s_0$ (e.g., of the order of $10^3$ km/s, see~\cite{bektalk}) could affect the dynamics of galaxy clusters (by boosting the modification with an effectively higher value of $a_0$) compared to the previous MOND equation, while keeping the less massive systems such as galaxies typically unaffected compared to usual MOND, while other (lower) values of $s_0$ could allow (modulo a renormalization of $a_0$) for a stronger modification in galaxy clusters as well as milder modification in subgalactic systems such as globular clusters, which, as we shall see hereafter could be interesting from a phenomenological point of view (see Sect.~6.6). However, the possibility of \textit{too} strong a modification should be carefully investigated, as well as, in a relativistic (see Sect.~7) version of the theory, the consequences on the dynamics of a scalar-field with a similar action.

\subsubsection{QUMOND}

Another way~\cite{qumond} of modifying gravity in order to reproduce Milgrom's law is to still keep the ``matter action'' unchanged $S_{\rm kin}+S_{\rm in}=\int \rho ({\bf v}^2/2 -  \Phi) d^3x \, dt$, thus ensuring that varying the action of a test particle with respect to the particle degrees of freedom leads to ${\rm d}^2\textbf{x}/{\rm d}t^2 = - \nabla \Phi$, but to invoke an auxiliary acceleration field ${\bf g}_N = -\nabla \Phi_N$ in the gravitational action instead of invoking an aquadratic Lagrangian in $|\nabla \Phi|$. The addition of such an auxiliary field can of course be done without modifying Newtonian gravity, by writing the Newtonian gravitational action in the following way\epubtkFootnote{This is similar to the Palatini formalism of GR, where the present auxiliary acceleration field is replaced by a connection}:
\begin{equation}
S_{{\rm grav} \, {\rm N}} = - \frac{1}{8 \pi G} \int (2 \nabla \Phi . {\bf g}_N - {\bf g}_N^{2}) \, d^3x \, dt .
\end{equation}
It gives, after variation over ${\bf g}_N$ (or over $\Phi_N$): ${\bf g}_N=-\nabla \Phi$. And after variation of the full action over $\Phi$: $-\nabla . {\bf g}_N=4 \pi G \rho$, i.e., Newtonian gravity. One can then introduce a MONDian modification of gravity by modifying this action in the following way, replacing ${\bf g}_N^{2}$ by a non-linear function of it and assuming that it derives from an auxiliary potential ${\bf g}_N = -\nabla \Phi_N$, so that the new degree of freedom is this new potential:
\begin{equation}
S_{{\rm grav} \, {\rm QUMOND}} \equiv - \frac{1}{8 \pi G} \int  [2 \nabla \Phi . \nabla \Phi_N - a_0^2 Q(|\nabla \Phi_N|^2/a_0^2)] \, d^3x \, dt .
\label{qumondact}
\end{equation}
Varying the total action with respect to $\Phi$ yields: $\nabla^2 \Phi_N = 4 \pi G \rho$. And varying it with respect to the auxiliary (Newtonian) potential $\Phi_N$ yields:
\begin{equation}
\nabla^2 \Phi = \nabla . \left[\nu \left( \frac{|\nabla \Phi_N|}{a_0}  \right) \nabla \Phi_N \right]
\label{QUMOND}
\end{equation}
where $\nu(y) = Q'(z)$ and $z=y^2$. The theory thus requires only to solve twice the Newtonian linear Poisson equation, with only one non-linear step in calculating the rhs term of Eq.~\ref{QUMOND}. For this reason, it is called the \textit{quasi-linear formulation of MOND} (QUMOND). In order to recover the $\nu$-function behavior of Milgrom's law (Eq.~\ref{inversemoti}), i.e., $\nu(y) \rightarrow 1$ for $y \gg 1$ and  $\nu(y) \rightarrow y^{-1/2}$ for $y \ll 1$, one needs to choose:
\begin{equation}
Q(z) \rightarrow z \; {\rm for} \; z \gg 1 \; {\rm and} \; Q(z) \rightarrow \frac{4}{3}z^{3/4} \; {\rm for} \; z \ll 1.
\end{equation}
The general solution of the system of partial differential equations is equivalent to Milgrom's law (Eq.~\ref{inversemoti}) up to a curl field correction, and is precisely equal to Milgrom's law in highly symmetric one-dimensional systems. However, this curl-field correction is \textit{different} from the one of AQUAL. This means that, outside of high symmetry, AQUAL and QUMOND cannot be precisely equivalent. An illustration of this is given in~\cite{virial3}: for a system with all its mass in an elliptical shell (in the sense of a squashed homogeneous spherical shell), the effective density of matter that would source the MOND force field in Newtonian gravity is uniformly zero in the void inside the shell for QUMOND, but nonzero for AQUAL.

The concept of the effective density of matter that would source the MOND force field in Newtonian gravity is extremely useful for an intuitive comprehension of the MOND effect, and/or for interpreting MOND in the dark matter language: indeed, subtracting from this effective density the baryonic density yields what is called the ``phantom dark matter'' distribution. In AQUAL, it requires deriving the Newtonian Poisson equation after having solved for the MOND one. On the other hand, in QUMOND, knowing the Newtonian potential yields direct access to the phantom dark matter distribution even before knowing the MOND potential. After choosing a $\nu$-function, one defines 
\begin{equation}
\tilde{\nu}(y) = \nu(y) - 1, 
\label{tildenu}
\end{equation}
and one has, for the phantom dark matter density,
\begin{equation}
\rho_{\rm ph} =  \frac{\nabla . (\tilde{\nu} \nabla \Phi_N )}{4 \pi G} .
\label{eq:phantom}
\end{equation}
This $\tilde{\nu}$-function appears naturally in an alternative formulation of QUMOND where one writes the action as a function of an auxiliary potential $\Phi_{\rm ph}$:
\begin{equation}
S_{{\rm grav} \, {\rm QUMOND}} = - \frac{1}{8 \pi G} \int  [ |\nabla \Phi|^2  - |\nabla \Phi_{\rm ph}|^2- a_0^2 H(|\nabla \Phi - \nabla \Phi_{\rm ph}|^2/a_0^2) ] \, d^3x \, dt ,
\label{qumondactbis}
\end{equation}
leading to a potential $\Phi_{\rm ph}$ obeying a QUMOND equation with $\tilde{\nu}(y) = H'(y^2)$, and $\Phi = \Phi_N + \Phi_{\rm ph}$.

Numerically, for a given Newtonian potential discretized on a grid of step $h$, the discretized phantom dark matter density is given on grid points $(i,j,k)$ by (see Figure~\ref{fig:discr} and cf. Eq.~\ref{eq:monddiscret}, see also~\cite{anguscosmo}):
\begin{eqnarray}
\label{numpdm}
&\ &\rho_{{\rm ph} \, (i,j,k)}= \\ \nonumber
&\ &[(\Phi_{N \, (i+1,j,k)}-\Phi_{N \, (i,j,k)})\tilde{\nu}_{M_1}-(\Phi_{N \, (i,j,k)}-\Phi_{N \, (i-1,j,k)})\tilde{\nu}_{L_1} \\ \nonumber
&\ &+(\Phi_{N \, (i,j+1,k)}-\Phi_{N \, (i,j,k)})\tilde{\nu}_{M_2}-(\Phi_{N \, (i,j,k)}-\Phi_{N \, (i,j-1,k)})\tilde{\nu}_{L_2} \\ \nonumber
&\ &+(\Phi_{N \, (i,j,k+1)}-\Phi_{N \, (i,j,k)})\tilde{\nu}_{M_3}-(\Phi_{N \, (i,j,k)}-\Phi_{N \, (i,j,k-1)})\tilde{\nu}_{L_3}] / (4 \pi G h^2) .
\end{eqnarray}
This means that \textit{any} N-body technique (e.g., treecodes or fast multipole methods) can be adapted to QUMOND (a grid being however necessary as an intermediate step). Once the Newtonian potential (or force) is locally known, the phantom dark matter density can be computed and then represented by weighted particles, whose gravitational attraction can then be computed in any traditional manner. An example is given in Figure~\ref{miyamoto}, where one considers a rather typical baryonic galaxy model with a small bulge and a large disk. Applying Eq.~\ref{numpdm} (with the $\nu$-function of Eq.~\ref{simplenu}) then yields the phantom density~\cite{lughausen}. Interestingly, this phantom density is composed of a round ``dark halo'' and a flattish ``dark disk'' (see~\cite{Milhalo} for an extensive discussion of how such a dark disk component comes about, see also~\cite{Bienayme} and Sect.~6.5.2 for observational considerations). Let us note that this phantom dark matter density can be \textit{slightly} separated from the baryonic density distribution in non-spherical situations~\cite{knebe}, and that it can be \textit{negative}~\cite{Milnegative, XufenMW}, contrary to normal dark matter. Finding the signature of such a local negative dark matter density could be a way of exhibiting a clear signature of MOND.

\epubtkImage{miyamotonagai.png}{%
\begin{figure}[htbp]
  \centerline{(a)\parbox[t]{6.6cm}{\vspace{0pt}\includegraphics[width=6.6cm]{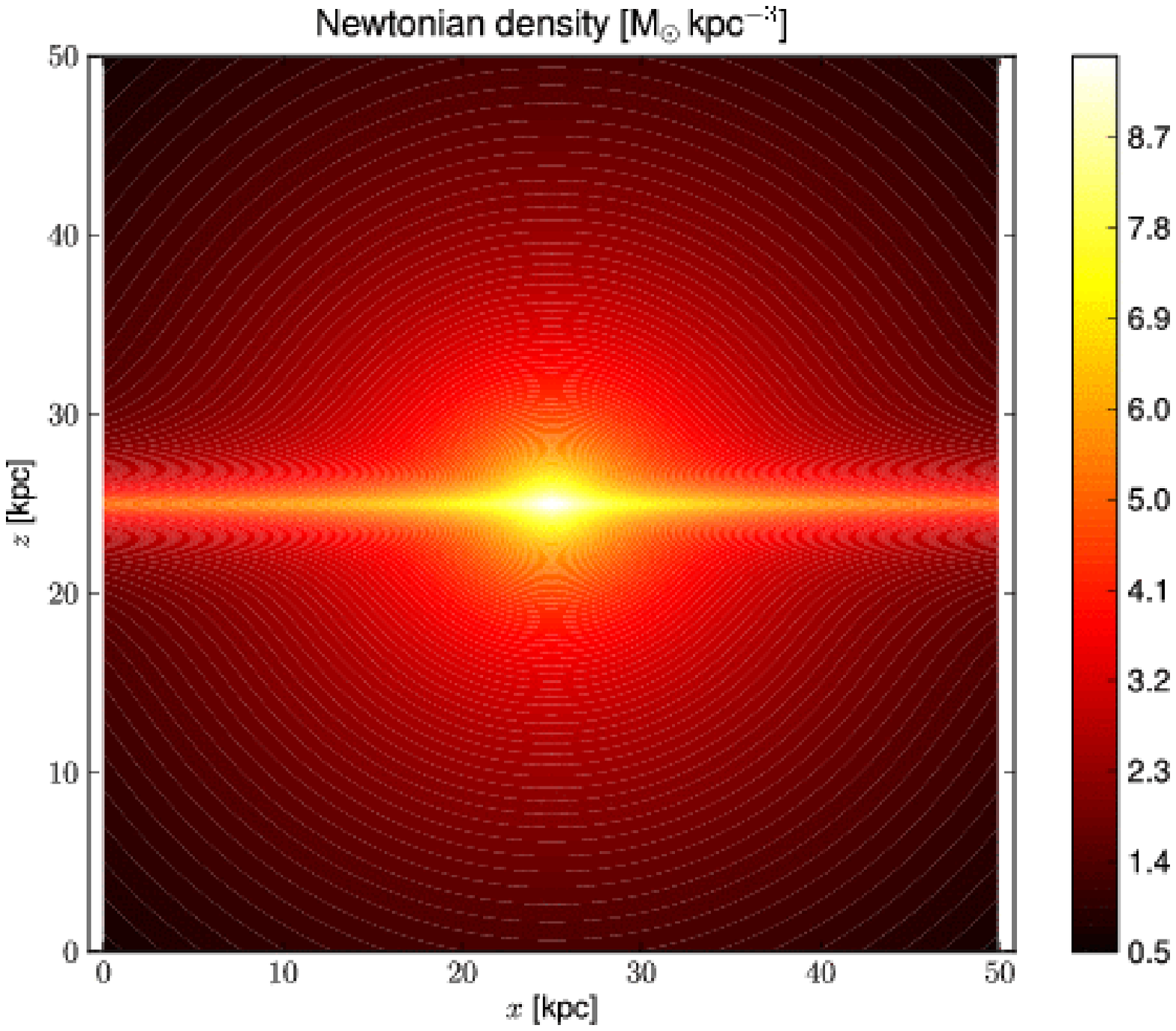}}\qquad
              (b)\parbox[t]{6.6cm}{\vspace{0pt}\includegraphics[width=6.6cm]{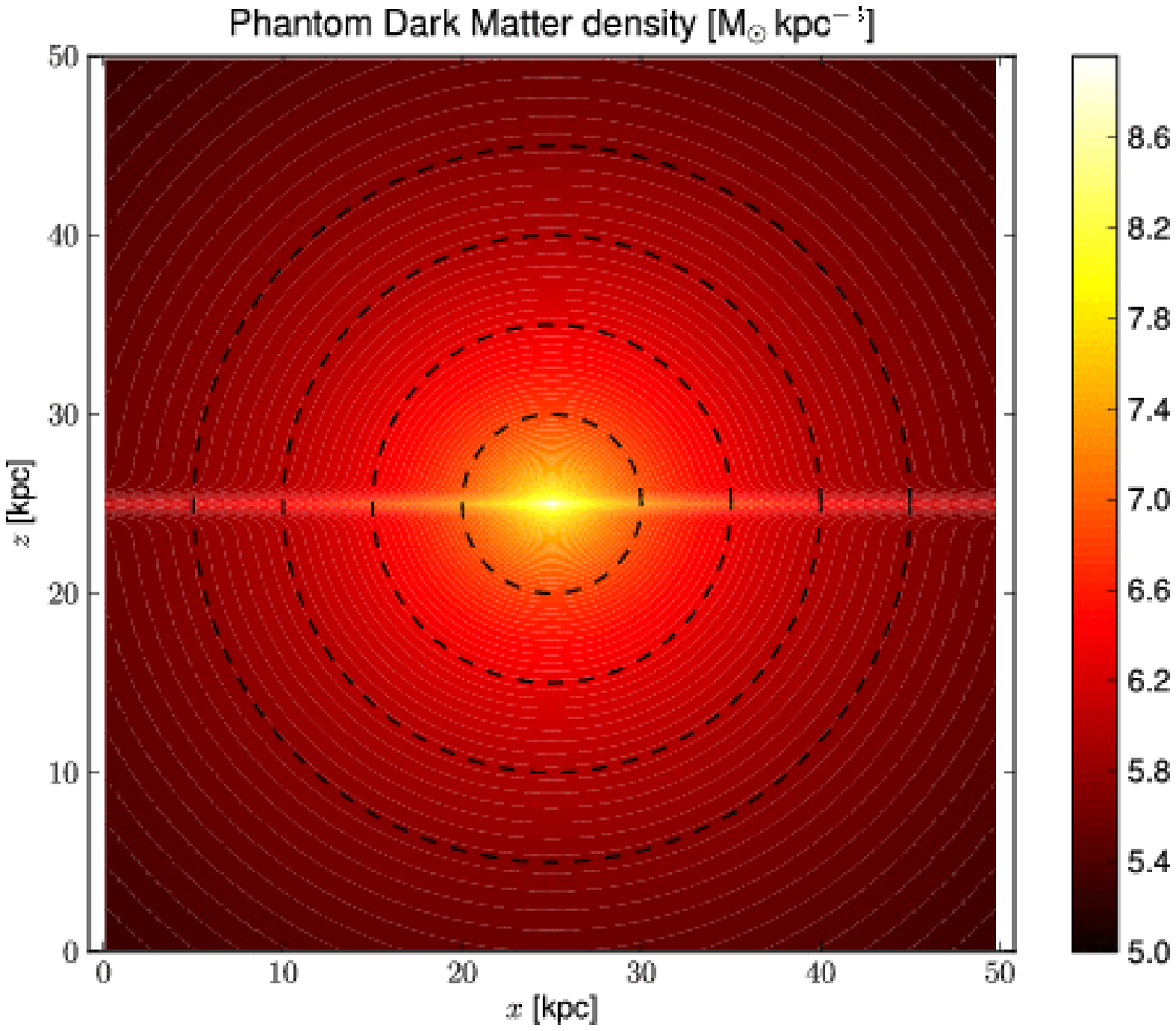}}}
  \caption{(a) Baryonic density of a model galaxy made of a small Plummer bulge with a mass of $2 \times 10^{8}\,M_{\odot}$ and Plummer radius of 185~pc, and of a Miyamoto--Nagai disk of $1.1 \times 10^{10}\,M_{\odot}$, a scale-length of 750~pc and a scale-height of 300~pc. (b) The derived phantom dark matter density distribution: it is composed of a spheroidal component similar to a dark matter halo, and of a thin disky component (Figure made by Fabian L\"ughausen~\cite{lughausen})}
  \label{miyamoto}
\end{figure}}

Finally, let us note that, as shown in~\cite{qumond, virial3}, (i) a system made of high-acceleration constituents, but with a low-acceleration center-of-mass, moves according to a low-acceleration MOND law, while (ii) the virial of a system is given by
\begin{equation}
W = -\frac{2}{3}\sqrt{GM^3a_0} - \frac{1}{4 \pi G} \int \left[ - \frac{3}{2} a_0^2 Q(|\nabla \Phi_N|^2/a_0^2) + 2 \nu(|\nabla \Phi_N|/a_0)|\nabla \Phi_N|^2 \right] d^3x ,
\end{equation}
meaning that for a system entirely in the extremely weak field limit where  $\nu(y) = y^{-1/2}$ and $Q(z) = (4/3) z^{3/4}$, the second term vanishes and we thus get $W=(-2/3)\sqrt{GM^3a_0}$, precisely like in Bekenstein--Milgrom MOND. This means that, although the curl-field correction is \textit{in general} different in AQUAL and QUMOND, the two-body force in the deep-MOND limit is the same~\cite{virial3}.

\subsection{The interpolating function}

The basis of the MOND paradigm is to reproduce Milgrom's law, Eq.~\ref{moti}, in highly symmetrical systems, with an interpolating function asymptotically obeying the conditions of Eq.~\ref{muasympt}, i.e.,  $\mu(x) \rightarrow 1$ for $x \gg 1$ and  $\mu(x) \rightarrow x$ for $x \ll 1$. Obviously, in order for the relation between $g$ and $g_N$ to be univoquely determined, another constraint is that $x\mu(x)$ must be a monotonically increasing function of $x$, or equivalently
\begin{equation}
\label{constrmu}
\mu(x) + x \mu'(x) > 0 ,
\end{equation}
or equivalently
\begin{equation}
\frac{d \, {\rm ln} \mu}{d \, {\rm ln} x} > -1 .
\end{equation}
Even though this leaves some freedom for the exact shape of the interpolating function, leading to the various families of functions hereafter, let us insist that it is already extremely surprising, from the dark matter point of view, that the MOND prescriptions for the asymptotic behavior of the interpolating function did predict all the aspects of the dynamics of galaxies listed in Sect.~5.

As we have seen in Sect.~6.1, an alternative formulation of the MOND paradigm relies on Eq.~\ref{inversemoti}, based on an interpolating function 
\begin{equation}
\nu(y)=1/\mu(x) \; {\rm where} \; y=x\mu(x).
\end{equation}
In that case, we also have that $y\nu(y)$ must be a monotonically increasing function of $y$.

Finally, as we shall see in details in Sect.~7, many MOND relativistic theories boil down to \textit{multifield} theories where the weak-field limit can be represented by a potential $\Phi = \sum_i \phi_i$, where each $\phi_i$ obeys a generalized Poisson equation, the most common case being
\begin{equation}
\label{sumpot}
\Phi = \Phi_N + \phi,
\end{equation}
where $\Phi_N$ obeys the Newtonian Poisson equation and the scalar field $\phi$ (with dimensions of a potential) plays the role of the phantom dark matter potential and obeys an equation of either the type of Eq.~\ref{BM} or of Eq.~\ref{QUMOND}. When it obeys a QUMOND type of equation (Eq.~\ref{QUMOND}), the $\nu$-function must be replaced by the $\tilde{\nu}$-function of Eq.~\ref{tildenu}. When it obeys a BM-like equation (Eq.~\ref{BM}), the classical interpolating function $\mu(x)$ acting on $x=|\nabla \Phi|/a_0$ must be replaced by another interpolating function $\tilde{\mu}(s)$ acting on $s=|\nabla \phi|/a_0$, in order for the total potential $\Phi$ to conform to Milgrom's law\epubtkFootnote{Confusing these 2 interpolating functions $\mu(x)$  and $\tilde{\mu}(s)$ can lead to serious mistakes~\cite{Wojtak}, as illustrated by~\cite{Bekrebut}}. In the absence of a renormalization of the gravitational constant, the two functions are related through~\cite{FGBZ}
\begin{equation}
\tilde{\mu}(s) = (x-s)s^{-1} \; {\rm where} \; s=x[1-\mu(x)] .
\end{equation}
For $x \ll 1$ (the deep-MOND regime), one has $s = x(1-x) \ll 1$ and $x \sim s(1+s)$, yielding $\tilde{\mu}(s) \sim s$, i.e., although it is generally different, $\tilde{\mu}$ has the same low-gravity asymptotic behavior as $\mu$.

In spherical symmetry, all these different formulations can be made equivalent by choosing equivalent interpolating functions, but the theories will typically slightly differ outside of spherical symmetry (i.e., the curl field will be slightly different). As an example, let us consider a widely used interpolating function~\cite{FB05,things,noordermeer,ZF06} yielding excellent fits in the intermediate to weak gravity regime of galaxies (but not in the strong gravity regime of the Solar system), known as the ``simple'' $\mu$-function (see Figure~\ref{fig:mu}):
\begin{equation}
\mu(x) = \frac{x}{1+x}.
\label{simplemu}
\end{equation}
This yields $y=x^2/(1+x)$, and thus $x=[y+(y^2+4y)]/2$, and $\nu=(1+x)/x$ yields the ``simple'' $\nu$-function:
\begin{equation}
\label{simplenu}
\nu(y) = \frac{1+(1+4y^{-1})^{1/2}}{2}.
\end{equation}
It also yields $s=x[1-\mu(x)]=x/(1+x)=\mu$, and hence $x=s/(1-s)$, yielding for the ``simple'' $\tilde{\mu}$-function:
\begin{equation}
\tilde{\mu}(s) = \frac{s}{1-s}.
\end{equation}
A more general family of $\tilde{\mu}$-functions is known as the $\alpha$-family~\cite{AFZ}, valid for $0 \leq \alpha \leq 1$ and including the simple function as the $\alpha=1$ case \footnote{In principle, $\alpha$ can be slightly larger, but if $\alpha \gg 1$, then in the range of gravities of interest for galaxy dynamics (between $0.1 a_0$ and a few times $a_0$) the scalar field contribution $s$ is too small to account for the MOND effect, or said in another way, the corresponding Milgrom $\mu$-function would deviate significantly from $\mu(x)=x$ (i.e., $\mu(x) > x$, so that there would be less modification to the Newtonian prediction).}:
\begin{equation}
\tilde{\mu}_{\alpha}(s) = \frac{s}{1-\alpha s}
\end{equation}
corresponding to the following family of $\mu$-functions:
\begin{equation}
\mu_{\alpha}(x) = \frac{2x}{1+ (2-\alpha) x + [(1-\alpha x)^2 + 4x]^{1/2}}
\label{alphafamily}
\end{equation}
The $\alpha=0$ case is sometimes referred to as ``Bekenstein's $\mu$-function'' (see Figure~\ref{fig:mu}) as it was used in~\cite{TeVeS}. The problem here is that all these $\mu$-functions approach $1$ quite slowly, with $\zeta \leq 1$ in their asymptotic expansion for $x \rightarrow \infty$, $\mu(x) \sim 1 - A x^{-\zeta}$. Indeed, since $s=x[1-\mu(x)]$, its asymptotic behavior is $s \sim A x^{-\zeta+1}$. So, if $\zeta >1$, $s \rightarrow 0$ for $x \rightarrow \infty$ as well as for $x \rightarrow 0$, which would imply that $x(s)=s\tilde{\mu}(s)+s$ would be a multivalued function, and that the gravity would be ill-defined. This is problematic because even for the extreme case $\zeta=1$, the anomalous acceleration does not go to zero in the strong gravity regime: there is still a constant anomalous ``Pioneer-like" acceleration $x[1-\mu(x)] \rightarrow A$, which is observationally excluded\footnote{In principle, one could make $A = \alpha^{-1}$ as small as desired in the $\alpha$-family, by not limiting $\alpha$ to the range between $0$ and $1$, but passing solar system constraints would require $\alpha>20$ which would cancel the MOND effect in the range of interest for galaxy dynamics.} from very accurate planetary ephemerides \cite{Fienga}. What is more, these $\tilde{\mu}$-functions, defined only in the domain $0<s<\alpha^{-1}$, would need very carefully chosen boundary conditions to avoid covering values of $s$ outside of the allowed domain when solving for the Poisson equation for the scalar field.

The way out to design $\tilde{\mu}$-functions corresponding to acceptable $\mu$-functions in the strong gravity regime is to proceed to a renormalization of the gravitational constant\cite{FGBZ}: this means that the bare value of $G$ in the Poisson and generalized Poisson equations ruling the bare Newtonian potential $\phi_N$ and the scalar field $\phi$ in Eq.~\ref{sumpot} is \textit{different} from the gravitational constant measured on Earth, $G_N$ (related to the true Newtonian potential $\Phi_N$). One can assume that the bare  gravitational constant $G$ is related to the measured one through 
\begin{equation}
G_N = \xi G, 
\end{equation}
meaning that $x=y+s$ where $x=\nabla \Phi/a_0$, $y=\nabla \phi_N/a_0 = \nabla \Phi_N/(\xi a_0)$, and $s \tilde{\mu}(s)=y$. We then have for Milgrom's law:
\begin{equation}
x\mu(x) = \xi (x - s) = \xi s \tilde{\mu}(s).
\end{equation}
In order to recover $\mu(x) \rightarrow 1$ for $x \rightarrow \infty$, it is straightforward to show~\cite{FGBZ} that it suffices that $\tilde{\mu}(s)  \rightarrow \tilde{\mu}_0$ for $s \rightarrow \infty$, and that $\xi = 1 + \tilde{\mu}_0^{-1}$. Then if $\zeta > 1$ in the asymptotic expansion $\mu(x) \sim 1 - x^{-\zeta}$, one has $s \sim (1 + \tilde{\mu}_0^{-1})^{-1}x^{-\zeta+1} + (1+\tilde{\mu}_0)^{-1} x$. This second linear term allows $s$ to go to infinity for large $x$ and thus $x(s)$ to be single-valued. On the other hand, for the deep-MOND regime, the renormalization of $G$ implies that $\tilde{\mu}(s) \rightarrow s/\xi$ for $s \ll 1$.

We can then use, even in multifield theories, $\mu$-functions quickly asymptoting to 1. For each of these functions, there is a one-parameter family of corresponding $\tilde{\mu}$-functions (labelled by the parameter $\tilde{\mu}(\infty)=\tilde{\mu}_0$), obtained by inserting $\mu(x)$ into $s=x[1-\xi^{-1}\mu(x)]$ and making sure that the function is increasing and thus invertible. A useful family of such $\mu$-functions asymptoting more quickly towards 1 than the $\alpha$-family is the $n$-family:
\begin{equation} 
\mu_n(x)={x\over (1+x^n)^{1/n}}.
\label{nfamily}
\end{equation}
The case $n=1$ is again the simple $\mu$-function, while the case $n=2$ has been extensively used in rotation curve analysis from the very first analyses~\cite{BBS, Kent}, to this day~\cite{SM02}, and is thus known as the ``standard'' $\mu$-function (see Figure~\ref{fig:mu}). The corresponding $\tilde{\mu}$-function for $n\geq 2$ has a very peculiar shape of the type shown in figure~3 of~\cite{Bruneton07} (which might be considered a fine-tuned shape, necessary to account for Solar System constraints). On the other hand, the corresponding $\nu$-function family is:
\begin{equation} 
\nu_n(y)=\left[{1+(1+4y^{-n})^{1/2}\over 2}   \right]^{1/n}.
\end{equation}
As the simple $\mu$-function ($\alpha=1$ or $n=1$) fits well galaxy rotation curves (see also Sect.~6.5.1) but is excluded in the Solar System (see also Sect.~6.4), it can be useful to define $\mu$-functions that have a gradual transition similar to the simple function in the low to intermediate gravity regime of galaxies, but a more rapid transition towards 1 than the simple function. Two such families are described in~\cite{rings} in terms of their $\nu$-function:
\begin{equation} 
\nu_{\beta}(y)=(1-e^{-y})^{-1/2} + \beta e^{-y}
\end{equation}
and
 \begin{equation}
\label{gammafamily}
\nu_{\gamma}(y)=(1-e^{-y^{\gamma/2}})^{-1/\gamma}+(1-\gamma^{-1})e^{-y^{\gamma/2}}.
\end{equation}
Finally, yet another family was suggested in~\cite{McGMW}, obtained by deleting the second term of the $\gamma$-family, and retaining the virtues of the $n$-family in galaxies, but approaching 1 more quickly in the Solar system:
\begin{equation}
\label{deltafamily}
\nu_{\delta} (y) = (1-e^{-y^{\delta/2}})^{-1/\delta}.
\end{equation}
To be complete, it should be noted that other $\mu$-functions considered in the literature include~\cite{unruh,zhaoneut} (see also Sect.~7.10):
\begin{equation}
\mu(x) = \frac{(1+4x^2)^{1/2}-1}{2x},
\label{unruhmu}
\end{equation}
and
\begin{equation}
\mu(x) = 1 - (1+x/3)^{-3}.
\end{equation}
This simply shows the variety of shapes that the interpolating function of MOND can in principle take\footnote{Note that, among the freedom of choice of that function, one could additionally even imagine that the $\mu$-function is not a scalar function but a ``tensor" $\mu_{ij}$ such that the modification becomes anisotropic and the modified Poisson equation becomes something like $\partial_i[\mu_{ij} g_j] = 4 \pi G \rho$}. Very precise data for rotation curves, including negligible errors on the distance and on the stellar mass-to-light ratios (or, in that case, purely gaseous galaxies) should allow to pin down its precise form, at least in the intermediate gravity regime and for ``modified inertia'' theories (Sect.~6.1.1) where Milgrom's law is exact for circular orbits. Nowadays, galaxy data still allow some, but not much, wiggle room: they tend to favor the $\alpha=n=1$ simple function~\cite{things} or some interpolation between $n=1$ and $n=2$~\cite{FB05}, while combined data of galaxies and the Solar System (see Sects.~6.4 and 6.5) rather tend to favor something like the $\gamma=\delta=1$ function of Eq.~\ref{gammafamily} and Eq.~\ref{deltafamily} (which effectively interpolates between $n=1$ and $n=2$, see Figure~\ref{fig:mu}), although slightlty higher exponents (i.e., $\gamma > 1$ or $\delta > 1$) might still be needed in the weak gravity regime in order to pass Solar system tests involving the external field from the Galaxy~\cite{BlanchetNovak}. Again, it should be stressed that the most salient aspect of MOND is however \textit{not} its precise interpolating function, but rather its successful predictions on galactic scaling relations and Kepler-like laws of galactic dynamics (Sect.~5.2), as well as its various beneficial effects on, e.g., disk stability (see Sect.~6.5), all predicted from its asymptotic form. The very concept of a pre-defined interpolating function should even in principle fully disappear once a more profound parent theory of MOND is discovered (see also, e.g.,~\cite{galileon}).

\epubtkImage{mufunctions.png}{%
\begin{figure}[htbp]
  \centerline{\includegraphics[width=10.5cm]{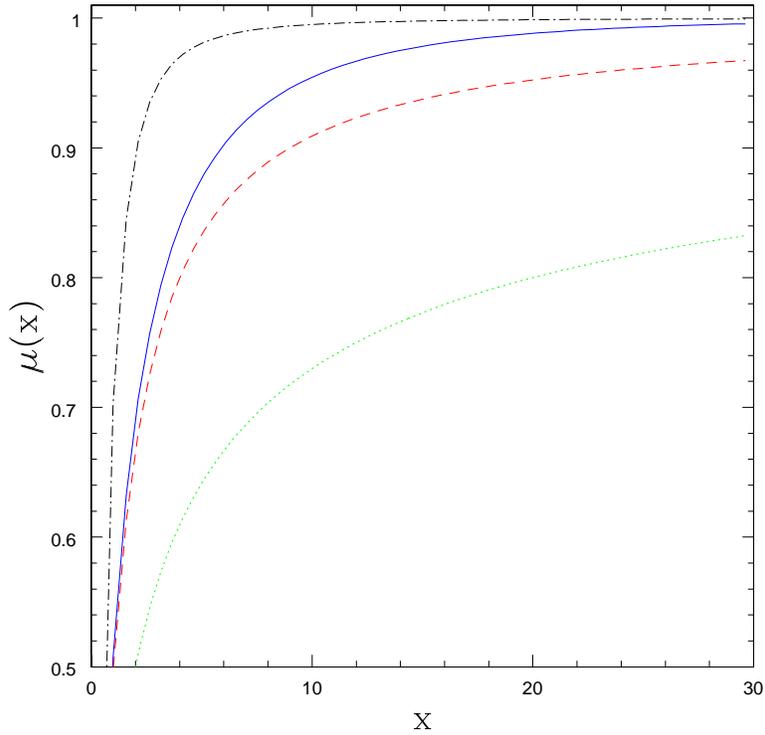}}
  \caption{Various $\mu$-functions. Dotted green line: the $\alpha=0$ ``Bekenstein'' function of Eq.~\ref{alphafamily}. Dashed red line: the $\alpha=n=1$ ``simple'' function of Eq.~\ref{alphafamily} and Eq.~\ref{nfamily}. Dot-dashed black line: the n=2 ``standard'' function of Eq.~\ref{nfamily}. Solid blue line: the $\gamma=\delta=1$ $\mu$-function corresponding to the $\nu$-function defined in Eq.~\ref{gammafamily}  and Eq.~\ref{deltafamily}. The latter function closely retains the virtues of the $n=1$ simple function in galaxies ( $x < \sim 10$ ), but approaches 1 much more quickly and connects with the $n=2$ standard function as $x \gg 10$.}
  \label{fig:mu}
\end{figure}}

To end this section on the interpolating function, let us stress that if the $\mu$-function asymptotes as $\mu(x)=x$ for $x \rightarrow 0$, then the energy of the gravitational field surrounding a massive body is infinite~\cite{BM84}. What is more, if the $\tilde{\mu}$ function of relativistic multifield theories asymptotes in the same way to zero before going to negative values for time-evolution dominated systems (see Sect.~9.1), then a singular surface exists around each galaxy, on which the scalar degree of freedom does not propagate, and can therefore not provide a consistent picture of collapsed matter embedded into a cosmological background. A simple solution\cite{FGBZ,Sanderseps} consists in assuming a modified asymptotic behavior of the $\mu$-function, namely of the form 
\begin{equation} 
\mu(x) \sim \varepsilon_0 + x  \; {\rm for} \; x \ll 1.
\end{equation} 
In that case there is a return to a Newtonian behavior (but with a very strong renormalized gravitational constant $G_N/\varepsilon_0$) at a very low acceleration scale $x \ll \varepsilon_0$, and rotation curves of galaxies are only approximately flat until the galactocentric radius 
\begin{equation} 
R \sim \frac{1}{\varepsilon_0}\sqrt{\frac{G_N M} {a_0}}.
\end{equation} 
One must thus have $\varepsilon_0 \ll 1$ to not affect the observed phenomenology in galaxies. Note that the $\mu$-function will thus never go to zero, also at the center of a system. Conversely, in QUMOND and the likes, one can modify the $\nu$-function in the same way:
\begin{equation}
\nu(y) \sim \frac{1}{\varepsilon_0+y^{1/2}}  \; {\rm for} \; y \ll 1.
\end{equation}

\subsection{The external field effect}

The above return to a rescaled Newtonian behavior at very large radii and in the central parts of isolated systems, in order to avoid theoretical problems with the interpolating function, would happen \textit{anyway}, even with the interpolating function going to zero, for any non-isolated system in the Universe (and this return to a Newtonian behavior could actually happen at much lower radii) because of a very peculiar aspect of MOND: the \textit{external field effect}, which appeared in its full significance already in the pristine formulation of MOND~\cite{original}.

In practice, no objects are truly isolated in the Universe and this has wider and
more subtle implications in MOND than in Newton--Einstein gravity. In the linear Newtonian dynamics, the internal dynamics of a subsystem (a cluster in a galaxy, or a galaxy in a galaxy cluster for instance) in the field of its mother system decouples. Namely, the internal dynamics is always the same independently of any external field (constant across the subsystem) in which the system is embedded (of course, if the external field varies across the subsystem, it manifests itself as tides). This has subsequently been built in as a fundamental principle of GR: the Strong Equivalence Principle (see also Sect.~7). But MOND \textit{has to} break this fundamental principle of GR. This is because, as it is an acceleration-based theory, what counts is the \textit{total} gravitational acceleration with respect to a pre-defined frame (e.g., the CMB frame\epubtkFootnote{It is interesting to note that different MOND theories offer (very) different answers to the generic question ``acceleration with respect to what?''. For instance, in the MOND-from-vacuum idea (see~\cite{unruh} and Sect.~7.10), the total acceleration is measured with respect to the quantum vacuum, which is well defined. In BIMOND (Sect.~7.8) it is the relative acceleration between the two metrics, which is also well defined through the difference of Christoffel symbols.}). The MOND effects are thus only observed in systems where the \textit{absolute} value of the gravity both internal, $g$, and external, $g_e$ (from a host galaxy, or astrophysical system, or large scale structure), is less than $a_0$ . If $g_e < g < a_0$ then we have standard MOND effects. However, if the hierarchy goes as $g < a_0 < g_e$, then the system is purely Newtonian\footnote{A Cavendish experiment in a freely falling satellite in Earth orbit would thus return a Newtonian result in MOND}, and if $g < g_e < a_0$ then the system is Newtonian with a renormalised gravitational constant. Ultimately, whenever $g$ falls below $g_e$ (which always happens at some point) the gravitational attraction falls again as $1/r^2$. This is most easily illustrated in a thought experiment where one considers MOND effects in one dimension. In Eq.~\ref{BM}, one has $\nabla \Phi = {\bf g} + {\bf g_e}$ and $4 \pi G \rho = \nabla . ({\bf g_{N}} + {\bf g_{Ne}})$, which in one dimension leads to the following revised Milgrom's law (Eq.~\ref{moti}) including the external field:
\begin{equation}
g \, \mu\left(\frac{g+g_e}{a_0}\right)  + g_e \, \left[ \mu\left(\frac{g+ge}{a_0} \right) - \mu \left( \frac{g_e}{a_0} \right) \right] = g_N ,
\label{motiefe}
\end{equation}
such that, when $g \rightarrow 0$, we have Newtonian gravity with a renormalized gravitational constant $G_{\rm norm}\approx G/[\mu_e (1+L_e)]$ where $\mu_e = \mu(g_e/a_0)$ and $L_e = (d \, {\rm ln} \mu/d \, {\rm ln} x)_{x=g_e/a_0}$, assuming as before that the external field only varies on a much larger scale than the internal system. Similarly, for QUMOND (Eq.~\ref{QUMOND}) in one dimension, one gets the equivalent of Eq.~\ref{inversemoti}:
\begin{equation}
g = g_N \, \nu\left(\frac{g_N+g_{Ne}}{a_0}\right) + g_{Ne} \, \left[ \nu\left(\frac{g_N+g_{Ne}}{a_0} \right) - \nu \left( \frac{g_{Ne}}{a_0} \right) \right]  
\label{inversemotiefe}
\end{equation}
When dealing in the future with very extended rotation curves whose last observed point is in the extreme weak-field limit, it could be interesting, as a first order approximation, to use the latter formulae\footnote{For instance, using the ``simple" function $\mu(x) = x/(1+x)$ in Eq.~\ref{motiefe} would lead to $g = [(g_N a_0 + g_N g_e - 2 g_e a_0 - g_e^2) + \sqrt{(2 g_e a_0 +g_e^2 -g_N a_0 - g_N g_e)+4 g_N (a_0 +g_e)^3}]/[2 (a_0+g_e)]$}, adding the external field as an additional parameter of the MOND fit to the external parts of the rotation curve. This would of course only be a first order approximation because this would neglect the three-dimensional nature of the problem and the direction of the external field.

Now, in three dimensions, the problem can be analytically solved only in the extreme case of the completely external-field dominated part of the system (where $g \ll g_e$) by considering the perturbation generated by a body of low mass $m$ inside a uniform external field, assumed along the $z$-direction, ${\bf g_e} = g_e {\bf 1_z}$. Eq.~\ref{BM} can then be linearized and solved with the boundary condition that the total field equals the external one at infinity~\cite{BM84} to yield:
\begin{equation}
\Phi(x,y,z) = - \frac{Gm}{\mu_e \tilde{r}},
\label{asympt_efe}
\end{equation}
with
\begin{equation}
\tilde{r} = r (1 + L_e (x^2 + y^2)/r^2)^{1/2},
\end{equation}
squashing the isopotentials along the external field direction. This is thus the asymptotic behavior of the gravitational field in any system embedded in a constant external field. Similarly, in QUMOND (Eq.~\ref{QUMOND}), one gets
\begin{equation}
\Phi(x,y,z) = - \frac{Gm \nu_e}{\tilde{r}},
\label{asymptqumond}
\end{equation}
with
\begin{equation}
\tilde{r} = r/[1 + (L_{Ne}/2) (x^2+y^2)/r^2 ] ,
\end{equation}
where $L_{Ne} = (d \, {\rm ln} \nu/d \, {\rm ln} y)_{y=g_{Ne}/a_0}$.

For the exact behavior of the MOND gravitational field in the regime where $g$ and $g_e$ are of the same order of magnitude, one again resorts to a numerical solver, both for the BM equation case and for the QUMOND case (see Eq~\ref{eq:monddiscret} and Eq.~\ref{numpdm}). For the BM case, one adds the three components of the external field (no longer assumed to be in the $z$-direction only) in the argument of $\mu_{M_1}$ which becomes $\{[(\Phi(B)-\Phi(A))/ h - g_{e_x}]^2 + [(\Phi(I)+\Phi(H)-\Phi(K)-\Phi(J))/(4h) - g_{e_y}]^2 +  [(\Phi(C)+\Phi(D)-\Phi(E)-\Phi(F)) / (4h) - g_{e_z}]^2 \}^{1/2}$, and similarly for the other $M_i$ and the $L_i$ points on the grid (Figure~\ref{fig:discr}). One also adds the respective component of the external field to the term estimating the force at the $M_i$ and $L_i$ points in Eq.~\ref{eq:monddiscret}. With $M_1$ for instance, one changes $(\Phi_{i+1,j,k}-\Phi_{i,j,k}) \rightarrow (\Phi_{i+1,j,k}-\Phi_{i,j,k} - h g_{e_x})$ in the first term of Eq.~\ref{eq:monddiscret}. One then solves this discretized equation with the large radius boundary condition for the Dirichlet problem given by Eq.~\ref{asympt_efe} instead of Eq.~\ref{potentielMondien}. Exactly the same is applicable to calculating the phantom dark matter component of QUMOND with Eq.~\ref{numpdm}, except that now the \textit{Newtonian} external field is added to the terms of the equation in exactly the same way.

This external field effect (EFE) is a remarkable property of MONDian theories, and because this breaks the strong equivalence principle, it allows us to derive properties of the gravitational field in which a system is embedded from its internal dynamics (and not only from tides). For instance the return to a Newtonian (Eq.~\ref{asympt_efe} or Eq.~\ref{asymptqumond}) instead of a logarithmic (Eq.~\ref{potentielMondien}) potential at large radii is what defines the escape speed in MOND. By observationally estimating the escape speed from a system (e.g., the Milky Way escape speed from our local neighbourhood, see discussion in Sect.~6.5.2), one can estimate the amplitude of the external field in which the system is embedded, and by measuring the shape of its isopotential contours at large radii, one can determine the direction of that external field, without resorting to tidal effects. It is also noticeable that the phantom dark matter has a tendency to become negative in ``conoidal" regions perpendicular to the external field direction (see figure 3 of \cite{XufenMW}): with accurate enough weak-lensing data, detecting these pockets of negative phantom densities could in principle be a smoking gun for MOND~\cite{XufenMW}, but such an effect would be extremely sensible to the detailed distribution of the baryonic matter. A final important remark about the EFE is that it prevents most possible MOND effects in Galactic disk open clusters or in wide binaries, apart from a possible rescaling of the gravitational constant. Indeed, for wide binaries located in the Solar neighbourhood, the galactic EFE (coming from the distribution of mass in our Galaxy) is about $1.5 \times a_0$. The corresponding rescaling of the gravitational constant then depends on the choice of the $\mu$-function, but could typically account for up to a 50\% increase of the effective gravitational constant. Although this is not properly speaking a MOND effect, it could still perhaps imply a systematic offset of mass for very long period binaries. However, any effect of the type claimed to be observed by \cite{binaries} would not be {\it a priori} expected in MOND due to the external field effect.

\subsection{MOND in the solar system}

The primary place to test modified gravity theories is of course the Solar System, where General Relativity has until now passed all the proposed tests. Detecting a deviation from  Einsteinian gravity in our backyard would actually be the holy grail of modified gravity theories, in the same sense as direct detection in the lab is the holy grail of the CDM paradigm. However, MOND anomalies typically manifest themselves only in the weak-gravity regime, several orders of magnitudes below the typical gravitational field exerted by the Sun on, e.g., the inner planets. But in the case of modified inertia (Sect.~6.1.1), the anomalous acceleration at any location depends on properties of the whole orbit (non-locality), so that anomalies may appear in the motion of Solar system bodies that are on highly eccentric trajectories taking them to large distances (e.g., long period comets or the Pioneer spacecraft), where accelerations are low~\cite{Milgromsun}. Such MOND effects have been proposed as a possible mechanism for generating the Pioneer anomaly~\cite{Milgromsun, Turyshev}, without affecting the motions of planets, whose orbits are fully in the high acceleration regime. On the other hand, in classical, non-relativistic modified gravity theories (Sects.~6.1.2 and 6.1.3), small effects could still be observable and would primarily probe two aspects of the theory: (i) the shape of the interpolating function (Sect.~6.2) in the regime $x \gg 1$, and (ii) the external Galactic gravitational field (Sect.~6.3) acting on the Solar system, testing the interpolating function in the regime $x \ll 1$. 

If, as a first approximation, one considers the Solar system as isolated, and the Sun as a point mass, the MOND effect in the inner Solar system appears as an anomalous acceleration field in addition to the Newtonian one. In units of $a_0$, the amplitude of the anomalous acceleration is given by $x[1-\mu(x)]$, which can be constrained from the motion of the inner planets, typically their perihelion precession and the (non)-variation of Kepler's constant~\cite{original,Sanderssun, Sereno}. These constraints typically exclude the whole $\alpha$-family of interpolating functions (Eq.~\ref{alphafamily}) that are natural for multi-field theories such as TeVeS (see Sect.~6.2 and Sect.~7) because they yield  $x[1-\mu(x)] > 1$ for $x \gg 1$ while it must be smaller than 0.04 at the orbit of Mars~\cite{Sanderssun}\footnote{See also \cite{Iorio} and constraints excluding such functions also from Lunar Laser Ranging \cite{Exirifad}, neglecting the external field effect from the Sun on the Earth-Moon system since it is 3 orders of magnitude below the internal gravity of the system.}. This of course does not mean that the $\mu$-function cannot be represented by the $\alpha$-family in the intermediate gravity regime characterizing galaxies, but it must be modified in the strong gravity regime\epubtkFootnote{This is why, although the ``simple'' $\alpha=1$ function is known to very well represent the gravitational field of spiral galaxies~\cite{FB05,things,noordermeer,ZF06}, we hereafter, in Sect.~6.5.1, rather use the the $\gamma=\delta=1$ function of Eq.~\ref{gammafamily} and Eq.~\ref{deltafamily} in order to fit spiral galaxy rotation curves.}. Another potential effect of MOND is anomalously strong tidal stresses in the vicinity of saddle points of the Newtonian potential, which might be tested with the LISA pathfinder~\cite{Mag1,Mag2,Mag3,Mag4}. The MOND bubble is typically quite big and clearly detectable, but some fine-tuned interpolating functions could still make the effect rather small and unobservable~\cite{Mag3,galianni}.

The approximation of an isolated Solar system being incorrect, it is also important to add the effect of the external field from the Galaxy. Its amplitude is typically of the order of $\sim 1.5 \times a_0$. From there, Milgrom~\cite{Milgromsun} has predicted (both analytically and numerically) a subtle anomaly in the form of a quadrupole field that may be detected in planetary and spacecraft motions (as subsequently confirmed by~\cite{BlanchetNovak, Hees}). This has been used to constrain the form of the interpolating function in the weak acceleration regime characteristic of the external field itself. Constraints have essentially been set on the $n$-family of $\mu$-functions from the perihelion precession of Saturn~\cite{BlanchetNovak2,Fienga}, namely that one must have $n > 8$ in order to fit these data\epubtkFootnote{For the $\gamma$-family  of Eq.~\ref{gammafamily} and $\delta$-family of Eq.~\ref{deltafamily}, it means that even slightly sharper transitions than $\gamma =1$ and $\delta = 1$ might still be needed.}.

It should however be noted that it is slightly incoherent to compare the \textit{classical} predictions of MOND with observational constraints obtained by a global fit of Solar System orbits using a \textit{fully relativistic} first-post-Newtonian model. Although the above constraints on classical MOND models are useful guides, proper constraints can thus only truly be set on the various \textit{relativistic} theories presented in Sect.~7, the first order constraints on these theories coming from their own post-newtonian parameters~\cite{Bonvin, Cliftonppn, Giannios05, Sagi09, Sanderssun, Tamaki08}. What is more, and makes all these tests perhaps unnecessary, it has recently been shown that it was possible to cancel \textit{any} deviation from General Relativity at small distances in most of these relativistic theories, independently of the form of the $\mu$-function~\cite{galileon}.

\subsection{MOND in rotationally supported stellar systems}

\subsubsection{Rotation curves of disk galaxies}
\label{section:disks}

The root and heart of MOND, as modified inertia or modified gravity, is Milgrom's formula (Eq.~\ref{moti}). Up to some small corrections outside of symmetrical situations, this formula yields (once $a_0$ and the form of the transition function $\mu$ are chosen) a unique prediction for the total effective gravity as a function of the gravity produced by the visible baryons. It is absolutely remarkable that this formula, devised 30 years ago, has been able to successfully predict an impressive number of galactic scaling relations (the ``Kepler-like'' laws of Sect.~5.2, backed by the modern data of Sect.~4.3) that were very unprecise and/or unobserved at the time, and which still are a puzzle to understand in the $\Lambda$CDM framework. What is more, this formula is not only predicting global scaling relations successfully, we show in this section that it also predicts the shape and amplitude of galactic rotation curves at \textit{all} radii with uncanny precision, and this for all disk galaxy Hubble types~\cite{3741,noordermeer}. Of course, the absolute exact prediction of MOND depends on the exact formulation of MOND (as modified inertia or some form or other of modified gravity), but the differences are small compared to observational error bars, and even compared with the differences between various $\mu$-functions.

\epubtkImage{RCcomp.png}{%
\begin{figure}[htbp]
  \centerline{\includegraphics[width=9cm]{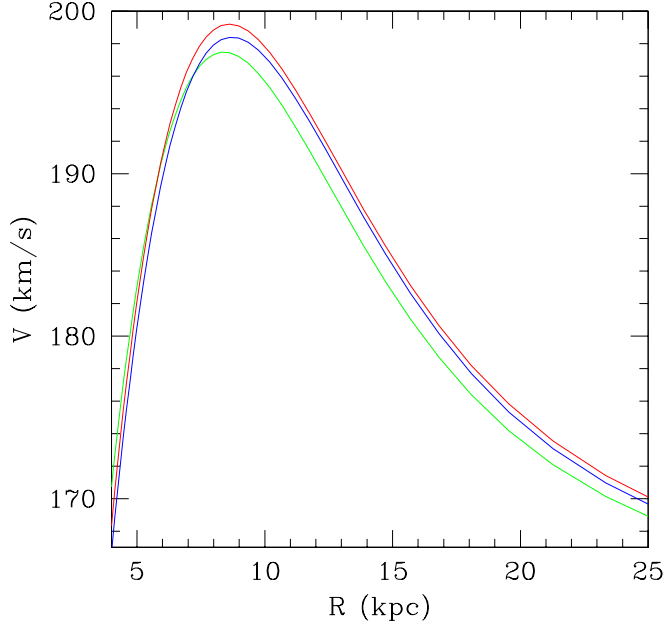}}
  \caption{Comparison of theoretical rotation curves for the inner parts (before the rotation curve flattens) of an HSB exponential disk~\cite{FGBZ}, computed with three different formulations of MOND. Green: Milgrom's formula; Blue: Bekenstein-Milgrom MOND (AQUAL); Red: TeVeS-like multi-field theory.}
  \label{fig:RCcomp}
\end{figure}}

\epubtkImage{n6946_LR_mond.png}{%
\begin{figure}[htbp]
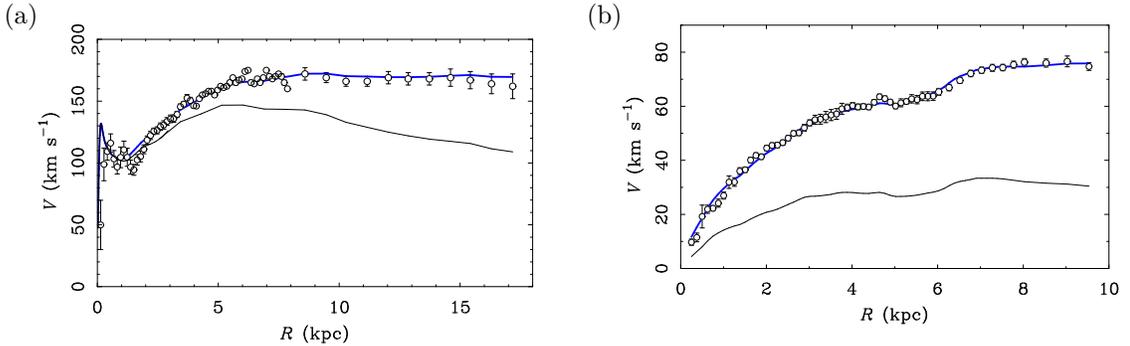

  \centerline{(a)\parbox[t]{6.6cm}{\vspace{0pt}\includegraphics[width=6.6cm]{n6946_LR_mond}}\qquad
              (b)\parbox[t]{6.6cm}{\vspace{0pt}\includegraphics[width=6.6cm]{n1560_LR_mond}}}
  \caption{Examples of detailed MOND rotation curve fits of the HSB and LSB galaxies of Figure~\ref{fig:4panel} (NGC~6946 on the left and NGC~1560 on the right).
  The black line represents the Newtonian contribution of stars and gas as determined by numerical
  solution of the Newtonian Poisson equation for the observed light distribution, as per Figure~\ref{fig:4panel}. The blue line is the MOND fit with the $\gamma=\delta=1$ function of Eq.~\ref{gammafamily} and Eq.~\ref{deltafamily}, the only free parameter being the stellar mass-to-light ratio.  In the $K$-band, the best fit value is $0.37\,M_{\odot}/L_{\odot}$
  for NGC 6946 and $0.18\,M_{\odot}/L_{\odot}$ for NGC 1560.  In practice, the best fit mass-to-light ratio can co-vary
  with the distance to the galaxy and $a_0$; here $a_0$ is held fixed ($1.2 \times 10^{-10}\mathrm{\ m\, s}^{-2}$)
  and the distance has been held fixed to the best observed value
  (5.9~Mpc for NGC~6946~\cite{Kara00} and 3.45~Mpc for
  NGC~1560~\cite{Kara03}).  Milgrom's formula provides an effective
  mapping between the rotation curve predicted by the observed baryons
  and the observed rotation, including the bumps and wiggles.}
  \label{figure:RCexamples}
\end{figure}}

\epubtkImage{m33gradient_LR.png}{%
\begin{figure}[htbp]
  \centerline{\includegraphics[width=14.5cm]{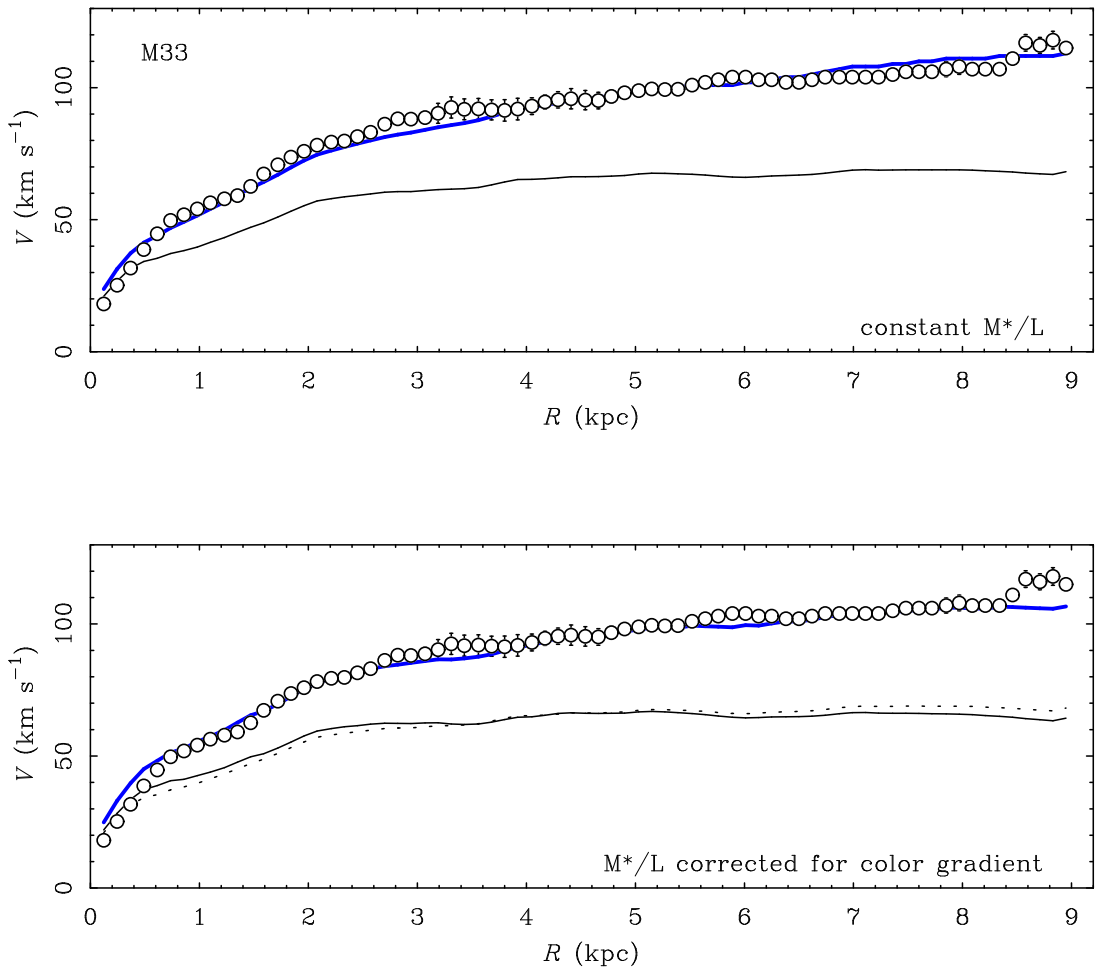}}
  \caption{The rotation curve~\cite{deulvdh} and MOND fit~\cite{sanders96} of the Local Group spiral M33 assuming a constant stellar
  mass-to-light ratio (top panel).  While the overall shape is a good match, there is a slight mismatch at $\sim$~3~kpc
  and above 7~kpc.  The observed color gradient implies a slight variation in the mass-to-light ratio, in the sense that
  the stars at small radii are slightly redder and heavier than those at large radii.  Applying stellar population models~\cite{Bel03} to the observed color gradient produces a slight adjustment of the Newtonian mass model.  The 
  dotted line in the lower panel reiterates the constant $M/L$ model from the top panel while the solid line
  has been corrected for the observed color gradient.  This slight adjustment to the baryonic mass distribution
  considerably improves the fit.  
  %Note that MOND fits assume circular motion.  One expects non-circlar motions to begin to become significant at small radii, so the modest over-fit there is not surprising.  The outermost points cannot be fit by any smooth and continuous mass distribution with purely circular motion in any theory.
   }
  \label{figure:m33gradient}
\end{figure}}

\epubtkImage{ALLresid_LR.png}{%
\begin{figure}[htbp]
  \centerline{\includegraphics[width=14.5cm]{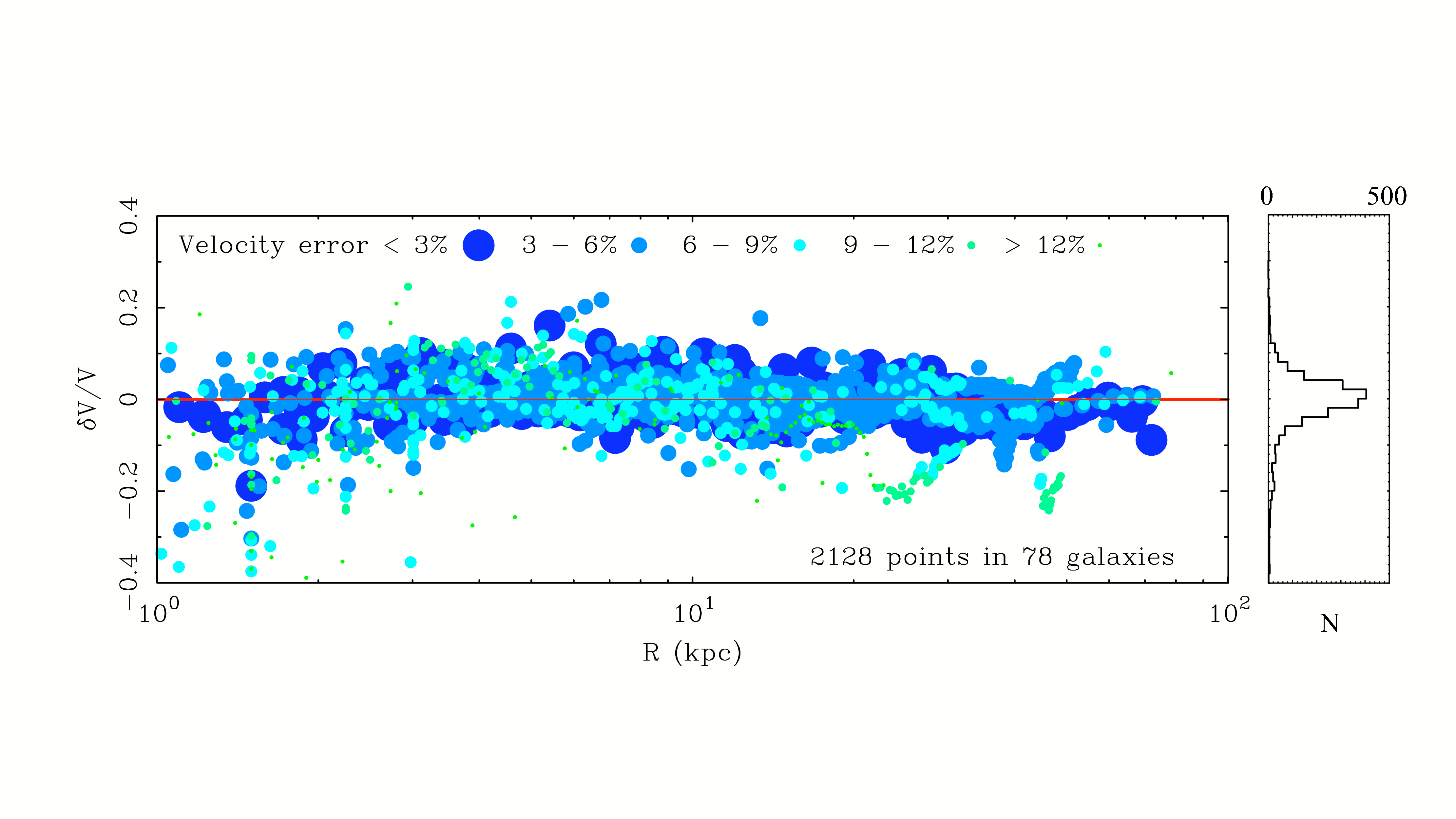}}
  \caption{Residuals of MOND fits to the rotation curves
  of 78 nearby galaxies (all data to which authors have access) including about two thousand individual resolved measurements. Data for 21 galaxies are either new or improved in terms of spatial resolution and velocity accuracy over those in \cite{SM02}.  More accurate points are illustrated with larger 
symbols. The histogram of residuals is plotted on the right panel, and is well fitted by a 
Gaussian of width $\Delta v/v \sim 0.04$. The bulk of the more accurate data are in good accord with MOND. There are a few deviant points, mostly at small radii where non-circular motions are ubiquitous and observational resolution (beam smearing) can be a challenge.  These are but a few trees outlying from a very clear forest.}
  \label{figure:RCresiduals}
\end{figure}}

In order to illustrate this, we plot in Figure~\ref{fig:RCcomp} the theoretical rotation curve of an HSB exponential disk (see~\cite{FGBZ} for exact parameters) computed with three different formulations of MOND\epubtkFootnote{Note that the rotation curves of Figure~\ref{fig:RCcomp} become flat only at larger radii than shown here, see~\cite{FGBZ}.}: Milgrom's formula (Eq.~\ref{moti}), representative of circular orbits in modified inertia, AQUAL (Eq.~\ref{BM}), and a multi-field theory (Eq.~\ref{sumpot}) representative of a whole class of relativistic theories (see Sects.~7.1 to 7.4), all with the $\alpha=n=1$ ``simple'' $\mu$-function of Eq.~\ref{alphafamily} and Eq.~\ref{nfamily}. One can see velocity differences of only a few percents in this case, while, in general, it has been shown that the maximum difference between formulations is of the order of 10\% for any type of disk~\cite{Bradaexact}. This justifies using Milgrom's formula as a proxy for MOND predictions on rotation curves, keeping in mind that, in order to constrain MOND within the modified gravity framework, one should actually calculate predictions of the various modified Poisson formulations of Sect.~6.1 for each galaxy model, and for each choice of galaxy parameters~\cite{angusrotcurves}.

\epubtkImage{2Giants.png}{%
\begin{figure}[htbp]
  \centerline{\includegraphics[width=0.9\textwidth]{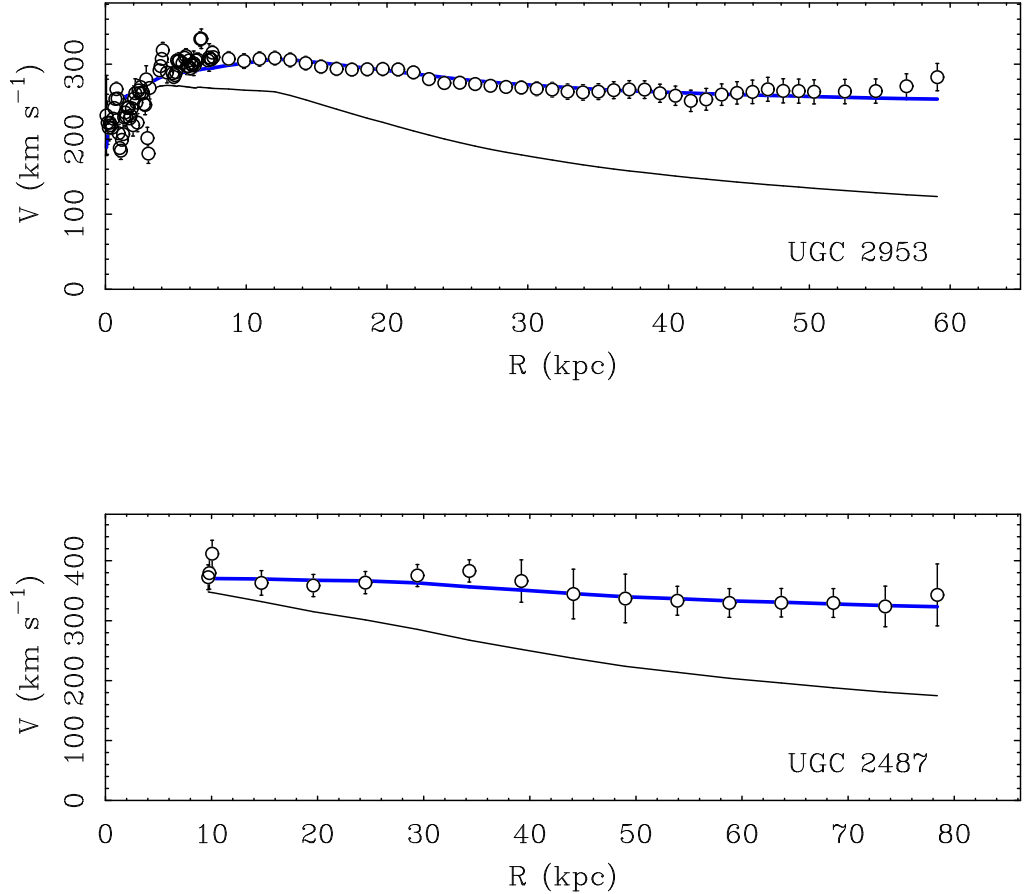}}
  \caption{Examples of MOND fits (blue lines, using Eq.~\ref{deltafamily} 
with $\delta = 1$) to two massive galaxies~\cite{noordermeer}. With baryonic masses in excess of $10^{11}\,M_{\odot}$, these are among the most massive, rapidly rotating disk galaxies known.  Stars dominate the mass, and Newtonian dynamics suffices to explain the innermost regions because of the high acceleration, but the mass discrepancy becomes apparent as the Keplerian decline (black lines) falls well below the data at the enormous radii spanned by these giant disks (the diameter of UGC~2487 spans half a million light-years).}
  \label{figure:2Giants}
\end{figure}}

\epubtkImage{2Dwarfs.png}{%
\begin{figure}[htbp]
  \centerline{\includegraphics[width=0.9\textwidth]{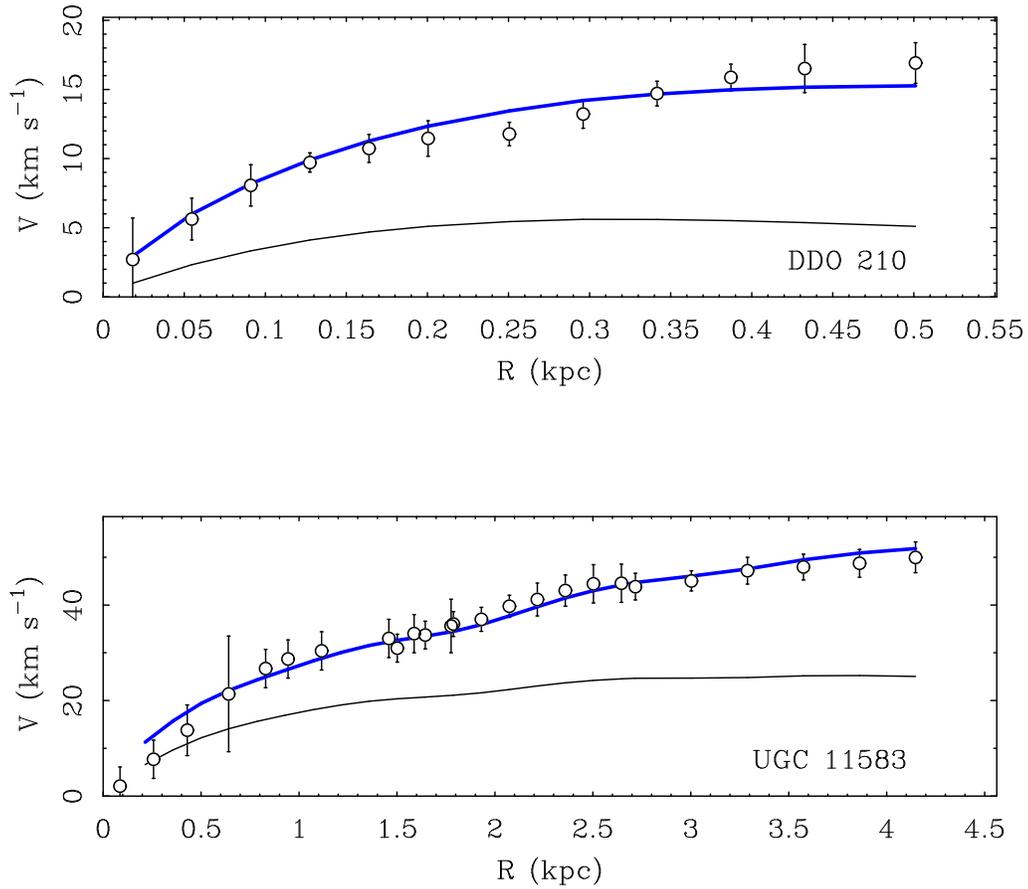}}
  \caption{Examples of MOND fits (blue lines) to two dwarf galaxies~\cite{Millowmass}. The data for DDO~210 come from~\cite{begumC04A}, and those for UGC~11583 (also known as KK98~250) are from~\cite{begumC04B} augmented with high resolution data from~\cite{MRdBlongslit,KdN06}.  The high gas content of these galaxies make them strong tests of MOND, as the one fit-parameter -- the mass-to-light ratio of the stars -- has only a minor impact on the fit. What is more, as they are deep in the MOND regime, the exact form of the interpolating function (Sect.~6.2) has also little impact on the fits, making them the cleanest tests of MOND, with essentially no wiggle room. Note that, with a mass of only a few million solar masses (comparable in mass to the largest globular clusters), the Local Group dwarf DDO~210 is the smallest galaxy known to show clear rotation ($V_f \sim 15 \,$km/s).  It is the lowest point in Figure~\ref{figure:btf}.}
  \label{figure:2Dwarfs}
\end{figure}}

\epubtkImage{RCfits_LR.png}{%
\begin{figure}[htbp]
  \centerline{\includegraphics[width=0.9\textwidth]{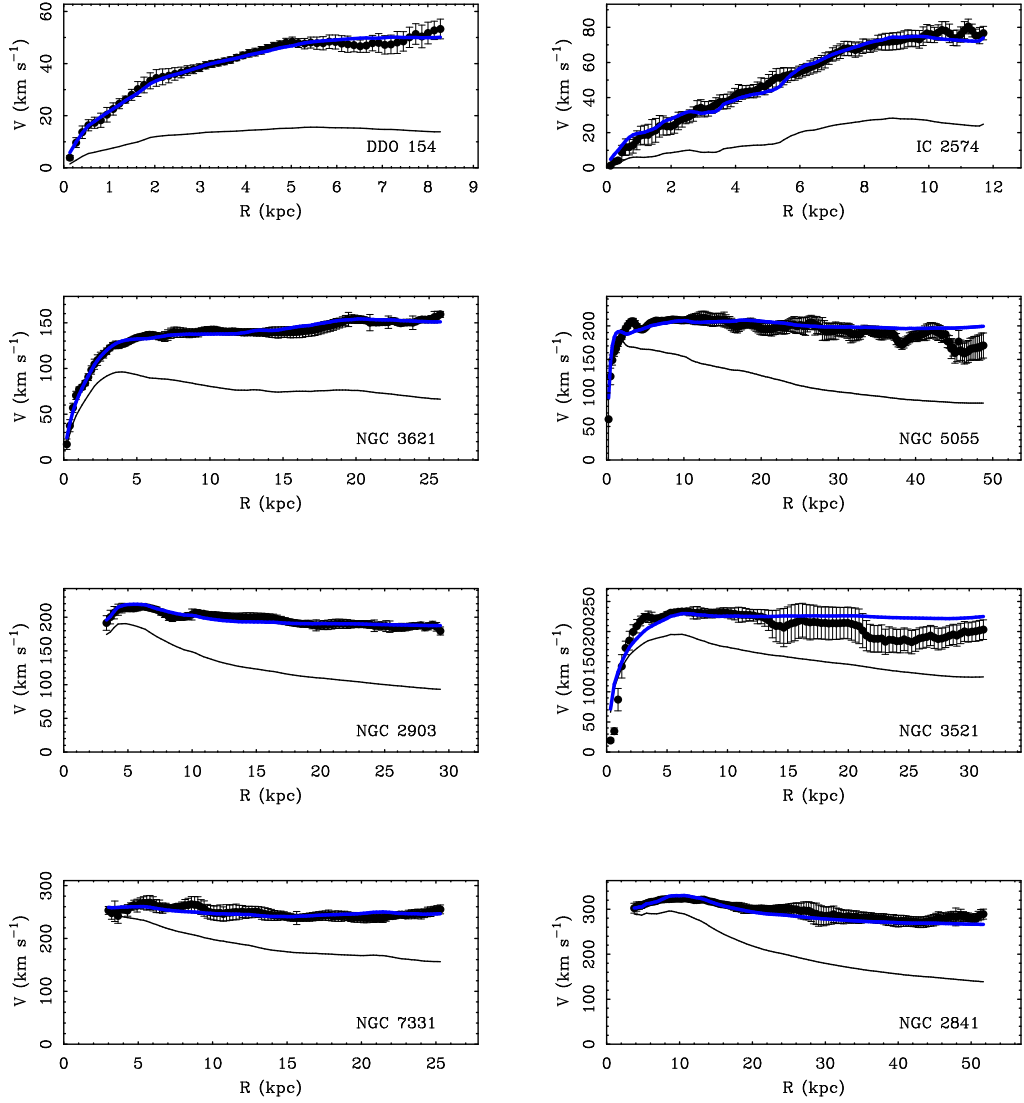}}
  \caption{MOND rotation curve fits for representative galaxies from the THINGS survey 
\cite{dBTHINGS,things,FTHINGS}.  Galaxies are chosen to illustrate a broad range of mass, from $M_b \sim 3 \times 10^8\,M_{\odot}$ to $\sim 3\times 10^{11}\,M_{\odot}$. All galaxies have high resolution interferometric 21~cm data for the gas and $3.6\mu$ photometry for mapping the stars.  The Newtonian baryonic mass model is shown as a black line and the MOND fit as a blue line (as in 
Figure~\ref{figure:RCexamples}).  The fits use the interpolating function of Eq.~\ref{deltafamily} with $\delta = 1$.}
  \label{figure:RCfits_panel}
\end{figure}}

\epubtkImage{LSBmondfits_LR.png}{%
\begin{figure}[htbp]
  \centerline{\includegraphics[width=0.9\textwidth]{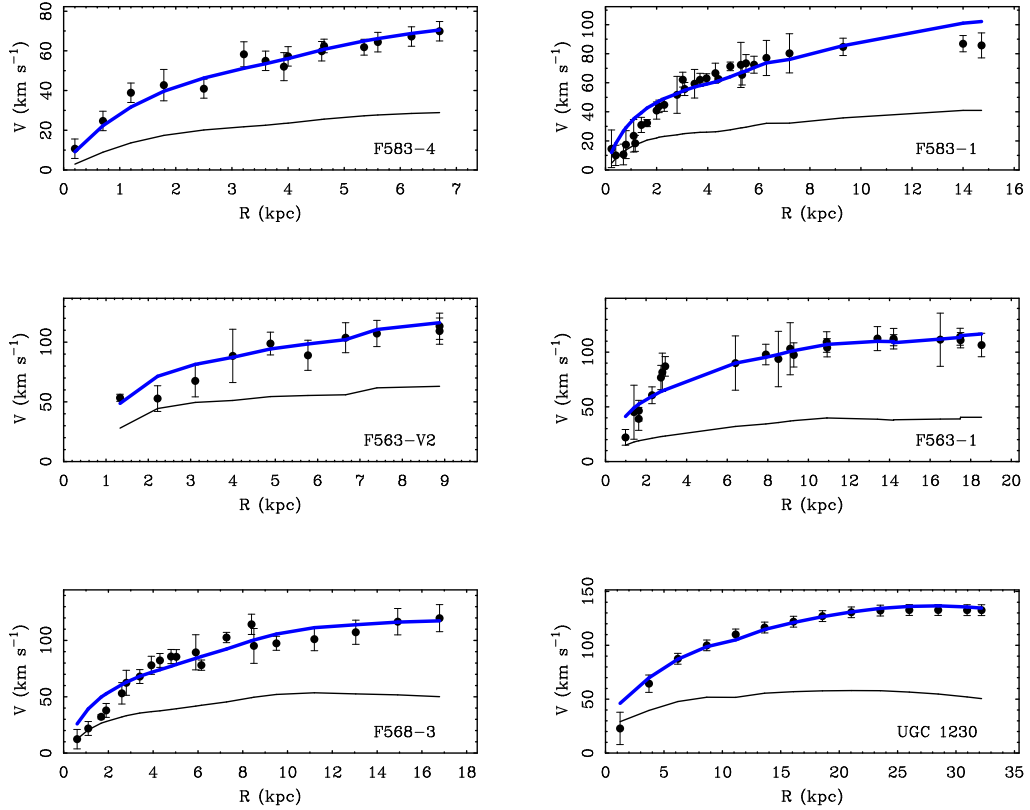}}
  \caption{MOND rotation curve fits for low surface brightness galaxies~\cite{dBM98} updated with 
high resolution H$\alpha$ data~\cite{KdN06, KdN08} and using Eq.~\ref{deltafamily} 
with $\delta = 1$. Low surface brightness galaxies are important tests of MOND because their low surface densities ($\Sigma \ll a_0/G$) place them well into the MOND regime everywhere, and the exact form of the interpolating function is rather unimportant. Their baryonic mass models fall well short of explaining the observed rotation at any but the smallest radii in Newtonian dynamics, and MOND nevertheless provides the necessary additional force everywhere (lines as per Figure~\ref{figure:RCexamples}).}
  \label{figure:LSBfits_panel}
\end{figure}}

\epubtkImage{MLcolor.png}{%
\begin{figure}[htbp]
  \centerline{\includegraphics[width=12.5cm]{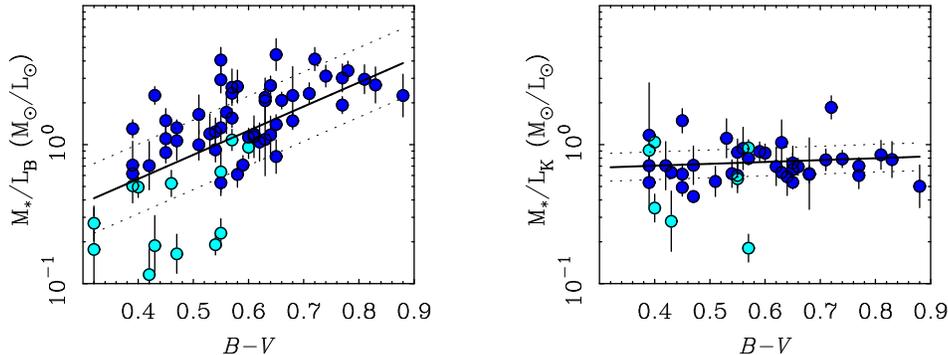}}
  \caption{A comparison of the mass-to-light ratios obtained
  from MOND rotation curve fits (points) with the independent expectations
  of stellar population synthesis models (lines)~\cite{Bel03}.  
  The mass-to-light ratio in the optical (blue $B$-band, left) and 
  near-infrared (2.2 $\mu$m $K$-band, right) are shown as a function
  of $B-V$ color (the ratio of blue to green light).  The one free parameter
  of MOND rotation curve fits reproduces the normalization, slope, and 
  scatter expected from what we know about stars.  Not all galaxies illustrated here have both $B$ and $K$-band data.  Some have neither, instead having photometry in some other bandpass (e.g., $V$ or 
$R$ or $I$).}
  \label{figure:MLcolor}
\end{figure}}

The procedure is then the following (see also Sect.~4.3.4 for more details). One usually assumes that light traces stellar mass (constant mass-to-light ratio, but see hereafter the counter-example M33), and one adds to this baryonic density the contribution of observed neutral hydrogen, scaled up to account for the contribution of primordial helium. The Newtonian gravitational force of baryons is then calculated via the Newtonian Poisson equation, and the MOND force is simply obtained via Eq.~\ref{moti} or Eq.~\ref{inversemoti}. First of all, an interpolating function must be chosen, then one can determine the value of $a_0$ by fitting, all at once, a sample of high-quality rotation curves with small distance uncertainties and no obvious non-circular motions. Then all individual rotation curve fits can be performed with the mass-to-light ratio of the disk as the single free parameter of the fit\footnote{If one assumes that a lot of dark baryons are present in the form of molecular gas, one can add another free parameter in the form of a factor multiplying the gas mass\cite{Tiretdarkbaryons}. Good MOND fits can then still be obtained but with a lower value of $a_0$.}. It turns out that using the simple interpolating function ($\alpha=n=1$, see Eqs.~\ref{alphafamily} and \ref{nfamily}) yields a value of $a_0 = 1.2 \times 10^{-10} {\rm m} {\rm s}^{-2}$, and excellent fits to galaxy rotation curves~\cite{things}. However, as already pointed out in Sects.~6.3 and 6.4, this interpolating function yields  too strong a modification in the Solar System, so we hereafter rather use the $\gamma =\delta = 1$ interpolating function of Eqs.~\ref{gammafamily} and \ref{deltafamily} (solid blue line on Figure~\ref{fig:mu}), very similar to the simple interpolating function in the intermediate to weak gravity regime.

Figure~\ref{figure:RCexamples} shows two examples of detailed MOND fits to rotation curves of Figure~\ref{fig:4panel}. The black line represents the Newtonian contribution of stars and gas and the blue line is the MOND fit, the only free parameter being the stellar mass-to-light ratio\epubtkFootnote{The mass-to-light ratio is not really a constant in galaxies, however. Figure~\ref{figure:m33gradient} thus gives an example of a rotation curve fit (to the Local Group galaxy M33), where the variation of the mass-to-light ratio according to the color-gradient has been included, even improving the MOND fit.}. Not only does MOND predict the general trend for low surface brightness (LSB) and high surface-brightness (HSB) galaxies, it also predicts the observed rotation curves in great detail. This procedure has been carried out for 78 nearby galaxies (all galaxy rotation curves to which the authors have access), and the residuals between the observed and predicted velocities, at every point in all these galaxies (thus about two thousand individual measurements), are plotted in Figure~\ref{figure:RCresiduals}. As an illustration of the variety and richness of rotation curves fitted by MOND, as well as of the range of magnitude of the discrepancies covered, we display in Figure~\ref{figure:2Giants} fits to rotation curves of extremely massive HSB early-type disk galaxies~\cite{noordermeer} with $V_f$ up to 400~km/s, and in Figure~\ref{figure:2Dwarfs} fits to very low mass LSB galaxies~\cite{Millowmass} with $V_f$ down to 15~km/s. In the latter, gas-rich, small galaxies, the detailed fits are insensitive to the exact form of the interpolating function (Sect.~6.2) and to the stellar mass-to-light ratio~\cite{3741,Millowmass}. We then display in Figure~\ref{figure:RCfits_panel} eight fits for representative galaxies from the latest high-resolution THINGS survey~\cite{things,FTHINGS}, and in Figure~\ref{figure:LSBfits_panel} six fits of yet other LSB galaxies (as these provide strong tests of MOND and depend less on the exact form of the interpolating function than HSB ones) from~\cite{dBM98}, updated with high resolution H$\alpha$ data~\cite{KdN06, KdN08}. The overall results for the whole 78 nearby galaxies (Figure~\ref{figure:RCresiduals}) are globally very impressive, although there are a few outliers among the 2000 measurements. These are but a few trees outlying from a very clear forest. It is actually only as the quality of the data decline~\cite{sanders96} that one begins to notice small disparities.  These are sometimes attributable to external disturbances that invalidate the assumption of equilibrium~\cite{SV98}, non-circular motions or bad observational resolution.  For targets that are intrinsically difficult to observe, minor problems become more common~\cite{dBM98,swatersmond}. These typically have to do with the challenges inherent in combining disparate astronomical data sets (e.g., rotation curves measured independently at optical and radio wavelengths) and constraining the inclinations. A single individual galaxy that can be considered as a bit problematic is NGC~3198~\cite{bottema,things}, but this could simply be due to a problem with the potentially too high Cepheids-based distance (reddening problem mentioned in~\cite{Macri}). Indeed, the adopted distance plays an important role in the MOND fitting procedure, as the value of the centripetal acceleration $V_c^2/R$ depends on the distance through the conversion of the observed angular radius in arcsec into the physical radius $R$ in kpc. Note that other galaxies such as NGC~2841 had historically posed problems to MOND but that these have largely gone away with modern data (see~\cite{things} and Figure~\ref{figure:RCfits_panel}).

We finally note that what makes all these rotation curve fits really impressive is that either (i) stellar mass-to-light ratios are unimportant (in the case of gas-rich galaxies) yielding excellent fits with essentially zero free parameters (apart from some wiggle room on the distance), or (ii) stellar mass-to-light ratios are important, and their best-fit value, obtained on purely dynamical grounds assuming MOND, vary with galaxy color as one would expect on purely astrophysical grounds from stellar population synthesis models~\cite{Bel03}. There is absolutely nothing built into MOND that would require that redder galaxies should have higher stellar mass-to-light ratios in the $B$-band, but this is what the rotation curve fits require. This is shown on Figure~\ref{figure:MLcolor} where the best-fit mass-to-light ratio in the $B$-band is plotted against $B-V$ color index (left panel), and the same for the $K$-band (right panel).

\subsubsection{The Milky Way}

Our own Milky Way galaxy (a HSB galaxy) is a unique laboratory within which present and future surveys will allow us to perform many precision tests of MOND (at a level of precision that might even discriminate between the various versions of MOND described in Sect.~6.1) that are not feasible with external galaxies. Concerning the rotation curve however, the test is at present not the most conclusive, as the outer rotation curve of the Milky Way is paradoxically much less precisely known than that of external galaxies (the forthcoming Gaia mission should allow to improve this situation, although the rotation curve will not be measured directly). Nevertheless, past studies of the inner rotation curve of the Milky Way~\cite{FB05,FB2,McGMW}, measured with the tangent point method, compared to the baryonic content of the inner Galaxy~\cite{Bissantz,Flynn}, have shown full agreement between the rotation curve and MOND, assuming as usual the simple interpolating function ($\alpha=n=1$ in Eqs.~\ref{alphafamily} and \ref{nfamily}) or the $\gamma =\delta = 1$ interpolating function (Eqs.~\ref{gammafamily} and \ref{deltafamily}). The inverse problem was also tackled, i.e., deriving the surface density of the inner Milky Way disk from its rotation curve (see Figure~\ref{fig:MWRC}): this exercise~\cite{McGMW} led to a derived surface density fully consistent with star count data, and also even reproducing the details of bumps and wiggles in the surface brightness (Renzo's rule, Sect.~4.3.4), while being fully consistent with the (somewhat imprecise) constraints on the outer rotation curve of the Galaxy~\cite{xuexue}.

\epubtkImage{MW_emp_LR.png}{%
\begin{figure}[htbp]
  \centerline{\includegraphics[width=14.5cm]{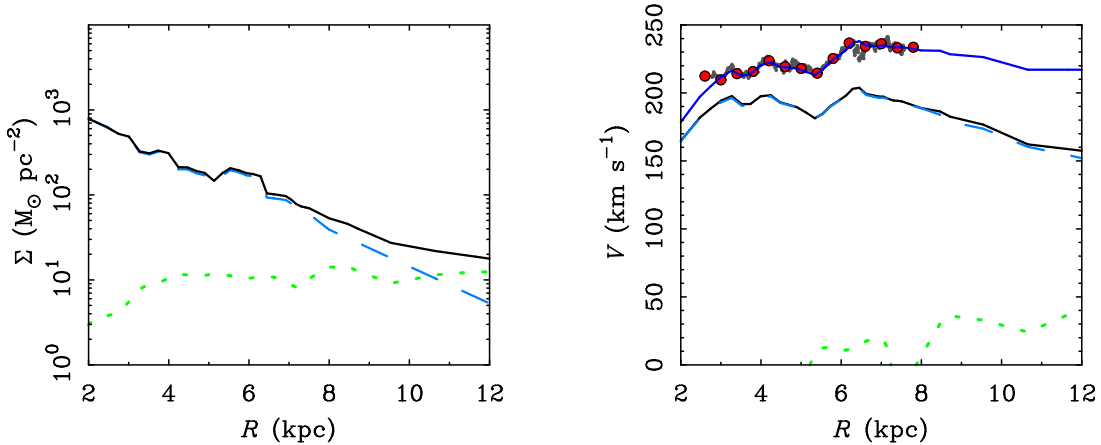}}
  \caption{The mass distribution of the Milky Way disk (left) inferred from fitting in MOND the observed bumps and wiggles in the
  rotation curve of the Galaxy (right)~\cite{McGMW}.  The Newtonian contributions of the stellar and gas disk are shown
  as dashed and dotted lines as per Figure~\ref{fig:4panel}.  The resulting model is consistent with independent 
  star count data~\cite{Flynn} and compares favorably to constraints on the rotation curve at radii beyond those 
  included in the fit~\cite{xuexue}. The prominent feature at $R \approx 6$ kpc  corresponds to the Centaurus spiral arm.}
  \label{fig:MWRC}
\end{figure}}

However, especially with the advent of present and future astrometric and spectroscopic surveys, the Milky Way offers a unique opportunity to test many other predictions of MOND. These include the effect of the ``phantom dark disk'' (see Figure~\ref{miyamoto}) on vertical velocity dispersions and on the tilt of the stellar velocity ellipsoid, the precise shape of tidal streams around the Galaxy, or the effects of the external gravitational field in which the Milky Way is embedded on fundamental parameters such as the local escape speed. All these predictions can however slightly vary depending on the exact formulation of MOND (mainly Bekenstein--Milgrom MOND, QUMOND, or multi-field theories, the predictions being anyway difficult to make in modified inertia versions of MOND when non-circular orbits are considered). Most of the predictions made until today and reviewed hereafter have been using the Bekenstein--Milgrom version of MOND (Eq.~\ref{BM}).

Based on the baryonic distribution from, e.g., the Besan\c{c}on model of the Milky Way~\cite{Robin}, one can compute the MOND gravitational field of the Galaxy by solving the BM-equation (Eq.~\ref{BM}). This has been done in~\cite{XufenMW}. Then one can apply the Newtonian Poisson equation to it, in order to find back the density distribution that would have yielded this potential within Newtonian dynamics~\cite{Bienayme,Bienayme2}. In this context, as already shown (Figure~\ref{miyamoto}), MOND predicts a disk of ``phantom dark matter'' allowing one to clearly differentiate it  from a Newtonian model with a dark halo:
\begin{itemize}
\item[(i)] By measuring the force perpendicular to the Galactic plane: at  the Solar  radius, MOND predicts a 60 percent enhancement  of the dynamical surface density at 1.1~kpc above the plane compared to the baryonic surface density, a value in agreement with current data (Table~1, see also~\cite{Nipotivert}).  The enhancement would become more  apparent at large galactocentric radii where the stellar disk mass density becomes negligible.
\item[(ii)]  By determining  dynamically the  scale length  of the  disk mass
 density distribution. This scale length is a factor $\sim$~1.25 larger
 than the  scale length of the  visible stellar disk  if Bekenstein--Milgrom MOND applies. Such a  test could  be applied  with existing  RAVE data ~\cite{DR3}, but the accuracy of available proper motions  still limits  the possibility  to explore  the gravitational forces too far from the Solar neighbourhood.
\item[(iii)]  By  measuring the  velocity  ellipsoid  tilt  angle within  the
 meridional  galactic plane.  This  tilt is  different within the MOND and Newton+dark halo cases in the inner part of  the Galactic disk. The tilt of about 6  degrees at z=1~kpc at  the Solar radius is  in agreement with
 the recent determination of  $7.3\pm 1.8$ degrees obtained by~\cite{S08}. The difference between MOND and a Newtonian model with a spherical halo becomes significant at z=2~kpc. Interestingly, recent data \cite{monithick} on the tilt of the velocity ellipsoid at these heights clearly favor the MOND prediction~\cite{Bienayme}.
\end{itemize}

Such tests of MOND could be applied with the first release of future Gaia data. To fix the ideas on the \textit{current} local constraints, the predictions of the Besan\c con MOND model are compared with the relevant observations in Table~\ref{tab1}. Let us however note that these predictions are \textit{extremely} dependent on the baryonic content of the model~\cite{Bissantz,Flynn,Robin}, so that testing MOND at the precision available in the Milky Way heavily relies on star counts, stellar population synthesis, census of the gaseous content (including molecular gas), and inhomogeneities in the baryonic distribution (clusters, gas clouds).

Another test of the predictions of MOND for the gravitational potential of the Milky Way is the thickness of the HI layer as a function of position in the disk (see also Sect.6.5.3): it has been found~\cite{Salcedo} that Bekenstein--Milgrom MOND and it phantom disk successfully accounts for the most recent and acurate flaring of the HI layer beyond 17~kpc from the center, but that it slightly underpredicts the scale-height in the region between 10 and 15~kpc. This could indicate that the local stellar surface density in this region should be slightly smaller than usually assumed, in order for MOND to predict a less massive phantom disk and hence a thicker HI layer. Another explanation for this discrepancy would rely on non-gravitational phenomena, namely ordered and small-scale magnetic fields and cosmic rays contributing to support the disk.

Yet another test would be the comparison of the observed Sagittarius stream~\cite{Ibata,Law} with the predictions made for a disrupting galaxy satellite in the MOND potential of the Milky Way. Basic comparisons of the stream with the orbit of a point mass has shown accordance at the zeroth order~\cite{Read}. In reality, such an analysis is not straightforward because streams do not delineate orbits, and because of the non-linearity of MOND. Combining a MOND N-body code with a Bayesian technique~\cite{Varghese} in order to efficiently explore the parameter space, it should however be possible to rigorously test MOND with such data in the near future, including for external galaxies, which will thus lead to an exciting battery of new observational tests of MOND.

Finally, a last test of MOND in the Milky Way involves the external field effect of Sect.~6.3. As explained there,  the return to a Newtonian (Eq.~\ref{asympt_efe} or Eq.~\ref{asymptqumond}) instead of a logarithmic (Eq.~\ref{potentielMondien}) potential at large radii is defining the escape speed in MOND. By observationally estimating the escape speed from a system (e.g., the Milky Way escape speed from our local neighbourhood), one can estimate the amplitude of the external field in which the system is embedded. With simple analytical arguments, it was found~\cite{FBZ} that with an external field of $0.01a_0$, the local escape speed at the Sun's radius was about 550~km/s exactly as observed (within the observational error range~\cite{SmithRAVE}). This was later confirmed by rigorous modeling in the context of Bekenstein--Milgrom MOND and with the Besan\c{c}on baryonic model of the Milky Way~\cite{Xufenstab}. This value of the external field, $10^{-2} \times a_0$, corresponds to the order of magnitude of the gravitational field exerted by Large Scale Structure, estimated from the acceleration endured by the Local Group during a Hubble time in order to attain a peculiar velocity of 600~km/s.

\begin{table}[htb]
\caption[Values predicted from the Besan\c{c}on model of the Milky Way
  in MOND as seen by a Newtonist (i.e., in terms of phantom dark
  matter contributions) compared to current observational constraints
  in the Milky Way, for the local dynamical surface density and the
  tilt of the stellar velocity ellipsoid.]{Values predicted from the
  Besan\c{c}on model of the Milky Way in MOND as seen by a Newtonist
  (i.e., in terms of phantom dark matter contributions) compared to
  current observational constraints in the Milky Way, for the local
  dynamical surface density and the tilt of the stellar velocity
  ellipsoid~\cite{Bienayme}. Predictions for a round dark halo without
  a dark disk are also compatible with the current constraints,
  though~\cite{HF04,S08}. The tilt at $z$~=~2~kpc should be more
  discriminating.}
\label{tab1}
\centering
\begin{tabular}{l c c }
\toprule
&  MOND predictions & Observations \\
\midrule
Surface density $\Sigma_{\odot} (z=1.1\mathrm{\ kpc})$ & $78\,M_{\odot}/\mathrm{pc}^2$ & $74\pm6\,M_{\odot}/\mathrm{pc}^2$~\cite{HF04} \\
\midrule
Velocity ellipsoid tilt at $z=1$~kpc & 6 degrees & $7.3\pm1.8$ degrees~\cite{S08} \\
\bottomrule
\end{tabular}
\end{table}

\subsubsection{Disk stability and interacting galaxies}

A lot of questions in galaxy dynamics require using N-body codes. This is notably necessary for studying stability of galaxy disks, the formation of bars and spirals, or highly time-varying configurations such as galaxy mergers. As we have seen in Sect.~6.1.2, the BM modified Poisson equation (Eq.~\ref{BM}) can be solved numerically using various methods~\cite{Bienayme,BradaM,Ciotticode,Feix,Llinares,Tiret}. Such a Poisson solver can then be used in particle-mesh N-body codes. More general codes based on QUMOND (Sect~6.1.3) are currently under development.

The main results obtained via these simulations are the following (the comparison with observations will be discussed below):
\begin{itemize}
\item[(i)] LSB disks are more unstable regarding bar and spiral instabilities in MOND than in the Newton+spherical halo equivalent case,
\item[(ii)] Bars always tend to appear more quickly in MOND than in the Newton+spherical halo equivalent, and are not slowed down by dynamical friction, leading to fast bars,
\item[(iii)] LSB disks can be both very thin and extended in MOND thanks to the effect of the ``phantom disk'', and vertical velocity dispersions level off at 8~km/s, instead of 2~km/s for Newtonian disks,
\item[(iv)] Warps can be created in apparently isolated galaxies from the external field effect of large scale structure in MOND,
\item[(v)] Merging time-scales are longer in MOND for interacting galaxies, 
\item[(vi)] Reproducing interacting systems such as the Antennae require relatively fine-tuned initial conditions in MOND, but the resulting galaxy is more extended and thus closer to observations, thanks to the absence of angular momentum transfer to the dark halo.
\end{itemize}

Concerning the first point (i), Brada \& Milgrom~\cite{BradaM}
investigated the important problem of stability of disk galaxies. They
demonstrated that MOND, as anticipated~\cite{Milstab}, has an effect
similar to a dark halo in stabilizing a rotationally supported disk,
thereby explaining the upper limit in surface density seen in the data
(Sect.~4.3.2), and also showing how it damps the growth-rate of
bar-forming modes in the weak gravitational field regime.  In a
comparison of MOND disks with the equivalent Newtonian+halo
counterpart (with identical rotation curves), they found that, as the
surface density of the disk decreases, the growth-rate of the
bar-forming mode decreases similarly in both cases. However, in the
limit of very low surface densities, typical of LSB galaxies, the MOND
growth rate stops decreasing, contrary to the Newton+dark halo case
(Figure~\ref{figure:growthrate}). This could provide a solution to the
stability challenge of Sect.~4.2, as observed LSBs do exhibit bars and
spirals, which would require an \textit{ad hoc} dark component within
the self-gravitating disk of the Newtonian system. One can also see on
this figure that if the surface density is typical of intermediate HSB
galaxies, the bar systematically forms quicker in MOND. 

\epubtkImage{diskstability.png}{%
\begin{figure}[htb]
  \centerline{\includegraphics[width=0.6\textwidth]{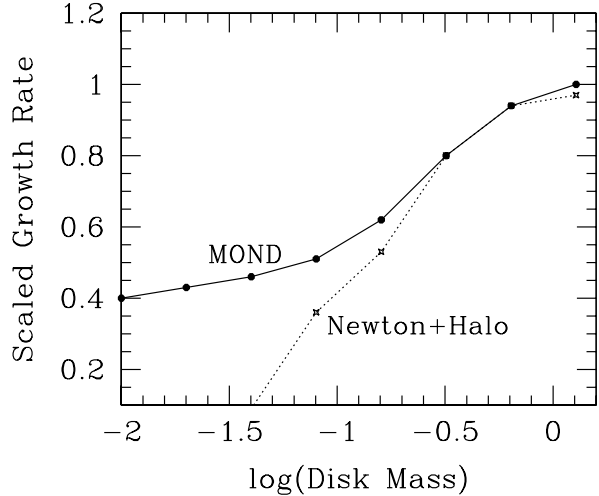}}
  \caption{The scaled growth-rate of the $m=2$ instability in Newtonian disks with a dark halo (dotted line) and MONDian disks (solid line) as a function of disk mass. In the MOND case, as the disk mass decreases, the surface density decreases and the disk sinks deeper into the MOND regime. At very low masses, however, the growth-rate saturates. In the equivalent Newtonian case, the rotation curve is maintained at the MOND level by supplementing the force with a round stabilizing dark halo which causes the growth-rate to crash~\cite{BradaM,SM02}. An \textit{ad-hoc} dark disk could help maintain the growth rate in the dark matter context.}
  \label{figure:growthrate}
\end{figure}}

This was confirmed in recent simulations~\cite{Combesgal, Tiret}, where it was additionally found that (ii) the bar is sustained longer, and is not slowed down by dynamical friction against the dark halo, which leads to fast bars, consistent with the observed fast bars in disk galaxies (measured through the position of resonances). When gas inflow and external gas accretion are included, however, a larger range of situations are met regarding pattern speeds in MOND, all compatible with observations~\cite{Tiretgas}. Since the bar pattern speed has a tendency to stay constant, the resonances remain at the same positions, and particles are trapped on these orbits more easily than in the Newtonian case, which leads to the formation of rings and pseudo-rings as observed (see Fig~\ref{fig:olivier2} and Figure~\ref{figure:olivier3}). All these results have been shown to be rather independent of the exact choice of interpolating $\mu$-function~\cite{Tiretgas}. 

What is more, (iii) LSB disks can be both very thin and extended in MOND thanks to the stabilizing effect of the ``phantom disk'', and vertical velocity dispersions level off at 8~km/s, as typically observed~\cite{BJBB,KdN08}, instead of 2~km/s for Newtonian disks with $\Sigma = 1\,M_{\odot}\,\mathrm{pc}^{-2}$ (depending on the thickness of the disk). However, the observed value is usually attributed to non-gravitational phenomena. Note that~\cite{MdB98} utilized this fact to predict that conventional analyses of LSB disks would infer abnormally high mass-to-light ratios for their stellar populations -- a prediction that was subsequently confirmed~\cite{fuchs,Saburova}. But let us also note that this stabilizing effect of the phantom disk, leading to very thin stellar and gaseous layers, could even be too strong in the region between 10 and 15~kpc from the galactic center in the Milky Way (see Sect.~6.5.2), and in external galaxies~\cite{Zari}, even though, as said, non-gravitational effects such as ordered and small-scale magnetic fields and cosmic rays could significantly contribute to the prediction in these regions.

Via these simulations, it has also been shown (iv) that the external field effect of MOND (Sect.~6.3) offers a mechanism other than the relatively weak effect of tides in inducing and maintaining warps~\cite{Bradawarp}. It was demonstrated that a satellite at the position and with the mass of the Magellanic clouds can produce a warp in the plane of the Galaxy with the right amplitude and form~\cite{Bradawarp}, and even more importantly, that isolated galaxies could be affected by the external field of large scale structure, inducing a differential precession over the disk, in turn causing a warp~\cite{Combesgal}. This could provide a new explanation for the puzzle of isolated warped galaxies.

Interactions and mergers of galaxies are (v) very important in the cosmological context of galaxy formation (see also Sect.~9.2). It has been found~\cite{ciottibin} from analytical arguments that dynamical friction should be much more efficient in MOND, for instance for bar slowing down or mergers occuring more quickly.  But simulations display exactly the opposite effect, in the sense of bars not slowing down and merger time-scales being much larger in MOND~\cite{Nipotimerge,Tiretinteract}. Concerning bars, Nipoti~\cite{nipotifriction} found that they were indeed slowed down more in MOND, as predicted analytically~\cite{ciottibin}, but this is because their bars were unrealistically small compared to observed ones. In reality, the bar takes up a significant fraction of the baryonic mass, and the reservoir of particles to interact with, assumed infinite in the case of the analytic treatment~\cite{ciottibin}, is in reality insufficient to affect the bar pattern speed in MOND. Concerning long merging time-scales, an important constraint from this would be that, in a MONDian cosmology, there should perhaps be less mergers, but longer ones than in $\Lambda$CDM, in order to keep the total observed amount of interacting galaxies unchanged. This is indeed what is expected (see Sect.~9.2). What is more, the long merging time-scales would imply that compact galaxy groups do not evolve statistically over more than a crossing time. In contrast, in the Newtonian+dark halo case, the merging time scale would be about one crossing time because of dynamical friction, such that compact galaxy groups ought to undergo significant merging over a crossing time, contrary to what is observed~\cite{Kroupa}. Let us also note that, in MOND, many passages in binary galaxies will happen before the final merging, with a starburst triggered at each passage, meaning that the number of observed starbursts as a function of redshift cannot be used as an estimate of the number of mergers~\cite{Combesgal}.

Finally, (vi) at a more detailed level, the Antennae system, the prototype of a major merger, has been shown to be nicely reproducible in MOND~\cite{Tiretinteract}. This is illustrated on Figure~\ref{figure:olivier4}. On the contrary, while it is well established that CDM models can result in nice tidal tails, it turns out to be difficult to simultaneously match the narrow morphology of many observed tidal tails with rotation curves of the systems from which they come~\cite{dubinski}. In MOND, reproducing the Antennae requires relatively fine-tuned initial conditions, but the resulting tidal tails are narrow and the galaxy is more extended and thus closer to observations than with CDM, thanks to the absence of angular momentum transfer to the dark halo (solution to the angular momentum challenge of Sect.~4.2).

\epubtkImage{olivier2a.png}{%
\begin{figure}[htbp]
  \centerline{(a)\parbox[t]{5.5cm}{\vspace{0pt}\includegraphics[width=5.5cm]{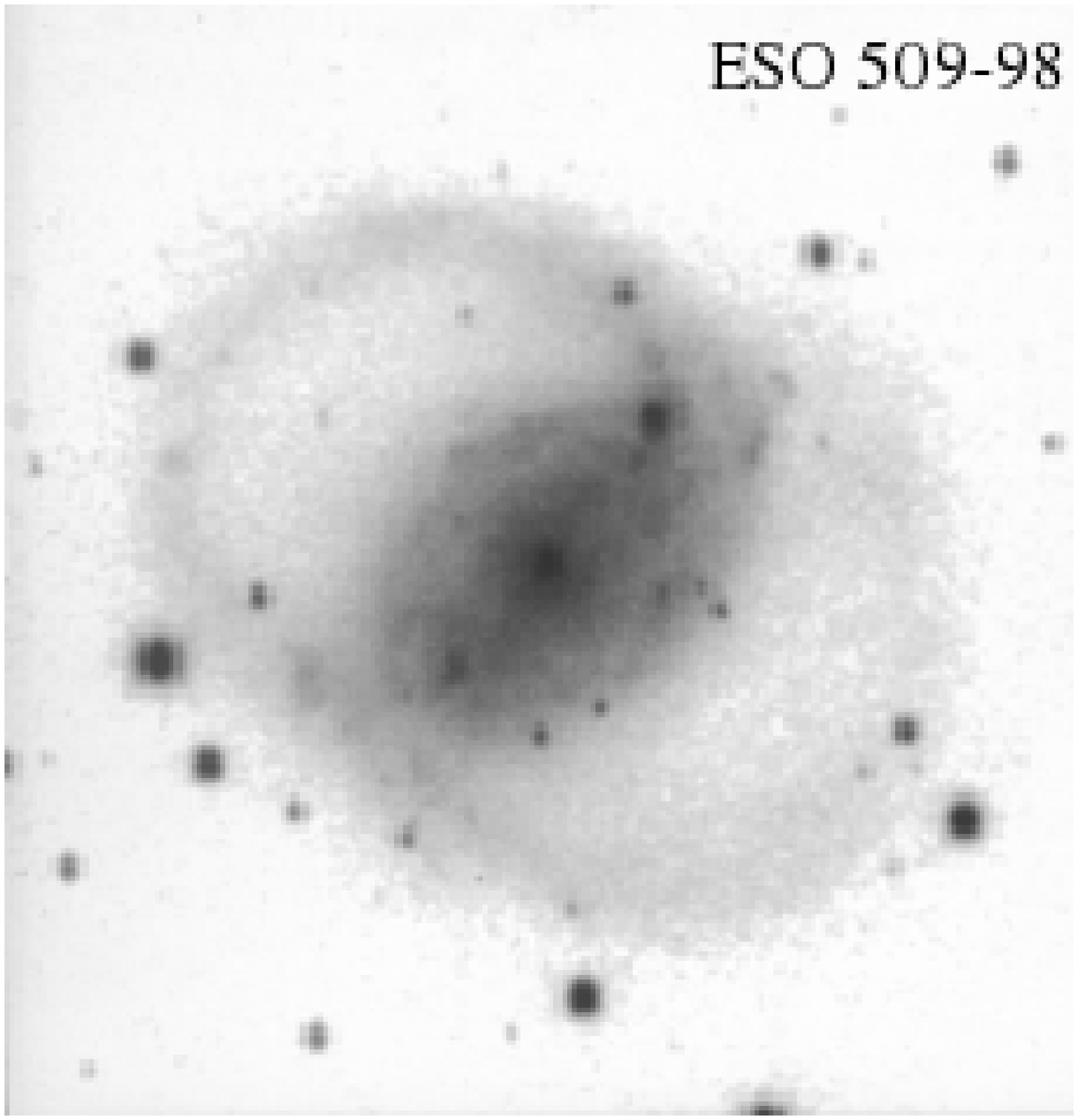}}\qquad
              (b)\parbox[t]{5.5cm}{\vspace{0pt}\includegraphics[width=5.5cm]{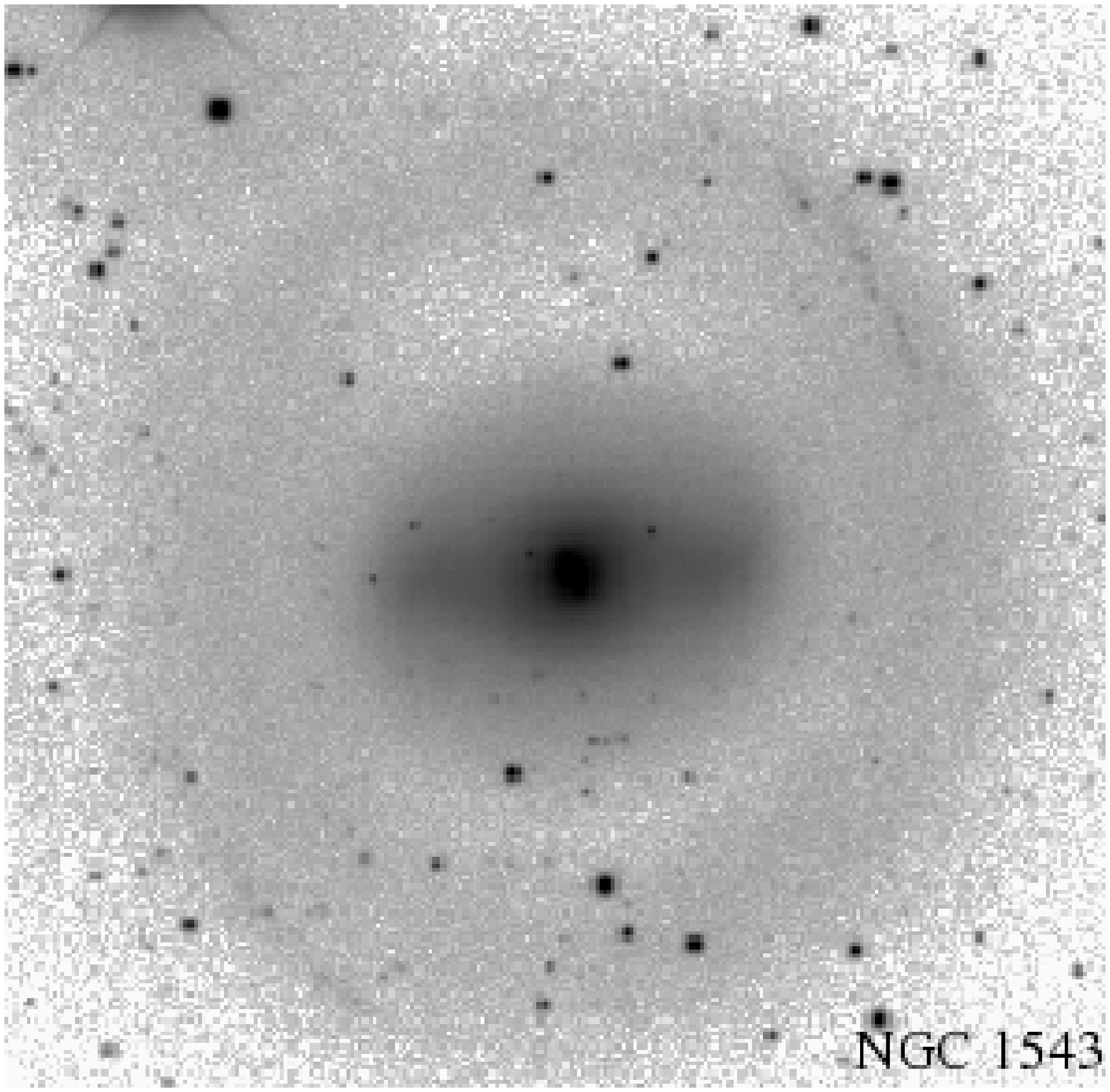}}}
  \caption{(a) The galaxy ESO 509-98   (b) The galaxy NGC 1543. These are two examples of galaxies that exhibit clear ring and pseudo-ring structures.}
  \label{fig:olivier2}
\end{figure}}

\epubtkImage{olivier3.png}{%
\begin{figure}[htbp]
  \centerline{\includegraphics[width=14cm]{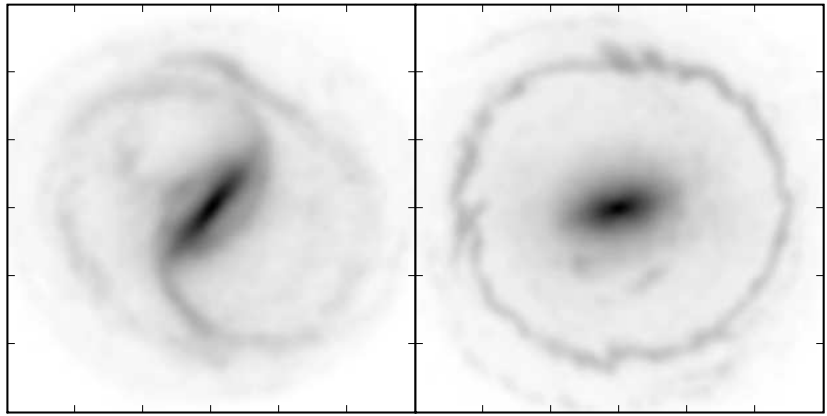}}
  \caption{Simulations of ESO 509-98 and NGC 1543 in MOND, to be compared with Figure~\ref{fig:olivier2}. Rings and pseudo-rings structures are well reproduced with modified 
gravity (Figure courtesy of O. Tiret).}
  \label{figure:olivier3}
\end{figure}}

\epubtkImage{olivier4.png}{%
\begin{figure}[htbp]
  \centerline{\includegraphics[width=10.5cm]{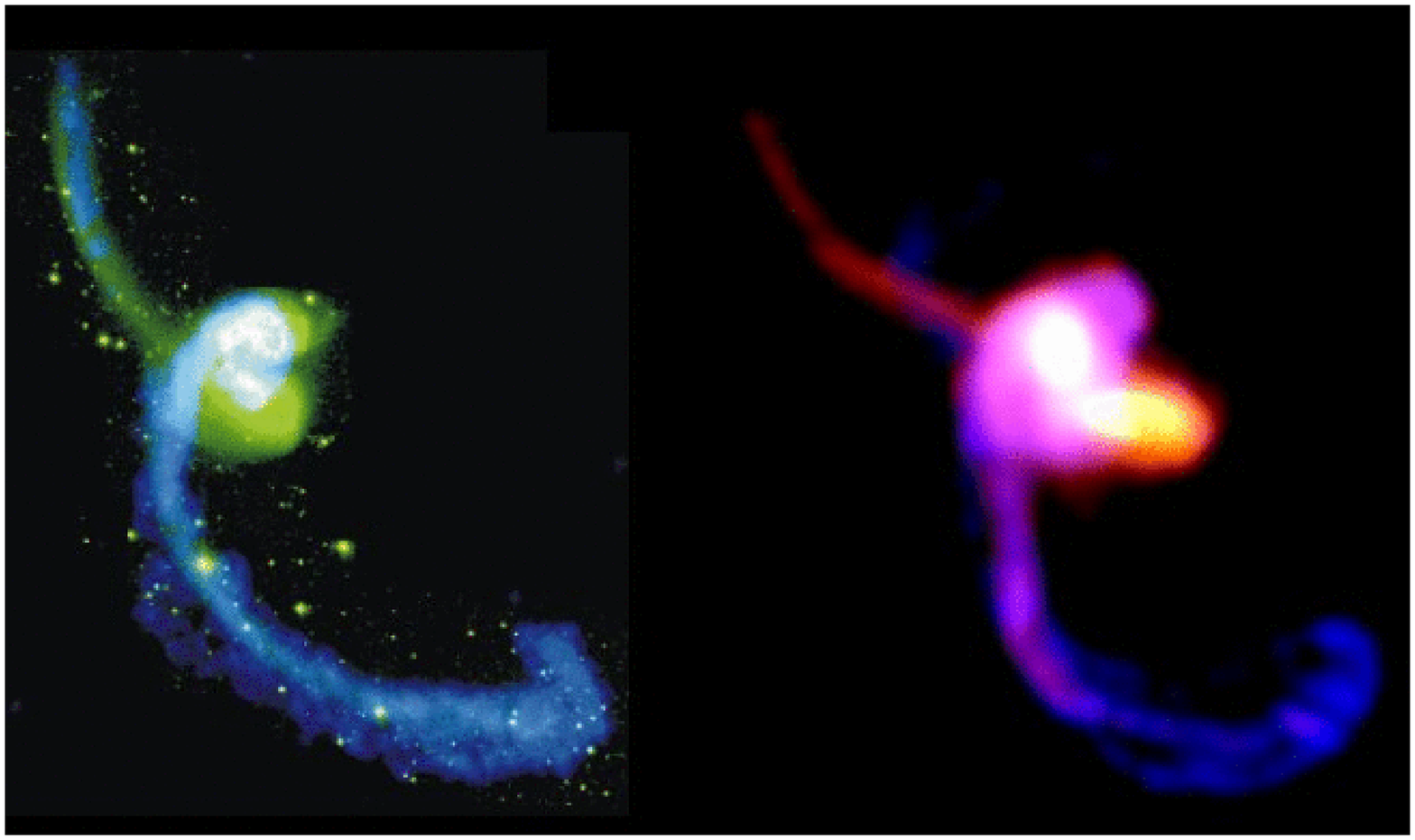}}
  \caption{Simulation of The Antennae with MOND (right,~\cite{Tiretinteract}) compared to the observations (left,~\cite{Hibbard}). In the observations, the gas is represented in blue and the stars in green. In the simulation the gas is in 
blue and the stars are in yellow/red. (Figure courtesy of O. Tiret)}
  \label{figure:olivier4}
\end{figure}}

\subsubsection{Tidal dwarf galaxies}

As seen in, e.g., Figure~\ref{figure:olivier4}, left panel, major mergers between spiral galaxies are frequently observed with dwarf galaxies at the extremity of their tidal tails, called Tidal Dwarf Galaxies (TDG). These young objects are formed through gravitational instabilities within the tidal tails, leading to local collapse of gas and star formation. These objects are very common in interacting systems: in some cases dozens of such condensations are seen in the tidal tails, with a few ones having a mass typical of other dwarf galaxies in the Universe. In the $\Lambda$CDM model, however, these objects are difficult to form, and require very extended dark matter distribution~\cite{Bournaud03}. In MOND simulations~\cite{Tiretinteract,Combesgal}, however, the exchange of angular momentum occurs within the disks, whose sizes are inflated. For this reason, it is much easier with MOND to form TDGs in extended tidal tails.

\epubtkImage{NGC5291.png}{%
\begin{figure}[htbp]
\centerline{\includegraphics[width=7.5cm]{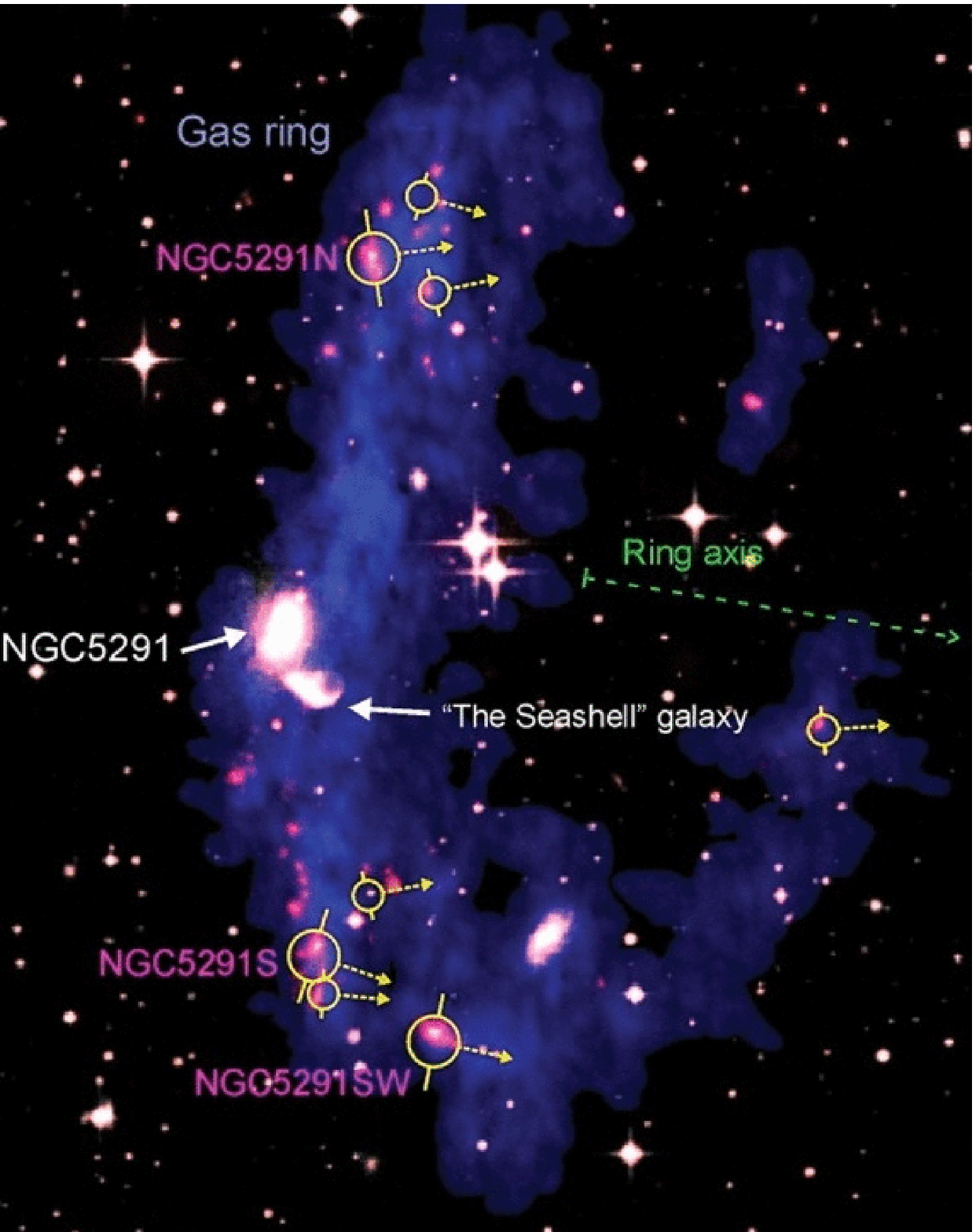}}
  \caption{The NGC 5291 system~\cite{Bournaud07}. VLA atomic hydrogen 21-cm map (blue) superimposed on an optical image (white). The UV emission observed by GALEX (red) traces dense star-forming concentrations. The most massive of these objects are rotating with the projected spin axis as indicated by dashed arrows. The three most massive ones are denoted as NGC5291N, NGC5291S, and NGC5291W. Figure courtesy of F. Bournaud.}
  \label{fig:5291}
\end{figure}}

\epubtkImage{RCtdg.png}{%
\begin{figure}[htbp]
  \centerline{\includegraphics[width=14.5cm]{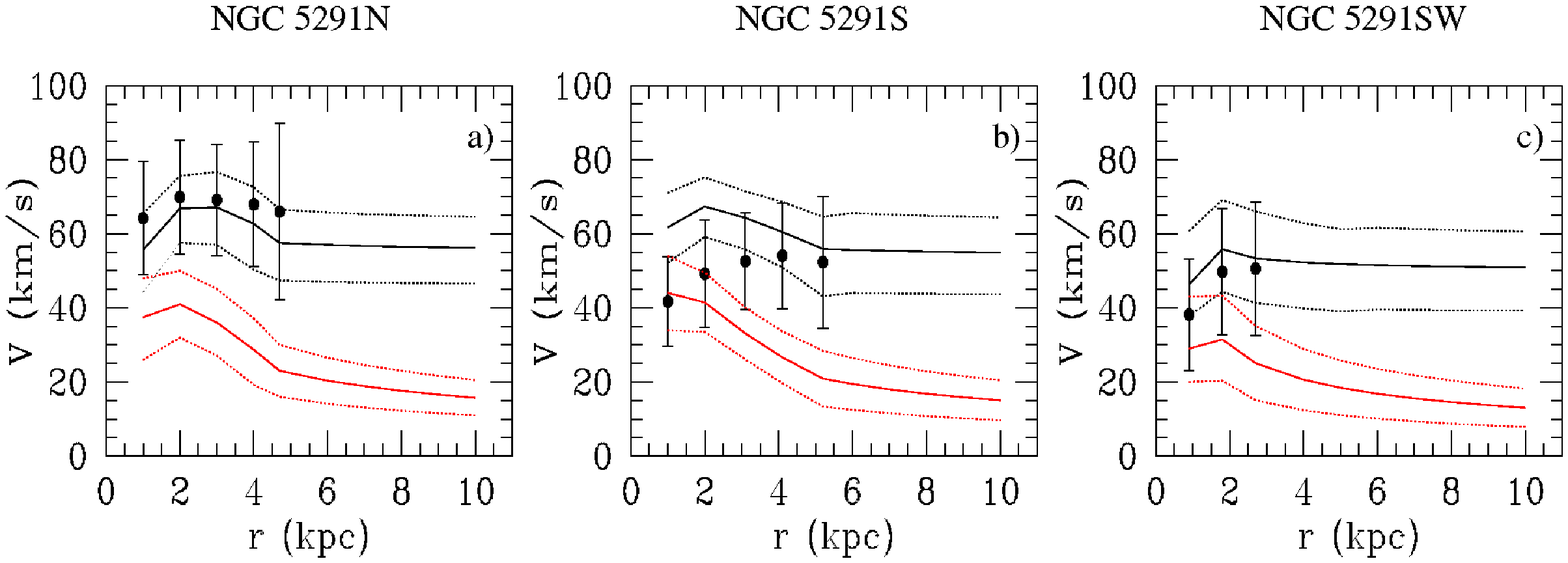}}
  \caption{Rotation curves of the three TDGs in the NGC 5291 system. In red: $\Lambda$CDM prediction (with no additional cold molecular gas), with the associated uncertainties. In black: MOND prediction with the associated uncertainties (prediction with zero free parameter, ``simple'' $\mu$-function assumed).}
  \label{figure:RCtdg}
\end{figure}}

What is more, in the $\Lambda$CDM context, these objects are not expected to drag CDM around them, the reason being that these objects are formed out of the material in the tidal tails, itself made of the dynamically cold, rotating, material in the progenitor disk galaxies. In these disks, the local ratio of dark matter to baryons is close to zero. For this reason, the $\Lambda$CDM prediction is that these objects should not exhibit a mass discrepancy problem. However, the first ever measurement of the rotation curve of three TDGs in the NGC~5291 ring system (Figure~\ref{fig:5291}) has rather revealed the presence of dark matter in these three objects~\cite{Bournaud07}. A solution to explain this in the standard picture could then be to resort to dark baryons in the form of cold molecular gas in the disks of the progenitor galaxies. However, it is very surprising that a very different kind of dark matter, in this case baryonic dark matter, would conspire to assemble itself precisely in the right way such as to put the three TDGs (see Sect.~4.3.1) on the baryonic Tully--Fisher relation (when this baryonic dark matter is not taken into account in the baryonic budget of the BTF). Another possibility, not resorting to baryonic dark matter, would be that, by chance, the three TDGs have been observed precisely edge-on. However, if we simply consider the most natural inclination coming from the geometry of the ring ($i=45^\circ$, see~\cite{Bournaud07}), and apply Milgrom's formula to the visible matter distribution with \textit{zero} free parameters~\cite{GentileTDG,Miltidal}, one gets very reasonable curves (Figure~\ref{figure:RCtdg}). Playing around a little bit with the inclinations allowed perfect fits to these rotation curves~\cite{GentileTDG}, while the influence of the external field effect has been shown not to significantly change the result.  We can therefore conclude that $\Lambda$CDM has severe problems with these objects, while MOND does exceedingly well in explaining their observed rotation curves.

However, the observations of only three TDGs are of course not enough, from a statistical point of view, in order for this result to be as robust as needed. Many other TDGs should be observed to randomize the uncertainties, and consolidate (or invalidate) this potentially extremely important result, that could allow to really discriminate between Milgrom's law being either a consequence of  some fundamental aspect of gravity (or of the nature of dark matter), or simply a mere recipe for how CDM organizes itself inside spiral galaxies. As a summary, since the internal dynamics of tidal dwarfs should \textit{not} be affected by CDM, they \textit{cannot} obey Milgrom's law for a statistically significant sample of TDGs if Milgrom's law is only linked to the way CDM assembles itself in galaxies. Observations of the internal dynamics of TDGs should thus be one of the observational priorities of the coming years in order to settle this debate.

Finally, let us note that it has been suggested~\cite{Kroupa}, as a possible solution to the satellites phase-space correlation problem of Sect.~4.2, that most dwarf satellites of the Milky Way could have been formed tidally, thereby being old tidal dwarf galaxies. They would then naturally appear in closely related planes, explaining the observed disk-of-satellites. While this scenario would lead to a missing satellites catastrophe in $\Lambda$CDM (see Sect.~4.2), it could actually make sense in a MONDian Universe (see Sect.~9.2).

\subsection{MOND in pressure supported stellar systems}

We have already outlined (Sect.~5.2) how Milgrom's formula accounts for general scaling relations of pressure-supported systems such as the Faber-Jackson relation (Figure~\ref{figure:faberjackson} and see~\cite{sanders10}), and that isothermal systems have a finite mass in MOND with the density at large radii falling approximately as $r^{-4}$~\cite{Milisotherm}. Note also that, in order to match the observed fundamental plane, MOND models must actually deviate somewhat from being strictly isothermal and isotropic: a radial orbit anisotropy in the outer regions is needed~\cite{sanders00,cardone11}. Here we concentrate on slightly more detailed predictions and scaling relations. In general, these detailed predictions are less obvious to make than in rotationally supported systems, precisely because of the new degree of freedom introduced by the anisotropy of the velocity distribution, very difficult to constrain observationally (as higher order moments than the velocity dispersions would be needed to constrain it). As we shall see, the successes of MOND are in general a bit less impressive in pressure-supported systems than in rotationally supported ones, and even in some cases really problematic (e.g., in the case of galaxy clusters, see Sect.~6.6.4). Whether this is due to the fact that predictions are less obvious to make, or whether this truly reflects a breakdown of Milgrom's formula for these objects (or the fact that certain theoretical versions of MOND would explicitly deviate from Milgrom's formula in pressure-supported systems, see Sect.~6.1.1) remains unclear.

\subsubsection{Elliptical galaxies}

Luminous elliptical galaxies  are dense bodies of old stars with very little gas and typically large internal accelerations. The age of the stellar populations suggest they formed early and all the gas has been used to form stars. To form early, one might expect the presence of a massive dark matter halo, but the study of, e.g.,~\cite{romanowsky} showed that actually, there is very little evidence for dark matter within the effective radius, and even several effective radii, in ellipticals. On the other hand, these are very high surface brightness objects and would thus not be expected to show a large mass discrepancy within the bright optical object in MOND. And indeed, the results of~\cite{romanowsky} were shown to be in perfect agreement with MOND predictions, assuming very reasonable anisotropy profiles~\cite{dearth}. On the theoretical side, it was also importantly shown that triaxial elliptical galaxies can be reproduced using the Schwarzschild orbit superposition technique~\cite{Wang}, and that these models are stable~\cite{XufenWang}\footnote{Separable models have also been investigated in \cite{Ciottistackel}}.

Interestingly, some observational studies circumvented the mass-anisotropy degeneracy by constructing non-parametric models of observed elliptical galaxies, from which equivalent circular velocity curves, radial profiles of mass-to-light ratio, and anisotropy profiles as well as high-order moments could be computed~\cite{gerhard01}. Thanks to these studies, it was e.g., shown ~\cite{gerhard01} that, although not much dark matter is needed, the equivalent circular velocity curves (see also~\cite{Weijmans} where the rotation curve could also be measured directly) tend to become \textit{flat} at much \textit{larger} accelerations than in thin exponential disk galaxies. This would seem to contradict the MOND prescription, for which flat circular velocities typically occur well below the acceleration threshold $a_0$, but not at accelerations of the order of a few times $a_0$ as in ellipticals. However, as shown in~\cite{RFGS}, if one assumes the simple interpolating function ($\alpha=n=1$ in Eq.~\ref{alphafamily} and Eq.~\ref{nfamily}), known to yield excellents fits to spiral galaxy rotation curves (see Sect.~6.5.1), one finds that MONDian galaxies exhibit a flattening of their circular velocity curve at high accelerations if they can be described by a Jaffe profile~\cite{Jaffe} in the region where the circular velocity is constant. Since this flattening at high accelerations is not possible for exponential profiles, it is thus remarkable that such flattenings of circular velocity curves at high accelerations are only observed in elliptical galaxies. What is more,~\cite{gerhard01}, as well as~\cite{T09}, derived from their models scaling relations for the configuration space and phase-space densities of dark matter in ellipticals, and these DM scaling relations have been shown~\cite{RFGS} to be in very good agreement with the MOND predictions on ``phantom DM'' (Eq.~\ref{eq:phantom}) scaling relations. This is displayed on Figure~\ref{figure:ellipticals_richtler}. Of course, some of these galaxies are residing in clusters, and the external field effect (see Sect.~6.3) could thus modify the predictions, but this was shown to be negligible for most of the analyzed sample, because the galaxies are far away from the cluster center~\cite{RFGS}. Note that when closer to the center of galaxy clusters, interesting behaviors such as lopsidedness caused by the external field effect could allow new tests of MOND in the near future~\cite{Xufenlops}. However, this would require modeling both the orbit of the galaxy in the cluster to take into account time-variations of the external field, as well as a precise estimate of the external field from the cluster itself, which can be tricky as the whole cluster should be modeled at once due to the non-linearity of MOND~\cite{Dai1,Matsuoscreening}.

\epubtkImage{ellipticals_richtler.png}{%
\begin{figure}[htbp]
  \centerline{\includegraphics[width=0.55\textwidth]{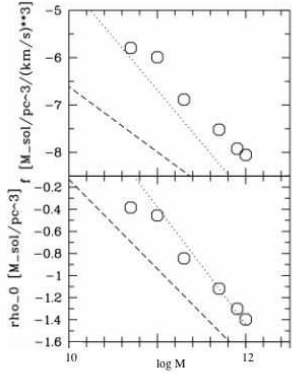}}
  \caption{MOND phantom dark matter scaling relations in ellipticals. The circles display central density $\rho_0$, and central phase space density $f$ of the phantom dark halos predicted by MOND for different masses of baryonic Hernquist profiles (with scale-radius $r_H$ related to the effective radius by $R_{\mathrm{eff}} = 1.815 \, r_H$). The dotted lines are the scaling relations of~\cite{gerhard01}, and the dashed lines those of~\cite{T09}, which exhibit a very large observational scatter in good agreement with the MOND prediction~\cite{RFGS}.}
  \label{figure:ellipticals_richtler}
\end{figure}}

At a more detailed level, precise full line-of-sight velocity dispersion profiles of individual ellitpticals, typically measured with tracers such as PNe or globular clusters populations, have been reproduced by solving Jeans equation in spherical symmetry:
\begin{equation}
 \frac{d \sigma^2}{dr} + \sigma^2 \frac{(2 \beta + \alpha)}{r} = - g(r)
\label{Jeans}
\end{equation}
where $\sigma$ is the radial velocity dispersion, $\alpha =  d ln \rho / d ln r$ is the slope of the tracer density $\rho$, and $\beta = 1-  (\sigma_\theta ^2 +  \sigma_\phi ^2) / 2 \sigma^2$ is the velocity anisotropy. Note that on the left-hand side, one uses the density and the velocity dispersion of the tracers only, which can be different from the density producing the gravity on the right-hand side if a specific population of tracers such as globular clusters is used. When the global kinematics of a galaxy is analyzed, we do expect in MOND that the gravity on the right-hand side of Eq.~\ref{Jeans} is generated by the observed mass distribution, so both should be fit simultaneously: Figure~\ref{fig:n7507} (provided by \cite{Sandersinprep}) shows an example. In general, it was found that field galaxies are all fit very naturally with MOND~\cite{Tiretellipt,4636} (see also~\cite{Weijmans}). On the other hand, the MOND modification has been found to slightly underpredict the velocity dispersions in large elliptical galaxies at the very center of galaxy clusters~\cite{1399}, which is just the small-scale equivalent of the problem of MOND in clusters, pointing towards missing baryons (see Sect.~6.6.4).

\epubtkImage{n7507prun.png}{%
\begin{figure}[htbp]
  \centerline{(a)\parbox[t]{5.5cm}{\vspace{0pt}\includegraphics[width=6.2cm]{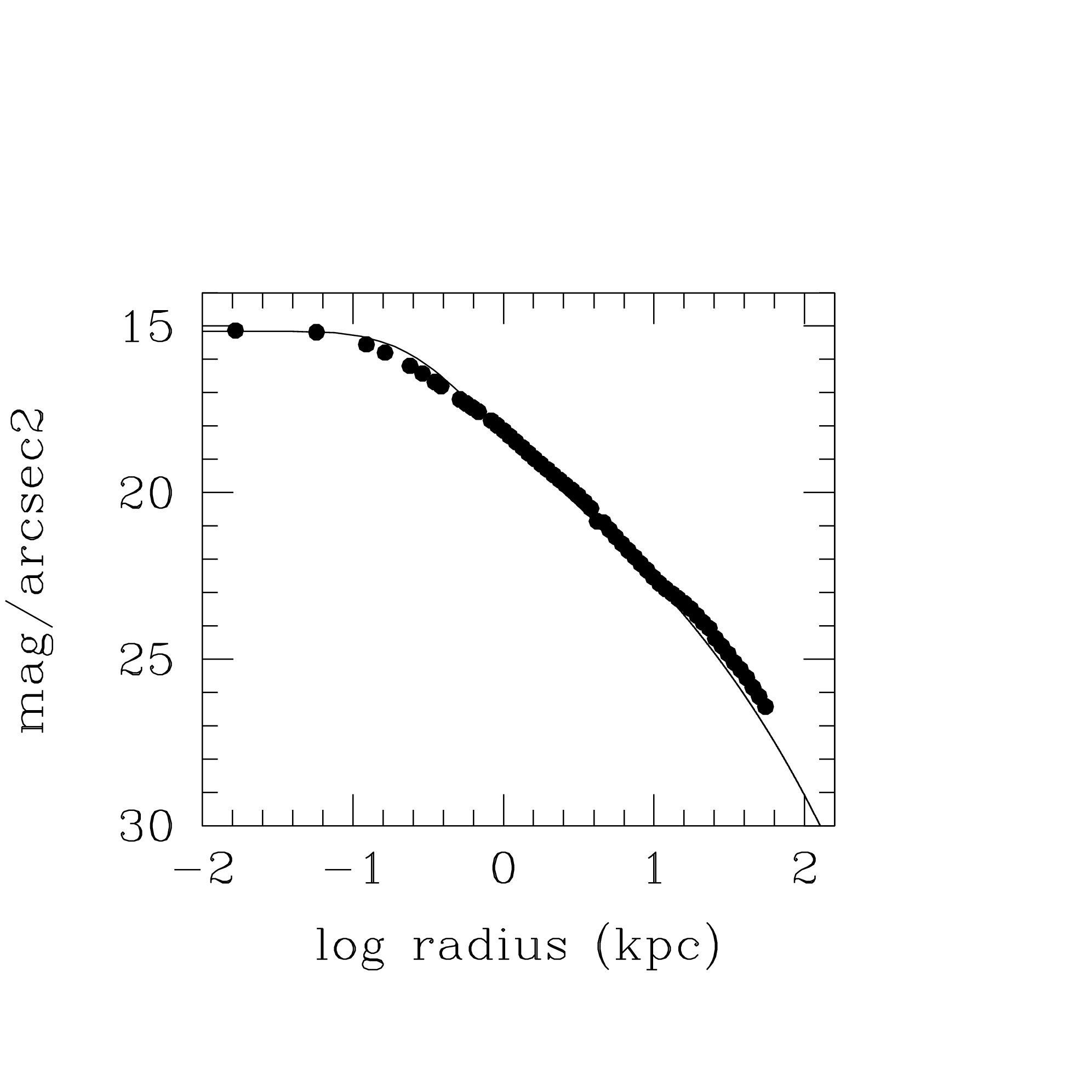}}\qquad
              (b)\parbox[t]{6cm}{\vspace{0pt}\includegraphics[width=5.5cm]{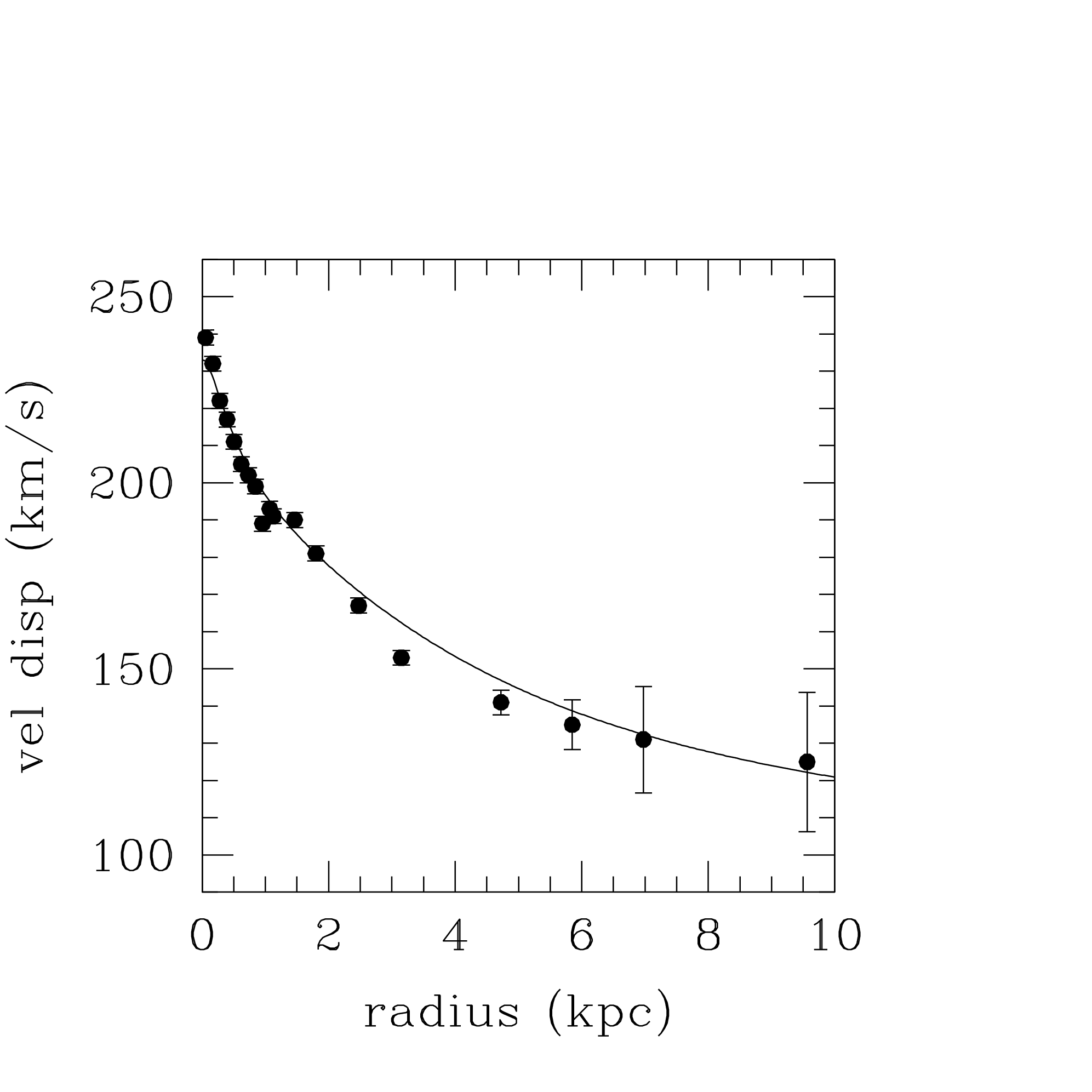}}}
  \caption{The surface brightness (a) and velocity dispersion (b) profiles of the elliptical 
galaxy NGC~7507 \cite{Salinas} fitted by MOND (lines \cite{Sandersinprep}). Elliptical 
galaxies can be approximated in MOND as high order polytropes with some radial orbit 
anisotropy \cite{sanders00}.  This particular case has a polytropic index of 14 with 
anisotropy of the Osipkov-Merritt form with an anisotropy radius of 5~kpc and maximum 
anisotropy $\beta=0.75$ at large radii \cite{Sandersinprep}.  The stellar mass-to-light 
ratio is $\Upsilon_*^B = 3.03\;M_{\odot}/L_{\odot}$.  This simple model captures the 
gross properties of both the surface brightness and velocity dispersion profiles. The galaxy is well-fitted by MOND, contrary to the claim of \cite{Salinas}.}
  \label{fig:n7507}
\end{figure}}

On the other hand, \cite{Klypin} used satellite galaxies of ellipticals to test MOND at distances of several 100~kpcs. They used the stacked SDSS satellites to generate a pair of mock galaxy groups with reasonably precise line-of-sight velocity dispersions as a function of radius across the group. When these systems were first analysed by~\cite{Klypin} they claimed that MOND was excluded by 10$\sigma$, but this was only for models that had constant velocity anisotropy. It was then found~\cite{Angusdss} that with varying anisotropy profiles similar to those found in simulations of formation of ellipticals by dissipationless collapse in MOND~\cite{Nipoticollapse}, excellent fits to the los velocity dispersions of both mock galaxies  could be found
are excellent and can be taken as strong evidence that MOND describes the dynamics in the surroundings of relatively isolated ellipticals very well.

Finally, let us note an intriguing possibility in a MONDian Universe (see also Sect.~9.2).  While massive ellipticals would form at $z \approx 10$~\cite{Sandersform} from monolithic dissipationless collapse~\cite{Nipoticollapse}, dwarf ellipticals could be more difficult to form. A possibility to form those would then be that tidal dwarf galaxies would be formed and survive more easily (see Sect.~6.5.4) in major mergers, and could then evolve to lead to the population of dwarf ellipticals seen today, thereby providing a natural explanation for the observed density-morphology relation~\cite{Kroupa} (more dwarf ellipticals in denser environments).

\subsubsection{Dwarf spheroidal galaxies}

Dwarf spheroidal (dSph) satellites of the Milky Way~\cite{SG,Walker09} exhibit some of the largest mass discrepancies observed in the Universe. In this sense, they are extremely interesting objects in which to test MOND. Observationally, let us note that there are essentially two classes of objects in the galactic stellar halo: globular clusters (see Sect.~6.6.3 hereafter) and dSph galaxies. These overlap in baryonic mass, but not in surface brightness, nor in age or uniformity of the stellar populations. The globular clusters are generally composed of old stellar populations, they are HSB objects and mostly exhibit no mass discrepancy problem, as expected for HSB objects in MOND. The dSphs, on the contrary, generally contain slightly younger stellar populations covering a range of ages, they are extreme LSB objects and exhibit, as said before, an extreme mass discrepancy, as generically expected from MOND. So, contrary to the case of $\Lambda$CDM where different formation scenarios have to be invoked (see also Sect.~6.6.3), the different mass discrepancies in these objects find a natural explanation in MOND. 

At a more detailed level, MOND should also be able to fit the whole velocity dispersion profiles, and not only give the right ballpark prediction. This analysis has recently been possible for the eight ``classical'' dSph around the Milky Way~\cite{Walker09}. Solving Jeans equation (Eq.~\ref{Jeans}), it was found~\cite{AngusdSph} that the four most massive and distant dwarf galaxies (Fornax, Sculptor, Leo~I and Leo~II) have typical stellar mass-to-light ratios, exactly within the expected range. Assuming equilibrium, two of the other four (smallest and most nearby) dSphs have mass-to-light ratios that are a bit higher than expected (Carina and Ursa Minor), and two have very high ones (Sextans and Draco). For all these dSphs, there is a remarkable correlation between the stellar $M/L$ inferred from MOND and the ages of their stellar populations~\cite{Hernandez}. Concerning the high inferred stellar $M/L$, note that it has been shown~\cite{BMdwarfs} that  a dSph will begin to suffer tidal disruption at distances from the Milky Way that are 4-7 larger in MOND than in CDM, Sextans and Draco could thus actually be partly tidally disrupted in MOND. And indeed, after subjecting the five dSphs with published data to an interloper removal algorithm~\cite{Serra}, it was found that Sextans was probably littered with unbound stars which inflated the computed $M/L$, while Draco's projected distance-l.o.s. velocity diagram actually looks as out-of-equilibrium as Sextans' one. Ursa Minor, on the other hand, is the typical example of an out-of-equilibrium system, elongated and showing evidence of tidal tails. In the end, only Carina thus has a suspiciously high $M/L$ ($>4$, see~\cite{Serra}). 

What is more, there is a possibility that, in a MONDian Universe, dSphs are not primordial objects but have been tidally formed in a major merger (see Sect.~9.2 as a solution to the phase-space correlation challenge of Sect.~4.2). In addition to the MOND effect, it would be possible that these objects \textit{never} really reach a stable equilibrium~\cite{KroupanoDM}, and exhibit artifically high $M/L$ ratio. This is even more true for the recently discovered ``ultra-faint'' dwarf spheroidals, that are also, due to to their extremely low-density, very much prone to tidal heating in MOND. Indeed, at face-value, if these ultrafaints are equilibrium objects, their velocity dispersions are much too high compared to what MOND predicts, and rule out MOND straightforwardly. However, unless this is due to systematic errors linked with the smallness of the velocity dispersion to measure (one must distinguish between $\sigma \approx 2\mathrm{\ km\, s}^{-1}$ and $\sigma \approx 5\mathrm{\ km\, s}^{-1}$), and/or to high intrinsic stellar $M/L$ ratios related to stochastic effects linked with the small number of stars \cite{Herultra}, it was also found~\cite{MWolf} that these objects are all close to filling their MONDian tidal radii, and that their stars can complete only a few orbits for every orbit of the satellite itself around the Milky Way (see Figure~\ref{fig:dSph}). As Brada \& Milgrom~\cite{BMdwarfs} have shown, it then comes as no surprise that they are displaying out-of-equilibrium dynamics in MOND (and even more so in the case of a tidal formation scenario~\cite{KroupanoDM}).

\epubtkImage{MONDtidalradius_LR.png}{%
\begin{figure}[htbp]
  \centerline{\includegraphics[width=13.5cm]{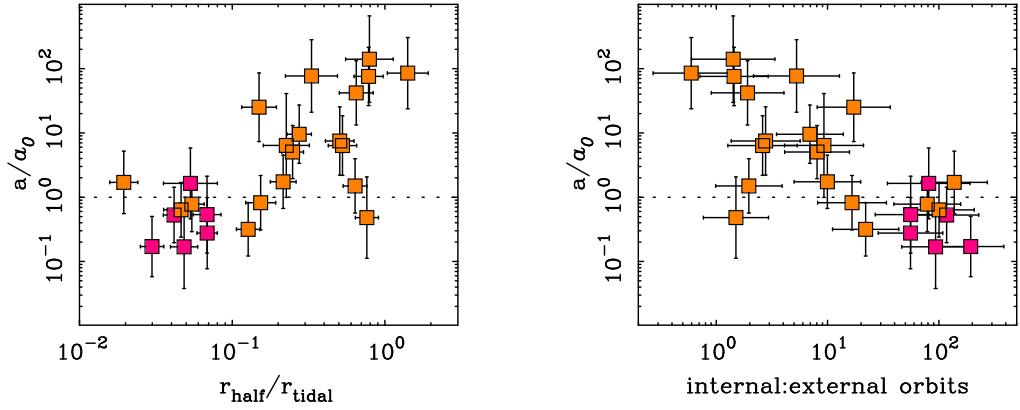}}
  \caption{The characteristic acceleration, in units of $a_0$, in the smallest galaxies known:  the dwarf
  satellites of the Milky Way (orange squares) and M31 (pink squares)~\cite{MWolf}.  The so-called classical dwarfs,
  with thousands of velocity measurements of individual stars~\cite{Walker09}, are largely consistent with MOND.
  The more recently discovered ``ultrafaint'' dwarfs, tiny systems with only a handful of stars~\cite{SG}, typically
  are not, in the sense that their measured  velocity dispersions and accelerations are too high.  This could be due to
  systematic uncertainties in the data~\cite{koposov}, as we must distinguish between 
  $\sigma \approx 2\mathrm{\ km\, s}^{-1}$ and $\sigma \approx 5\mathrm{\ km\, s}^{-1}$.
  Nevertheless, there may be a good physical reason for the non-compliance of the ultrafaint galaxies in the context of MOND.
  The deviation of these objects only occurs in systems where the stars are close to filling their MONDian tidal radii:
  the left panel shows the half light radius relative to the tidal radius.  Such systems may not be in equilibrium.
  Brada \& Milgrom~\cite{BMdwarfs} note that systems will no longer respond adiabatically to the influence of their
  host galaxy when a star in a satellite galaxy can complete only a few orbits for every orbit the satellite makes
  about its host.  The deviant dwarfs are in this regime (right panel).}
  \label{fig:dSph}
\end{figure}}

\subsubsection{Star clusters}

Star clusters come in two types: open clusters and globular clusters. Most observed open clusters are in the inner parts of the Milky Way disk, and for that reason, the prediction of MOND is that their internal dynamics is Newtonian~\cite{original} with, perhaps, a slightly renormalized gravitational constant and slightly squashed isopotentials, due to the external field effect (Sect.~6.3). The possibility of distinguishing Newtonian dynamics from MOND in these objects would therefore require extreme precision. On the other hand, globular clusters are mostly HSB halo objects (see Sect.~6.6.2), and are consequently predicted to be Newtonian, and most of those that are fluffy enough to display a MONDian behavior are close enough to the Galactic disk to be affected by the external field effect (Sect.~6.3), and so are Newtonian, too. Interestingly, MOND thus provides a natural explanation for the dichotomy between dwarf spheroidals and globular clusters. In $\Lambda$CDM, this dichotomy is rather explained by the formation history~\cite{Kravtsov,Sanders2419}: globular clusters are supposedly formed in primordial disk-bound supermassive molecular clouds with high baryon-to-dark matter ratio, and later become more spheroidal due to subsequent mergers. In MOND, it is of course not implied that the two classes of objects have necessarily the same formation history, but the different dynamics are qualitatively explained by MOND itself, not by the different formation scenarios.

However, there exist a \textit{few} globular clusters (roughly, less than $\sim$~10 compared to a total number of $\sim$~150) both fluffy enough to display typical internal accelerations well below $a_0$, and far away enough from the Galactic plane to be more or less immune from the external field effect~\cite{Baumgardt,Haghi1,Haghi2,Sollima1}. These should thus in principle display a MONDian mass discrepancy. They include, e.g., Pal~14 and Pal~3, or the large fluffy globular cluster NGC~2419. Pal~3 is interesting, because it indeed tends to display a larger than Newtonian global velocity dispersion, broadly in agreement with the MOND prediction (Baumgardt \& Kroupa, private communication). However, it is difficult to draw too strong a conclusion from this (e.g., on excluding Newtonian dynamics), since there are not many stars observed, and one or two outliers would be sufficient to make the dispersion grow artificially, while a slightly higher than usual mass-to-light ratio could reconcile Newtonian dynamics with the data. Other clusters such as NGC~1851 and NGC~1904 apparently display the same MONDian behavior~\cite{Scarpaglob} (see also \cite{Herglob}. On the other hand, Pal~14 displays exactly the opposite behavior: the measured velocity dispersion is Newtonian~\cite{Jordi}, but again the number of observed stars is too small to draw a statistically significant conclusion~\cite{GentilePal14}, and it is still possible to reconcile the data with MOND assuming a slightly low stellar mass-to-light ratio~\cite{SollimaPal14}. Note that if the cluster is on a highly eccentric orbit, the external gravitational field could vary very rapidly both in amplitude and direction, and it is possible that the cluster could take some time to accomodate this by still displaying a Newtonian signature in its kinematics after a sudden decrease of the external field. 

NGC~2419 is an interesting case, because it allows not only for a measure of the global velocity dispersion, but also of the detailed velocity dispersion profile~\cite{Rodrigo2419}. And, again, like in the case of Pal~14 (but contrary to Pal~3), it displays Newtonian behavior. More precisely, it was found, solving Jeans equations (Eq.~\ref{Jeans}), that the best MOND fit, although not extremely bad in itself, was 350 times less likely than the best Newtonian fit without DM~\cite{Rodrigo2419,Rodrigopoly}. The stability~\cite{Nipotiradial} of this best MOND fit has however not been checked in detail. These results are, however, heavily debated as they rely on the small quoted measurement errors on the surface density, and even a slight rotation of only the outer parts of this system near the plane of the sky (which would not show up in th velocity data) would make a considerable difference in the right direction for MOND \cite{2419part2}. However, these observations, together with the results on Pal~14, although not ruling out any theory, are not a resounding success for MOND. It could however perhaps indicate that globular clusters are generically on highly eccentric orbits, and out of equilibrium due to this (however, the effect would have to be opposite to that prevailing in ultra-faint dwarfs, where the departure from equilibrium would boost the velocity dispersion instead of decreasing it). A stronger view on these results could indicate that MOND as formulated today is an incomplete paradigm (see, e.g., Eq.~\ref{Bekmod}), or that MOND is an effect due to the fundamental nature of the DM fluid in galaxies (see Sects.~7.6 and 7.9), which is absent from globular clusters. Concerning NGC~2419, it is however perhaps useful to remind that it is very plausibly \textit{not} a globular cluster. It is part of the Virgo stream and is thus most probably the remaining nucleus of a disrupting satellite galaxy in the halo of the Milky Way, on a generically highly eccentric orbit. Detailed N-body simulations of such an event, and of the internal dynamics of the remaining nucleus, would thus be the key to confront MOND with observations in this object. All in all, the situation regarding MOND and the internal dynamics of globular clusters thus remains unclear.

On the other hand, it has been noted that MOND seems to overpredict the Roche lobe volume of globular clusters~\cite{Zhaoglob1,Zhaoglob2,Zhaoglob3}. Again, the fact that globular clusters could generically be on highly eccentric orbits could come to the rescue here. What is more, it was shown that, in MOND, globular clusters can have a cutoff radius which is unrelated to the tidal radius when non-isothermal~\cite{Sanders2419}. In general the cutoff radii of dwarf spheroidals, which have comparable baryonic masses, are larger than those of the globular clusters, meaning that those may well extend to their tidal radii because of a possibly different formation history than globular clusters. 

Finally, a last issue for MOND related to globular clusters~\cite{nipotifriction,sanchez06} is the existence of five such objects surrounding the Fornax dwarf spheroidal galaxy. Indeed, under similar environmental conditions, dynamical friction occurs on significantly shorter timescales in MOND than standard dynamics~\cite{ciottibin}, which could cause the globular clusters to spiral in and merge within at most 2~Gyrs~\cite{sanchez06}. However, this strongly depends on the orbits of the globular clusters, and in particular their initial radius~\cite{Angusfornax}, which can allow for a Hubble time survival of the orbits in MOND.

\subsubsection{Galaxy groups and clusters}
\label{clusters}

As pointed out earlier (3rd Kepler-like law of Sect.~5.2), it is a  
natural consequence of Milgrom's law that  at the effective baryonic  
radius of the system, the typical acceleration $\sigma^2/R$ is always  
observed to be of the order of $a_0$, thereby naturally explaining the  
linear relation between size and temperature for galaxy  
clusters~\cite{Mohr,Sandersneut}. However, one of the main predictions  
of Milgrom's formula is the so-called baryonic Tully--Fisher relation  
(circular velocity vs. baryonic mass, Figure~\ref{figure:btf}), and  
its equivalent for isotropic pressure supported systems, the  
Faber-Jackson relation (stellar velocity dispersion vs.\ baryonic  
mass, Figure~\ref{figure:faberjackson}), both for their slope and  
normalization. For systems such as galaxy clusters, where the hot  
intra-cluster gas is the major baryonic component, this relation can  
also be translated into a ``gas temperature vs. baryonic mass''  
relation $M_b \propto T^2$, plotted on  
Figure~\ref{figure:clustersbtf}, as the line $\log (M_b/M_{\odot}) = 2  
\log (T/\mathrm{keV}) +12.9$ (note that this differs slightly  
from~\cite{sandersclusters} where solar metallicity gas is assumed).  
Note on this figure that observations are closer to the MOND predicted  
slope than to the conventional prediction of $M \propto T^{3/2}$ in  
$\Lambda$CDM, without the need to invoke preheating (a need that may  
arise as an artifact of the mismatch in slopes).

\epubtkImage{cluster_LR.png}{%
\begin{figure}[htb]
   \centerline{\includegraphics[width=0.75\textwidth]{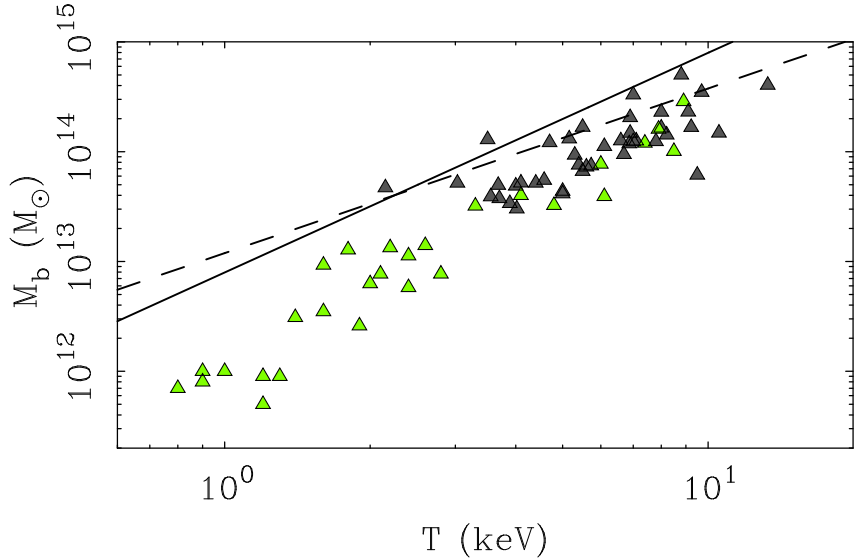}}
   \caption{The baryonic mass--X-ray temperature relation for rich  
clusters (gray triangles~\cite{Reiprich02,sandersclusters})
   and groups of galaxies (green triangles~\cite{angfambuo}).  The  
solid line indicates the {\it a priori} prediction of MOND: the data  
are reasonably consistent with the slope ($M \propto T^2$), but not  
with the normalization.  This is the residual missing baryon problem  
in MOND:  there should be roughly twice as much mass (on average) as  
observed. Also shown is the scaling relation {\it a priori} expected  
in $\Lambda$CDM (dashed line \cite{EMN96}).  This is in better (if not  
perfect) agreement with the normalization of the data for rich  
clusters, but not the slope.  The difference is sometimes attributed  
to preheating of the gas \cite{YB07},
which might also occur in MOND.}
   \label{figure:clustersbtf}
\end{figure}}

\epubtkImage{missingbaryons_LR.png}{
\begin{figure}[!ht]
   \centerline{\includegraphics[width=0.75\textwidth]{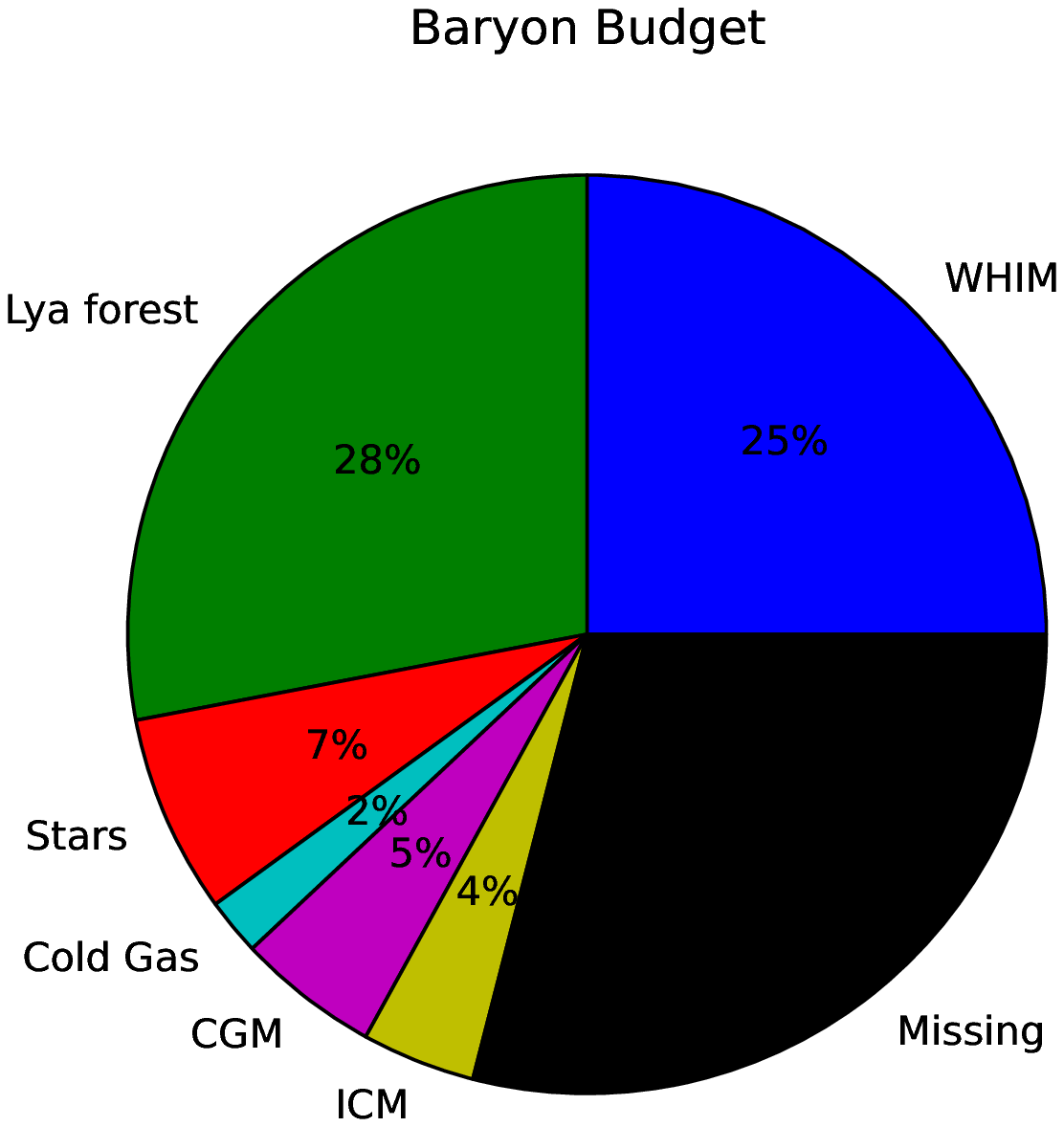}}
   \caption{The baryon budget in the low redshift Universe adopted  
from \cite{SSDmissingbaryons}.
The census of baryons includes the detected Warm-Hot Intergalactic  
Medium (WHIM), the
Lyman$\alpha$ forest, stars in galaxies, detected cold gas in galaxies  
(atomic HI and molecular
H$_2$), other gas associated with galaxies (the Circumgalactic Medium,  
CGM), and the
Intracluster Medium (ICM) of groups and clusters of galaxies.  The sum  
of known baryons
falls short of the density of baryons expected from Big Bang  
Nucleosynthesis: $\sim 30\%$
are missing.  These missing baryons presumably exist in some as yet  
undetected (i.e.,
dark) form.  If a fraction of these dark baryons reside in clusters  
(an amount roughly
comparable to that in the ICM) it would suffice to explain the  
residual mass discrepancy
problem MOND suffers in galaxy clusters.}
   \label{fig:missingbaryons}
\end{figure}}

So, interestingly, the data are still reasonably consistent with the  
slope predicted by MOND~\cite{Sanders94}, but not with the  
normalization. There is roughly a factor of two of residual missing  
mass in these  
objects~\cite{Gerbal,Pointecout,Sanders99,sandersclusters,Sandersneut,The}.  
This conclusion, reached from applying the hydrostatic equilibrium  
equation to the temperature profile of the X-ray emitting gas of these  
objects, has also been reached for low mass X-ray emitting  
groups~\cite{angfambuo}. This is essentially because, contrary to the  
case of galaxies, there is observationally a need for ``Newtonian''  
missing mass in the central parts\footnote{The conventional baryon fraction of clusters
increases monotonically with radius\cite{SciSimionescu},
only obtaining the cosmic value of 0.17 at or beyond the virial radius.
One might therefore infer the presence of dark baryons in cluster cores
in $\Lambda$CDM as well as in MOND.} of clusters, where the observed  
acceleration is usually slightly larger than $a_0$, meaning that the  
MOND prescription is not enough to explain the observed discrepancy  
between visible and dynamical mass there. For this reason, the  
residual missing mass in MOND is essentially concentrated in the  
central parts of clusters, where the ratio of MOND dynamical mass to  
observed baryonic mass reaches a value of 10, to then only decrease to  
a value of roughly $\sim 2$ in the very outer parts, where almost no  
residual mass is present. The profile of this residual mass would thus  
consist of a large constant density core of about 100\,--\,200~kpc in  
size (depending on the size of the group/cluster in question),  
followed by a sharp cutoff.

The need for this residual missing mass in MOND might be taken in one  
of the five following ways:
\begin{itemize}
\item[(i)] Practical falsification of MOND,
\item[(ii)] Evidence for missing baryons in the central parts of clusters,
\item[(iii)] Evidence for non-baryonic dark matter (existing or exotic),
\item[(iv)] Evidence that MOND is an incomplete paradigm,
\item[(v)] Evidence for the effect of additional fields in the parent  
relativistic theories of MOND, not included in Milgrom's formula.
\end{itemize}
If (i) is correct, one still needs to explain the success of MOND on galaxy scales
with $\Lambda$CDM.  Such an explanation has yet to be offered.
Thus, tempting as case (i) is, it is worth giving closer inspection of the four other
possibilities. 

The second case (ii) would be most in line with the
elegant absence of need for any non-baryonic mass in MOND (however,
see the ``dark fields'' invoked in Sect.~7).  It has happened
before that most of the baryonic mass was in an unobserved component.
From the 1930s when Zwicky first discovered the missing mass problem in
clusters till the 1980s, it was widely presumed that the stars in the observed
galaxies represented the bulk of baryonic mass in clusters.
Only after the introduction of MOND (in 1983) did it become widely appreciated
that the diffuse X-ray emitting intracluster gas (the ICM) greatly outweighed the stars.
That is to say, some of the missing mass problem in clusters was due to optically dark baryons ---
instead of the enormous mass discrepancies implied by cluster dynamical mass to optical light
ratios in excess of 100 \cite{BLD95}, the ratio of dark to baryonic mass is only $\sim 8$ 
conventionally \cite{giodini,MdB98dm}.
So we should not be too hasty in presuming we now have a complete census of
baryons in clusters.  Indeed, in the 
global baryon inventory of the Universe, $\sim 30\%$ of the baryons produced
during big bang nucleosynthesis (BBN) are missing  
(Figure~\ref{fig:missingbaryons}), and presumably reside in some as yet undetected
(dark) form.  It is estimated \cite{Fukugita,SSDmissingbaryons}
that the observed baryons in clusters only account for about 4\% of
those produced during BBN (Figure~\ref{fig:missingbaryons}).
This is much less than the 30\% of baryons that are still missing.
Consequently, only a modest fraction of the dark baryons need
to reside in clusters to solve
the problem of missing mass in the central regions of clusters
in MOND. It should be highlighted that this missing mass only appears
in MOND for systems with a high abundance of ionised gas and X-ray
emission. Indeed,  for  even smaller galaxy groups, devoid of gas, the
MOND predictions for the velocity dispersions of individual galaxies
are again perfectly in line with the
observations~\cite{Milgroup1,Milgroup2}. It is then no stretch of the
imagination to surmise that these gas rich systems, where the residual
missing baryons problem have equal quantities of molecular hydrogen or
other molecules. Milgrom~\cite{marriage} has, e.g., proposed that the
missing mass in MOND could entirely be in the form of cold, dense gas
clouds. There is an extensive literature discussing searches for cold
gas in the cores of galaxy clusters but what is usually meant there is
quite different from what is meant here, since those searches
consisted in trying to find the signature of \textit{diffuse} cold
molecular gas at a temperature of $\sim$~30~K. The proposition of
Milgrom~\cite{marriage} rather relies on the work of Pfenniger \&
Combes~\cite{Pfen}, where dense gas clouds with a temperature of only
a few Kelvins ($\sim$~3~K), Solar System size, and of a Jupiter mass,
were considered to be possible candidates for both galactic and
extragalactic dark matter. These clouds would behave in a
\textit{collisionless} way, just like stars. Since the dark mass
considered in the context of MOND cannot be present in galaxies, it is
however not subject to the galactic constraints on such gas
clouds. Note that the total sky covering factor of such clouds in the
core of the clusters would be of the order of only $10^{-4}$, so that
they would only occult a minor fraction of the X-rays emitted by the
hot gas (and it would be a rather constant fraction). For the same
reason, the chances of a given quasar having light absorbed by them is
very small. Still, \cite{marriage} notes that these clouds could be
probed through X-ray flashes coming out of individual collisions
between them. Of course, this speculative idea also raises a number of
questions, the most serious one being how these clumps form and
stabilize, and why they form only in clusters, X-ray emitting groups
and some ellipticals at the center of these groups and clusters, but
not in individual spiral galaxies. As noted above, the fact that
missing mass in MOND is necessarily associated with an abundance of
ionised gas could be a hint at a formation and stabilization process
somehow linked with the presence of hot gas and X-ray emission
themselves. Then, there is the issue to know whether the clouds
formation would be prior or posterior to the cluster formation. We
note that a rather late formation mechanism could help increase the
metal abundance, solving the problem of small-scale variations of
metallicity in clusters when the clouds are
destroyed~\cite{morris}. Milgrom~\cite{marriage} also noted that these
clouds could alleviate the cooling flow conundrum, because whatever
destroys them (e.g., cloud-cloud collisions and dynamical friction
between the clouds and the hot gas) is conducive to heating the core
gas, and thus preventing it from cooling too quickly. Such a heating
source would not be transient and would be quite isotropic, contrary
to AGN heating.

Another possibility (iii) would be that this residual missing mass in  
clusters is in the form of non-baryonic matter. There is one obviously  
existing form of such matter: neutrinos.  If $m_{\nu} \approx  
\sqrt{\Delta m^2}$~\cite{neutrinoDM2}, then the neutrino mass is too  
small to be of interest in this context. But there is nothing that  
prevents it from being larger (note that the ``cosmological''  
constraints from structure formation in the $\Lambda$CDM context obviously do not apply in MOND). 
Actual \textit{model-independent} experimental limits on the  
electron neutrino mass from the Mainz/Troitsk experiments, counting  
the highest energy electrons in the $\beta$-decay of  
Tritium~\cite{Mainz} are $m_\nu<2.2\mathrm{\ eV}$.  Interestingly, the  
KATRIN experiment (the KArlsruhe TRItium Neutrino experiment, under  
construction) will be able to falsify these 2eV electron neutrinos at  
95\% confidence. If the neutrino mass is substantially larger than the  
mass differences, then all types have about the same mass, and the  
cosmological density of three left-handed neutrinos and their  
antiparticles~\cite{Sandersneut} would be
\begin{equation}
\Omega_\nu = 0.062 m_\nu,
\end{equation}
where  $m_\nu$ is the mass of a single neutrino type in eV. If one  
assumes that clusters of galaxies respect the baryon-neutrino  
cosmological ratio, and that the MOND missing mass is mostly made of  
neutrinos as suggested by~\cite{sandersclusters,Sandersneut}, then the  
mass of neutrinos must indeed be around 2~eV. Combined with the effect  
of additional degrees of freedom in relativistic MOND theories  
(Sect.~7), it has been shown that the CMB anisotropies could also be  
reproduced (see Sect.~9.2 and~\cite{SMFB}), while this hot dark matter  
would obviously free-stream out of spiral galaxies and would thus not  
perturb the MOND fits of Sect.~6.5.1. The main limit on the neutrino  
ability to condense in clusters comes from the Tremaine--Gunn  
limit~\cite{tg79}, stating that the phase space density must be  
preserved during collapse. This is a density level half the quantum  
mechanical degeneracy level in phase-space:
\begin{equation}
f_{\max} = \frac{1}{2} \sum_{i=1}^{i=6} \frac{m_{\nu_i}^4}{h^3}.
\end{equation}
Converting this into configuration space, the maximum density for a  
cluster of a given temperature, $T$, is defined for a given mass of  
one neutrino type as~\cite{tg79}:
\begin{equation}
\label{tg}
\frac{ \rho_{\nu}^{\max}}{7\times 10^{-5}\,M_{\odot}  
pc^{-3}}=\left(\frac{T}{1keV}\right)^{1.5}\left(\frac{m_{\nu}}{2eV}\right)^4.
\end{equation}
Assuming the temperature of the neutrino fluid as being equal (due to
violent relaxation) to the mean emission weighted temperature of the
gas, Sanders~\cite{sandersclusters} showed that such 2~eV neutrinos at
the limit of experimental detection could indeed account for the bulk
of the dynamical mass in his sample of galaxy clusters of $T >
4\mathrm{\ keV}$ (see also Sect.~8.3 for gravitational lensing
constraints). This has the great advantage of naturally reproducing
the proportionality of the electron density in the cores of clusters
to $T^{3/2}$, as observed~\cite{Sandersneut}. However, looking at the
central region of low-temperature X-ray emitting galaxy groups, it was
found~\cite{angfambuo} that the needed central density of missing mass
far exceeded this limit by a factor of several hundred. One would need
one neutrino species with $m \sim 10\mathrm{\ eV}$ to reach the
required densities. One exotic possibility is then the idea of
right-handed eV-scale \textit{sterile} neutrinos~\cite{AFD}: as
strange as this sounds, this mass for sterile neutrinos could also
provide a good fit to the CMB acoustic peaks (see Sect.~9.2). This
could indeed sound as the strangest and most complicated Universe
possible, combining true non-baryonic (hot) dark matter with a
modification of gravity, but if this is what it takes to
simultaneously explain the Kepler-like laws of galactic dynamics and
the extragalactic evidence for dark matter, it is useful to remember
that there are \textit{a priori} both good reasons for there being
more particles than those of the standard model of particle physics
and that there is \textit{a priori} no reason that General Relativity
should be valid over a wide range of scales where it has never been
tested. In any case, experiments that can address the existence of
such a $\sim 10\mathrm{\ eV}$-scale sterile neutrino would thus be
very interesting, as this kind of particle could provide the dark
matter candidate \textit{only} in a modified gravity framework, since
such a hot dark matter particle would be unable to form small
structures and to provide the dark matter that would be needed in
galaxies.

Yet another possibility (iv) would be that MOND is incomplete, and  
that a new scale should be introduced, in order to effectively enhance  
the value of $a_0$ in galaxy clusters, while lowering it to its  
preferred value in galaxies. There are several ways to implement such  
an idea. For instance Bekenstein~\cite{bektalk} proposed adding a  
second scale in order to allow for effective variations of the  
acceleration constant as a function of the deepness of the potential  
(Eq.~\ref{Bekmod}). This idea should be investigated more in the  
future, but it is not clear that such a simple rescaling of $a_0$  
would account for the exact spatial distribution of the residual  
missing mass in MOND clusters, especially in cases where it is  
displaced from the baryonic distribution (see Sect.~8.3). However, as  
even Gauss' theorem would not be valid anymore in spherical symmetry,  
the high non-linearity might provide non-intuitive results, and it  
would thus clearly be worth investigating this suggestion in more  
detail, as well as developing similar ideas with other additional  
scales in the future (such as, for instance, the baryonic matter  
density, see~\cite{BLSF,FamaeyNZ} and Sect.~7.6).

Finally, as we shall see in the next section (Sect.~7), parent  
relativistic theories of MOND often require additional degrees of  
freedom in the form of ``dark fields'', which can nevertheless be  
globally subdominant to the baryon density, and thus do not  
necessarily act precisely as true ``dark matter''. The last  
possibility (v) is thus that these fields, obviously not included in  
Milgrom's formula, are responsible for the cluster missing mass in  
MOND. An example of such fields are the vector fields of TeVeS  
(Sect.~7.4) and Generalized Einstein-Aether theories (Sect.~7.7). It  
has been shown (see Sect.~9.2) that the growth of the spatial part of  
the vector perturbation in the course of cosmological evolution can  
successfully seed the growth of baryonic structures, just as dark  
matter does. If these seeds persist, it was shown~\cite{Daibullet}  
that they could behave in very much the same way as a dark matter halo  
in relatively unrelaxed galaxy clusters. However, it remains to be  
seen whether the spatially concentrated distribution of missing mass  
in MOND would be naturally reproduced in all clusters. In other  
relativistic versions of MOND (see, e.g., Sect.~7.6 and Sect.~7.9),  
the ``dark fields'' are truly massive and can be thought of as true  
dark matter (although more complex than simple collisionless DM),  
whose energy density outweighs the baryonic one, and could provide the  
missing mass in clusters. However, again, it is not obvious that the  
centrally concentrated distribution of residual missing mass in  
clusters would be naturally reproduced. All in all, there is no obviously
satisfactory explanation for the problem of residual missing mass in the center 
of galaxy clusters, which remains one of the most serious problems facing MOND.

\newpage

\section{Relativistic MOND Theories}
\label{sec:covariant}

In Sect.~6, we have considered the classical theories of MOND and their predictions in a vast number of astrophysical systems. However, as already stated at the beginning of Sect.~6, these classical theories are only toy-models until they become the weak-field limit of a relativistic theory (with invariant physical laws under differentiable coordinate transformations), i.e., an extension of General Relativity (GR) rather than an extension of Newtonian dynamics. Here, we list the various existing relativistic theories boiling down to MOND in the quasi-static weak-field limit. It is useful to restate here that the motivation for developing such theories is not to get rid of dark matter but to explain the Kepler-like laws of galactic dynamics predicted by Milgrom's law (see Sect.~5). As we shall see, many of these theories include new fields, so that dark matter is often effectively replaced by ``dark fields'' (although, contrary to dark matter, their energy density can be subdominant to the baryonic one; note that, even more importantly, in a static configuration these dark fields are fully determined by the baryons, contrary to the traditional dark matter particles which may in principle be present independently of baryons).

These theories are great advances because they enable us to calculate the effects of gravitational lensing and the cosmological evolution of the Universe in MOND, which are beyond the capabilities of  classical theories. However, as we shall see, many of these relativistic theories still have their limitations, ranging from true theoretical or observational problems to more aesthetical problems such as the arbitrary introduction of an interpolating function (Sect.~6.2) or the absence of an understanding of the $\Lambda \sim a_0^2$ coincidence. What is more, the new fields introduced in these theories have no counterpart yet in microphysics, meaning that these theories are, \textit{at best}, only effective. So, despite the existing effective relativistic theories presented here, the quest for a more profound relativistic formulation of MOND continues. Excellent reviews of existing theories can also be found in, e.g.,~\cite{Bek06,Bek2010,Bruneton07,Clifton,Esposito,Halle,MDorDM,Skordis09,SkorZlos}.

The heart of GR is the \textit{equivalence principle(s)}, in its weak (WEP), Einstein (EEP) and strong (SEP) form. The WEP states the universality of free fall, while the EEP states that one recovers special relativity in the freely falling frame of the WEP. These equivalence principles are obtained by assuming that all known matter fields are universally and minimally coupled to one single metric tensor, the physical metric. It is perfectly fine to keep these principles in MOND, although certain versions can involve another type of (dark) matter not following the same geodesics as the known matter, and thus effectively violating the WEP. Additionally, note that the local Lorentz invariance of special relativity could be {\it spontaneously} violated in MOND theories. The SEP, on the other hand, states that all laws of physics, including gravitation itself, are fully independent of velocity and location in spacetime. This is obtained in GR by making the physical metric itself obey the Einstein--Hilbert action. This principle \textit{has} to be broken in MOND (see also Sect.~6.3).  We now recall below how GR connects with Newtonian dynamics in the weak-field limit, which is actually the regime in which the modification must be set in order to account for the MOND phenomenology of the ultra-weak field limit. The action of GR writes as the sum of the matter action and the Einstein--Hilbert (gravitational) action\epubtkFootnote{If the action has the units of $\hbar$ the factor in front of the gravitational action is rather $c^3/16\pi G$. And if one wishes to include a cosmological constant $\Lambda$, the integral then rather reads $\int{d^4 x\sqrt{-g}\, (R-2\Lambda)}$}:
\begin{equation}\label{eq:SGR}
S_{\mathrm{GR}} \equiv \vphantom{\int}S_{\mathrm{matter}}[\mathrm{matter},
g_{\mu\nu}] + \frac{c^4}{16\pi G}
\int{d^4 x\sqrt{-g}\, R},
\end{equation}
where $g$ denotes the determinant of the metric tensor $g_{\mu\nu}$ with $(-,+,+,+)$ signature\epubtkFootnote{With this signature the proper-time is defined by $d \tau^2 = -g_{\mu\nu}\, dx^\mu dx^\nu$}, and $R=R_{\mu\nu}g^{\mu\nu}$ is its scalar curvature, $R_{\mu\nu}$ being the Ricci tensor (involving second derivatives of the metric). The matter action is a functional of the matter fields, depending on them and their first derivatives. For instance, the matter action of a free point particle $S_{\rm pp}$ writes:
\begin{equation}\label{eq:Spp}
S_{\rm pp} \equiv - \int m c\, ds =
-\int m c\sqrt{-g_{\mu\nu}(x)\,v^\mu v^\nu}\, dt,
\end{equation}
depending on the positions $x$ and on their time-derivatives $v^\mu$. Varying the matter action w.r.t. matter fields degrees of freedom yields the equations of motion, i.e., the geodesic equation in the case of a point particle:
\begin{equation}
\label{geo}
\frac{d^2 x^\mu}{d\tau^2} = -\Gamma^\mu_{\alpha \beta} \frac{dx^\alpha}{d\tau} \frac{d x^\beta}{d\tau},
\end{equation}
where the proper time $\tau = s$ is approximately equal to $ct$ for slowly moving non-relativistic particles, and $\Gamma^\mu_{\alpha \beta}$ is the Christoffel symbol involving first derivatives of the metric. On the other hand, varying the total action w.r.t. the metric yields Einstein field equations:
\begin{equation}
\label{Ein}
R_{\mu\nu} - \frac{1}{2} R g_{\mu\nu} = \frac{8 \pi G}{c^4} T_{\mu\nu},
\end{equation}
where $T_{\mu\nu}$ is the stress-energy tensor defined as the variation of the Lagrangian density of the matter fields over the metric. 

In the static weak-field limit, the metric writes (up to third order corrections in $1/c^3$)\footnote{Note that, at 1PN, this weak-field metric can also be written as $g_{00} = -e^{2\Phi/c^2} , \; g_{ij} = e^{2\Psi/c^2} \delta_{ij}$. Note also that Taylor expanding Eq.~\ref{eq:Spp} yields $S_{\rm pp} = \int m ({\bf v}^2/2 -  \Phi_N - c^2) \, dt$, so that the sum of the classical kinetic and internal actions for a point particle (see Eq.~\ref{Newtonac}) are now lumped together into the matter action.}:
\begin{equation}
\label{metric}
g_{0i}=g_{i0}=0 \; , \; g_{00} \underbrace{=}_{\mathrm{Taylor}} -1 -  \frac{2\Phi}{c^2} \; , \; g_{ij} = \underbrace{=}_{\mathrm{Taylor}} \left(1 +  \frac{2\Psi}{c^2} \right)  \delta_{ij},
\end{equation}
where, in GR,
\begin{equation} 
\Phi=\Phi_N \; \; {\rm and} \; \; \Psi=-\Phi_N, 
\end{equation}
and $\Phi_N$ is the Newtonian gravitational potential. From the $(0,0)$ components of the weak-field metric, one gets back Newton's second law for massive particles $d^2x^i/dt^2 = - \Gamma^i_{00} = -\partial \Phi_N/dx^i$ from the geodesic equation (Eq.~\ref{geo}). On the other hand, Einstein (Eq.~\ref{Ein}) equations give back the Newtonian Poisson equation $\nabla^2 \Phi_N=4 \pi G \rho$. The metric thus plays the role of the gravitational potential, and the Christoffel symbol plays the role of acceleration. Note however that if time-like geodesics are determined by the $(0,0)$ component of the metric, this is not the case for null geodesics. While the gravitational redshift for light-rays is solely governed by the $g_{00}$ component of the metric too, the deflection of light is on the other hand also governed by the $g_{ij}$ components (more specifically by $\Phi-\Psi$ in the weak-field limit). This means that, in order for the anomalous effects of any modified gravity theory respectively on lensing and dynamics to correspond to a similar\epubtkFootnote{The derived lensing and dynamical masses are typically very close to each other but the data are not yet precise enough to ascertain that they are exactly identical.} amount of ``missing mass'' in GR, it is crucial that $\Psi \simeq -\Phi$ in Eq.~\ref{metric} for such a theory.

\subsection{Scalar-tensor k-essence}

MOND is an acceleration-based modification of gravity in the ultra-weak-field limit, but since the Christoffel symbol, playing the role of acceleration in GR, is not a tensor, it is in principle not possible to make a general relativistic theory depend on it. Another natural way to account for the departure from Newtonian gravity in the weak-field limit and to account for the violation of the SEP inherent to the external field effect is to resort to a scalar-tensor theory, as first proposed by~\cite{BM84}. The added scalar field can play the role of an auxiliary potential, and its gradient then has the dimensions of acceleration and can be used to enforce the acceleration-based modification of MOND.

The relativistic theory of~\cite{BM84} depends on two fields, an ``Einstein metric'' $\tilde{g}_{\mu\nu}$ and a scalar field $\phi$. The physical metric $g_{\mu\nu}$ entering the matter action is then given by a conformal transformation of the Einstein metric\epubtkFootnote{The frame associated to the Einstein metric is called the ``Einstein frame'' as opposed to the ``matter frame'' or ``Jordan frame'', associated to the physical metric.} through an exponential coupling function:
\begin{equation}
\label{conformal}
g_{\mu\nu} \equiv e^{2\phi} \tilde{g}_{\mu\nu}.
\end{equation}
In order to recover the MOND dynamics, the Einstein--Hilbert action (involving the Einstein metric) remains unchanged ($\int{d^4 x\sqrt{-\tilde{g}}\, \tilde{R}}$), and the dimensionless scalar field is given a so-called k-essence action, with no potential and a non-linear, aquadratic, kinetic term\epubtkFootnote{k-essence fields have also recently been reintroduced as possible dark energy fluids, that could also drive inflation~\cite{Damour, Stein, Chiba}. This name comes from the fact that their dynamics is dominated by their kinetic term $f(X)$ (in the case of RAQUAL, there is no potential at all), contrary to other dark energy models such as quintessence, in which the scalar field potential plays the crucial role.} inspired by the AQUAL action of Eq.~\ref{aqualaction}:
\begin{equation}
\label{scalaraction}
S_\phi \equiv - \frac{c^4}{2 k^2 l^2 G}\int{d^4 x\sqrt{-\tilde{g}}\, f(X)},
\end{equation}
where $k$ is a dimensionless constant, $l$ is a length-scale, $X=kl^2\tilde{g}^{\mu\nu}\phi,_{\mu} \phi,_{\nu}$, and $f(X)$ is the ``MOND function''. Since the action of the scalar field is similar to that of the potential in the Bekenstein--Milgrom version of classical MOND, this relativistic version is known as the \textit{Relativistic Aquadratic Lagrangian} theory, RAQUAL.

Varying the action w.r.t. the scalar field yields, in a \textit{static} configuration, the following modified Poisson's equation for the scalar field: 
\begin{equation}
\protect\label{eqn:poissonkessence} 
c^2 \nabla.\left[\nabla \phi f'(k l^2 |\nabla \phi|^2)\right] = k G \rho,
\end{equation}
and the $(0,0)$ component of the physical metric is given by $g_{00}=-e^{2(\Phi_N+c^2\phi)/c^2}$, leading us precisely to the situation of Eq.~\ref{sumpot} in the weak-field, with $\Phi=\Phi_N+c^2\phi$, with 
\begin{equation}
s=(c^2/a_0)|\nabla\phi| = (Xc^4/kl^2a_0^2)^{1/2}
\end{equation}
 and 
\begin{equation}
\tilde{\mu}(s) = (4 \pi c^2/k) f'(X), 
\label{tildemu}
\end{equation}
whose finely tuned relation with the $\mu$-function of Milgrom's law is extensively described in Sect.~6.2. We note that the standard choice for $X \ll 1$ is $f'(X) \sim (X/3)^{1/2}$, meaning that in order to recover $\tilde{\mu}(s)=s/\xi$ for small $s$, where $\xi = G_N/G$ (see Sect.~6.2), one must define the length-scale as 
\begin{equation}
l \equiv (c^2 \sqrt{3k})/(4 \pi \xi a_0).
\end{equation}

It was immediately realized~\cite{BM84} that a k-essence theory such as RAQUAL can exhibit superluminal propagations whenever $f''(X) >0$~\cite{JPB06}. Although it does not threaten causality~\cite{JPB06}, one has to check that the
Cauchy problem is still well-posed for the field equations. It has been shown~\cite{JPB06,Rendall} that it requires the otherwise free
function
$f$ to satisfy the following properties, $\forall X$: 
\begin{eqnarray} 
f'(X) > 0 \\
f'(X) + 2 X f''(X) >0, 
\end{eqnarray}
which is the equivalent of the constraints of Eq.~\ref{constrmu} on Milgrom's $\mu$-function. 

However, another problem was immediately realized at an observational level~\cite{BM84, BekSan}. Because of the conformal transformation of Eq.~\ref{conformal}, one has that $\Psi \neq -\Phi$ in the RAQUAL equivalent of Eq.~\ref{metric}. In other words, as it is well-known that gravitational lensing is insensitive to conformal rescalings of the metric, apart from the contribution of the stress-energy of the scalar field to to the source of the Einstein metric~\cite{BekSan, Bruneton07}, the ``non-Newtonian'' effects of the theory respectively on lensing and dynamics do not \textit{at all} correspond to similar amounts of ``missing mass''. This is also considered a generic problem with any local pure metric formulation of MOND~\cite{SoussaWoodard}.

\subsection{Stratified theory}

A solution to the above gravitational lensing problem due to the conformal rescaling of the metric in RAQUAL has been presented in~\cite{stratified}. Inspired by ``stratified'' theories of gravity~\cite{Ni}, Sanders~\cite{stratified} suggested, in addition to the scalar field $\phi$ of RAQUAL, the use of a \textit{non-dynamical} timelike vector field $U_\mu=(-1,0,0,0)$  with unit-norm $U^2=-1$ (in terms of the Einstein metric), in order to enforce a \textit{disformal} relation between the Einstein and physical metrics:
\begin{equation}
\label{disformal}
g_{\mu\nu} \equiv e^{-2\phi} \tilde{g}_{\mu\nu} - 2\sinh(2\phi) U_\mu U_\nu.
\end{equation}
The second term only affects the $g_{00}$ component, and it then appears immediately that  $\Psi = -\Phi$ in the weak-field limit (rhs terms of Eq.~\ref{metric}), and the problem of lensing is cured. However, the prescription that a 4-vector points in the time direction is not a covariant one, and the theory should involve strong preferred frames effects, although these can now be fully suppressed, as well as any deviation from GR at small distances, with an appropriate additional ``Galileon'' term in addition to the asymptotic deep-MOND k-essence term in the action of the scalar field~\cite{galileon} (the other advantage being that the interpolating function then does not have to be inserted by hand). In any case, endowing the vector field with covariant dynamics of its own has thus been the next logical step in developing relativistic MOND theories.

\subsection{Original Tensor-Vector-Scalar theory}

The idea of the Tensor-Vector-Scalar theory of Bekenstein~\cite{TeVeS}, dubbed TeVeS, is to keep the disformal relation of Eq.~\ref{disformal} between the Einstein metric $\tilde{g}_{\mu\nu}$ and the physical metric $g_{\mu\nu}$ to which matter fields couple, but to replace the above non-dynamical vector field by a dynamical vector field $U_\mu$ with an action ($K$ being a dimensionless constant):
\begin{equation}
S_U \equiv - \frac{c^4}{16 \pi G}\int{d^4 x \sqrt{-\tilde{g}}\, \left[ \frac{K}{2} \tilde{g}^{\alpha\beta} \tilde{g}^{\mu\nu} U_{[\alpha,\mu]} U_{[\beta,\nu]} - \lambda (\tilde{g}^{\mu\nu}U_\mu U_\nu + 1) \right] },
\label{Bekoriginal}
\end{equation}
akin to that of the electromagnetic 4-potential vector field ($U_{[\mu,\nu]}$ playing the role of the Faraday tensor), but without the coupling term to the 4-current, and with a constraint term forcing the unit norm $U^{\mu} U_\nu =\tilde{g}^{\mu \nu}U_\mu U_\nu = -1$ ($\lambda$ being a Lagrange multiplier function, to be determined as the equations are solved). The first term in the integrand takes care of approximately aligning $U_\mu$ with the 4-velocity of matter (when simultaneously solving for (i) the Einstein-like equation of the Einstein metric $\tilde{g}_{\mu\nu}$ and for (ii) the vector equation obtained by varying the total action with respect to $U_\mu$). 

Finally, the k-essence action for the scalar field is kept as in RAQUAL (Eq.~\ref{scalaraction}), but with 
\begin{equation}
X_{\rm teves}=kl^2(\tilde{g}^{\mu\nu}-U^\mu U^\nu) \phi,_{\mu} \phi,_{\nu}.
\label{XTEVES}
\end{equation} 
Contrary to RAQUAL, this scalar field exhibits no superluminal propagation modes.  However,~\cite{Bruneton07} noted that such superluminal propagation might have to be re-introduced in order to avoid excessive Cherenkov radiation and suppression of high-energy cosmic rays (see also~\cite{MilCher}).

The static weak-field limit equation for the scalar field is precisely the same as Eq.~\ref{eqn:poissonkessence}, and the scalar field enters the static weak field metric Eq.~\ref{metric} as $\Phi =-\Psi = \Xi \Phi_N + c^2 \phi$ meaning that lensing and dynamics are compatible, with $\Xi$ being a factor depending on $K$ and on the cosmological value of the scalar field~(see Eq. 58 of \cite{TeVeS}). This can be normalized to yield $\Xi=1$ at redshift zero. Again, all the relations between the free function $f$ and Milgrom's $\mu$-function can be found in Sect.~6.2 (see also~\cite{FGBZ, SkorZlos}).

This theory has played a true historical role as a \textit{proof of concept} that it was possible to construct a fully relativistic theory both enhancing dynamics and lensing in a coherent way and reproducing the MOND phenomenology for static configurations with the dynamical 4-vector pointing in the time direction. However, the question remained whether these static configurations would be stable. What is more, although a classical Hamiltonian\epubtkFootnote{Expressed in terms of $U_\mu$ and its congugate momenta $P^{\mu} = \partial L / \partial \dot{U}_\mu$} unbounded from below in flat spacetime would not \textit{necessarily} be a concern at the classical level (and even less if the model is only ``phenomenological"), it would inevitably become a worry for the existence of a stable quantum vacuum (see however~\cite{horwitz}). And indeed, it was shown in~\cite{Clayton} that models with such ``Maxwellian" vector fields having a TeVeS-like Lagrange multiplier constraint in their action have a corresponding Hamiltonian density that can be made arbitrarily large and negative (see also Sect. IV.A of~\cite{Bruneton07}). What is more, even at the classical level, it has been shown that spherically symmetric solutions of TeVeS are heavily unstable~\cite{Seifert1,Seifert2}, and that this type of vector field causes caustic singularities~\cite{Contaldi}, in the sense that the integral curves of the vector are timelike geodesics meeting each other when falling into gravity potential wells. Another form was thus needed for the action of the TeVeS vector field.

\subsection{Generalized Tensor-Vector-Scalar theory}

The generalization of TeVeS was proposed by Skordis~\cite{genTeVeS}. Inspired by the fact that Einstein-Aether theories~\cite{EA,Jacobson} also present instabilities when the unit-norm vector field is ``Maxwellian" as above, it was simply proposed to use a more general Lagrangian density for the vector field, akin to that of Einstein-Aether theories:
\begin{equation}
\label{EAaction}
S_U \equiv - \frac{c^4}{16 \pi G}\int{d^4 x \sqrt{-\tilde{g}}\, \left[ K^{\alpha \beta \mu \nu} U_{\beta,\alpha} U_{\nu,\mu} - \lambda (\tilde{g}^{\mu\nu}U_\mu U_\nu + 1) \right] },
\end{equation}
where
\begin{equation}
\label{K}
K^{\alpha \beta \mu \nu} = c_1 \tilde{g}^{\alpha \mu} \tilde{g}^{\beta \nu} + c_2 \tilde{g}^{\alpha \beta} \tilde{g}^{\mu \nu}+ c_3 \tilde{g}^{\alpha \nu} \tilde{g}^{\beta \mu} + c_4 U^\alpha U^\mu \tilde{g}^{\beta \nu}
\label{ci}
\end{equation}
for a set of constants $c_1, c_2, c_3, c_4$. Interestingly, spherically symmetric solutions depend only on the combination $c_1-c_4$, not on $c_2$ and $c_3$ that can in principle be chosen to avoid the instabilities of the original TeVeS theory. The original unstable theory is of course also included in this generalization through a specific combination of the four $c_i$ (see, e.g.,~\cite{SkorZlos}).

This generalized version is thus the current ``working version'' of what is now called TeVeS: a tensor-vector-scalar theory with an Einstein-like metric, an Einstein-Aether-like unit-norm vector field, and a k-essence-like scalar field, all related to the physical metric through Eq.~\ref{disformal}. It has been extensively studied, both in its original and generalized form. It has for instance been shown that, contrary  to many gravity theories with a scalar sector, the theory evidences no cosmological evolution of the Newtonian gravitational constant $G$ and only minor evolution of Milgrom's constant $a_0$~\cite{FGBZ, BekSagi}. The fact that the latter is still put in by hand through the length-scale of the theory $l \sim c^2/a_0$, and has no dynamical connection with the Hubble or cosmological constant is however perhaps a serious conceptual shortcoming, together with the free function put by hand in the action of the scalar field (but see~\cite{galileon} for a possible solution to the latter shortcoming). The relations between this free function and Milgrom's $\mu$ can be found in~\cite{FGBZ, SkorZlos} (see also Sect.~6.2), the detailed structure of null and timelike geodesics of the theory in~\cite{SkorZlos}, the analysis of the parametrized post-Newtonian coefficients (including the preferred-frame parameters quantifying the local breaking of Lorentz invariance) in~\cite{Giannios05, Sagi09, Sanderssun, Tamaki08}, solutions for black holes and neutron stars in~\cite{Lasky09, Lasky11, Lasky08, Lasky10, Sagi08,Sotani1,Sotani2}, and gravitational waves in~\cite{Kahya1,Kahya2,Kahya3,Sagi10}. It is important to remember that TeVeS is \textit{not} equivalent to GR in the strong regime, which is why it can be tested there, e.g. with binary pulsars or with the atomic spectral lines from the surface of stars~\cite{dedeo}, or other very strong field effects\epubtkFootnote{It is also important to remember that some interpolation functions (Sect.~6.2) are already excluded by Solar system tests, and it is thus useless to exclude these over and over again.}. However, these effects can always generically be suppressed (at the price of introducing a Galileon type term in the action \cite{galileon}), and such tests would never test MOND as a paradigm. It is by testing gravity in the weak field regime that MOND can really be put to the test.

Finally, let us note that TeVeS (and its generalization) has been shown to be expressible (in the ``matter frame'') \textit{only} in terms of the physical metric $g_{\mu\nu}$, and the vector field $U_\mu$~\cite{Zlosnik06}, the scalar field being eliminated from the equations through the ``unit-norm'' constraint in terms of the Einstein metric $\tilde{g}^{\mu \nu}U_\mu U_\nu = -1$, leading to $g^{\mu \nu}U_\mu U_\nu = -e^{-2 \phi}$. In this form, TeVeS is sometimes thought of as GR with an additional ``dark fluid'' described by a vector field~\cite{ZhaoTeV}. 

\subsection{Bi-Scalar-Tensor-Vector theory}

In TeVeS~\cite{TeVeS}, the ``MOND function'' $f(X_{\rm teves})$ of Eq.~\ref{scalaraction}, where $X_{\rm teves} \sim (\tilde{g}^{\mu\nu}-U^\mu U^\nu) \phi,_{\mu} \phi,_{\nu}$, could also be expressed as a potential $V$ of a non-dynamical scalar field $q$, i.e., a scalar action for TeVeS of the form:
\begin{equation}
S_\phi \propto - \int{d^4 x\sqrt{-\tilde{g}}\, \left[ \frac{1}{2}q^2 X_{\rm teves} + V(q) \right] }
\end{equation}
After variation of the action w.r.t. this non-dynamical field, one gets $qX=-V'(q)$, and variation w.r.t. to $\phi$ yields the usual BM Poisson equation for $\phi$ (Eq.~\ref{BM}), with $q^2 \propto \tilde{\mu}(\sqrt{X})$. Inspired by an older theory (Phase Coupling Gravity~\cite{PCG1,PCG2}) devised in 
a partially successful attempt to eliminate superluminal propagation from RAQUAL (but plagued with the same gravitational lensing problem as RAQUAL, and with additional instabilities), Sanders~\cite{BSTV} proposed to make this field dynamical by adding a kinetic term $\tilde{g}^{\mu\nu} q,_{\mu} q,_{\nu}$ in the action, leading to the following very general action for the scalar fields $\phi$ and $q$:
\begin{equation}
S_{(\phi \, q)} \propto - \int{d^4 x\sqrt{-\tilde{g}}\, \left[ \frac{1}{2}(\tilde{g}^{\mu\nu} q,_{\mu} q,_{\nu} + H(q) (\tilde{g}^{\mu\nu}+U^\mu U^\nu) \phi,_{\mu} \phi,_{\nu}) - F(q) U^\mu U^\nu \phi,_{\mu} \phi,_{\nu} + V(q) \right] }.
\end{equation}
In this theory (dubbed BSTV for bi-scalar-tensor-vector theory), the physical metric has the same \textit{a priori} form as in TeVeS, meaning that $\phi$ is the matter-coupling scalar field, while $q$ only influences the strength of that coupling. A remarkable achievement of the theory is that the quasi-static field equation for $\phi$ can be obtained only in a cosmological context, and thereby naturally explains the connection between $a_0$ and $H_0$~\cite{BSTV}. What is more, oscillations of the $q$ field around its expectation value can be considered as massive dark matter, and is allowing an explanation of the peaks of the angular power spectrum of the Cosmic Microwave Background~\cite{BSTV}. Unfortunately, various instabilities and a Hamiltonian unbounded by below have been evidenced in Sect. IV.A of~\cite{Bruneton07}, thus most likely ruling out this theory, at least in its present form.

\subsection{Non-minimal scalar-tensor formalism}

As a consequence of the inability of RAQUAL (the scalar-tensor k-essence of Sect.~7.1) to enhance gravitational lensing, all other attempts reviewed so far (Sect.~7.2 to Sect.~7.5) have been plagued with an aesthetically unpleasant growth of additional fields and free parameters. This has led Bruneton \& Esposito-Far\`ese~\cite{Bruneton07} to consider models with fewer additional fields. They first considered pure metric theories in which matter is not only coupled to the metric but also non-minimally to its curvature (Eqs 5.1 and 5.2 of~\cite{Bruneton07}). While they showed that such models can indeed reproduce the MOND dynamics, they also concluded that they are generically unstable if locality is to be preserved (but see Sect.~7.10). They then considered models in which at most \textit{one} scalar field is added, without any additional vector field, but where this field is coupled non-minimally to matter, in the sense that the matter-coupling depends on the scalar field itself but also on its first derivatives. In other words, the gradient of the scalar field is replacing the dynamical vector field of TeVeS. The simple scalar field action is just the normal action of a massive scalar field:
\begin{equation}
S_\phi = - \frac{c^4}{8 \pi G l^2}\int{d^4 x\sqrt{-\tilde{g}}\, [X + 2V(\phi)]},
\end{equation}
with $X=l^2 \tilde{g}^{\mu\nu}\phi,_{\mu} \phi,_{\nu}$ and $V(\phi) = l^2 m^2 \phi^2 /2$.  The physical metric $g_{\mu\nu}$ is then disformally related to the Einstein metric through (see Eq.~5.11 of~\cite{Bruneton07}):
\begin{equation}
g_{\mu\nu} \equiv  A^2 \tilde{g}_{\mu\nu} + B \phi,_{\mu} \phi,_{\nu},
\end{equation}
with the functionals 
\begin{equation}
A(\phi, X)=e^{\eta \phi} - \phi h(Y) Y/\eta \; , \;  B(\phi,X)=-4 \phi \eta^{-1} Y/X,
\end{equation}
where $Y=(\eta a_0)^{1/2}c^{-1} X^{-1/4}$. The free function $h(Y)$ is the ``MOND function'' playing the role of Milgrom's $\mu$. An alternative formulation of the model is obtained by separating the matter action into a normal matter action and an ``interaction term'' between the scalar field, the metric and the matter fields~\cite{BLSF}. Considering the massive scalar field as a dark matter fluid, this model can thus be interpreted as non-standard baryon-dark matter interaction leading to the MOND behavior. If the scalar mass $m$ is small enough, it is a pure MOND theory, but if it is higher, it can lead to a ``DM+ MOND'' behavior, especially noteworthy in regions of high gravity such as the center of galaxy clusters (see Sect.~\ref{clusters} and discussions in~\cite{BLSF}). Let us note that, while this theory exhibits superluminal propagations outside of matter, it is in principle not a problem for causality~\cite{JPB06}. It has also been possible to study the behavior of the theory \textit{within} matter, e.g., within the dilute HI gas inside galaxy disks (an analysis which is mostly too difficult to perform in other models reviewed so far): this led to a deadly problem, i.e., that the Cauchy problem becomes ill-posed and the solutions to field equations ill-defined. A possible solution was proposed in~\cite{BLSF}, namely to make the matter coupling (or, equivalently, the baryon-scalar DM interaction) depend on the local density of matter\epubtkFootnote{A characteristic matter density $\rho_0$ thus becomes an additional order parameter, in the spirit of the velocity scale $s_0$ of~\cite{bektalk}, see Eq.~3.1 of~\cite{BLSF}}: this can also lead to an interesting phenomenology, where only gas-rich systems behave according to Milgrom's law, while others would behave in a CDM way~\cite{FamaeyNZ}. A lot remains to be studied within this framework.

\subsection{Generalized Einstein-Aether theories} 

All theories reviewed so far are best expressed in the ``Einstein frame'', and involve an \textit{a priori} form for the physical metric to which matter couples (an \textit{a priori} form expressed as a function of the Einstein metric and of the other additional fields). However, the work of~\cite{Zlosnik06} has shown that, for instance, TeVeS (Sects. 7.3 and 7.4) is expressible as a pure Tensor-Vector theory in the matter frame, and that the physical metric then both satisfies the Einstein--Hilbert action and couples \textit{minimally} to the matter fields, just like in GR. In fact, the modification of gravity in TeVeS thus only comes from the coupling of the physical metric to the vector field. The idea of Zlosnik et al.~\cite{GEA} was then that a similar, but simpler, modification of gravity could be obtained by devising a simple Tensor-Vector theory in the matter frame, with no \textit{a priori} on the geometry of the physical metric. Starting from the extensively studied Einstein-Aether theories~\cite{EA,Jacobson}, with a vector action of the type of Eq.~\ref{EAaction}, the idea is to make the k-essence free function $f(X)$ (the ``MOND function'' of Eq.~\ref{scalaraction}) act directly on the vector field rather than on an additional scalar field. This thus leads to \textit{vector k-essence}, or \textit{Generalized Einstein-Aether} (GEA) theories (also called non-canonical Einstein-Aether theories), in which the Einstein--Hilbert and matter actions remain as in GR, but with an additional unit-norm vector field with the following action~\cite{SkorZlos,GEA}:
\begin{equation}
\label{GEAaction}
S_U \equiv - \frac{c^4}{16 \pi G l^2}\int{d^4 x \sqrt{-g}\, \left[ f(X_{\rm gea}) - l^2 \lambda (g^{\mu\nu}U_\mu U_\nu + 1) \right] },
\end{equation}
where (see Eq.~\ref{K} and replacing $\tilde{g}^{\mu \nu}$ by $g^{\mu \nu}$)
\begin{equation}
X_{\rm gea} = l^2 K^{\alpha \beta \mu \nu} U_{\beta,\alpha} U_{\nu,\mu}.
\label{XGEA}
\end{equation}
The unit-norm constraint fixes the vector field in terms of the metric, and from there we have that, in the weak-field limit, $X_{\rm gea} \propto - |\nabla \Phi|^2$, with $\Phi$ defined as in Eq.~\ref{metric}. The Einstein equation in the weak-field limit then yields a BM type of Poisson equation (Eq.~\ref{BM}) for the full gravitational potential $\Phi$, with $\mu = f' + (1-f')/(1-C/2)$ and $C=c_1-c_4$~\cite{SkorZlos}. In the deep-MOND limit, the usual choice for $f$ is of the type $f(X_{\rm gea}) \propto (-X_{\rm gea})^{3/2}+2X_{\rm gea}/C$, and the length-scale must be fixed as:
\begin{equation}
l \equiv  \frac{(2-C)c^2}{3/2 C^{3/2} a_0}.
\end{equation}
Let us note that this weak-field limit of GEA theories is different from that of RAQUAL or TeVeS, where only the scalar field $\phi$ obeys a BM-like equation governed by an interpolating function $\tilde{\mu}(s)$, and where the total potential is given by Eq.~\ref{sumpot}. 

The remarkable feature of GEA theories allowing for the desired enhancing of gravitational lensing without any apriori on the form of the physical metric is that, writing the metric as in Eq.~\ref{metric}, it can be shown~\cite{SkorZlos} that in the limit $X_{\rm gea} \rightarrow 0$ the action of Eq.~\ref{GEAaction} is only a function of $\Upsilon = \Phi + \Psi$ and is thus \textit{invariant} under disformal transformations $[\Phi \rightarrow \Phi + \beta(r) \, ; \,  \Psi \rightarrow \Psi - \beta(r)]$, of the type of Eq.~\ref{disformal}. These GEA theories are currently extensively studied, mostly in a cosmological context (see Sect.~9), but also for their parametrized post-Newtonian coefficients in the Solar system~\cite{Bonvin} or for black hole solutions~\cite{Tamaki}.

Interestingly, it has been shown that all these vector field theories (TeVeS, BSTV, GEA) are all part of a broad class of theories studied in~\cite{Halle}. Yet other phenomenologically interesting theories exist among this class, such as, for instance, the $V\Lambda$ models considered by Zhao \& Li~\cite{ZhaoLi1, ZhaoLi2, ZhaoLi3} with a \textit{dynamical norm} vector field, whose norm obeys a potential (giving it a mass) and has a non-quadratic kinetic term {\` a}-la-RAQUAL, in order to try reproducing both the MOND phenomenology and the accelerated expansion of the Universe, while interpreting the vector field as a fluid of neutrinos with varying mass~\cite{zhaoneut2,zhaoneut}. This has the advantage of giving a microphysics meaning to the vector field. Such vector fields have also been argued to arise naturally from dimensional reduction of higher dimensional gravity theories~\cite{Bek06, Mavromatos}, or, more generally, to be necessary from the fact that quantum gravity could need a preferred rest frame~\cite{EA} in order to protect the theory against instabilities when allowing for higher derivatives to make the theory renormalizable (e.g., in Ho{\v r}ava gravity~\cite{BPS,Horava}). Inspired by this possible need of a preferred rest frame in quantum gravity, relativistic MOND theories boiling down to particular cases of GEA theories in which the vector field is hypersurface-orthogonal have, for instance, been proposed in~\cite{BlanchetMarsat,SandersHo}.

\subsection{Bimetric theories}

In the previous theories, the acceleration-dependence of MOND enters the equations through a free ``MOND function'' $f(X)$ acting either on the contracted gradient of an added scalar field, with dimensions of acceleration (Eq.~\ref{XTEVES}), or on a scalar formed with the first derivatives of a vector field (Eq.~\ref{XGEA}) with a unit-norm constraint relating it to the gradient of the potential in the physical metric. The ``MOND function'' could not act directly on the Christoffel symbol because this is not a tensor, and such a theory would thus violate general covariance. However, if there is more than one metric entering gravitation, the difference between the associated Christoffel symbols \textit{is} a tensor, and one can construct from it a scalar with dimensions of acceleration, on which the ``MOND function'' can act. Such theories  in which there are two dynamical rank-2 symmetric tensor fields are called \textit{bimetric theories}~\cite{Isham1, Isham2, Rosen}. Milgrom~\cite{bimond,bimond2} proposed to construct a whole parametrized class of bimetric MOND theories (dubbed BIMOND), involving an auxiliary metric, with various phenomenological behaviors in the weak-field limit, ranging from Bekenstein--Milgrom MOND to QUMOND as well as a mix of both (see~\cite{MDorDM}). As one example (parameters $\alpha=-\beta=-1$ in the general class of BIMOND theories, for which we refer the reader to the review~\cite{MDorDM}), the auxiliary metric $\hat{g}_{\mu \nu}$ can e.g. be introduced precisely in the same way as the auxiliary potential $\Phi_{\rm ph}$ in the QUMOND classical action of Eq.~\ref{qumondactbis} :
\begin{equation}
S \equiv \vphantom{\int}S_{\rm m}[{\rm matter},
g_{\mu\nu}] + \vphantom{\int}S_{\rm m}[{\rm twin \, matter},
\hat{g}_{\mu\nu}] + \frac{c^4}{16\pi G} \int{d^4 x\sqrt{-g}\, [R - \hat{R} - 2 l^{-2} f(X_{\rm bimond})]},
\label{bimondact}
\end{equation}
where $l \equiv c^2/a_0$, and
\begin{equation}
X_{\rm bimond} = l^2 g^{\mu \nu} (C^{\alpha}_{\mu \beta} C^{\alpha}_{\nu \beta} - C^{\alpha}_{\mu \nu}C^{\beta}_{\beta \alpha}),
\end{equation}
where  $C^{\alpha}_{\mu \nu} = \Gamma^{\alpha}_{\mu \nu} -  \hat{\Gamma}^{\alpha}_{\mu \nu}$. The MONDian modification of gravity is thus introduced through the interaction between the space-time on which matter lives and the auxiliary space-time (on which some ``twin matter'' might live). This modification is acceleration-based since the interaction involves the difference of Christoffel symbols, playing the role of acceleration. By varying the action w.r.t. both metrics, we obtain two sets of Einstein-like equations, which boil down in the static weak-field limit to $\hat{\Phi}=-\hat{\Psi}$ and $\Phi=-\Psi$ in Eq.~\ref{metric} (so this yields the correct amount of gravitational lensing for normal photons w.r.t. the ``matter metric" $g_{\mu\nu}$), as well as the following generalized Poisson equations:
\begin{equation}
\nabla^2 \Phi = 4 \pi G \rho + \nabla . [f'(|\nabla(\Phi - \hat{\Phi})|^2/a_0^2) \nabla(\Phi - \hat{\Phi})] \; {\rm and} \; \nabla^2 \hat{\Phi} = 4 \pi G \hat{\rho} + \nabla . [f'(|\nabla(\Phi - \hat{\Phi})|^2/a_0^2) \nabla(\Phi - \hat{\Phi})].
\end{equation}
or, equivalently,
\begin{equation}
\nabla^2 (\Phi -  \hat{\Phi}) = 4 \pi G (\rho - \hat{\rho}) \; {\rm and} \; \nabla^2 \Phi = 4 \pi G \rho + \nabla . [f'(|\nabla(\Phi - \hat{\Phi})|^2/a_0^2) \nabla(\Phi - \hat{\Phi})] .
\end{equation}
This is equivalent to QUMOND (Eq.~\ref{QUMOND}) if the matter and twin matter are well separated (which is natural if they repel each other), the function $f$ playing the role of $H$ in Eq.~\ref{qumondactbis}, with $f'(X_{\rm bimond}) \rightarrow 0$ for $X_{\rm bimond} \gg 1$ and $f'(X_{\rm bimond}) \rightarrow X_{\rm bimond}^{-1/4}$ for $X_{\rm bimond} \ll 1$. Note that the existence of this putative twin matter is far from being necessary (putting $\hat{\rho} = 0$ everywhere yields exactly QUMOND), but it might be suggested by the existence of the auxiliary metric within the theory. Again, it is mandatory to stress that the formulation of BIMOND sketched above is actually far from unique and can be suitably parametrized to yield a whole class of BIMOND theories with various phenomenological behaviors~\cite{bimond, bimond2,MDorDM}. For instance, in matter-twin matter symmetric versions of BIMOND ($\alpha=\beta=1$, see~\cite{MDorDM}), and within a fully symmetric matter-twin matter system, a cosmological constant is given by the zero-point of the MOND function, naturally of the order of 1, thereby naturally leading to $\Lambda \sim a_0^2$ for the large-scale Universe. Matter and twin matter would not interact at all in the high-acceleration regime, and would repel each other in the MOND regime (i.e., when the acceleration difference of the two sectors is small compared to $a_0$), thereby possibly playing a crucial role in the Universe expansion and structure formation~\cite{bimond3}.

This promising broad class of theories should be carefully theoretically investigated in the future, notably against the existence of ghost modes~\cite{Boulanger}. At a more speculative level, this class of theories can be interpreted as a modification of gravity arising from the interaction between a pair of membranes: matter lives on one membrane, twin matter on the other, each membrane having its own standard elasticity but coupled to the other one. The way the shape of the membrane is affected by matter then depends on the combined elasticity properties of the double membrane, but matter response depends only on the shape of its home membrane. Interestingly, bimetric theories have also been advocated~\cite{Manrique} to be a useful ingredient for the renormalizability of quantum gravity (although they currently considered theories with only metric interactions, not derivatives like in BIMOND).

\subsection{Dipolar dark matter}
\label{sec:dipolar}

As we have seen, many relativistic MOND theories do invoke the existence of new ``dark fields'' (scalar or vector fields), which, if massive, can even sometimes truly be thought of as ``dark matter'' enjoying non-standard interactions with baryons\epubtkFootnote{In the case of TeVeS and GEA theories, the dark fields do not really count as dark matter because their energy density is subdominant to the baryonic one.} (Sect.~7.6 and~\cite{BLSF}). The bimetric version of MOND (Sect.~7.8) also invokes the existence of a new type of matter, the ``twin matter''. This clearly shows that, contrary to common misconceptions, MOND is not necessarily about ``getting rid of dark matter'' but rather about reproducing the success of Milgrom's law in galaxies. It might require adding new fields, but the key point is that these fields, very massive or not, would \textit{not} behave simply as collisionless particles.

In a series of papers, Blanchet \& Le Tiec~\cite{Blanchet1,Blanchet2,Blanchet3,Blanchet4,Blanchet5,Blanchet} have pushed further the idea that the MOND phenomenology could arise from the fundamental properties of a form of dark matter itself, by suggesting that dark matter could carry a space-like\epubtkFootnote{This is to be contrasted with the time-like nature of TeVeS and GEA vector fields in the static weak-field limit} four-vector gravitational dipole moment $\xi^{\mu}$, following the analogy between Milgrom's law and Coulomb's law in a dieletric medium proposed by \cite{Blanchet2} (see Eq.~\ref{coulomb}) or between the Bekenstein--Milgrom modified Poisson equation and Gauss' law in terms of free charge density (see Eq.~\ref{BM}). The dark matter medium is described as a fluid with mass current $J^\mu = \rho u^\mu$ (where $\rho$ is the equivalent of the mass density of the atoms in a dielectric medium, i.e., it is the ordinary mass density of a pressureless perfect fluid, and $u^\mu$ is the four-velocity of the fluid\footnote{And the current $J^\mu$ is conserved, i.e., $\nabla_\mu J^\mu = 0$}) endowed with the dipole moment vector $\xi^\mu$ (which will affect the total density in addition to the above mass density $\rho$), with the following action~\cite{Blanchet}:
\begin{equation}
S_{\rm DM} \equiv \int{d^4 x\sqrt{-g}\, [ c^2 (J_\mu \dot{\xi}^\mu -\rho) - W(P) ]},
\label{ddmact}
\end{equation}
where $P$ is the norm of the projection {\it perpendicular} to the four-velocity (not the norm of the polarization field\epubtkFootnote{It can be shown that only the projection perpendicular to the four-velocity enters the field equations deduced from the action of Eq.~\ref{ddmact}. Thus the dipole moment is always fully space-like.}) of the polarization field $P^\mu = \rho \xi^\mu$, and where the dot denotes the covariant proper time derivative. The specific dynamics of this dark matter fluid will thus arise from the coupling between the current and the dipolar field (analogue to the coupling to an external polarization field in electromagnetism), as well as from the internal non-gravitational force acting on the dipolar dark particles and characterized by the potential $W(P)$. Let us note that the normal matter action and the gravitational Einstein--Hilbert action are just the same as in GR.

The equations of motion of the dark matter fluid are then gotten by varying the action w.r.t. the dipole moment variable $\xi^\mu$ and w.r.t. to the current $J^\mu$, boiling down in the non-relativistic limit to:
\begin{equation}
\frac{d{\bf v}}{dt} = {\bf g} - {\bf f},
\label{fluidmotion}
\end{equation}
\begin{equation}
\frac{d^2 \mbox{\boldmath$\xi$}}{dt^2} = {\bf f} + \frac{1}{\rho}\nabla [W(P) - P W'(P)] + ({\bf P}  \nabla){\bf g} , 
\label{dipolemotion}
\end{equation}
where ${\bf v}$ is the ordinary velocity of the fluid, ${\bf g} = -\nabla \Phi$ is the gravitational field, and ${\bf f} = -({\bf P}/P) W'/\rho$ is the internal non-gravitational force field making the dark particles motion non-geodesic. What is more, the Poisson equation in the weak-field limit is recovered as:
\begin{equation}
-\nabla . ({\bf g} - 4 \pi {\bf P}) = 4 \pi G (\rho_b + \rho).
\label{poissondipolar}
\end{equation}
In order to then reproduce the MOND phenomenology in galaxies, the next step is the so-called ``weak-clustering hypothesis'', namely the fact that, in galaxies, the dark matter fluid does not cluster much ($\rho \ll \rho_b$) and is essentially at rest (${\bf v} = 0$) because the internal force of the fluid precisely balances the gravitational force, in such a way that the polarization field ${\bf P}$ is precisely aligned with the gravitational one ${\bf g}$, and $g \propto -W'(P)$. The potential thus plays the role of the ``MOND function'', and e.g. choosing to determine it up to third order in expansion as 
\begin{equation}
W(P) \propto \Lambda/(8 \pi) + 2 \pi P^2 + 16 \pi^2 P^3/(3a_0) + {\cal O}(P^4)
\end{equation}
then yields the desired MOND behavior in Eq.~\ref{poissondipolar}, with the $n=1$ ``simple'' $\mu$-function (see Eqs.~\ref{simplemu} and \ref{nfamily}).

This model has many advantages. The monopolar density of the dipolar atoms $\rho$ will play the role of CDM in the early Universe, while the minimum of the potential $W(P)$ naturally adds a cosmological constant term, thus making the theory precisely equivalent to the $\Lambda$CDM model for expansion and large scale structure formation. The dark matter fluid behaves like a perfect fluid with zero pressure at first order cosmological perturbation around a FLRW background and thus reproduces CMB anisotropies. Let us also note that, if the potential $W(P)$ defining the internal force of the dipolar medium is to come from a fundamental theory at the microscopic level, one expects that the dimensionless coefficients in the expansion all be of order unity after rescaling by $a_0^2$, thus naturally leading to the coincidence $\Lambda \sim a_0^2$. 

However, while the weak clustering hypothesis and stationarity of the dark matter fluid in galaxies are suppported by an exact and stable solution in spherical symmetry~\cite{Blanchet4}, it remains to be seen whether such a configuration would be a natural outcome of structure formation within this model. The presence of this stationary DM fluid being necessary to reproduce Milgrom's law in stellar systems, this theory looses a bit of the initial predictability of MOND, and inherits a bit of the flexibility of CDM, inherent to invoking the presence of a DM fluid. This DM fluid could, e.g., be absent from some systems such as the globular clusters Pal 14 or NGC 2419 (see Sect.~6.6.3), thereby naturally explaining their apparent Newtonian behavior. However, the weak clustering hypothesis in itself might be problematic for explaining the missing mass in galaxy clusters, due to the fact that the MOND missing mass is essentially concentrated in the central parts of these objects (see Sect.~6.6.4). 

\subsection{Non-local theories and other ideas}

All the models hereabove somehow invoke the existence of new ``dark fields'', notably because for \textit{local} pure metric theories, the Hamiltonian is generically unbounded from below if the action depends on a finite number of derivatives~\cite{Bruneton07, Esposito, SoussaWoodard}. A somewhat provocative solution would thus be to consider \textit{non-local} theories. A non-local action could, e.g., arise as an effective action due to quantum corrections from super-horizon gravitons~\cite{Soussa2}. Deffayet, Esposito-Far\`ese \& Woodard~\cite{DEW} have notably exhibited the form that a pure metric theory of MOND could take in order to yield MONDian dynamics and MONDian lensing for a static, spherically symmetric baryonic source.

In such a static spherically symmetric geometry, the Einstein--Hilbert action of Eq.~\ref{eq:SGR} can be rewritten in the weak-field expansion as~\cite{DEW}:
\begin{equation}
S_{\rm EH} = {\rm surfaceterm} +  \frac{c^4}{16\pi G} \int{d^4 x [-rab' +a^2/2 + {\cal O}(a^3,b^3)]},
\end{equation}
where $(1+a)$ and $-(1+b)$ are the weak-field $g_{rr}$ and $g_{00}$ components of the static weak-field metric, respectively. The MOND modification to this action implies to obtain as a solution in the deep-MOND limit $a=rb'=2(GMa_0)^{1/2}/c^2$, where the first equality ensures that lensing and dynamics are consistent, leading to the following tentative action in the ultra-weak-field limit~\cite{DEW}:
\begin{equation}
S_{\rm MOND} \sim \frac{c^4}{16\pi G} \int{d^4 x \left[ lr^2 \left(\frac{\alpha}{3} \left(b' - \frac{a}{r} \right) - \frac{b'^3}{6} + {\cal O}(a^4,b^4) \right) \right]} ,
\end{equation}
where $l \equiv c^2/a_0$ and $\alpha$ is an arbitrary constant. While it is \textit{impossible} to express this form of the action as a local functional of a general metric, Deffayet et al.~\cite{DEW} showed that it was entirely possible to do so in a non-local model, making use of the non-local inverse d'Alembertian and of a TeVeS-like vector field, introduced not as an additional ``dark field'', but as a non-local functional of the metric itself (by e.g. normalizing the gradient of the volume of the past light-cone). A whole class of such models is constructible, and a few examples are given in~\cite{DEW}, for which stability analyses are still needed, though. 

As already mentioned in Sect.~6.1.1, this non-locality was also inherent to classical toy models of ``modified inertia''. In GR, this would mean making the matter action of a point particle (Eq.~\ref{eq:Spp}) depend on all derivatives of its position, but such models are very difficult to construct~\cite{Mil94} and no fully-fledged theory exists along these lines. A few interesting heuristic ideas have however been proposed in this context. For instance, Milgrom~\cite{unruh} proposed that the inertial force in Newton's second law could be defined to be proportional to the difference between the Unruh temperature and the Gibbons--Hawking one. It is indeed well-known that, in Minkowski space-time, an accelerated observer sees the vacuum as a thermal bath with a temperature proportional to the observer's acceleration $T_U = a h/(4 \pi^2 k c)$~\cite{unruh2,unruh1} where $h$ is the Planck constant and $k$ the Boltzmann constant. On the other hand, a constant-accelerated observer in de~Sitter space-time (curved with a positive cosmological constant $\Lambda$) sees a non-linear combination of that vacuum radiation and of the Gibbons--Hawking radiation (with temperature $T_{\mathrm{GH}}= (\Lambda/3)^{1/2} h/(4 \pi^2 k)$~\cite{GibbonsHawking}) due to the cosmological horizon in the presence of a positive $\Lambda$.  Namely, the Unruh temperature of the radiation seen by such an accelerated observer in de~Sitter spacetime is~\cite{GibbonsHawking} $T_U =  (a^2 + c^2 \Lambda/3)^{1/2} h/(4 \pi^2 k c)$. The idea of Milgrom~\cite{unruh} is to then define the right-hand side of the norm of Newton's second law as being proportional to the difference between the two temperatures:
\begin{equation}
|{\bf F}|= \frac{4 \pi^2 m k c}{h} (T_U - T_{\mathrm{GH}}),
\end{equation}
which trivially leads to $F = m \mu(a/a_0) a$ with $a_0 \equiv c (\Lambda/3)^{1/2}$ (which is however observationally too large by a factor $2 \pi$) and the interpolating function $\mu(x)$ having the exact form of Eq.~\ref{unruhmu}. In short, observers experiencing a very small acceleration would see an Unruh radiation with a small temperature close to the Gibbons--Hawking one, meaning that the inertial resistance defined by the difference between the two radiation temperatures would be smaller than in Newtonian dynamics, and thus the corresponding acceleration would be larger. However, no relativistic version (if at all possible) of this approach has been developed yet: a few difficulties arise due to the direction of the acceleration, or by the fact that stars in galaxies are free-falling objects along geodesics, and not accelerated by a non-gravitational force as in the case of basic Unruh radiation. It was interestingly noted~\cite{modinproc} that the de~Sitter space-time could be seen as a 4-dimensional pseudo-sphere embedded in a 5-dimensional flat Minkowski space, and that the acceleration of a constant-accelerated observer in this flat space would be exactly $a_5 = (a^2 + c^2 \Lambda/3)^{1/2}$. Then , MOND could arise from symmetry arguments in this 5-dimensional space similar to those leading to special relativity in Minkowski space~\cite{modinproc}. Interestingly, arguments very similar to this whole vacuum radiation approach have also recently been made in the context of entropic gravity~\cite{Ho1,Ho2,Klink,Verlinde}. Finally, another interesting idea to get MOND dynamics has been the tentative modification of special relativity, making the Planck length and the length $l=\Lambda^{-1/2} \sim c^2/a_0$ two new invariants in addition to the speed of light, an attempt known as Triply Special Relativity~\cite{smolin}. In any case, despite all these attempts, there is still no fully-fledged theory of MOND at hand which would derive from first principles, and the quest for such a formulation of MOND continues.

\newpage

\section{Gravitational Lensing in Relativistic MOND}
\label{sec:lensing}

The viable MOND theories from the previous section, although still mostly effective, have the great advantage of proving that constructing relativistic MOND theories \textit{is} possible, and that it is thus possible to calculate from them the effects of gravitational lensing. But the non-uniqueness of the theories of course means that there is not really a unique prediction for gravitational lensing, especially in heavily time-dependent configurations, or when the predictions of the theories for the expansion history of the Universe deviate from the concordance model. As we have seen, some theories also deviate slightly from classical MOND predictions for dynamics of quasi-static systems, due to the presence of massive dark fields, and the same would of course happen for gravitational lensing. However, at the zeroth order, and in static weak-field configurations, we \textit{can} make predictions for all theories whose expansion history would be similar to that of $\Lambda$CDM (see Sect.~9.1) and whose static weak-field limit is represented by a physical metric\footnote{This equality $\Psi=-\Phi$ in the weak-field metric is put in by hand in all TeVeS-like theories (Sects. 7.2 to 7.6) through a disformal relation between the Einstein and physical metrics, and is a generic prediction of GEA (Sect.~7.7), BIMOND (Sect.~7.8) and DDM (Sect.~7.9) theories} with $\Psi=-\Phi$ in Eq.~\ref{metric} ($\Phi$ obeying Eq.~\ref{BM}). In this case, the way the light propagates on the null geodesics of this metric is exactly the same in all these theories once $\Phi$ is known. What differs from GR is only the relation between $\Phi$ and the underlying mass distribution of the lens.

\subsection{Strong lensing by galaxies}

When multiple images of a background source are produced by a gravitational lens, one talks about \textit{strong lensing}. In that case, most of the light bending occurs within a small range around the lens compared to the lens-source distance $D_{ls}$ and the observer-source distance $D_s$ (where the distances are the usual luminosity distances in cosmology). In this so-called \textit{thin-lens approximation}\epubtkFootnote{By this, we however do \textit{not} mean that the MOND lensing can be computed from the projected surface density on the lens-plane as in GR, because the convergence parameter (Eq.~\ref{kappa} below) is \textit{not} a measure of the projected surface density anymore. This is also sometimes referred to as the ``thin-lens approximation" in GR, and is {\it not} valid in MOND: two lenses with the same projected surface density can have different convergence parameters, because lensing also depends strongly on the distribution of the source mass along the line-of-sight in MOND.}, the resulting deflection angle can be written as:
\begin{equation}
\mbox{\boldmath$\alpha$} = \frac{2}{c^2}\int_{-\infty}^{\infty} \nabla_{\bot}\Phi dz,
\label{deflection}
\end{equation}
where $\Phi=-\Psi$ is the non-relativistic gravitational potential of Eq.~\ref{metric} (obeying a MONDian Poisson equation), and $\nabla_{\bot}$ denotes the two-dimensional gradient operator perpendicular to light propagation. The lens equation then relates the observed two-dimensional angular position of the source in the lens plane $\mbox{\boldmath$\theta$}$ to its original angular position in the source plane $\mbox{\boldmath$\beta$}$ through:
\begin{equation}
\mbox{\boldmath$\theta$} = \mbox{\boldmath$\beta$} + \frac{D_{ls}}{D_s} \mbox{\boldmath$\alpha$},
\label{lenseq}
\end{equation}
where it appears clearly that the expansion history will play an important role in converting redshifts to distances. It is also convenient to make the deflection angle $\mbox{\boldmath$\alpha$}$ derive from a deflection potential $\Upsilon$ in the lens-plane:
\begin{equation}
\Upsilon(\mbox{\boldmath$\theta$}) = \frac{2D_{ls}}{c^2D_{s}D_{l}}\int_{-\infty}^{\infty}\Phi(D_{l}\mbox{\boldmath$\theta$},z)dz,
\end{equation}
If a source is much smaller than the angular scale on which the lens properties change, the lens equation Eq.~\ref{lenseq} can locally be linearized as:
\begin{equation}
\mbox{\boldmath$\beta$}(\mbox{\boldmath$\theta$}) = \mbox{\boldmath$\beta$}_0 + \mathcal{A}(\mbox{\boldmath$\theta$}) (\mbox{\boldmath$\theta$}-\mbox{\boldmath$\theta$}_0),
\end{equation}
where the inverse magnification matrix is
\begin{equation}
\mathcal{A}(\mbox{\boldmath$\theta$}) = \frac{\partial\mbox{\boldmath$\beta$}}{\partial\mbox{\boldmath$\theta$}}, \, {\rm where} \; \mathcal{A}_{11} = 1 - \kappa - \gamma_1 , \; \mathcal{A}_{12} = \mathcal{A}_{21} = -\gamma_2 , \; \mathcal{A}_{22} = 1 - \kappa + \gamma_1
\end{equation}
The convergence $\kappa$ is directly given by the Laplacian of the deflection potential $\Upsilon$:
\begin{equation}
\kappa = \frac{1}{2} \nabla^2 \Upsilon .
\label{kappa}
\end{equation}
The so-called Einstein radius is the radius within the lens-plane within which the mean convergence is $\langle \kappa \rangle=1$. The existence of a region where $\kappa$ is of that order is sufficient to produce multiple images and is the definition of strong lensing. On the other hand, the shear components $\gamma_{1},\gamma_{2}$ are given by
\begin{equation}
\gamma_{1} = \frac{1}{2}\left(\frac{\partial^{2}\Upsilon}{\partial\theta_{1}^{2}}-\frac{\partial^{2}\Upsilon}{\partial\theta_{2}^{2}}\right), \; \; \gamma_{2} = \frac{\partial^{2}\Upsilon}{\partial\theta_{1}\partial\theta_{2}}.
\label{eq:shear}
\end{equation}
Due to Liouville's theorem, gravitational lensing preserves the surface brightness, but it changes the apparent solid angle of a source. The resulting flux ratio between image and source can be expressed in terms of the magnification $M$,
\begin{equation}
M^{-1} = (1-\kappa)^{2}-\gamma_1^2 - \gamma_2^2.
\end{equation}
The flux ratio between two images A and B is $f_{AB}=A_A/A_B$. Let us finally note that (i) the time-delay between the different images can be deduced directly from the lensing potential and depends on the Hubble constant and convergence at the Einstein radius, and that (ii) points in the lens plane where $M^{-1}=0$ (infinite magnification) form closed curves called the \textit{critical curves}. Their corresponding curves located in the source plane are called caustics. The location of the source with respect to caustics determines the number of images, a source outside of the outermost caustic producing only one image while each caustic crossing changes the number of images by a factor of two. Spherically symmetric models of galaxy lenses can never produce observed quadruple-imaged systems because the innermost caustic of spherical models degenerates into a point.

As outlined hereabove, what differs from GR in \textit{all} the relativistic MOND theories is the relation between the non-relativistic potential $\Phi$ and the underlying mass distribution of the lens $\rho$. However, different theories yield slightly different relations between $\Phi$ and $\rho$ in the weak-field limit (see especially Sect.~6.1 and Sect.~6.2). For instance, while GEA theories (Sect.~7.7) boil down to Eq.~\ref{BM} in the static weak-field limit, TeVeS (Sect.~7.4) leads to the situation of Eq.~\ref{sumpot}, and BIMOND (Sect.~7.8) to Eq.~\ref{QUMOND}. However, like in the case of rotation curves (see Figure~\ref{fig:RCcomp}), the differences are only minor outside of spherical symmetry (and null in spherical symmetry), and the global picture can be obtained by assuming a relation given by the BM equation (Eq.~\ref{BM}). 

\epubtkImage{q2237.png}{%
\begin{figure}[htbp]
  \centerline{(a)\parbox[t]{5.5cm}{\vspace{0pt}\includegraphics[width=5.5cm]{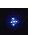}}\qquad
              (b)\parbox[t]{6cm}{\vspace{0pt}\includegraphics[width=6cm]{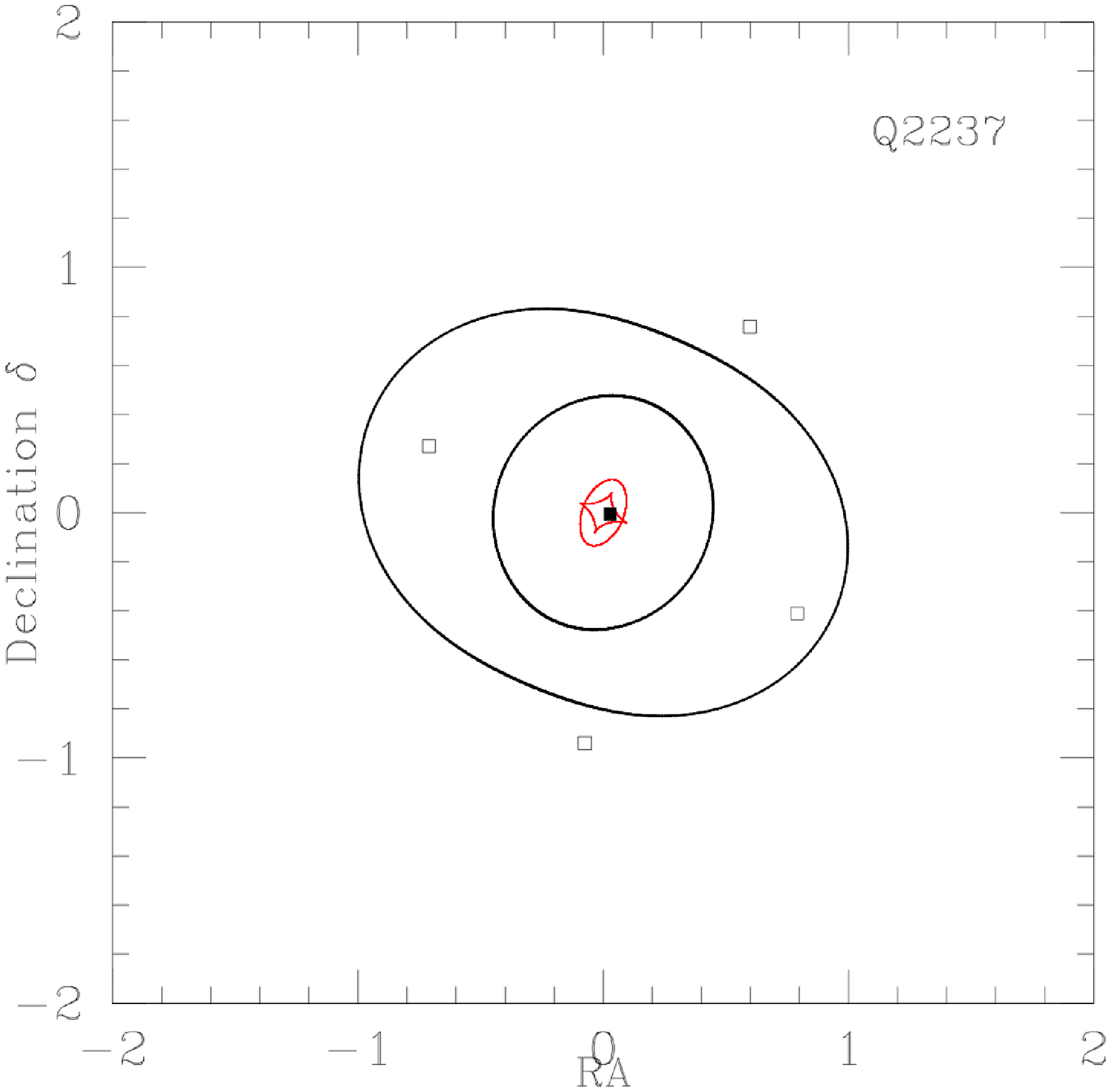}}}
  \caption{(a) The four images of the quasar Q2237+030 (known as the Einstein cross), gravitationally lensed by an isolated bulge-disk galaxy known as Huchra's lens~\cite{Huchra}. \copyright\ ESA's faint object camera on HST.  (b) The empty squares denote the four observed positions of the images, and the filled square denotes the MOND-fit unique position of the source~\cite{Shan}. The critical curves for which $M^{-1}=0$ in the lens plane are displayed in black, and their corresponding caustics in the source plane in red.}
  \label{fig:q2237}
\end{figure}}

The first studies of strong lensing by galaxies in relativistic MOND theories~\cite{Chiu1, Zhaolens2, Zhaolens1} made use of the CfA-Arizona Space Telescope Lens Survey (CASTLES) and made a one parameter-fit of the lens mass to the observed size of the Einstein radius, both for point-mass models and for Hernquist spheres (with observed core radius). Zhao et al.~\cite{Zhaolens1} also compared the predicted and observed flux ratios $f_{AB}$. They used the $\alpha=0$ $\mu$-function of Eq.~\ref{alphafamily}, and concluded that reasonably good fits could be obtained with a lens mass corresponding to the expected baryonic mass of the lens. Shan et al.~\cite{Shan} then improved the modeling method by considering analytic non-spherical models with locally spherically symmetric isopotentials on both sides of the symmetry plane $z=0$, implying no curl field correction (${\bf S} = 0$) in Eq.~\ref{curl}. The MOND non-relativistic potential $\Phi$ can then analytically be written, and using Eq.~\ref{deflection}, one can analytically compute the two components $\alpha_1$ and $\alpha_2$ of the deflection angle vector $\mbox{\boldmath$\alpha$}$ as a function of the three parameters of the model, namely the lens-mass and two scale-lengths controlling the extent and flattening of the lens (see Eq.~18 of~\cite{Shan}). Using the lens equation hereabove (Eq.~\ref{lenseq}), one can then trace back light-rays for each observed image to the source plane and fit the lens parameters as well as its inclination in order for the source position to be the same for each image. The quality of the fit is thus quantified by the squared sum of the source position differences. This notably allowed~\cite{Shan} to fit in MOND the famous quadruple-imaged system Q2237+030 known as the Einstein cross (see Figure~\ref{fig:q2237}), a quasar gravitationally lensed by an isolated bulge-disk galaxy~\cite{Huchra}. For three other quadruple-imaged systems of the CASTLES survey, the fits were however less successful mostly because of the intrinsic limitations of the analytic model of Shan et al.\cite{Shan} at reproducing at the same time both a large Einstein radius and a large shear. What is more it does not take into account the effects of the environment in the form of an external shear, which is also often needed in GR to fit quadruple-imaged systems. For 10 isolated double-imaged systems in the CASTLES survey, the fits were much more succesful\epubtkFootnote{Note that, in order for the problem to be well constrained, a regularization method was used in order to penalize solutions deviating from the fundamental plane as well as face-on solutions and solutions with an anomalous flux ratio or M/L ratio (see Eq.~21 of~\cite{Shan}.}. For non-isolated systems however, especially for those lenses residing in groups or clusters, the need for an external shear might be coupled to a need for dark mass on galaxy group scales (see Sect.~6.6.4 and Sect.~8.3).

Due to the fact that all the above models were using the so-called Bekenstein $\mu$-function ($\alpha=0$ in Eq.~\ref{alphafamily}), and that this function has a tendency of slightly underpredicting stellar mass-to-light ratios in galaxy rotation curves fits~\cite{FGBZ}, it was claimed that this was a sign for a MOND missing mass problem in galaxy lenses~\cite{Ferreras1,Ferreras2,Yusaf}. While such a missing mass is indeed possible, and even corroborated by some dynamical studies~\cite{1399} of galaxies residing inside clusters (i.e., the small-scale equivalent of the problem of MOND in clusters), for \textit{isolated} systems with well-constrained stellar mass-to-light ratio, the use of the simple $\mu$-function ($\alpha=1$ in Eq.~\ref{alphafamily}) has on the contrary been shown to yield perfectly acceptable fits~\cite{Chiu2} in accordance with the lensing fundamental plane~\cite{Sanderslens}.

Finally, the probability distribution of the angular separation of the two images in a sample of lensed quasars has been investigated by Chen~\cite{Chen2,Chen1}. This important question has proved somewhat troublesome for the $\Lambda$CDM paradigm, but is well explained by relativistic MOND theories~\cite{Chen2}. 

\subsection{Weak lensing by galaxies}

A gravitational lens does not only produce multiple images close to caustics, but also weakly distorted images (arclets) of other background sources. The weak and noisy signals from several individual arclets (not necessarily detected by eye, but rather numerically exploited with the help of image analysis) can be averaged by statistical techniques to get the shear components $\gamma_1$ and $\gamma_2$ in Eq.~\ref{eq:shear} from the mean ellipticity of the images. One can then get the convergence $\kappa$ from the azimuthal average of the tangential component of the shear. This is what is known as \textit{weak lensing}. In the case of galaxy-galaxy weak lensing, since the gravitational distortions induced by an individual lens are too small to be detected, one has to resort to the study of the ensemble averaged signal around a large number of lenses. This has been investigated in the context of MOND for a sample of relatively isolated galaxy-lenses, stacked by luminosity ranges~\cite{Tianlens}. The derived MOND masses were obtained by fitting a point mass model to the lensing data within a distance of 200~kpc from the lens. While the MOND masses are perfectly compatible with the baryonic masses in all galaxies less luminous than $10^{11} \, L_\odot$, it was found that the required MOND mass-to-light ratios tended to be slightly too high ($M/L \simeq 10$) for the most massive an luminous galaxies ($L > 10^{11} \, L_\odot$). However, this whole result is dictated by only one data point which ``pulls up" the result and make all the data points lie  below the ``best fit', and the curve is "pulled up" strongly by only the first point. The mass-to-light ratios could thus easily be scaled down by a factor of two, making these galaxies in perfect agreement with MOND. But it is also worth noting that due to the very large distances probed, the presence of some weakly clustering residual mass (hot dark matter, or some sort of ``dark field'' in the relativistic MOND theories) could start playing a role at these distances. While ordinary neutrinos are still too weakly clustering, a slightly more massive fermion such as a $10 \,$eV-scale sterile neutrino could cluster on these scales, and of course, the presence of baryonic dark matter in the form of dense molecular gas clouds could also be present around these very massive objects (see Sect.~6.6.4).

Also related to weak lensing, it is important to recall that the ``phantom dark matter'' of MOND (Eq.~\ref{eq:phantom}) can sometimes become negative in cones perpendicular to the direction of the external gravitational field in which a system is embedded: with accurate enough weak-lensing data, detecting these pockets of negative phantom densities around a sample of non-isolated galaxies could in principle be a smoking gun for MOND~\cite{XufenMW}, but such an effect would be extremely sensitive to the detailed distribution of the baryonic matter, and finding a sample of galaxies with similar gravitational environments would also be extremely difficult.

\subsection{Strong and weak lensing by galaxy clusters}

Gravitational lensing is a complementary technique to the hydrostatic equilibrium of the X-ray emitting gas (Sect.~6.6.4) to probe the mass distribution of galaxy clusters. Since clusters are the most recently formed structures, they could be slightly out of equilibrium, which makes gravitational lensing extremely interesting as this technique is fully independent from the relaxed or unrelaxed nature of the lens. A famous example of such a clearly unrelaxed object is the cluster 1E0657-56, known as the Bullet Cluster (Figure~\ref{figure:bullet1}). It is actually a pair of clusters which collided at high-speed ($>3100$~km/s) at $z=0.3$. In the collision, the dissipational hot X-ray emitting gas which dominates the baryonic matter was separated from the negligible and collisionless galaxies and any presumed collisionless dark matter. Using background galaxies to map the shear field, the convergence map of the cluster was provided by~\cite{Clowe}, a convergence very conspiciously centered where the collisionless dark matter should be \footnote{Note, however, that this is not {\it always} the case in colliding clusters: Abell 520 actually provides a counter-example to the bullet cluster in which the mass peaks indicated by weak lensing do {\it not} behave as collisionless matter should \cite{JeeMahdavi}.}. It would appear difficult to reconstruct such a configuration merely by modifying gravity, but the non-linearity of MOND does not guarantee that the convergence from a two-center baryonic distribution would be indeed centered on the two centers. Indeed, while the linear relation between the matter density and the gravitational potential implies that the convergence parameter is a direct measurement of the projected surface density in the weak-field limit of GR, this is not the case anymore in MOND due to the non-linearity of the modified Poisson equation. Actually, it has been shown that, in MOND, it is possible to have a non-zero convergence along a line of sight where there is zero projected matter~\cite{AFZ}. What is more, the gravitational environment might play an important role on the internal gravitational field too~\cite{Dai1,Matsuoscreening}, and the additional degrees of freedom of the various relativistic theories might play a non-negligible role, especially in non-static situations~\cite{Daibullet}. Neglecting possible effects of the gravitational environment and non-trivial features of the additional fields of the relativistic theories out of equilibrium, i.e., simply assuming that the physical metric is given by $\Psi=-\Phi$ in Eq.~\ref{metric}, and that $\Phi$ obeys Eq.~\ref{BM}), a MOND model of the bullet cluster was produced~\cite{angbul}, in which a parametrized potential was fitted to the convergence map to then determine the underlying mass distribution from Eq.~\ref{BM}. The result is displayed on Figure~\ref{figure:bullet}, and exactly the same conclusion was reached by going from the baryonic density to the convergence map~\cite{Feix}. The main conclusions are that (i) the amount of residual missing mass needed to account for the convergence map of the bullet cluster is the same as in all other clusters (Sect.~6.6.4 and~\cite{Takahashi}), but that (ii) if it is made of dark baryons, they must be in a \textit{collisionless} form, since the residual missing mass is centered on the collisionless galaxies and not on the dissipational hot gas. The dense molecular gas clouds proposed by Milgrom~\cite{marriage} (see discussion in Sect.~6.6.4) satisfy this criterion, and would mostly behave like individual stars. Like in most clusters with $T > 4 \, {\rm keV}$, ordinary neutrinos with a $2 \,$eV mass would be broadly sufficient to account for the missing mass deduced from weak lensing (and, obviously, heavier exotic hot dark matter particles such as $10 \,$eV sterile neutrinos would do the job too). 

For  TeVeS (Sect.~7.4) and GEA (Sect.~7.7), the growth of the spatial part of the vector perturbation in the course of cosmological evolution can successfully seed the growth of baryonic structures, just as dark matter does, and it is possible to reconstruct the gravitational field of the bullet cluster without any extra matter but with a substantial contribution from the vector field. However, why the dynamical evolution of the vector field perturbations would lead to precisely such a configuration remains unclear. Similarly, the massive scalar field of Sect.~7.6 or the monopolar part of the dipolar DM of Sect.~7.9 could in principle provide the off-centered missing mass too, but again, why they would appear distributed as they do remains unclear, especially in the case of dipolar DM which is supposed to cluster only very weakly, and in principle not to appear as densely clustered. Whether the twin matter of BIMOND (Sect.~7.8) could help providing the right convergence map also remains to be seen, while for non-local models (Sect.~7.10), there is a strong dependence on the past light-cone, meaning that recently disturbed systems such as the Bullet may be far from the static MOND limit (but in that case, it would not be clear why all the other clusters from Sect.~6.6.4 exhibit the same amount of residual missing mass). So, while the bullet cluster clearly does not represent the MOND-killer that it was supposed to be, explaining its convergence map remains an outstanding challenge for all MOND theories. However, the bullet cluster also represents an outstanding challenge to $\Lambda$CDM (see Sect.~4.2), due to its high collision speed~\cite{Lee}. In that respect, MOND is much more promising~\cite{Angusbullet}.

On the other hand, a comprehensive weak lensing mass reconstruction of the rich galaxy cluster Cl0024+17 at $z=0.4$~\cite{J07} has been argued to have revealed the first dark matter structure that is offset from \textit{both} the gas and galaxies in a cluster. This structure is ringlike, located between $r \sim 60''$ and $r \sim 85''$. It was, again, argued to be the result of a collision of two massive clusters 1\,--\,2~Gyr in the past, but this time along the line-of-sight. It has also been argued~\cite{J07} that this offset was hard to explain in MOND. Assuming that this ringlike structure is real and not caused by instrumental bias or spurious effects in the weak lensing analysis (due, e.g., to the unification of strong and weak-lensing or to the use of spherical/circular priors), and that cluster stars and galaxies do not make up a high fraction of the mass in the ring (which would be too faint to observe anyway), it has been shown that, for certain interpolating functions with a sharp transition, this is actually \textit{natural} in MOND~\cite{rings}. A peak in the phantom dark matter distribution generically appears close to the transition radius of MOND $r_t=(GM/a_0)^{1/2}$, especially when most of the mass of the system is well-contained inside this radius (which is the case for the cluster Cl0024+17). This means that the ring in  Cl0024+17 could be the first manifestation of this pure MOND phenomenon, and thus be a resounding success for MOND in galaxy clusters. However, the sharpness of this phantom dark matter peak strongly depends on the choice of the $\mu$-function, and for some popular ones (such as the ``simple'' $\mu$-function) the ring cannot be adequately reproduced by this pure MOND phenomenon. In this case, a collisional scenario would be needed in MOND too, in order to explain the feature as a peak of cluster dark matter. Indeed, we already know that there is a mass discrepancy in MOND clusters, and we know that this dark matter must be in collisionless from (e.g., neutrinos or dense clumps of cold gas). So the results of the simulation with purely collisionless dark particles~\cite{J07} would surely be very similar in MOND gravity. Again, it was shown that the density of missing mass was compatible with 2~eV ordinary neutrinos, like in most clusters with $T > 4\mathrm{\ keV}$~\cite{wedding}. Finally, let us note that strong lensing was also recently used as a robust probe of the matter distribution on scales of 100~kpc in galaxy clusters, especially in the cluster Abell~2390~\cite{straightarc}. A residual missing mass was again found, compatible with the densities provided by fermionic hot dark matter candidates only for masses of $\sim 10 \,$eV and heavier. All in all, the problem posed by gravitational lensing from galaxy clusters is thus very similar to the one posed by the temperature profiles of their X-ray emitting gas (Sect.~6.6.4), and remains one of the two main current problems of MOND, together with its problem at reproducing the CMB anisotropies (see Sect.~9.2).

Finally, let us note in passing that another (non-lensing) test of relativistic MOND theories in galaxy clusters has been performed by analysing the gravitational redshifts of galaxies in 7800 galaxy clusters \cite{Wojtak}, which were originally found to be difficult to reconcile with MOND: however, this original analysis assumed a distribution of residual missing mass in MOND by simply scaling down the Newtonian dynamical mass represented by a NFW halo by a factor 0.8, and the analysis confused the interpolating functions $\mu(x)$  and $\tilde{\mu}(s)$ (see Sect.~6.2). A subsequent analysis \cite{Bekrebut} showed that these gravitational redshifts were in accordance with relativistic MOND when the correct residual mass and acceptable $\mu$-functions were used.

\epubtkImage{bullet1.png}{%
\begin{figure}[htbp]
\centerline{\includegraphics[width=10.5cm]{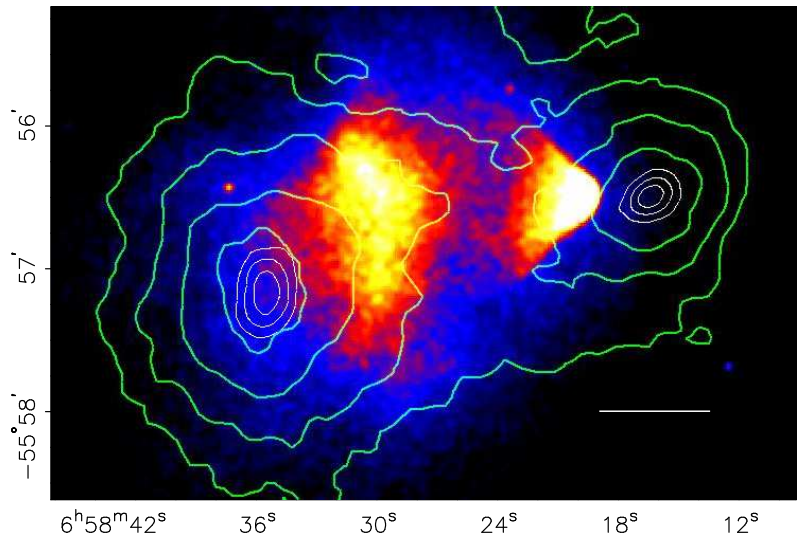}}
  \caption{The bullet cluster 1E0657-56. The hot gas stripped from both subclusters after the collision is colored red-yellow. The green and white curves are the isocontours of the lensing convergence parameter $\kappa$ (Eq.~\ref{kappa}). The two peaks of $\kappa$ do not coincide with those of the gas which makes up most of the baryonic mass, but are skewed in the direction of the galaxies. The white bar corresponds to 200~kpc. Figure courtesy of D.~Clowe.}
  \label{figure:bullet1}
\end{figure}}

\epubtkImage{bullet.png}{%
\begin{figure}[htbp]
\centerline{\includegraphics[width=0.9\textwidth]{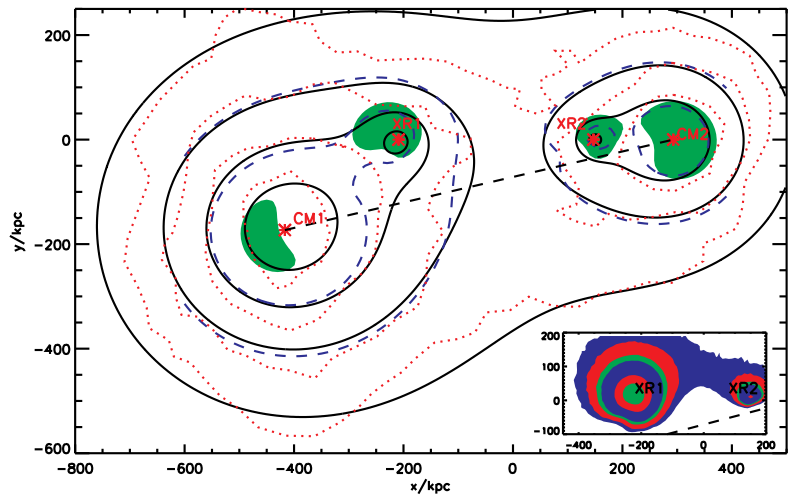}}
  \caption{A MOND model of the bullet cluster~\cite{angbul}. The fitted $\kappa$-map (solid black lines) is overplotted on the convergence map of~\cite{Clowe} (dotted red lines). The four centres of the parametrized potential used are the red stars. Also overplotted (blue dashed line) are two contours of surface density. Note slight distortions compared to the contours of $\kappa$. The green shaded region corresponds to the clustering of 2~eV neutrinos. Inset: The surface density of the gas in the model of the bullet cluster.}
  \label{figure:bullet}
\end{figure}}

\clearpage
\subsection{Weak lensing by large-scale structure}

The weak-lensing method can also be applied on larger scales, i.e., mapping the shear-field induced by large-scale structures. On these scales, the metric of the expanding Universe forming structure is well represented by a \textit{Newtonianly perturbed} Friedmann--Lema\^itre--Robertson--Walker (FLRW) metric:
\begin{equation}
\label{ngaugemetric}
g_{0i}=g_{i0}=0 \; , \; g_{00} = - \left( 1 +  2\Phi \right) \; , \; g_{ij} = a(t)^2 \left(1 +  2\Psi \right)  \delta_{ij},
\end{equation}
where $a(t)$ is the scale factor. Like in the static weak-field case (Eq.~\ref{metric}), $\Phi$ is the non-relativistic potential in units of $c^2$, \textit{but} the equality $\Psi=-\Phi$ in Eq.~\ref{metric} does \textit{not} necessarily imply the equality in Eq.~\ref{ngaugemetric}. In GR, this equality is actually respected for both cases (apart for perturbations around a FLRW background sourced by anisotropic stress), but the relativistic MOND theories, which have been constructed in order to yield the equality for the static weak-field limit in Eq.~\ref{metric}, do \textit{not} harbor this equality in the perturbed FLRW case, and the quantity $\Phi+\Psi$ is referred to as the {\it gravitational slip}. For instance, in the TeVeS (Sects. 7.3 and 7.4) and GEA (Sect.~7.7) theories, based on unit-norm vector fields, the equality is broken due to the growth of vector perturbations in the course of cosmological evolution (see e.g.~\cite{dodelson} and Sect.~9.2). 

Like in the static case, weak gravitational lensing from large-scale structure will actually depend on $\Phi-\Psi$, whereas galaxy clustering will arise only from the non-relativistic potential $\Phi$. By combining information on the matter overdensity at a given redshift (obtained from measuring the peculiar velocity field) and on the weak lensing maps, Zhang et al.~\cite{zhang} proposed a clever method to observationally estimate $\Phi-\Psi$. This allowed Reyes et al.~\cite{reyesnat} to use luminous red galaxies in the SDSS survey in order to exclude one model from the original TeVeS theory (Sect.~7.3) with the original $f(X)$ function of~\cite{TeVeS}, thus explicitly showing how such measurements could be a possible future smoking-gun for all theories based on dynamical vector fields. But note that other MOND theories such as BIMOND would not be affected by such measurements.

Let us however finally note a caveat in the interpretation of the weak lensing shear map in the context of relativistic MOND. While intercluster filaments negligibly contribute to the weak lensing signal in GR, a single filament inclined by $\pi/4$ from the line of sight can cause substantial distortion of background sources pointing toward the filament's axis in relativistic MOND theories~\cite{filaments}. Since galaxies are generally embedded in filaments or are projected on such structures, this contribution should be taken into account when interpreting weak lensing data. This additional difficulty for interpreting weak-lensing data in MOND is not only true for filaments, but more generally for all low-density structure such as sheets and voids.

\newpage

\section{MOND and Cosmology}
\label{sec:cosmology}

\subsection{Expansion history}

A viable theory of modified gravity, including dark fields or not, should not only be able to reproduce observations in quasi-stationary galactic and extragalactic systems, but also to reproduce all of the major probes of observational cosmology, including (i) the Hubble diagram out to large $z$, (ii) the anisotropies in the cosmic microwave background (CMB), and (iii) the matter power spectrum on large scales. The first requires a detailed knowledge of FLRW cosmology, and the last two a knowledge of cosmological perturbations on a FLRW background.

Concerning the first point, the FLRW solutions have been extensively studied for TeVeS (Sects.~7.3 and 7.4, see e.g.~\cite{Bourliot}) and GEA (Sect.~7.7, see e.g.~\cite{Zlosnikcosmo}) theories, for BIMOND (Sect.~7.8, see e.g.~\cite{cliftonbimond}), and for theories based on dipolar dark matter (Sect.~7.9, see e.g.~\cite{Blanchet}). In the latter case, the theory~\cite{Blanchet4,Blanchet} has been shown to be strictly equivalent to $\Lambda$CDM out to first order cosmological perturbations (but very different in the galaxy formation regime), together with a natural explanation for $\Lambda \sim a_0^2$. For the other theories, it has been shown that the contribution of the extra-fields to the overall expansion is subdominant to the baryonic mass and does not affect the overall expansion~\cite{Ferreira09}. Such theories can predict an extremely wide range of cosmological behavior, ranging from accelerated expansion to contraction on a finite time scale~\cite{Bourliot}. The key point is that the expansion history mainly depends on the form of the ``MOND function'' $f(X)$ for the \textit{unconstrained} domain $X<0$ in any of these theories. 

\epubtkImage{mus_ZF.png}{%
\begin{figure}[htbp]
  \centerline{\includegraphics[width=10.5cm]{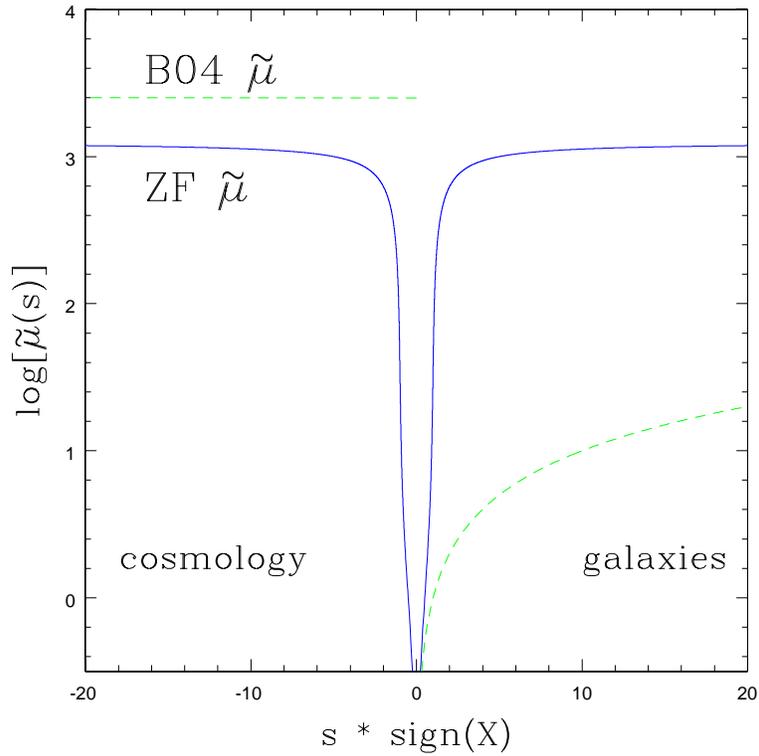}}
  \caption{In solid blue, the Zhao--Famaey~\cite{ZF06} $\tilde{\mu}(s)$-function (Eq.~\ref{tildemu}) of TeVeS (Sect. 7.3 and 7.4), compared to the original Bekenstein one (dashed green) with a discontinuity at $s=0$~\cite{TeVeS}. The ZF function provides a more natural transition from static systems (the positive side) to cosmology (the negative side).}
  \label{fig:mus_ZF}
\end{figure}}

For instance in TeVeS, $X \propto (\nabla \phi)^2 > 0$ in static configurations (see Eq.~\ref{XTEVES}), and $X \propto -2 (\partial \phi/\partial t)^2$ in evolving homogeneous and isotropic configurations such as the expanding Universe. The form of $f(X)$ is clearly constrained from the MOND phenomenology \textit{only} for $X>0$, meaning that \textit{a lot} of freedom exists for $X<0$. Exactly the same is true in GEA and BIMOND theories~\cite{cliftonbimond}. For instance, Bekenstein~\cite{TeVeS} originally proposed for TeVeS a $f'$-function (corresponding to $\tilde{\mu}$, see Eq.~\ref{tildemu}) with a discontinuity at $X=0$ (the B04 function on Figure~\ref{fig:mus_ZF}) not enabling galaxies to collapse continuously out of the Hubble expansion. Afterwards, Zhao \& Famaey proposed an improved ``mirror-function'' $f'(X)$ such that the corresponding $\tilde{\mu}$-function reproduces the simple $\mu$-function ($\alpha=1$ in Eq.~\ref{alphafamily}) for $X>0$, and $f(X) = f(-X)$ for the cosmological regime $X<0$ (see Figure~\ref{fig:mus_ZF}, leading to an acceptable expansion history. However, when connecting a static galaxy to the expanding Universe, the limit $\tilde{\mu}(0)=0$ would predict the existence of a singular surface around each galaxies on which the scalar degree of freedom does not propagate, meaning that it is better to reconnect the two sides at $\tilde{\mu}(0)=\varepsilon$ (see Sect.~6.2). In addition, the integration constant $f(0)$ can play the role of the cosmological constant~\cite{Hao} to drive accelerated expansion, but even some $f(0)=0$ models can drive late-time acceleration~\cite{Diaz}, which is not surprising since k-essence scalar fields were also introduced to address the dark energy problem. In the case of BIMOND (see Sect.~7.8), a symmetric matter-twin matter early Universe yields a cosmological constant through the zero-point of the MOND function, thereby naturally leading to $\Lambda \sim a_0^2$. 

All in all, with the additional freedom of a hypothetical dark component in the matter sector, in the form of e.g. ordinary or sterile neutrinos, playing with the form of $f(X)$ for $X<0$ in TeVeS, GEA and BIMOND always allows one to reproduce an expansion history and a Hubble diagram almost precisely identical to $\Lambda$CDM, justifying the assumption made in Sect.~8 to assume this expansion history for gravitational lensing in relativistic MOND. However, it is important to note that MOND theories are not providing a unique prediction on this.

\subsection{Large scale structure and Cosmic Microwave Background}
\label{subsection:CMB}

Modified gravity theories should of course not only produce a reasonable Hubble expansion but also reproduce the observed anisotropies in the CMB, and the matter power spectrum. Taken at face value, these require not only dark matter, but non-baryonic cold dark matter. Any alternative theory must account for these, just as dark matter models need to explain galaxy scale phenomenology.

Using the hypothesis that the Universe is filled with some form of cold dark matter, it is possible to simultaneously fit observations of the CMB~\cite{Komatsu11} and provide an elegant picture for the growth of large scale structure~\cite{SFW}. An obvious question is thus how MOND fares with these subjects. Of course, as we have seen, there is no unique existing MOND theory (Sect.~7), and the basic theory underlying MOND as a paradigm is probably yet to be found. Nevertheless, we can make a few general considerations about how any MOND theory should behave, and then look in more details at specific predictions from existing relativistic theories. The general picture is that, in some ways MOND does surprisingly well, in others it clearly gives no real unique prediction by now, and in still others it appears to fail outright. 

If one alters the force law as envisioned by MOND, the effective long range force becomes stronger. Though details will of course depend on the specific relativistic theory, we can speculate about the consequences of a MOND-like force in cosmology. Note however that most of what follows cannot be rigorously justified at the moment for lack of a compelling unique underlying theory. But obviously, because of the stronger force, dynamical measures of the cosmic mass density will be overestimated, just as in galaxies.
Applying MOND to the peculiar motions of galaxies yields $\Omega_m \approx \Omega_b$~\cite{MdB98}.
There are large uncertainties in estimating the extragalactic peculiar acceleration field, so this merely shows that MOND might alleviate the need for non-baryonic dark matter inferred conventionally from $\Omega_m > \Omega_b$. 

The stronger effective gravitational attraction of MOND would change the growth rate of perturbations.  Instead of adding dark mass to speed the growth of structure, we now rely on the modified force law to do the work. While it is obvious that MOND will form structures
more rapidly than conventional gravity with the same source perturbation, we immediately encounter a challenge posed by the non-linear nature of the theory, precluding an easy linear perturbation analysis. One can nevertheless sketch a naive overview of how structure might form under the influence of MOND.
The following picture emerges from numerical calculations of particles interacting under MOND in an assumed background \cite{sandersstructure,nusser,knebe,Llinares}, and is thus obviously slightly (or very) different from the various relativistic MOND theories of Sect.~7 and from those yet to be found, especially from those MONDian theories involving the existence of some form of dark matter (twin matter, dipolar dark matter, etc.). In the early Universe, perturbations cannot grow because the baryons are coupled to the photon fluid.  The mass density
is lower, so matter domination occurs later than in $\Lambda$CDM.  Consequently, MOND structure formation initially has to lag behind
$\Lambda$CDM at very high redshift ($z > 200$).  However, as the influence of the photon field declines and perturbations begin to enter the MOND regime, structure formation rapidly speeds up.  Large galaxies may form by $z \approx 10$ and clusters by $z \approx 2$~\cite{anguscosmo,sandersstructure}, considerably earlier than in $\Lambda$CDM.  By $z=0$, the voids have become more empty than in $\Lambda$CDM, but otherwise simulations (of collisionless particles, which is of course not the best representation of the baryon fluid) show the same qualitative features of the cosmic web~\cite{knebe,Llinares}.
This similarity  is not surprising since MOND is a subtle alteration of the force law.  The chief difference is in the timing of when structures of a given mass appear, it being easier to assemble a large mass early in MOND. This means that MOND is promising in addressing many of the challenges of Sect.~4.2, namely the high-z clusters challenge~\cite{anguscosmo} and Local Void challenge, as well as the bulk flow challenge and high collisional velocity of the bullet cluster~\cite{Angusbullet,Llinaresbullet}, again due to the much larger than Newtonian MOND force in the structure formation context. What is more, it could allow large massive galaxies to form early ($z \approx 10$) from monolithic dissipationless collapse~\cite{Sandersform}, with well-defined relationships between the mass, radius and velocity dispersion. Consequently, there would be less mergers than in $\Lambda$CDM at intermediate redshifts, in accordance with constraints from interacting galaxies (see Sect.~6.5.3), which could explain the observed abundance of large thin bulgeless disks unaffected by major mergers (see Sect.~4.2), and in those rare mergers between large spirals, tidal dwarf galaxies would be formed and survive more easily (see Sect.~6.5.4). This could lead to the intriguing possibility that most dwarf galaxies are not primordial but have been formed tidally in these encounters~\cite{Kroupa}. These populations of satellite galaxies, associated with globular clusters that formed along with them, would naturally appear in (more than one) closely related planes (because a gas-rich galaxy pair undergoes many close encounters in MOND before merging, see Sect.~6.5.3), thereby perhaps providing a natural solution to the Milky Way satellites phase-space correlation problem of Sect.~4.2. What is more, the density-morphology relation for dwarf ellipticals (more dE galaxies in denser environments~\cite{Kroupa}), observed in the field, in galaxy groups and in galaxy clusters could also find a natural explanation.

Actually, the chief problem seems not to be forming structure in MOND, but the danger of over-producing it~\cite{nusser,SM02}.  
The amplitude of the power spectrum is well measured at $z = 1091$ in the CMB and at $z \approx 0$ by surveys like
the Sloan Digital Sky Survey.  Simulations normalized to the CMB overproduce the structure at $z = 0$ by a factor
of $\sim$~2.  Given the uncertainty in the parent relativistic theory and hence the appropriate form of the expansion history,
this seems remarkably close.  Given the non-linear nature of the theory, MOND could easily have been wrong by many
orders of magnitude in this context.  Nevertheless, it may be necessary to somehow damp the growth of structure at
late times~\cite{SM02}.  In this regard, a laboratory measurement of the ordinary neutrino mass might be relevant.  Conventional
structure cannot form in $\Lambda$CDM if $m_{\nu} > 0.2\mathrm{\ eV}$~\cite{Komatsu11}.  In contrast, some modest damping from
a non-trivial neutrino mass might be desirable in MOND, and is also relevant to the CMB and clusters of galaxies (see Sect.~6.6.4). 

In addition to mapping the growth factor as a function of redshift, one would also like to predict the power spectrum of mass fluctuations as a function of scale at a given epoch.  It is certainly possible to match the power spectrum of galaxies at $z=0$
\cite{SM02}, but because of MOND's non-linearity and the uncertainty in the background cosmology, it is rather harder to know 
if such a match faithfully represents a viable theory.  Indeed, a natural prediction of baryon dominated cosmologies is the presence of strong baryon acoustic oscillations in the matter power spectrum at $z=0$ \cite{McGBAO,Dodprob}.  Dodelson \cite{Dodprob} portrays this as a problem, but as already pointed out in \cite{McGBAO}, the non-linearity of MOND can lead to mode mixing that washes out the initially strong signal by $z=0$.  A more interesting test would be provided by the galaxy power spectrum at high redshift ($z \sim 5$).  This is a challenging observation, as one needs both a large survey volume and high resolution in $k$-space.  The latter requirement arises because the predicted features in the power spectrum are very sharp.  The window functions necessarily employed in the analysis of large scale structure data are typically wider than the predicted features.  Convolution of the predicted power spectrum with the SDSS analysis procedure \cite{kspacerez} shows that essentially all the predicted features wash out, with the possible exception of the strongest feature on the largest scale.  This means that the BAO signal detected by SDSS and consistent with $\Lambda$CDM \cite{LCDMBAO} could also be interpreted as a confirmation of the \textit{a priori} prediction \cite{McGBAO} of such features\footnote{Even if BAO features are present at high redshift in MOND, it is not clear that low redshift structures will correlate with the ISW in the CMB as they should in conventional cosmology because of the late time non-linearity of MOND.} in MOND.  However, there is no definitive requirement that the BAO appears at the same scale as observed, or that it survives at all.  In relativistic theories such as TeVeS (Sect.~7.3 and 7.4), damping of the baryonic oscillations can be taken care of by parameters of the theory such as $K$ in original TeVeS (Eq.~\ref{Bekoriginal}, see figure~3 of \cite{SMFB}) or the $c_i$ coefficients in generalized TeVeS (Eq.~\ref{ci}). In any case, as in standard cosmology, the angular power spectrum of the CMB should be a cleaner probe.

\epubtkImage{CMB_LR.png}{%
\begin{figure}[htbp]
  \centerline{\includegraphics[width=0.75\textwidth]{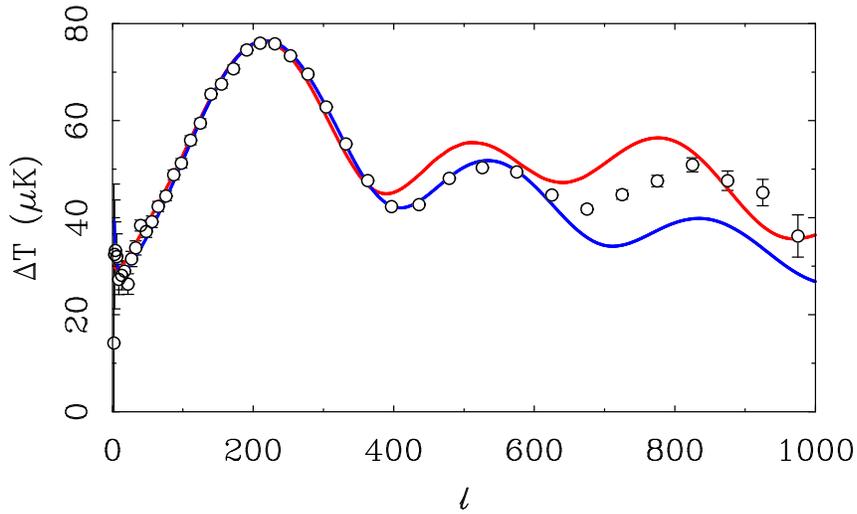}}
  \caption{The acoustic power spectrum of the cosmic microwave
    background as observed by WMAP~\cite{Komatsu11} together with the
    \emph{a priori} predictions of $\Lambda$CDM (red line) and no-CDM
    (blue line) as they existed in 1999~\cite{McGCMB1} prior to
    observation of the acoustic peaks.  $\Lambda$CDM correctly
    predicted the position of the first peak (the geometry is very
    nearly flat) but over-predicted the amplitude of both the second
    and third peak.  The most favorable \emph{a priori} case is shown;
    other plausible $\Lambda$CDM parameters~\cite{turner} predicted an
    even larger second peak.  The most important parameter adjustment
    necessary to obtain an \emph{a posteriori} fit is an increase in
    the baryon density $\Omega_b$ above what had previously been
    expected form big bang nucleosynthesis.  In contrast, the no-CDM
    model \emph{ansatz} made as a proxy for MOND successfully
    predicted the correct amplitude ratio of the first to second peak
    with no parameter adjustment~\cite{McGCMB2,McGCMB3}.  The no-CDM
    model was subsequently shown to under-predict the amplitude of the
    third peak~\cite{WMAP3}.}
  \label{fig:CMB1}
\end{figure}}

A first attempt to address the CMB was made before the existence of relativistic theories with a simple \textit{ansatz}~\cite{McGCMB1}: just as MOND returns precisely Newton in high accelerations, so any parent theory should contain GR (almost exactly, although this is not precisely the case for, e.g., TeVeS) in the appropriate strong-field limit. An obvious first assumption is that MOND effects do not yet appear in the very early Universe, so that pure GR suffices for calculations concerning the CMB. The chief difference between $\Lambda$CDM and a MONDian cosmology is then just the presence or absence of non-baryonic cold dark matter. With this \textit{ansatz}, we can make one robust prediction:  the shape of the acoustic power spectrum should follow pure baryonic diffusion damping.  There is no net forcing term, as provided by the extra degree of freedom of non-baryonic cold dark matter. With nothing but baryons, each acoustic peak should thus be lower than the previous one~\cite{silkdamping} as part of a simple damping tail (Figure~\ref{fig:CMB1}).  In contrast, there must be evidence of forcing present in a power spectrum where CDM outweighs the baryons.

The density of both the baryons and the non-baryonic cold dark matter are both critical to the shape of the acoustic
power spectrum.  For a given baryon density, models with CDM with have a larger second 
peak than models without it.  Similarly, the third peak is always lower than the second in purely baryonic models,
while it can be either higher or lower in CDM models, depending on the mix of each type of mass.
Moreover, both parameters were well constrained prior to observation of the CMB~\cite{turner}: $\Omega_b$ from 
BBN~\cite{BBNcomp1} and $\Omega_m$ from a variety of methods~\cite{DavisOm}.  It therefore seemed like a
straightforward exercise to predict the difference one should observe. The most robust prediction that could be made was the ratio of the amplitude of the first to second acoustic peak~\cite{McGCMB1}.
For the range of baryon and dark matter densities allowed at the time, $\Lambda$CDM predicted a range in this ratio
anywhere from 1.5 to 1.9.  That is, the first peak should be almost but not quite twice as large as the second, with the precise
value containing the information necessary to much better constrain both density parameters.  For the same baryon densities
allowed by BBN but no dark matter, the models fell in a distinct and much narrower range:  2.2 to 2.6, with the most plausible value
being 2.4.  The second peak is smaller (so the ratio of first to second higher) because there is no driving term to counteract
baryonic damping.  In this limit, the small range of relative peak heights follows directly from the narrow range in $\Omega_b$
from BBN.

\epubtkImage{BBN.png}{%
\begin{figure}[htbp]
  \centerline{\includegraphics[width=0.75\textwidth]{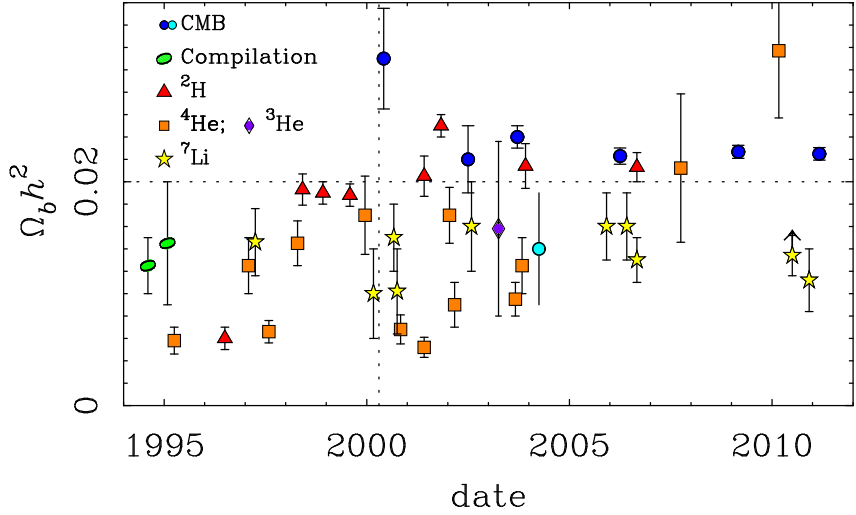}}
  \caption{Estimates of the baryon density $\Omega_b h^2$ [where 
  $h = H_0/(100\mathrm{\ km\, s^{-1}\, Mpc}^{-1})$]
  over time (updated~\cite{halobyhalo} from~\cite{McGCMB3}).  
  Big bang nucleosynthesis was already a well established field prior to 1995;
  earlier contributions are summarized by compilations (green ovals~\cite{BBNcomp1,BBNcomp2}) that gave the long-lived
  standard value $\Omega_b h^2 = 0.0125$~\cite{BBNcomp1}.  More recent estimates from individual isotopes are shown 
  as triangles ($^2\mathrm{H}$), squares ($^4\mathrm{He}$), diamonds ($^3\mathrm{He}$), and stars 
  ($^7\mathrm{Li}$).  Estimates of the baryon density based 
  on analyses of the cosmic microwave background are shown by circles (dark blue for $\Lambda$CDM; light blue for
  no-CDM).  No measurement of any isotope suggested a value greater than $\Omega_b h^2 = 0.02$ prior to observation
  of the acoustic peaks in the microwave background (dotted lines), which might be seen as a possible illustration of confirmation bias.  
  Fitting the acoustic peaks in $\Lambda$CDM requires $\Omega_b h^2 > 0.02$.  
  More recent measurements of $^2\mathrm{H}$ and $^4\mathrm{He}$ have
  migrated towards the $\Lambda$CDM
  CMB value, while $^7\mathrm{Li}$ remains persistently problematic~\cite{cyburt}.  It has been suggested that
  turbulent mixing might result in the depletion of primordial lithium necessary to reconcile lithium
  with the CMB (upward pointing arrow~\cite{LiMelendez}) while others~\cite{LiSbordone} argue that this would merely reconcile 
  some discrepant stars with the bulk of the data defining the Spite plateau, which persists in giving a $^7\mathrm{Li}$ 
  abundance discrepant from the $\Lambda$CDM CMB value.
  In contrast, the amplitude of the second peak of the microwave background
  is consistent with no-CDM and $\Omega_b h^2 = 0.014 \pm 0.005$~\cite{McGCMB3}.  
  Consequently, from the perspective of MOND, the CMB, lithium, deuterium, and helium all
  give a consistent baryon density given the uncertainties.}
  \label{fig:BBN}
\end{figure}}

The BOOMERanG experiment~\cite{boomerang} provided the first data capable of testing this prediction,
and was in good agreement with the no-CDM prediction~\cite{McGCMB2}.
This result was subsequently confirmed by WMAP, which measured a ratio $2.34 \pm 0.09$~\cite{WMAPpeaks}.
This is in good quantitative agreement with the \textit{a priori} prediction of the no-CDM \textit{ansatz}, and outside the range first expected in $\Lambda$CDM.  $\Lambda$CDM can nevertheless provide a good fit to the CMB power spectrum.
The chief parameter adjustment required to obtain a fit is the baryon density, which must be increased:  this is the reason for the
near doubling of the long-standing value $\Omega_b h^2 = 0.0125$~\cite{BBNcomp1} to the more recent $\Omega_b h^2 = 0.02249$~\cite{Komatsu11}. 

A critical question is whether the baryon density required by $\Lambda$CDM is consistent with the independently measured abundances of the light isotopes.  This question is explored in Figure~\ref{fig:BBN}.  Historically, no isotope
suggested a value $\Omega_b h^2 > 0.02$ prior to fits to the CMB requiring such a high value.  This is an important fact
to bear in mind, since historically cosmology has a long tradition of confirmation bias\epubtkFootnote{Perhaps the most
famous modern example of confirmation bias is in measurements 
of the Hubble constant~\cite{Trimble} where over many years de Vaucouleurs persistently found 
$H_0 \approx 100\mathrm{\ km\, s^{-1}\, Mpc}^{-1}$ while Sandage persistently found
$H_0 \approx 50\mathrm{\ km\, s^{-1}\, Mpc}^{-1}$.  Then, as now, there was a
conflation of data with theory:  the lower value of $H_0$ was more widely accepted because it was required for 
cosmology to be consistent with the ages of the oldest stars.}.
More recent measurements of deuterium and helium are consistent with the
high baryon density required by $\Lambda$CDM fits to the CMB.  Lithium persistently suggests a lower baryon density,
consistent with pre-CMB values.  If we are convinced of the correctness of $\Lambda$CDM, then it is easy to dismiss this
as some peculiarity of stars -- if exposed to the high temperatures in the cores of stars by turbulent mixing, lithium
might be depleted from it primordial value.  If we are skeptical of $\Lambda$CDM, then it is no surprise that measurements
of the primordial lithium abundance return the same value now as they did before.  From the perspective of the no-dark matter MOND view, 
the CMB, lithium, deuterium, and helium all give a consistent baryon density given the uncertainties. 

However, the no-CDM \textit{ansatz} must fail at some point.  It could fail outright if the parent MOND theory deviates substantially from
GR in the early Universe.  However, the more obvious~\cite{McGCMB1} points of failure are rather due to the anticipated early structure formation in MOND discussed above.  This should lead, in a true MOND theory, to early
re-ionization of the Universe and an enhancement of the integrated Sachs--Wolfe effect.  Evidence for both these effects
are present in the WMAP data~\cite{McGCMB3}.  Indeed, it turns out to be rather easy, and perhaps too easy, to
enhance the integrated Sachs--Wolfe (ISW) effect in theories like TeVeS or GEA~\cite{SMFB,Zuntz}.  Nevertheless, early re-inioniaztion is an especially natural consequence of MOND structure formation that was predicted \textit{a priori}~\cite{McGCMB1}.  In contrast, structure is expected to build up more
slowly in $\Lambda$CDM such that obtaining the observed early re-ionization implies that the earliest objects to collapse were
$\sim$~50 times as efficient at converting mass to ionizing photons as are collapsed objects at the present time~\cite{Sokasian04}.

One prediction of the no-CDM \textit{anzatz} that should not obviously fail is that the third peak should be smaller than the
second peak of the acoustic power spectrum of the CMB.  In a Universe governed by MOND rather than cold dark matter,
there is \textit{a priori} no obvious non-baryonic mass that is decoupled from the photon-baryon fluid.  It is therefore a strong expectation
that we observe only baryonic damping in the power spectrum, and each peak should be smaller in amplitude than the previous one.
Contrary to this expectation, WMAP observes the third peak to be nearly equal in amplitude to the second~\cite{WMAP3,Komatsu11}. This approximate equality of the second and third peaks falsifies the simple no-CDM \textit{anzatz}.  

\epubtkImage{garry.png}{%
\begin{figure}[htb]
  \centerline{\includegraphics[width=10.5cm]{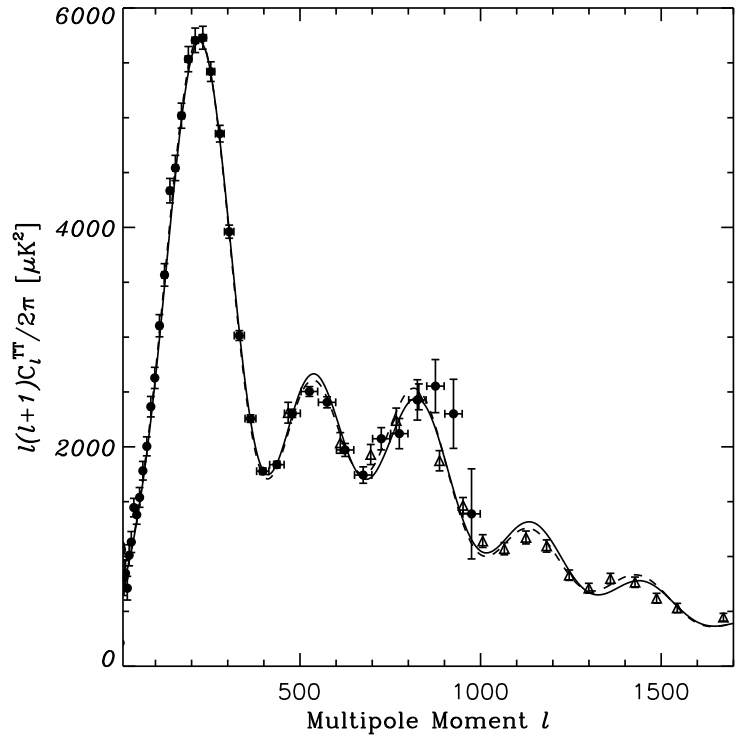}}
  \caption{CMB data as measured by the WMAP satellite year five data release (filled circles) and the ACBAR 2008 data release (triangles). Dashed line: $\Lambda$CDM fit. Solid line: HDM fit with a sterile neutrino of mass 11~eV (Figure courtesy of G.~Angus)}
  \label{CMBangus}
\end{figure}}

The PLANCK mission should soon report a new and much higher resolution measurement of the CMB acoustic power spectrum.  It is conceivable\footnote{At $\ell \approx 800$, the third peaks is only marginally resolved by WMAP.  This scale is comparable to a single  (frequency dependent) beam size, and as such is extraordinarily sensitive to corrections for the instrumental point spread function \cite{SSWMAPsys}.} that improved data will reveal a different power spectrum.  A third peak as low as that expected in the no-CDM \textit{anzatz} would be one of the few observations capable of clearly falsifying the existence of cosmic non-baryonic  dark matter.  A more likely result is basic confirmation of existing observations with only minor tweaks to the exact power spectrum.  Such a result would have little impact on the discussion here as it would simply confirm the need for some degrees of freedom in relativistic MOND theories that can play a role analogous to CDM.  However, the uncertainties on the best fit cosmological parameters may become negligibly small.  Precise as current data are, cosmology (with the exception of BBN) is still far from being over-constrained.  Hopefully PLANCK data will be sufficiently accurate that they either agree or clearly do not agree\footnote{Determining agreement between independent observations requires that we believe not just the result (e.g., the value of $H_0$ from direct distance measurements) but also its uncertainty.  The latter has always been challenging in astronomy, and the history of cosmology is replete with examples of results that were simply wrong.  While we may have entered the era of precision cosmology, we have yet to reach an era when data are so accurate that we can hope to challenge cosmology with falsification if, for example, PLANCK data require $H_0 < 60 \, {\rm km} \, {\rm s}^{-1} \, {\rm Mpc}^{-1}$ while galaxy distances require $H_0 > 70 \, {\rm km} \, {\rm s}^{-1} \, {\rm Mpc}^{-1}$.} with a host of other observations.

Presuming nothing substantial changes in the CMB data, we must understand the net forcing term in the acoustic oscillations leading to a high third peak.  This might be taken in one of three ways: 
\begin{itemize}
\item[(i)] Practical falsification of MOND, 
\item[(ii)] Proof of the existence of some form of non-baryonic matter particles, 
\item[(iii)] An indication of some necessary additional freedom in relativistic parent theories of MOND, playing the role of the non-baryonic mass in the CMB\epubtkFootnote{The third possibility actually means either non-local effects in non-local theories (Sect.~7.10), or the effect of additional fields in local modified gravity theories. The important difference with CDM is that these fields are \textit{not} simply representative of collisionless massive particles, that their behaviour is determined by the baryons in static configurations, and that they can be subdominant to the baryonic density. In theories where their energy density dominates that of baryons, these new fields then really act as dark matter in the early Universe, which is also a possibility (see Sect. 7.6 and 7.9)}. 
\end{itemize}
Tempting as the first case (i) is~\cite{silkboom}, we cannot know whether the CMB falsifies MOND until we have exhaustively explored the predictions of relativistic parent theories (Sect.~7).  The possibility of true non-baryonic mass (ii) \textit{a priori} seems unelegant, although a modification of gravity and the existence of non-baryonic dark matter are not mutually exclusive concepts.  What is more, there is one obviously existing form of non-baryonic mass
that may be relevant on cosmic scales:  neutrinos.  If $m_{\nu} \approx \sqrt{\Delta m^2}$~\cite{neutrinoDM2}, then the neutrino mass is
too small to be of interest in this context.  However, as discussed above, a modest neutrino mass may help to prevent MOND 
from over-predicting the growth of structure.  Independently, a mass $m_{\nu} \approx 1\mathrm{\ eV}$ to 2~eV for the three neutrino species provides a good match to the width of the acoustic peaks of the CMB~\cite{McGCMB3}, which are otherwise too wide in a purely baryonic Universe. Note that it provides as well a match to the missing mass in galaxy clusters of $T > 4\mathrm{\ keV}$ (see Sect.~6.6.4).  However, this neutrino mass is inadequate to explain the relatively high third peak in the no-DM ansatz.  Obtaining a match to that rather requires a  neutrino mass (for only one species) of $\sim$~10~eV~\cite{angussterilenu}.  Such a large mass violates experimental constraints on the ordinary neutrino mass~\cite{Mainz}, but it may be possible to have a sterile neutrino with a mass in that ballpark~\cite{Barry}.  As strange as this sounds, it provides a good fit to the CMB (Figure~\ref{CMBangus}), and it may provide the unseen mass in all clusters and groups (see Sect.~6.6.4~\cite{AFD,anguscosmo}).  Experiments that can address the existence of such a particle would thus be very interesting \cite{Mention}, although in between it is perhaps best to view it merely as the encapsulation of our ignorance about cosmology in modified gravity theories, much as dark energy currently plays the same role in conventional cosmology. The fit of Figure~\ref{CMBangus}~\cite{angussterilenu} is at least a proof of concept that \textit{cold} DM is definitely not required by the CMB alone.

Perhaps the most intriguing possibility is (iii), that the height of the third peak is providing a glimpse of some new aspect of modified gravity theories.  As we have seen, generalizations of GR seeking to incorporate MONDian phenomenology must, per force, introduce either non-locality (Sect.~7.10), or new degrees of freedom in local theories.  It is at least conceivable that these new degrees of freedom result in the net driving of the acoustic oscillations that is implied by the departure from pure baryonic damping.  For instance, Dodelson \& Liguori~\cite{dodelson} have shown that in TeVeS (Sects. 7.3 and 7.4) or GEA (Sect.~7.7) theories, based on unit-norm vector fields, the growth of the spatial part of the vector perturbation in the course of cosmological evolution is acting as an additional seed akin to non-baryonic dark matter\footnote{In TeVeS, the perturbations of the scalar field also play an important role in generating enhanced growth \cite{Feixperturb}.} (but unlike dark matter, its energy density is subdominant to the baryonic mass). Actually, it has been shown that, with the help of this effect prior to baryon-photon decoupling, it \textit{is} actually possible\epubtkFootnote{P.~Ferreira's talk, Alternative Gravities and Dark Matter Workshop, Edinburgh, April 2006.} to produce as high a third peak as the second one in TeVeS and GEA theories without non-baryonic dark matter, but at the cost of leading to unacceptably high temperature anisotropies in the CMB on large angular scales, due to an over-enhanced ISW effect~\cite{SMFB,Zuntz}. Indeed, when making the effect of the growth of the perturbed vector modes large, one also generates~\cite{Ferreira09,schmidt,zhang} a large \textit{gravitational slip} (see Sect.~8.4) in the perturbed FLRW metric (Eq.~\ref{ngaugemetric}), which in turn leads to enhanced ISW\epubtkFootnote{The ISW effect can be casted as the integral of $-(\Phi+\Psi)' + 2\Phi'$, thus involving both a gravitational slip part and a growth rate part.}. For this reason, acceptable fits to the CMB in TeVeS or GEA still need to appeal to non-baryonic mass~\cite{SMFB}. In this case, \textit{ordinary} neutrinos within their model-independent mass-limit~\cite{Mainz} are sufficient, though\epubtkFootnote{Note that the presence of non-baryonic matter in the form of massive neutrinos also helps damping the baryonic acoustic oscillations~\cite{Dodprob} at $z=0$, as can be seen on Figure~4 of~\cite{SMFB}.}. The gravitational slip could however be able to soon exclude at least some of these models from combined information on the matter overdensity and weak lensing~\cite{reyesnat,zhang}. However, an important caveat  is that all of the above arguments are based on adiabatic initial conditions\footnote{This means that, on surfaces of constant temperature, the densities of the various components (e.g. baryons, neutrinos, additional dark fields) are uniform, and that these components share a common velocity field.}. While initial isocurvature perturbations are basically ruled out in the GR context, this is not necessarily true for modified gravity theories, so that correlated mixtures of adiabatic and isocurvature modes could perhaps lower the ISW effect and/or raise the third peak \cite{Skordis09}.

Of course,  when the additional ``dark fields'' of relativistic MOND theories are truly massive (as is the case in some theories), they can be thought of as true ``dark matter'', whose energy density outweighs the baryonic one in the early Universe: this is the case for the second scalar field of BSTV (Sect.~7.5), the scalar field of Sect.7.6, and of course the dipolar dark matter of Sect.~7.9. In all these cases, reproducing the acoustic peaks of the CMB is, by construction, \textit{not} a problem at all (nor erasing the baryon acoustic oscillations in the matter power spectrum contrary to~\cite{Dodprob}), while the MOND phenomenology is still nicely recovered in galaxies. In the case of BIMOND (Sect.~7.8), the possible appeal to twin matter could also have important consequences on the growth of structure~\cite{bimond3} and of course on the CMB acoustic peaks too, although the latter analysis is still lacking. In an initially matter-twin matter symmetric Universe, if the initial quantum fluctuations are not identical in the two sectors, matter and twin matter would still segregate efficiently, since density differences grow much faster that the sum~\cite{bimond3}. The inhomogeneities of the two matter types would then develop, eventually, into mutually avoiding cosmic webs, and the tensors coming from the variation of the interaction term between the two metrics with respect to the matter metric can then act precisely as the energy-momentum tensor of cosmological dark matter~\cite{bimond3}, besides its contribution to the cosmological constant (see Sect.~9.1). Finally, the most thought-provoking and interesting possibility would perhaps be to explain all these cosmological observations through non-local effects (Sect.~7.10). In any case, it is likely that MOND will not be making truly clear predictions regarding cosmology until a more profound theory, based on first principles and underlying the MOND paradigm, will be found.

\newpage

\section{Summary and Discussion}
\label{sec:summary}

In this review, after briefly presenting the currently favored $\Lambda$CDM model of cosmology (which clearly works overwhelmingly well on large scales despite its slightly unelegant mixture of currently unknown elements, Sects.~2 and 3), we reviewed the few most outstanding challenges that this model is still facing (Sect.~4), which will have to be addressed one way or the other in the coming years. These include coincidences at $z=0$ between the scale of the energy density in dark energy, dark matter, and baryonic matter, as well as a common natural scale for the behavior of the dark matter and dark energy sectors. What is more, as far as galaxy formation is concerned, many predictions made by the model (keeping in mind that baryon physics could modify these predictions) were ruled out by observations: these include many observations indicating that structure formation should take place earlier than predicted, the low number of observed satellites around the Milky Way (especially the missing satellites at the low and high mass ends of the mass function), the phase-space correlation of satellite galaxies of the Milky Way as opposed to their predicted isotropic distribution, the apparent presence of constant DM density cores in the central parts of galaxies instead of the predicted cuspy dark halos, the over-abundance of large bulgeless thin disk galaxies that are extremely difficult to produce in simulations, or the presence of spiral arms in disks that should be immune to such instabilities. But even more challenging is the appearance (Figure~\ref{btflongbaseline}) of an acceleration constant $a_0 \simeq 10^{-10}\mathrm{\ m\, s}^{-2}$ (i.e., the common scale of the dark matter and dark energy sectors as $a_0 \sim \Lambda^{1/2}$ in natural units) in many \textit{a priori} unrelated scaling relations for DM and baryons in galaxies. These scaling relations involve a possibly devastating amount of fine-tuning for all collisionless dark matter models (Sect.~4.3), and can all be summarized by Milgrom's empirical formula (Sect.~5), meaning that the observed gravitational field in galaxies is mimicking a universal force law generated by the baryons alone. 

\epubtkImage{btfdeviationslongbaseline.png}{%
\begin{figure}[htbp]
  \centerline{\includegraphics[width=\textwidth]{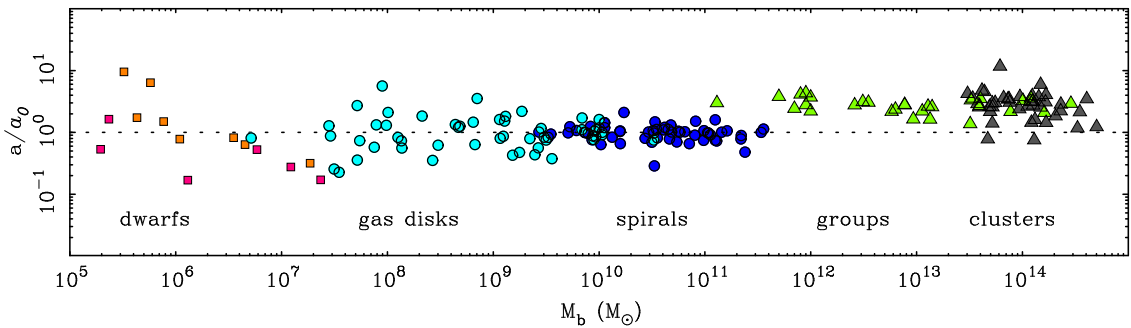}}
  \caption{The acceleration parameter $\sim V_f^4/(GM_b)$ of extragalactic systems, spanning ten decades in baryonic mass $M_b$. X-ray emitting galaxy groups and clusters are visibly offset from smaller systems, but by a remarkably modest amount over such a long baseline. The characteristic acceleration scale $a_0 \sim \sqrt{\Lambda}$ is in the data, irrespective of the interpretation. And it actually plays various other independent roles in observed galaxy phenomenology. This is natural in MOND (see Sect.~5.2), but not in $\Lambda$CDM (see Sect.~4.3).}
  \label{btflongbaseline}
\end{figure}}

With inert, collisionless and dissipationless DM, making Milgrom's law emerge requires a huge, and perhaps even unreasonable, amount of fine-tuning in the expected feedback from the baryons. Indeed, the relation between the distribution of baryons and DM should \textit{a priori} depend on the various different histories of formation, intrinsic evolution, and interaction with the environment of the various different galaxies, whereas Milgrom's law provides a sucessful unique and \textit{history-independent} relation. Given this puzzle, the central idea of Modified Newtonian Dynamics (MOND) is to rather explore the possibility that the force law is indeed effectively modified (Sect.~6). The main motivation for studying MOND is thus a fully \textit{empiricist} one, as it is driven by the \textit{observed} phenomenology on galaxy scales, and not by an aesthetic wish of getting rid of DM. The corollary is that it is \textit{a priori} not a problem for a theory designed to reproduce the uncanny successes of the MOND phenomenology to replace CDM by ``dark fields'' (see Sect.~7) or more exotic forms of DM, different from simple collisionless DM particles, contrary to the common belief that this would be against the spirit of the MOND paradigm (although it is true that it would be more elegant to avoid too many additional degrees of freedom). It is perhaps more important that, if MOND is correct in the sense of the acceleration $a_0$ being a \textit{truly} fundamental quantity, the strong equivalence principle \textit{cannot} hold anymore, and local Lorentz invariance could perhaps be spontaneously violated too.

At this juncture, it is worthwhile to summarize the general predictions of MOND, as a paradigm, and their observational tests (Table~\ref{table:mondtests}). As a mathematical description of the effective force law, MOND works remarkably well in individual galaxies.  As a modified gravity theory (at the classical level), it makes some predictions that are both unique and challenging to reproduce in the context of the $\Lambda$CDM paradigm.  However, MOND faces sharp challenges, particularly with cosmology and in rich clusters of galaxies, which will not be conclusively addressed without a viable parent theory (Sect.~7), based on first principles and underlying the MOND paradigm (if such a theory exists at all). In any case, in his series of papers introducing the idea in 1983, Milgrom~\cite{Milgrom1} made a few very \textit{explicit} predictions, which we quote hereafter, and compare with modern observational data (see also the Kepler-like laws of galactic dynamics in Sect.~5.2):

\begin{itemize}
\item \textit{``Velocity curves calculated with the modified dynamics on the basis of the observed mass in galaxies should agree with the observed curves.''}

It is now well established that MOND provides good fits to the rotation curves of galaxies (Figure~\ref{figure:RCresiduals}~\cite{SM02,things}), including bumps and wiggles associated to a baryonic counterpart (Figure~\ref{figure:RCexamples}, Kepler-like law n$^\circ$10 in Sect.~5.2).  These fits are obtained with a single free parameter per galaxy, the mass-to-light ratio of the stars. What makes them most impressive is that the best-fit mass-to-light ratios, obtained on purely dynamical grounds assuming MOND, vary with galaxy color exactly as one would expect from stellar population synthesis models~\cite{Bel03}, that are based on astronomers' detailed understanding of stars. Note that the rotation curves of galaxies are predicted to be \textit{asymptotically} flat, even though this flatness is not always attained at the last observed point (see Kepler-like law n$^\circ$1 in Sect.~5.2, and last explicit prediction hereafter).

\item \textit{``The relation between the asymptotic velocity and the mass of the galaxy is an absolute one.''}

This is the Baryonic Tully--Fisher relation with $M_b = a_0 G V_f^4$ (see Kepler-like laws n$^\circ$2 in Sect.~5.2).  It appears to hold quite generally~\cite{M05}, even for galaxies that we would conventionally expect to deviate from it~\cite{GentileTDG,MdB98,McGgasrich}.

\item \textit{``Analysis of the $z$-dynamics in disk galaxies using the modified dynamics should yield surface densities
which agree with the observed ones.''}

This states that in addition to the radial force giving the rotation curve, the motions of stars perpendicular to the disk
must also follow from the source baryons (see Sect.~6.5.3).  This proves to be a remarkably challenging observation, and such data for
external galaxies are dear to obtain~\cite{bershady}.  To make matters still more difficult, the radial acceleration usually dominates
the vertical ($V^2/r \gg \sigma_z^2/z$).  This has the consequence that the distinction between MOND and conventional 
dynamics is not pronounced in regions that are well observed, becoming pronounced only at
rather low baryonic surface densities~\cite{MdB98}.  The vertical velocity dispersions in low
surface density regions (see Sect.~6.5.3) is typically $\sim 8$~km/s~\cite{BJBB,KdN08}.  This exceeds the nominal Newtonian expectation (typically $\sim 2$~km/s for 
$\Sigma = 1\,M_{\odot}\,\mathrm{pc}^{-2}$, depending on the thickness of the disk), and is more in accordance with MOND.  
However, it would require a considerably more detailed analysis to consider this a test, let alone a success, of MOND. The Milky Way (Sect.~6.5.2) may provide an excellent test for this prediction~\cite{Bienayme,Salcedo} as more precision data become available.

\item \textit{``Effects of the modification are predicted to be particularly strong in (low surface brightness) dwarf galaxies.''}

The dwarf spheroidal satellite galaxies of the Milky Way have very low surface densities of stars, so (see Kepler-like law n$^\circ$8 in Sect.~5.2) are far into the MOND regime. As expected, these systems exhibit large mass discrepancies~\cite{Walker09,SG}.  Detailed fits to the better observed 
``classical'' dwarfs~\cite{AngusdSph} are satisfactory in most cases (see Sect.~6.6.2). The so-called ``ultrafaint'' dwarfs appear more problematic~\cite{MWolf}, in the sense that their velocity dispersions are higher
than expected.  This might be an indication of the MOND-specific external field effect (see Sect.~6.3 and~\cite{BMdwarfs}), as the field of the Milky Way dominates the internal fields of the ultrafaint dwarfs.  If so, these objects are not in dynamical equilibrium, which considerably
complicates their analysis.

\item \textit{Locally measured mass-to-light ratios should show no indication of hidden mass when $V_c^2/R \gg a_0$, but rise beyond the radius where $V_c^2/R \approx a_0$.}

We have paraphrased this prediction for brevity (see also  Kepler-like law n$^\circ$7 in Sect.~5.2). The test of this prediction is shown in Figures~\ref{figure:MDRA}, \ref{fig:mdasolsys}, and \ref{figure:MLRSd}. The predicted effect is obvious in the data with populations synthesis mass-to-light ratios for the stars~\cite{Bel03}, or with dynamical mass-to-light ratios~\cite{MdB98} that make no assumption about stellar mass.  In HSB spirals, there is no obvious need for dark matter in the inner regions, with the mass discrepancy only appearing at large radii as the acceleration drops below $a_0$ (Figure~\ref{figure:MDRA}).

\item \textit{``Disk galaxies with low surface brightness provide particularly strong tests.''}

Low surface brightness means low stellar surface density, which in turns means low acceleration.
LSB galaxies are thus predicted to be well into the modified regime (see also  Kepler-like law n$^\circ$8 in Sect.~5.2). This was a strong \textit{a priori} prediction, because few bona-fide examples of such objects were known at the time. Indeed, in 1983, when these predictions were published, it was widely thought that nearly all disk galaxies shared a common high surface brightness. One specific consequence of MOND for LSB galaxies is that they should lie on the same BTFR, with the
same normalization, as high surface brightness spirals.  This was subsequently observed to be the case~\cite{ZwaanTF,sprayTF}. There is \textit{no} systematic deviation from the BTFR with surface brightness (Figure~\ref{figure:btfresiduals}), thus contrary to what is naturally expected in conventional dynamics~\cite{MdB98,CRix}. Another consequence of low surface density is that the acceleration is low ($< a_0$) everywhere. As a result, the mass discrepancy appears at a smaller radius in low surface brightness galaxies, and is larger in amplitude than in high surface brightness galaxies.  This effect was subsequently observed (Figure~\ref{figure:MLRSd}~\cite{MdB98}).

\item \textit{``We predict a correlation between the value of the average surface density of a galaxy and the steepness with which the rotational velocity rises to its asymptotic value.''}

MOND does \textit{not} simply make rotation curves flat.  It predicts that high surface brightness galaxies have rotation curves that rise rapidly before becoming flat, and may even fall towards asymptotic flatness.  In contrast, low surface brightness galaxies should have slowly rising rotation curves that only gradually approach asymptotic flatness (see also  Kepler-like law n$^\circ$8 in Sect.~5.2).  Both morphologies are observed (Figure~\ref{figure:RCshape}).  The expected connection between dynamical acceleration and the surface density of the source baryons is illustrated in Figures~\ref{figure:XiSd} and \ref{figure:ARSd}.

\end{itemize}

The original predictions listed above cover many situations, but not all.  Indeed, once one writes a specific force law, its application must be completely general.  Such a hypothesis is readily subject to falsification, provided sufficiently accurate data to test it -- a perpetual challenge for astronomy. Table~\ref{table:mondtests} summarizes the tests discussed here.  By and large, tests of MOND involving rotationally supported disk galaxies are quite positive, as largely detailed above (see Sect.~6.5).  By construction, there is no cusp problem (solution to the challenge n$^\circ$6 of Sect.~4.2), and no missing baryons problem (solution to the challenge n$^\circ$10 of Sect.~4.2), as the way the dynamical mass-to-light ratio systematically varies with the circular velocity is a direct consequence of Milgrom's law (Kepler-like law n$^\circ$4 of Sect.~5.2). There does appear to be a relation between the quality of the data and the ease with which a MOND fit to the rotation curve is obtained, in the sense that fits are most readily obtained with the best data~\cite{BBS}.  As the quality of the data decline~\cite{sanders96}, one begins to notice small disparities.  These are sometimes
attributable to external disturbances that invalidate the assumption of equilibrium~\cite{SV98}.  For targets that are intrinsically
difficult to observe, minor problems become more common~\cite{dBM98,swatersmond}.  These typically have to do with the
challenges inherent in combining disparate astronomical data sets (e.g., rotation curves measured independently at optical
and radio wavelengths) and constraining the inclinations of low surface brightness galaxies (bear in mind that all velocities
require a $\sin(i)$ correction to project the observed velocity into the plane of the disk, and mass in MOND scales as the fourth
power of velocity).  Given the intrinsic difficulties of astronomical observations, it is remarkable that the success rate of MOND
fits is as high as it is:  of the 78 galaxies that have been studied in detail (see Sect.~6.5.1), only a few cases (most notably NGC~3198~\cite{bottema,things}) appear to pose challenges.  Given the predictive and quantitative success of the majority of the fits, it would seem unwise to ignore the forest and focus only on the outlying trees.

One rotationally supported system that is very familiar to us is the Solar system (see Sect.~6.4).  The Solar system is many orders of magnitude removed from the MOND regime (Figure~\ref{fig:mdasolsys}), so no strong effects are predicted.  However, it is of course possible
to obtain exquisitely precise data in the Solar system, so it is conceivable that some subtle effect may be observable~\cite{Sanderssun}.  Indeed, the lack of such effects on the inner planets already appears to exclude some slowly varying interpolation functions~\cite{BlanchetNovak}. Other tests may yet prove possible~\cite{Mag1,Milgromsun}, but, as they are strong-field gravity tests by nature, they all depend strongly on the parent relativistic theory (Sect.~7) and how it converges towards GR~\cite{galileon}. So in Table~\ref{table:mondtests}, we list the status of Solar system tests as unclear, depending on the parent relativistic theory.

An important aspect of galactic disks is their stability (see Sect.~6.5.3).  Indeed, the need to stabilize disks was one of the early motivations for invoking dark matter~\cite{OstPeeb}.  MOND appears able to provide the requisite stability~\cite{BradaM}.  Indeed, it gives good reason~\cite{Milstab} for the observed maximum in the distribution of disk galaxy surface densities at $\sim \Sigma_{\dagger} = a_0/G$ (Freeman's limit: Figure~\ref{fig:freeman} and  Kepler-like law n$^\circ$6 in Sect.~5.2).  Disks with surface densities below this threshold are in the low acceleration limit and can be stabilized by MOND.  Higher surface density disks would be purely in the Newtonian regime and subject to the usual instabilities. Going beyond the amount of stability required for existence, another positive aspect of MOND is that it does not over-stabilize disks. Features like bars and spiral arms are a natural result of disk self-gravity.  Conventionally, large halo-to-disk mass ratios suppress the growth of such features, especially in low surface brightness galaxies~\cite{Mihos}.  Yet such features are present\epubtkFootnote{\cite{MdB98} utilized this fact to predict that conventional analyses of low surface brightness disks
would infer abnormally high mass-to-light ratios for their stellar populations -- a prediction that was subsequently
confirmed~\cite{fuchs,Saburova}.}.  The 
suppression is not as great in MOND~\cite{BradaM}, and numerical simulations appear to do a good job of reproducing the range
of observed morphologies of spiral galaxies (solution to the challenge n$^\circ$9 of Sect.~4.2, see~\cite{Tiretgas}). Bars tend to appear more quickly and are fast, while warps can also be naturally produced (Sect.~6.5.3). There appears to be no reason why this should not extend to thin and bulgeless disks, whose ubiquity poses a challenge to galaxy formation models in $\Lambda$CDM.  This particular point of creating large bulgeless disks (challenge n$^\circ$8 of Sect.~4.2) can actually be solved thanks to early structure formation followed by a low galaxy-interaction rate in MONDian cosmology (see Sect.~9.2), but this definitely warrants further investigation, so we mark this case as merely promising in Table~\ref{table:mondtests}.

Interacting galaxies are by definition non-stationary systems in which the customary assumption of equilibrium does not generally
hold.  This renders direct tests of MOND difficult.  However, it is worth investigating whether commonly observed morphologies
(e.g., tidal tails) are even possible in MOND.  Initially, this seemed to pose a fundamental difficulty~\cite{MdB98}, as dark matter
halos play a critical role in absorbing the orbital energy and angular momentum that it is necessary to shed if passing galaxies
are to not only collide, but stick and merge. Nevertheless, recent
numerical simulations appear to do a nice job of reproducing observed morphologies~\cite{Tiretinteract}.  This is no trivial feat.
While it is well established that dark matter models can result in nice tidal tails, it turns out to be difficult to simultaneously
match the narrow morphology of many observed tidal tails with rotation curves of the systems from which they come~\cite{dubinski}.
Narrow tidal tails appear to be natural in MOND, as well as more extended resulting galaxies, thanks to the absence of angular momentum transfer to the dark halo (solution to the challenge n$^\circ$7 of Sect.~4.2).  Additionally, tidal dwarfs that form in these tails clearly have characteristics closer to those observed (see Sect.~6.5.4) than those from dark matter simulations~\cite{GentileTDG,Miltidal}.

Spheroidal systems also provide tests of MOND (Sect.~6.6).  Unlike the case of disk galaxies, where orbits are coplanar and nearly circular so that the centripetal acceleration can be equated with the gravitational force, the orbits in spheroidal systems are generally eccentric and randomly oriented.  This introduces an unknown geometrical factor usually subsumed into a parameter that characterizes the anisotropy of the orbits.  Accepting this, MOND appears to perform well in the classical dwarf spheroidal galaxies, but implies that the ultrafaint dwarfs are out of equilibrium (see Sect.~6.6.2).  For small systems like the ultrafaint dwarfs and star clusters (Sect.~6.6.3) within the Milky Way, the external field effect (Sect.~6.3) can be quite important.  This means that
star clusters generally exhibit Newtonian behavior by virtue of being embedded in the larger Galaxy.  Deviations from purely Newtonian behavior are predicted to be subtle and are fodder for considerable debate~\cite{Rodrigo2419,Sanders2419}, rendering the present status unclear (Table~\ref{table:mondtests}).  At the opposite extreme of giant elliptical galaxies (Sect.~6.6.1), the data accord well with MOND~\cite{dearth}.  Indeed, bright elliptical galaxies are sufficiently dense that their inner regions are well into the Newtonian regime.  In the MONDian context, this is the reason that it has historically been difficult to find clear evidence for mass discrepancies in these systems.  The apparent need for dark matter does not occur until radii where the accelerations become low.  That only spheroidal stellar systems appear to exist at surface densities in excess of $\Sigma_{\dagger}$ is the corollary of Freeman's limit:  such dense systems could not exist as stable disks, so must per force become elliptical galaxies, regardless of the formation mechanism that made them so dense.  That populations of elliptical galaxies should obey the Faber-Jackson relation (Kepler-like law n$^\circ$3 in Sect.~5.2, Figure~\ref{figure:faberjackson}) is also very natural to MOND~\cite{Sanders94,sanders10}.

The largest gravitationally bound systems are also spheroidal systems: rich clusters of galaxies.  The situation here is quite problematic for MOND (Sect.~6.6.4).  Applying MOND to ascertain the dynamical mass routinely exceeds the observed baryonic mass by a factor of 2 to 3.  In effect, MOND requires additional \textit{dark matter} in galaxy clusters. The need to invoke unseen mass is most unpleasant for a theory that otherwise appears to be a viable alternative to the existence of unseen mass.  However, one should remember that the present-day motivation for studying MOND is driven by the observed phenomenology on galaxy scales, summarized above, and not by an aesthetic wish of getting rid of DM. What is more, parent relativistic theories of MOND might well involve additional degrees of freedom in the form of ``dark fields''. But in any case, one must be careful not to conflate the rather limited missing mass problem that MOND suffers in clusters with \textit{the} non-baryonic collisionless cold dark matter required by cosmology.  There is really nothing about the cluster data that requires the excess mass to be non-baryonic, as long as it behaves in a collisionless way.  There could for instance be baryonic mass in some compact non-luminous form (see Sect.~6.6.4 for an extensive discussion). This might seem to us unlikely, but it does have historical precedent.  When Zwicky~\cite{zwicky} first identified the dark matter problem in clusters, the mass discrepancy was of order $\sim 100$.  That is, unseen mass outweighed the visible stars by two orders of magnitude.  It was only decades later that it was recognized that baryons residing in a hot intracluster gas greatly outweighed those in stars.  In effect, there were [at least] \textit{two} missing mass problems in clusters.  One was the hot gas, which reduces the conventional discrepancy from a factor of $\sim 100$ to a factor of $\sim 8$~\cite{giodini} in Newtonian gravity.  From this perspective, the remaining factor of two in MOND seems modest.  Rich clusters of galaxies are rare objects, so the total required mass density can readily be accommodated within the baryon budget of big bang nucleosynthesis.  Indeed, according to BBN, there must still be a lot of unidentified baryons lurking somewhere in the Universe. But the excess dark mass in clusters need not be baryonic, even in MOND.  Massive ordinary neutrinos~\cite{sandersclusters,Sandersneut} and light sterile neutrinos~\cite{angussterilenu,AFD} have been suggested as possible forms of dark matter that might provide an explanation for the missing mass in clusters.  Both are non-baryonic, but as they are \textit{hot} DM particle candidates, neither can constitute \textit{the} cosmological non-baryonic \textit{cold} dark matter. At this juncture, all we can say for certain is that we do not know what the composition of the unseen mass is. It could even just be an evidence for the effect of additional ``dark fields'' in the parent relativistic formulation of MOND, such as massive scalar fields, vector fields, dipolar dark matter, or even subtle non-local effects (see Sect.~7).

There are other aspects of cluster observations that are more in line with MOND's predictions.  Clusters obey a mass--temperature
relation that parallels the $M \propto T^2 \propto V^4$ prediction of MOND (Figures~\ref{figure:clustersbtf} and \ref{btflongbaseline})
more closely than the conventional prediction of $M \propto T^{3/2}$ expectation in $\Lambda$CDM, without the need to invoke
preheating (a need that may arise as an artifact of the mismatch in slopes).  Indeed, Figure~\ref{btflongbaseline} shows clearly
both the failing of MOND in the offset in characteristic acceleration between clusters and lower mass systems, and its successful 
prediction of the slope (a horizontal line in this figure).  A further test which may be important is the peculiar and bulk velocity of clusters.  For example, the collision velocity of the 
bullet cluster is so large\epubtkFootnote{The observed shock velocity of $\sim 4700$~km/s is thought to
be enhanced by hydrodynamical effects.  The collision velocity is improbable \textit{after} a substantial
($\sim 1700$~km/s) correction for this~\cite{Mastropietro}.}
as to be highly improbable in $\Lambda$CDM (occurring with a probability of $\sim 10^{-10}$~\cite{Lee}).
In contrast, large collision velocities are natural to MOND~\cite{Angusbullet}.  Similarly, the large scale peculiar velocity of clusters is observed to be $\sim 1000$~km/s~\cite{Kash}, well in excess of the expected 
$\sim 200\;\mathrm{km}\,\mathrm{s}^{-2}$.  Ongoing simulations with MOND~\cite{anguscosmo} show some promise to produce large peculiar velocities for clusters.  In general, one would expect high speed collisions to be more ubiquitous in MOND than $\Lambda$CDM.

An important line of evidence for mass discrepancies in the Universe is gravitational lensing in excess of that expected from
the observed mass of lens systems.  Lensing is an intrinsically relativistic effect that requires a generally covariant theory
to properly address.  This necessarily goes beyond MOND itself into specific hypotheses for its parent theory (Sect.~7), so is somewhat different than the tests discussed above. Broadly speaking, tests involving strong gravitational lensing fare tolerable well (Sect.~8.1), whereas weak lensing tests, that are sensitive to larger scale mass distributions, are more problematic (Sects.~8.2, 8.3, and 8.4) or simply crash into the usual missing mass problem of MOND in clusters. Note that weak lensing in relativistic MOND theories produces the same amount of lensing as required from dynamics, so this is not the problem. The problematic fact is just that some tests seem to require more dark matter than the effect of MOND provides.

On larger (cosmological) scales, MOND, as a modification of classical (non-covariant) dynamics, is simply unsatisfactory or mute.  MOND itself has no cosmology, providing analogs for neither the
Friedmann equation for the dynamics of the Universe, nor the Robertson-Walker metric for its geometry.
For these, one must appeal to specific hypotheses for the relativistic parent theory of MOND (Sect.~7), which is far from unique, and theoretically not really satisfactory, as none of the present candidates emerges from first principles.  At this juncture, it is not clear whether a compelling candidate cosmology will ever emerge. But on the other hand, there is \textit{nothing} about MOND as a paradigm that contradicts \textit{per se} the empirical pillars of the hot big bang:  Hubble expansion, big bang nucleosynthesis, and the relic radiation field (Sect.~9). The formation of large scale structure is one of the strengths of conventional theory, which can be approached with linear perturbation theory.  This leads to good fits of the power spectrum both at early times ($z \approx 1000$ in the cosmic microwave background) and at late times (the $z = 0$ galaxy power spectrum~\cite{Tegmark04}).  In contrast, the formation of structure in MOND is intrinsically non-linear.  It is therefore unclear whether MOND-motivated relativistic theories will inevitably match the observed galaxy power spectrum, a possible problem being how to damp the baryon acoustic oscillations~\cite{Dodprob,SMFB}. At this stage, a unique prediction does not exist. Nevertheless, there are two aspects of structure formation in MOND that appear to be fairly generic and distinct from $\Lambda$CDM. The stronger effective long range force in MOND speeds the growth rate, but has less mass to operate with as a source. Consequently, radiation domination persists longer and structure formation is initially inhibited (at redshifts of hundreds). Once structure begins to form, the non-linearity of MOND causes it to proceed more rapidly than in GR with CDM.  Three observable consequences would be (i) the earlier emergence of large objects like galaxies and clusters in the cosmic web (as well as the associated low interaction rate at smaller redshifts) providing a possible solution to challenge n$^\circ$2 of Sect.~4.2~\cite{anguscosmo}, (ii) the more efficient evacuation of large voids (possible solution to challenge n$^\circ$3 of Sect.~4.2), and (iii) larger peculiar (and collisional~\cite{Angusbullet}) velocities of galaxy clusters (solution to challenge n$^\circ$1 of Sect.~4.2). The potential downside to rapid structure formation in MOND, however, is that it may overproduce structure by redshift zero~\cite{nusser,Llinares}.

The final entries in Table~\ref{table:mondtests} regard the cosmic microwave background, discussed in more detail in Sect.~\ref{subsection:CMB}.  The third peak of the acoustic power spectrum of the CMB poses perhaps the most severe challenge to a MONDian interpretation of cosmology.  The amplitude of the third peak measured by WMAP is larger than expected in a Universe composed solely of baryons~\cite{WMAP3}.  This implies some substance that does not oscillate with the baryons. Cold dark matter fits this bill nicely.  In the context of MOND, we must invoke some other massive substance (i.e., non-baryonic dark matter such as, e.g., light sterile neutrinos~\cite{angussterilenu}) that plays the role of CDM, or rely on additional degrees of freedom in the relativistic parent theory of MOND (see Sect.~7) that would have the same net result (see the extensive discussion in Sect.~9.2), or even combine non-baryonic dark matter with these additional degrees of freedom~\cite{SMFB}.  While these are real possibilities, neither are \textit{a priori} particularly appealing, any more than it is to invoke CDM with complex fine-tuned feedback to explain rotation curves that apparently require only baryons as a source.

The missing baryon problem that MOND suffers in rich clusters of galaxies and the third peak of the acoustic power spectrum of the CMB are thus the most serious challenges presently facing MOND.  But even so, the interpretation of the acoustic power spectrum is not entirely clear cut.  Though there is no detailed fit to the power spectrum in MOND (unless we invoke $10 \,$eV-scale sterile neutrinos~\cite{angussterilenu}),
MOND did motivate the \textit{a priori} prediction~\cite{McGCMB1} of two aspects of the CMB that were surprising in $\Lambda$CDM (see Sect.~9.2). The amplitude ratio of the first-to-second peak in the acoustic power spectrum was outside the bounds expected ahead of time by $\Lambda$CDM for $\Omega_b$ from big bang nucleosynthesis as it was then known (see Sect.~\ref{subsection:CMB}). In contrast, the first:second acoustic peak ratio that is now well measured agrees well with the quantitative value predicted in 
advance for the case of the absence of cold dark matter~\cite{McGCMB2,McGCMB3}.  Similarly, the rapid formation of structure expected in MOND leads naturally to an earlier epoch of re-ionization than had been anticipated in $\Lambda$CDM~\cite{McGCMB1,McGCMB3}.  Thus, while the amplitude of the third peak is clearly problematic and poses a severe challenge for any MOND-inspired theories, the overall interpretation of the CMB is debatable.  While the existence of non-baryonic cold dark matter is \textit{a priori} the most obvious explanation of the third peak indeed, it is not at all obvious that straightforward CDM -- in the form of rather simple massive inert collisionless particles -- is uniquely required. 

Science is in principle about theories or models that are falsifiable, and thus that are presently either falsified or not. But in practice it does not (and cannot) really work that way: if a model that was making good predictions up to a certain point suddenly does not work anymore (i.e., does not fit some new data), one obviously first tries to adjust it to make it fit the observations rather than throwing it away immediately. This is what one calls the requisite ``compensatory adjustments" of the theory (or of the model): Popper himself drew attention to these limitations of falsification in \textit{The Logic of Scientific Discovery}~\cite{Popper}. In the case of the $\Lambda$CDM model of cosmology, which is mostly valid on large scales, the current main trend is to find the ``compensatory adjustments'' to the model to make it fit in galaxies, mainly by changing (or mixing) the mass(es) of the dark matter particles, and/or through artificially fine-tuned baryonic feedback in order to reproduce the success of MOND. Incidentally, exactly the same is true for MOND, but for the opposite scales: MOND works remarkably well in galaxies but apparently needs compensatory adjustments on larger scales to effectively replace CDM. Now does that mean that falsification is impossible? That all models are equal? Surely not. In the end, a theory or a model is really falsified once there are too many compensatory adjustments (needed in order to fit too many discrepant data), or once these become too twisted (like Tycho Brahe's geocentric model for the Solar system). But there is  obviously no truly quantitative way of ascertaining such global falsification. How one chooses to weigh the evidence presented in this review necessarily informs one's opinion of the relative merits of $\Lambda$CDM and MOND.  If one is most familiar with cosmology and large scale structure, $\Lambda$CDM is the obvious choice, and it must seem rather odd that anyone would consider an alternative as peculiar as MOND, needing rather bizarre adjustments to match observations on large scales. But if one is more concerned with precision dynamics and the observed phenomenology in a wide swath of galaxy data, it seems just as strange to invoke non-baryonic cold dark matter together with fine-tuned feedback to explain the appearance of a single effective force law that appears to act with only the observed baryons as a source. Perhaps the most important aspect before one throws away any model is to have a ``simpler'' model at hand, that still reproduces the successes of the earlier favored model but also naturally explains the discrepant data. In that sense, right now, it is absolutely fair to say that there is \textit{no} alternative which really does better \textit{overall} than $\Lambda$CDM, and in favor of which Ockham's razor would be. It would however probably be a mistake to persistently ignore the fine-tuning problems for dark matter and the related uncanny successes of the MOND paradigm on galaxy scales, as they could very plausibly point at a hypothetical better new theory. It is also important to bear in mind that MOND, as a paradigm or as a modification of Newtonian dynamics, is not itself generally covariant. Attempts to construct relativistic theories that contain MOND in the appropriate limit (Sect.~7) are correlated but distinct efforts, and one must be careful not to conflate the two. For example, some theories, like TeVeS (Sect.~7.4), might make predictions that are distinct from GR in the strong-field regime.  Should future tests falsify these distinctive predictions of TeVeS while confirming those of GR, this would perhaps falsify TeVeS as a viable parent theory for MOND, but would have no bearing on the MONDian phenomenology observed in the weak-field regime, nor indeed on the viability of MOND itself. It would perhaps simply indicate the need to continue to search for a deeper theory. It would for instance be extremely alluring if one would manage to find a physical connection between the dark energy sector and the possible breakdown of standard dynamics in the weak-field limit, since both phenomena would then simply reflect discrepancies with  the predictions of GR when $\Lambda \sim a_0^2$ is set to zero (see, e.g., Sect.~7.10). It is of course perfectly conceivable that such a deep theory does not exist, and that the apparent MONDian behavior of galaxies will be explained through small compensatory adjustments of the current $\Lambda$CDM paradigm, but one has yet to demonstrate how this will occur, and it will inevitably involve a substantial amount of fine-tuning that will have to be explained naturally. In any case, the existence of a characteristic acceleration $a_0$ (Figure~\ref{btflongbaseline}) playing various different roles in  many seemingly independent galactic scaling relations (see Sects.~4.3 and 5.2) is by now an empirically established fact, and it is thus mandatory for \textit{any} successful model of galaxy formation and evolution to explain it. The future of this field of research might thus still be full of exciting surprises for astronomers, cosmologists, and theoretical physicists.

\begin{table}[htbp]
\caption{Observational tests of MOND.}
\label{table:mondtests}
\centering
{\small
\begin{tabular}{lcccc}
  \toprule
Observational Test & Successful & Promising & Unclear & Problematic \\
  \midrule
  \textbf{Rotating Systems} & & & & \\
    solar system & & & X & \\
    galaxy rotation curve shapes & X & & & \\
    surface brightness $\propto \Sigma \propto a^2$  & X & & & \\
    galaxy rotation curve fits & X & & & \\
    fitted M$_*$/L & X & & & \\
  \midrule
  \textbf{Tully--Fisher Relation} & & & & \\
    baryon based & X & & & \\
    slope & X & & & \\
    normalization & X & & & \\
    no size nor $\Sigma$ dependence & X & & & \\
    no intrinsic scatter & X & & & \\
  \midrule
  \textbf{Galaxy Disk Stability} & & & & \\
    maximum surface density  & X & & & \\
    spiral structure in LSBGs   & X & & & \\
    thin \& bulgeless disks & & X & & \\
  \midrule
  \textbf{Interacting Galaxies} & & & & \\
    tidal tail morphology  & & X & & \\
    dynamical friction & & & X & \\
    tidal dwarfs & X & & & \\
  \midrule
  \textbf{Spheroidal Systems} & & & & \\
    star clusters & & & X & \\
    ultrafaint dwarfs & & & X & \\
    dwarf Spheroidals & X & & & \\
    ellipticals & X & & & \\
    Faber-Jackson relation & X & & & \\
  \midrule
  \textbf{Clusters of Galaxies} & & & & \\
    dynamical mass  & & & & X \\
    mass--temperature slope & X & & & \\
    velocity (bulk \& collisional)  & & X & & \\
  \midrule
  \textbf{Gravitational Lensing} & & & & \\
    strong lensing & X & & &  \\
    weak lensing (clusters \& LSS) & & & X &  \\
  \midrule
  \textbf{Cosmology} & & & & \\
    expansion history & & & X & \\
    geometry  & & & X & \\
    big bang nucleosynthesis & X & & & \\
  \midrule
  \textbf{Structure Formation} & & & & \\
    galaxy power spectrum & & & X & \\
    empty voids & & X & & \\
    early structure & & X & & \\
  \midrule
  \textbf{Background Radiation} & & & & \\
    first:second acoustic peak & X & & & \\
    second:third acoustic peak  & & & & X \\
    detailed fit  & & & & X \\
    early re-ionization & X & & & \\
  \bottomrule
\end{tabular}
}
\end{table}

\newpage

\section*{Acknowledgements}
\label{section:acknowledgements}

The authors are grateful for stimulating conversations about the dark
matter problem and MOND over the years, as well as for comments and
help from: Garry Angus, Jean-Philippe Bruneton, Martin Feix, Gianfranco Gentile,
HongSheng Zhao, Moti Milgrom, Rodrigo Ibata, Pavel Kroupa, Olivier
Tiret, Dominique Aubert, Jacob Bekenstein, Olivier Bienaym\'e, James
Binney, Luc Blanchet, Christian Boily, Greg Bothun, Laurent Chemin, Fran\c{c}oise
Combes, J\"org Dabringhausen, Erwin de Blok, Gilles Esposito-Far\`ese, Filippo
Fraternali, Andr\'e F\"uzfa, Alister Graham, Hosein Haghi, Ana\"elle Halle, 
Xavier Hernandez, Jean Heyvaerts, Alex Ignatiev, Alain Jorissen, Frans Klinkhamer, 
Joachim K\"oppen, Rachel Kuzio de Naray, Claudio Llinares, Fabian L\"ughausen, Jo\~{a}o
Magueijo, Mario Mateo, Chris Mihos, Ivan Minchev, Mustapha Mouhcine,
Carlo Nipoti, Adi Nusser, Marcel Pawlowski, Jan Pflamm-Altenburg, Tom
Richtler, Paolo Salucci, Bob Sanders, James Schombert, Ylva Schuberth,
Jerry Sellwood, Arnaud Siebert, Christos Siopis, Kristine Spekkens,
Rob Swaters, Marc Verheijen, Matt Walker, Joe Wolf, Herv\'e Wozniak,
Xufen Wu, and many others. We also thank Fr\'ed\'eric Bournaud, Chuck
Bennett, Douglas Clowe, Andrew Fruchter, Tom Jarrett, and, again,
Garry Angus and Olivier Tiret, for allowing us to make use of their
figures. We finally thank Clifford Will for inviting us to write this
review. We acknowledge the support of the CNRS, the AvH foundation,
and the NSF grant AST~0908370, and both acknowledge hospitality at
Case Western Reserve University, where a substantial part of this
review has been written.

\newpage

\bibliography{refs}

\end{document}